\documentclass[]{aa}     
\usepackage[usenames,dvipsnames]{color}                     
\usepackage{graphicx}       
\usepackage[varg]{txfonts}        
\usepackage{lscape}
\usepackage{rotating}

\usepackage{natbib}
\bibpunct{(}{)}{;}{a}{}{,} 
\begin{document}        
        
\def\ang{{\rm\AA}}        
        
\def\n{\noindent}        
        
\def\ph{\phantom}         
   
\newcommand{\fuse}{\emph{FUSE}}
\newcommand{\stis}{\emph{STIS}}
\newcommand{\cosp}{\emph{COS}}
\newcommand{\hst}{\emph{HST}}
\newcommand{\eso}{\emph{ESO}}
\newcommand{\uves}{\emph{UVES}}
\newcommand{\aat}{\emph{AAT}}
\newcommand{\flames}{\emph{FLAMES}}
\newcommand{\ergs}{ergs\,cm$^{-2}$\,s$^{-1}$\,\AA$^{-1}$}
\newcommand{\halpha}{\mbox{H$\alpha$}}
\newcommand{\hbeta}{\mbox{H$\beta$}}
\newcommand{\hgamma}{\mbox{H$\gamma$}}
\newcommand{\hdelta}{\mbox{H$\delta$}}
\newcommand{\bralpha}{\mbox{Br$\alpha$}}
\newcommand{\brgamma}{\mbox{Br$\gamma$}}
\newcommand{\lyalpha}{\mbox{Ly$\alpha$}}
\newcommand{\lybeta}{\mbox{Ly$\beta$}}
\newcommand{\lygamma}{\mbox{Ly$\gamma$}}
\newcommand{\lydelta}{\mbox{Ly$\delta$}}
\newcommand{\htwo}{\mbox{H$_{2}$}}
\newcommand{\lb}{$\lambda$}
\newcommand{\mdot}{$\dot{M}$}
\newcommand{\gcmcube}{g\,cm$^{-3}$}
\newcommand{\gcmcarre}{g\,cm$^{-2}$}
\newcommand{\vsini}{$v\sin\,i$}
\newcommand{\vrad}{$v_{\rm rad}$}
\newcommand{\vmc}{$v_{\rm MC}$}
\newcommand{\flux}{ergs\,cm$^{-2}$\,s$^{-1}$}
\newcommand{\teff}{$T_{\rm eff}$}
\newcommand{\logg}{$\log g$}
\newcommand{\loggc}{$\log g_{\rm c}$}
\newcommand{\met}{[Fe/H]}
\newcommand{\vturb}{$v_{\rm turb}$}
\newcommand{\vmax}{$v_{\rm max}$}
\newcommand{\vmin}{$v_{\rm min}$}
\newcommand{\vcl}{$v_{\rm cl}$}
\newcommand{\vinf}{$\rm v_{\infty}$}
\newcommand{\finf}{$f_{\infty}$}
\newcommand{\vt}{$v_{turb}$}
\newcommand{\vrot}{$v_{rot}$}
\newcommand{\radius}{$R_{*}/R_{\odot}$}
\newcommand{\rstar}{$R_{*}$}
\newcommand{\mstar}{$M_{*}$}
\newcommand{\zsol}{${\rm Z}_{\odot}$}
\newcommand{\lsol}{${\rm L}_{\odot}$}
\newcommand{\zzsol}{${\rm Z/Z}_{\odot}$}
\newcommand{\rsol}{$R_{\odot}$}
\newcommand{\msol}{$M_{\odot}$}
\newcommand{\tlusty}{{\sc Tlusty}}
\newcommand{\synspec}{{\sc Synspec}}
\newcommand{\cmfgen}{{\sc CMFGEN}}
\newcommand{\fastwind}{{\sc FASTWIND}}
\newcommand{\wmbasic}{{\sc WM-BASIC}}
\newcommand{\Av}{$A_{\rm v}$}
\newcommand{\ebv}{$\rm E(B-V)$}
\newcommand{\Rv}{$R_{\rm v}$}
\newcommand{\Mv}{$M_{\rm v}$}
\newcommand{\msp}{$M_{\rm spec}$}
\newcommand{\mev}{$M_{\rm evol}$}
\newcommand{\grid}{{\sc Ostar2002}}
\newcommand{\kms}{km\,s$^{-1}$}
\newcommand{\cun}{cm$^{-1}$\ }
\newcommand{\cdeux}{cm$^{-2}$\ }
\newcommand{\ctrois}{cm$^{-3}$\ }
\newcommand{\ci}{\mbox{C~{\sc i}}}
\newcommand{\cii}{\mbox{C~{\sc ii}}}
\newcommand{\ciii}{\mbox{C~{\sc iii}}}
\newcommand{\civ}{\mbox{C~{\sc iv}}}
\newcommand{\cv}{\mbox{C~{\sc v}}}
\newcommand{\hi}{\mbox{H~{\sc i}}}
\newcommand{\hyd}{\mbox{H}}
\newcommand{\he}{\mbox{He}}
\newcommand{\hei}{\mbox{He~{\sc i}}}
\newcommand{\heii}{\mbox{He~{\sc ii}}}
\newcommand{\oii}{\mbox{O~{\sc ii}}}
\newcommand{\oiii}{\mbox{O~{\sc iii}}}
\newcommand{\oiv}{\mbox{O~{\sc iv}}}
\newcommand{\ov}{\mbox{O~{\sc v}}}
\newcommand{\ovi}{\mbox{O~{\sc vi}}}
\newcommand{\ovii}{\mbox{O~{\sc vii}}}
\newcommand{\nii}{\mbox{N~{\sc ii}}}
\newcommand{\niii}{\mbox{N~{\sc iii}}}
\newcommand{\niv}{\mbox{N~{\sc iv}}}
\newcommand{\nv}{\mbox{N~{\sc v}}}
\newcommand{\fe}{\mbox{Fe~}}
\newcommand{\neon}{\mbox{Ne~}}
\newcommand{\p}{\mbox{P~}}
\newcommand{\piv}{\mbox{P~{\sc iv}}}
\newcommand{\pv}{\mbox{P~{\sc v}}}
\newcommand{\pvi}{\mbox{P~{\sc vi}}}
\newcommand{\siii}{\mbox{Si~{\sc ii}}}
\newcommand{\siiv}{\mbox{Si~{\sc iv}}}
\newcommand{\siv}{\mbox{S~{\sc iv}}}
\newcommand{\sv}{\mbox{S~{\sc v}}}
\newcommand{\svi}{\mbox{S~{\sc vi}}}
\newcommand{\feiv}{\mbox{Fe~{\sc iv}}}
\newcommand{\fev}{\mbox{Fe~{\sc v}}}
\newcommand{\fevi}{\mbox{Fe~{\sc vi}}}
\newcommand{\msolyr}{M$_{\odot}$\,yr$^{-1}$}
\newcommand{\pcyg}{P~Cygni}
\newcommand{\vesc}{$v_{esc}$}
\newcommand{\xit}{$\xi_{t}$}
\newcommand{\us}{$\times\,10^{-7}$}
\newcommand{\un}{$\times\,10^{-9}$}
\newcommand{\epv}{$\times\,10^{5}$}
\newcommand{\ev}{$\times\,10^{-5}$}
\newcommand{\evi}{$\times\,10^{-6}$}
\newcommand{\evii}{$\times\,10^{-7}$}
\newcommand{\eviii}{$\times\,10^{-8}$}
\newcommand{\eix}{$\times\,10^{-9}$}
\newcommand{\ex}{$\times\,10^{-10}$}
\newcommand{\logL}{$\log L/L_{\odot}$}
\newcommand{\lognHI}{$\log N$(H~{\sc i})}
\newcommand{\lognHII}{$\log N({\rm H}_{2})$}
\newcommand{\logLX}{$\log L_{\rm X}/L_{\rm bol}$}

\title{Massive stars at low metallicity}
\subtitle{Evolution and surface abundances of O dwarfs in the SMC
\thanks{Based on observations made with the NASA-ESA {\sl Hubble Space Telescope\/} (program GO 11625),
obtained at STScI, which is operated by AURA, Inc., under NASA contract NAS 5-26555.}
\thanks{Based on observations collected at the European Southern Observatory {\sl Very Large Telescope\/}, program 079.D-0073.}
}
\author{J.-C. Bouret\inst{1,2},  
               T. Lanz\inst{3},   
              F. Martins\inst{4}, W. L .F. Marcolino\inst{5},
               D. J. Hillier\inst{6}, E. Depagne\inst{7},  I. Hubeny\inst{8}}         
 
 \offprints{J.-C. Bouret}

 \institute{Laboratoire d'Astrophysique de Marseille - LAM, Universit\'e d'Aix-Marseille \& CNRS, UMR7326, 38 rue F. Joliot-Curie, 13388 Marseille Cedex 13, France , \email{Jean-Claude.Bouret@oamp.fr}
         \and
         NASA/Goddard Space Flight Center, Greenbelt MD 20771, USA
         \and
        Laboratoire J.-L. Lagrange, UMR 7293, Universit\'e de Nice-Sophia Antipolis, CNRS, Observatoire de la C\^ote d'Azur
        B.P. 4229, 06304 Nice Cedex 4, France, \email{Thierry.Lanz@oca.eu}
                \and
                 LUPM--UMR5299, Universit\'e Montpellier II \& CNRS, Place Eug\`ene Bataillon, F-34095 Montpellier Cedex 05, France, \email{fabrice.martins@univ-montp2.fr}
             \and
          Universidade Federal do Rio de Janeiro, Observat\'orio do Valongo. Ladeira Pedro Ant\^onio, 43, CEP 20080-090,
           Rio de Janeiro, Brazil, \email{wagner@astro.ufrj.br}
  \and
     Department of Physics and Astronomy \& Pittsburgh Particle physics, Astrophysics, and Cosmology Center (PITT PACC), University of Pittsburgh,  Pittsburgh, PA 15260, USA, \email{hillier@pitt.edu}     
        \and
  Leibniz Institut f\"ur Astrophysik Potsdam, An der Sternwarte 16, 14482 Potsdam, Germany, \email{edepagne@aip.de}
  \and
  Steward Observatory, University of Arizona, AZ, USA, \email{hubeny@as.arizona.edu}
    }

\date{Received 27/11/2012 ; Accepted 09/04/2013 }        

\abstract{}             
{We aim to study the properties of massive stars at low metallicity, with an emphasis on their evolution, rotation, and surface abundances. We focus on O-type dwarfs 
in the Small Magellanic Cloud. These stars are expected to have weak winds that do not remove significant amounts of their initial angular 
momentum.}
   {We analyzed the UV and optical spectra of twenty-three objects using the NLTE stellar atmosphere code \cmfgen\ and derived
    photospheric and wind properties.}
 {The observed binary fraction of the sample is $\approx$ 26\%, which is consistent with more systematic studies, if one considers that the actual binary fraction 
 is potentially larger owing to low-luminosity companions and that the sample was biased because it excluded obvious spectroscopic binaries.
 The location of the fastest rotators in the Hertzsprung-Russell (H-R) diagram built with fast-rotating evolutionary models and isochrones indicates that these could be several Myr old. 
The offset in the position of these fast rotators compared with the other stars confirms the predictions of evolutionary models that fast-rotating stars tend to evolve more vertically in the H-R diagram.
Only one star of luminosity class Vz, expected to best characterize extreme youth, is located on the ZAMS, the other two stars are more evolved. 
We found that the distribution of O and B stars in the $\epsilon (N)$ -- \vsini\ diagram is the same, which suggests that the mechanisms 
responsible for the chemical enrichment of slowly rotating massive stars depends only weakly on the star's mass. 
We furthermore confirm that the group of slowly rotating N--rich 
stars is not reproduced by the evolutionary tracks.
Even for more massive stars and faster rotators, our results call for stronger mixing in the models to explain the range of observed N abundances. 
All stars have an N/C ratio  as a function of stellar luminosity that match the predictions of the stellar evolution models well. 
More massive stars have a higher N/C ratio than the less massive stars. 
Faster rotators show on average a higher N/C ratio than slower rotators, again consistent with the expected trend of stronger mixing as rotation increases. 
When comparing the N/O versus N/C ratios with those of stellar evolution models, the same global qualitative agreement is reached.
The only discrepant behavior is observed for the youngest two stars of the sample, which both show very strong signs of mixing, which is unexpected for their evolutionary status.  }
{}

\keywords{Stars: early-type  -- Stars: fundamental parameters -- Stars: rotation -- Stars: abundances -- Galaxies: Magellanic Clouds}

\authorrunning{Bouret et al.}        
\titlerunning{Massive stars at low metallicity}        
        
\maketitle        
     
        
\section{Introduction}        
\label{intro_sect} 
During the past three decades, much effort has been invested in understanding the effect of essential processes
such as stellar winds (and the resulting mass loss) and stellar rotation on the evolution of massive stars.
Mass-loss rates are a crucial parameter of stellar evolution models because O-type stars loose a sizable
fraction of their total mass during their lifetime. 
Progress has been achieved in the parameterization of mass-loss rates on basic physical parameters, such as
luminosity and metallicity, from the theoretical and the observational sides \citep{vink01, krticka06, mokiem07, muijres12}. 
Although the general paradigm is successful in explaining the global behavior of stellar winds, several important questions remain open, 
including the presence of structures in the wind (clumping) and the so-called weak winds in late-O dwarfs \citep[see ][for a review]{puls08}. 

Likewise, major progress in stellar rotation theory has been achieved in the past decade. The effects of rotation
include an increase of the mass-loss rate, a change in evolutionary tracks in the H-R diagram, a lowering of the effective gravity,
an extension of the main-sequence phase, and the
mixing of CNO-cycle processed material up to the stellar surface \citep{maeder00, heger00}. 
Qualitatively, stellar evolution models that include rotation predict a surface enrichment of helium and nitrogen with an associated carbon and oxygen depletion during the main-sequence evolution. 
Quantitatively, stellar evolution models with rotation predict that fast-rotating stars undergo more mixing, and hence a
greater nitrogen (and helium) surface enrichment should be observed. Because of the sensitivity of evolutionary tracks
in the H-R diagram with rotation, the location of massive main-sequence stars in the H-R diagram does not provide
a unique determination of age and mass. Rotation thus introduces some degeneracy, but surface abundance
patterns could help resolve this degeneracy \citep{brott11a, kohler12}.

Hot-star winds are driven by momentum transfer from the radiation field to the gas via photon absorption
in spectral lines of metal elements. A dependence of the mass-loss rates on the stellar metallicity is thus predicted
($\dot{M} \propto Z^{0.69 \pm 0.10}$ -- Vink et al. 2001) and observed ($\dot{M} \propto Z^{0.83 \pm 0.16}$ -- Mokiem et al. 2007). In addition to mass, the stellar winds take
angular momentum away, and therefore the removal  of angular momentum during the star's evolution
is significantly reduced at lower metallicities, increasing the effects of rotation \citep{maeder01}. The
interplay between stellar rotation and the stellar wind, which blurs the stars' initial rotational velocity properties
within only a few million years in the case of Galactic massive stars, is thus expected to be reduced with decreasing
metal content. The initial conditions of rotational velocity remain better preserved during the main-sequence life of 
O-type stars at low metallicity. Massive stars in the Magellanic Clouds (MCs) are therefore
the appropriate targets to study the effects of rotation on the evolution of massive stars.

Several studies have tackled this topic, though from different perspectives. From the analysis of four
extreme O-type supergiants, \cite{crowther02} found that the \mbox{CNO} abundances differ greatly
from those inferred from nebular and stellar studies in the MCs. Nitrogen was strongly enhanced, with carbon
(and oxygen) moderately depleted, which is suggestive of mixing of unprocessed and \mbox{CNO}-processed material at their
surfaces.
These results were supported by \cite{hillier03}, who focused on two
giant O stars in the Small Magellanic Cloud (SMC) that have very similar effective temperatures and luminosities, but whose 
wind spectral signatures differ markedly. Nitrogen is substantially enhanced at the surface of one star,
while the N/C abundance ratio in the other is below solar which agrees with SMC nebular abundances. 
A consistent picture of the evolution of these two stars may be obtained if they both had a similar initial mass, 
$M \approx$~ 40~\msol, one star being a slow rotator and the other a fast rotator. 
\cite{bouret03} analyzed six O dwarfs in the SMC giant H II region \object{NGC 346} and found that a majority of main-sequence O stars show 
enhanced nitrogen at their surface, indicating either fast stellar rotation and/or very efficient mixing processes. 

The \cite{hillier03} and \cite{bouret03} results were extended to the complete sample of 18 O stars in the SMC 
(HST GO Program 7437, PI: D. Lennon) in \cite{heap06}. The program SMC O-type
stars were chosen for their narrow-lined spectra to avoid problems with line blending in the analysis.
\cite{heap06} found that the surface composition of a high percentage of the
stars (about 80\% of the sample) show surface nitrogen enrichment during their lifetime on the main sequence.
Furthermore, there is no enrichment in helium and no depletion in carbon or oxygen, although the latter is
poorly constrained, because we lack reliable abundance diagnostics. 
N/C enrichment factors of 10 -- 40 were found, which is significantly higher than the theoretical predictions of
factors of 5 -- 10 \citep{maeder01}. In the few cases where nitrogen surface abundance has
not yet been affected by mixing, \cite{heap06} also obtained \mbox{CNO} abundances
consistent with SMC nebular abundances.
Most stars in their sample show the so-called mass discrepancy between spectroscopic and evolutionary mass
estimates. Along with nitrogen enrichment, \cite{heap06} interpreted this discrepancy as a result of rapid rotation
that lowers the measured effective gravities. Therefore, most stars in their sample should be fast rotators with \vrot $\ge$ 200 \kms,
the low measured \vsini\ thus implying that most stars are viewed almost pole-on (note again that their sample was biased
toward stars with sharp spectral lines, hence low \vsini).

More recently, \cite{hunter07, hunter08, hunter09} investigated the link between surface abundance 
patterns and rotation on a larger, unbiased (against rotation) sample of Galactic and MC B-type stars observed
for the VLT-FLAMES survey \citep{evans05, evans06}.
They concluded that up to 40\% of their sample had nitrogen abundances that could not be explained by 
the theory of rotational mixing as implemented in stellar evolution models. 
In particular, they found populations of non-enriched fast rotators, highly enriched slow rotators, and
B-supergiants that are highly enriched compared with normal core-hydrogen-burning objects.

The results above have been used to calibrate the amount of convective overshooting and the 
efficiency of rotationally induced mixing in stellar evolution models \citep{brott11a}. 
More observational constraints were obtained using single-star population-synthesis and an extended grid 
of evolution models, to simulate population of stars with masses, ages, and rotational velocity distributions consistent 
with those from the VLT-FLAMES sample \citep{brott11b}. 
Although the observed fraction of stars without significant nitrogen enrichment could be reproduced, the simulations also
predicted too many rapid rotators with enhanced nitrogen. Other problems included the failure to account for populations 
of slowly rotating, nitrogen-enriched objects on one hand, and of rapidly rotating non-enriched objects on the other hand.
\cite{brott11b} concluded that in addition to rotational mixing, other mechanisms should be accounted for, including the effects of binarity and magnetic fields on stellar evolution. 

It is therefore fair to state that the relative influence and interplay of mass loss, metallicity, age, rotation, and binarity on the global evolution of massive stars are not yet fully understood \citep{maeder09}. 
To answer these questions, we must accurately place stars on the theoretical H-R diagram and examine their other properties, such as rotation 
rates, masses, mass-loss rates, and surface \mbox{CNO} abundances. Many of these quantities can be accurately determined from spectral diagnostics in the UV range. This spectral range contains all major UV resonance wind lines and numerous other metal lines, providing the essential diagnostics of the photospheric and wind properties 
of the stars \citep[see e.g.][]{hillier03, bouret03}. 

We present in this paper the first results of an extensive analysis of the UV spectra of 23 O-type dwarf stars in the SMC, with a special emphasis on the evolution and chemical properties. 
In Sect.~\ref{obs_sect}, we introduce the HST observing program, and the additional optical data used in this work.
Sect.~\ref{analysis_sect} presents the modeling tool and strategy, while
general results are presented in Sect.~\ref{sect_result}. We discuss the properties of the sample stars 
in Sect.~\ref{disc_sect} and summarize in Sect.~\ref{conclu_sect}. Atomic data used in the modeling are disccused in Appendix~A. 
Properties of individual stars are presented in Appendix~B.

\section{Target selection and observations}
\label{obs_sect}

\subsection{General presentation of the FUV program}
We aimed at mapping all spectral types and luminosity classes in the O-type range. For this aim, we
secured high-resolution far-ultraviolet (FUV) spectra of 22 O-type stars in the SMC with the {\sl Cosmic Origin
Spectrograph\/ (COS)} onboard the {\sl Hubble Space Telescope\/} (HST Program GO 11625). Combined with a second
set of FUV spectra obtained earlier with {\sl HST\/}/STIS (Program GO 7437, PI: D. Lennon), the whole sample contains 39 O stars. 
The sample was selected mainly from the UBVR CCD photometric survey of the Magellanic Clouds by \cite{massey02}. This survey is photometrically complete to V $\approx$ 15.7, although some stars could have been missed in crowded regions. Table~6 in \cite{massey02} provides cross-identifications with previous prism-objective surveys. We also added four O dwarfs that are members of the NGC 346 cluster and are included
in the ESO/VLT-FLAMES survey by \cite{evans06}.

In a few instances, we included two or three stars of the same spectral type and luminosity class to directly compare stars that were presumably similar, but had different apparent rotational velocities (as indicated by spectral qualifiers, for instance "n"). In addition to investigating the effects of rotation, this approach provides a direct way to distinguish
between general trends and individual anomalies.  Rotation velocity, however, was not a selection criterion {\it per se} because of the lack of prior information -- but for the spectral qualifier ''n'" -- for many selected stars. While the distribution of \vsini\ for OB stars in the SMC peaks around 150-180 \kms\  \citep{mokiem06}, only one star has  \vsini\ $\geq$ 120 \kms\ (\object{AzV 80}) in the STIS sample \citep{heap06}. This sample therefore shows a preponderance of stars with narrow-lined spectra, which may suggest a bias toward slower rotators (Heap et~al. argued, however, that the sample was biased toward stars seen mostly pole-on). Our selection of the COS sample aimed at offsetting this potential bias. We stress that  the profiles of UV wind lines retain the information on wind properties even for fast-rotating stars with broadened line profiles, because of the high
terminal velocities of O-star winds.

Finally, color excesses and bolometric corrections were derived from the Massey photometry and from theoretical colors \citep{lanz03}.  From this list, we selected stars with the lowest extinction to maximize the expected FUV flux, and in turn
the expected signal-to-noise ratio (S/N) in a given UV exposure.  For each selected target, we checked also
that there was no other bright nearby star that might be in the COS aperture.

In this paper, we focus on 23 stars that are classified as dwarfs, because they are presumably closer to the ZAMS and less
evolved than the giants and supergiants. 
One possible exception is \object{MPG~355}, classified as ON2III (f*) by \cite{walborn04} but also as O2V by \cite{massey09}. 
Taken at face value, this indicates that a non-ambiguous identification of luminosity classes from spectral morphology  becomes difficult for early-O-type stars. 
Problems with the luminosity class criterion (based on the morphology of \heii\ \lb4686) may also become more severe
at low-Z environments such as the SMC \citep[see the very interesting discussion about the specific case of early-O-type stars in low-Z environments in][]{rivero12}. 
Therefore, some early-O-type stars whose physical properties (visual luminosities and surface gravities) suggest they are giants, might actually be dwarfs according to their (spectroscopically determined) 
luminosity classes. Overall, dwarf and giant stars occupy nearly the same location in the H-R diagram at the hot end of the main sequence \citep[which is known to be the case for \object{MPG 355} from][]{bouret03}.  
Accordingly, we elected to include \object{MPG~355} is our study which will thus sample the full range from the earliest to the latest O stars.

O-type stars spend most of their lifetime in the dwarf phase, while later stages
are much shorter. We intend to investigate the effect of rotation on the surface abundances in this phase.
Their stellar winds are also weaker. An analysis of the more luminous objects 
(luminosity classes III and I) will be presented in a separate paper, with a stronger emphasis on mass loss.

\subsection{UV observations and optical data}

We observed each target in a sequence of four science exposures with {\sl HST\/}/COS, using gratings G130M and G160M with central wavelengths (\lb1291, \lb1327) and (\lb1577, \lb1623) for the adopted settings respectively. 
This sequence provides coverage of the FUV spectrum without a gap between 1132 and 1798~\AA. Overlapping segments were co-added to improve the final signal-to-noise ratio. 
Thanks to the high efficiency of COS in the FUV, S/N $\approx$ 20 - 30 per resolution element were typically achieved in a single {\sl HST\/} orbit for each star. For SMC targets, the visibility is 60 min per orbit. Accounting for various observation overheads, only 29~min were left for the science exposures. We therefore allocated 300~sec for the G130M exposures and
570~sec for the G160M exposures in order to achieve the best possible S/N over the whole spectral range. 

Some stars were observed previously with {\sl HST\/}/STIS  through the 0$\arcsec$.2 - 0$\arcsec$.2 aperture, using the far-UV MAMA detector in the E140M mode.
A spectral interval from 1150 \AA\ to 1700 \AA\ is covered in a single exposure at an effective spectral resolving power of R = 46,000. With one exception,  two or more observations per target were obtained. 
The exposure times were set according to the apparent magnitudes of the stars to equalize the S/Ns of the spectrograms. The total exposure times ranged from 45 minutes for MPG 355 (one observation) 
to 245 minutes for MPG 012 (five observations). At  1300 \AA\ (the wavelength with the highest sensitivity), the S/N per binned data point ranges from 55 to 110; at 1600 \AA\, it ranges from 25 to 50.
A more detailed description of the STIS observations and reduction can be found in \cite{walborn00} and  \cite{heap06}.

We supplemented the FUV spectra with optical spectra for most stars. First, we obtained spectra with the UVES spectrograph on the {\sl VLT} \citep{Dekker2000} under program 079.D-0073(A).
The spectra have a resolving power of about 45,000,  with five pixels per resolution element. The data were reduced using the UVES context within 
MIDAS \citep[see][]{Ballester2000}. The reduction process includes flat-fielding,  bias and sky-substraction, and a relative wavelength calibration. Cosmic-ray removal was performed 
with an optimal extraction method for each spectrum.
An additional {\sl VLT}/UVES spectrum was obtained for \object{AzV 388} in program 074.D-0164, and was kindly provided by P.~Crowther. 
The optical spectrum of \object{AzV 177}, obtained at {\sl CTIO} and presented in \cite{massey09}, was kindly provided by P. Massey. 
For those stars with {\sl HST\/}/STIS spectra, optical spectra were available from observations with  {\sl AAT}/UCLES or  {\sl ESO}/CASPEC. They were extensively presented in  \cite{walbornSMC} and  \cite{heap06}, to which we refer for
more details. 
Optical spectroscopy was also supplemented from the ESO {\sl VLT}/FLAMES survey of massive stars in the SMC
\citep{evans06}. See \cite{evans05, evans06}  for a general description of the properties of the FLAMES data and their reduction. 
For a few more objects (see Table \ref{tabone}), we used lower-resolution spectra obtained for the {\sl AAT}/2dF program in the SMC in \cite{evans04}.  These authors provide details about the data and their reduction. In total, we secured optical spectra for 18 stars out of the 23 stars of our sample.

The optical coverage is especially important in determining the stellar surface gravities, because this can only be poorly constrained from the UV spectrum\footnote{It is well known that UV wind features for O stars correlate with luminosity class \citep[e.g.][]{walborn84, walborn87}. Because we did not derive \mdot\ and \vinf\ from first principals, we cannot use this effect to further constrain the photospheric parameters}.
However, optical spectroscopy presents us
with some particular difficulties in determining the stellar parameters, in particular because of nebular emission features
in optical spectra \citep{heap06, mokiem06}. 
This nebular masking, for the Balmer and \hei\ lines that are  used to constrain \teff,  is especially critical for some stars in \object{NGC 346}.
Although the nebular features are clearly resolved in VLT/FLAMES spectra, an efficient and accurate substraction from photospheric line cores remains problematic \citep{evans05}. It is difficult to assess the actual level of filling that is left after removal of the nebular emission, and the \hei\ lines may seem weaker than they actually are, which would bias the \teff\ estimates toward higher values. This problem is also relevant for field stars observed at lower resolution (2dF spectra). 
We discuss our methodology of determining effective temperatures and surface gravities in more details in \S\ref{param_sect}.

\begin{sidewaystable*}
\centering \caption{Basic data for the stellar sample ordered by spectral type.}
\begin{tabular}{lcccllccllc}
\hline\hline
Star ID			&  R. A		& Dec.		& Spec.			&  \multicolumn{2}{l}{Spectroscopic Data}		&  $V$	&   $B-V$		&   \multicolumn{2}{c}{$E(B-V)$}	& Refs. \\
				&  (2000)		&  (2000)		&Type			&         FUV               &    Optical					&		&			&  Galactic	&	SMC			&              \\
\hline
\object{MPG 355}	&  00 59 00.75	&  -72 10 28.2	&  ON2III (f$^{*}$)	&   {\sl HST}/STIS  &  {\sl AAT}/UCLES, {\sl VLT}/UVES		&  13.50	&  $-0.23$		&  0.032		&   0.040			& 1                \\        
\object{AzV 177}	&  00 56 44.17	&  -72 03 31.3	&  O4V ((f))		&   {\sl HST}/COS  &  {\sl CTIO\/}				&   14.53	&  $-0.21$		&   0.040		&   0.040			& 2         \\ 
\object{AzV 388}	&  01 05 39.45	&  -72 29 26.8	&  O4V			&   {\sl HST}/COS  & {\sl VLT}/UVES				&   14.09	&  $-0.21$		&   0.032		&   0.050			& 3                \\ 
\object{MPG 324}	&  00 58 57.56	&  -72 10 35.7	&  O4V((ff$^{+}$))	&   {\sl HST}/STIS  &  {\sl AAT}/UCLES, {\sl VLT}/FLAMES	&   14.07	&  $-0.28$		&   0.037		&   0.030			& 4			                \\ 
\object{MPG 368}	&  00 59 01.81	&  -72 10 31.3	&  O6V			&   {\sl HST}/STIS  &  {\sl AAT}/UCLES, {\sl VLT}/UVES		&   14.18   &  $-0.23$	&   0.037		&   0.030			& 1                \\ 
\object{MPG 113}	&  00 58 31.77	&  -72 10 57.9	&  OC6Vz 			&   {\sl HST}/STIS  &  {\sl ESO}/CASPEC, {\sl VLT}/FLAMES	&   15.01	&  $-0.26$		&   0.037		&   0.050			& 4               \\ 
\object{AzV 243}	&  01 00 06.78	&  -72 47 18.7	&  O6V			&   {\sl HST}/COS  &  {\sl VLT}/FLAMES				&   13.84   &  $-0.17$	&   0.037		&   0.060			& 5                 \\ 
\object{AzV 446}	&   01 09 25.46	&  -73 09 29.7	&  O6.5V			&   {\sl HST}/COS  &  ... 						&   14.59	&  $-0.24$		&   0.035           	&   0.030			& 3                 \\ 
\object{MPG 356}	&   00 59 00.91	&  -72 11 09.3	&  O6.5V			&   {\sl HST}/STIS  &  {\sl VLT}/UVES				&   15.50 	&  $-0.26$		&   0.037    	&   0.060  			& 6              \\ 
\object{AzV 429}	&   01 07 59.87	&  -72 00 53.9	&  O7V             		&   {\sl HST}/COS  &  ... 						&   14.63   &  $-0.20$  	&   0.037		&   0.055 			& 3    \\ 
\object{MPG 523}	&   00 59 08.68	&  -72 10 14.1	&  O7Vz			&   {\sl HST}/COS  &  {\sl VLT}/FLAMES				&   15.51   &  $-0.26$  	&   0.027		&   0.020			& 4                \\ 
\object{NGC346-046} 	&   00 59 31.84	&  -72 13 35.2	&  O7Vn 		&   {\sl HST}/COS  &  {\sl VLT}/FLAMES  		&   15.44	&  $-0.28$		&   0.020 		&   0.020			& 4                \\ 
\object{NGC346-031}	&   00 59 54.04	&  -72 04 31.3	&  O8Vz		&   {\sl HST}/COS  &  {\sl VLTFLAMES\/}			&   15.02   &  $-0.26$  	&   0.040		&   0.050			& 4                \\ 
\object{AzV 461}	&   01 11 25.53	&  -72 09 48.5	&  O8V			&   {\sl HST}/COS  &  {\sl AAT/2dF\/}						&   14.57   &  $-0.23$	&   0.020		&   0.013			& 3                 \\ 
\object{MPG 299}	&   00 58 55.22	&  -72 09 06.6	&  O8Vn			&   {\sl HST}/COS  &  {\sl VLT}/FLAMES				&   15.50   &  $-0.30$	&   0.037		&   0.010			& 4                \\ 
\object{MPG 487}	&   00 59 06.71	&  -72 10 41.3	&  O8V			&   {\sl HST}/STIS  &  {\sl AAT}/UCLES, {\sl VLT}/UVES		&   14.53   &  $-0.22$	&   0.020		&   0.057			& 1                \\ 
\object{AzV 267}	&   01 01 15.68	&  -72 06 35.4	&  O8Vn			&   {\sl HST}/COS  &  ...						&   14.84	&  $-0.26$		&   0.037		&   0.010			& 3                 \\ 
\object{AzV 468}	&   01 12 05.82	&  -72 40 56.2	&  O8.5V			&   {\sl HST}/COS  &  ...						&   15.11   &  $-0.25$	&   0.030		&   0.010			& 3                 \\ 
\object{AzV 148}	&   00 53 42.19	&  -72 42 35.2	&  O8.5V			&   {\sl HST}/COS  &  ...						&   14.12   &  $-0.20$	&   0.037		&   0.050			& 3                 \\ 
\object{MPG 682}	&   00 59 18.59	&  -72 11 09.9	&  O9V			&   {\sl HST}/COS  &  {\sl VLT}/FLAMES				&    14.91	&  $-0.26$		&   0.037		&   0.030			& 4                \\ 
\object{AzV 326}	&   01 03 09.26	&  -72 25 56.5	&  O9V			&   {\sl HST}/COS  &  {\sl AAT/2dF\/}						&   13.92   &  $-0.12$	&   0.020		&   0.040			& 3                 \\ 
\object{AzV 189}	&   00 57 32.51	&  -72 28 51.0	&  O9V			&   {\sl HST}/COS  &  {\sl AAT/2dF\/}						&   14.37   &  $-0.11$	&   0.037		&   0.100			& 3                 \\ 
\object{MPG 012}	&   00 58 14.10	&  -72 10 44.2	&  B0IV			&   {\sl HST}/STIS  &  {\sl ESO}/CASPEC, {\sl VTL}/FLAMES	&   14.98	&  $-0.14$		&   0.037             &   0.097			& 4                \\ 
\hline
\end{tabular}
 \tablefoot{
The primary star identifications are ``MPG numbers'' from \cite{massey89} or ``NGC346-XXX numbers'' from \cite{evans06} for stars in \object{NGC 346}.
We use AzV numbers \citep{azzopardi75, azzopardi82} for SMC field stars. 
The color excess, $E(B-V)$, is divided into two contributions from the Galactic foreground 
and the SMC field and is derived from this study (see Sect. \ref{param_sect}).
The following sources were used for coordinates, spectral classification, and photometry: 
1. \cite{massey09}; 2. \cite{massey05}; 3. \cite{massey02}; 4. \cite{evans06}; 5. \cite{mokiem06}; 6. \cite{massey89}
}
\label{tabone}
\end{sidewaystable*}

\section{Spectroscopic analysis}
\label{analysis_sect}
\subsection{Model atmosphere}
\label{mod_sect}
We performed the analysis using the model atmosphere code \cmfgen\  \citep{hillier98, hillier03}. 
This code computes NLTE line-blanketed model atmospheres, solving the radiative transfer and statistical equilibrium equations in the comoving frame of the fluid in a spherically-symmetric outflow.
To facilitate the extensive inclusion of line blanketing, a formalism of super-levels was adopted.  
This approach allowed us to incorporate many energy levels from ions of many different species in the model atmosphere calculations. 
More specifically, we included in this work ions of H, He, C, N, O, Ne, Mg, Si, S, Ar, Ca, and Fe, which represent up to $\approx$ 14,000 individual levels.  
After convergence, a formal solution of the radiative transfer equation was computed  in the observer's frame \citep{busche05}, thus providing the synthetic spectrum for a comparison to observations. 

The photospheric density structure was initially described with an hydrostatic density structure from a \tlusty\ model atmosphere \citep{hubeny95, lanz03}  in the quasi-static deep layers,
while we used a standard $\beta$-velocity law in the wind. The $\beta$-velocity law is connected to the hydrostatic density structure just above the sonic point  (at approximately 15 \kms). 
The mass-loss rate, density, and velocity are related via the continuity equation.

In more advanced models we iterated on the density structure so that the hydrostatic equation,
\begin{equation}
{dP/dr} = - g_{\rm grav} + g_{\rm rad},
\end{equation}
where
\begin{equation}
g_{\rm grav} = GM/r^{2}
\end{equation}
and the radiative acceleration $g_{\rm rad}$
\begin{equation}
g_{\rm rad} = \frac{4\pi}{c}\rho\int\chi_{\nu}H_{\nu}d\nu
\end{equation}

\noindent was satisfied. We solved for the density structure by integrating the hydrostatic equation using the Runge-Kutta method. To facilitate the integration, 
the Rosseland LTE opacity was computed as a function of temperature and electron number density for the model abundances prior to the \cmfgen\ run. 
In a typical \cmfgen\ model, a hydrostatic iteration is performed after the first model iteration and then every $n$ (with $n\sim 8$) iterations. Generally three to four hydrostatic iterations are needed to obtain convergence to better than 5\% everywhere. 

Model atmospheres were calculated assuming that the winds were clumped, for all stars showing obvious spectral signatures of a wind (i.e., P~Cygni line profiles). We adopted
a parametric treatment of optically thin clumps with a void interclump medium that is described by two parameters, a volume filling factor $f$ and the velocity where clumping starts $v_{\rm cl}$. 
This simple approach has some limitation such as a monotonic description, while more elaborate theoretical simulations by \cite{runacres02, puls06} 
instead suggest that clumping is non-monotonic. 
The radial distribution of the clumping factor was constrained in \cite{puls06} and \cite{najarro11}, from simultaneous
modeling of \halpha, infrared, millimeter, and radio observations. Nevertheless, the actual variation is still uncertain, and we therefore adopted this reasonable simple approach. 
When no clear wind signatures are observed, we restricted ourselves to calculating homogeneous wind models. Since \mdot$_{\rm hom}$ = \mdot$_{\rm clump}/\sqrt f$, 
the mass-loss rates that we derived must be considered potentially as upper limits (and these mass-loss rates remain quite uncertain).

Finally, we accounted for the effects of shock-generated X-ray emission in the model atmospheres. We adopted the standard Galactic luminosity ratios log$\rm L_{X}/L_{bol} \approx -7$ \citep{sana06}, 
with the bolometric luminosities determined in this study. 

\subsection{Stellar and wind parameters}
\label{param_sect}
To determine the stellar and wind parameters and surface abundances, we followed the same methodology as in previous papers.
We discuss here the main points only, and refer to \cite{bouret12} for a detailed description of our procedure and
diagnostics in more general cases. 

\begin{itemize}
\item {\bf Stellar luminosity}:
Since the \cosp\  and \stis\ spectra are flux-calibrated, we may use them together with optical and NIR photometry from \cite{bonanos10} (and references therein) to compute the luminosity and interstellar extinction, by fitting the observed spectral energy distribution (SED) from the UV to the near-IR range. We used two corrections for reddening, one for the Galactic foreground 
using the \cite{cardelli89} data and one for the SMC interstellar medium \citep{witt00}. 
The corresponding values are listed in Table \ref{tabone}.
Both extinction curves are valid over the range 1150 \AA\ to 30,000 \AA. The ratio of total-to-selective extinction was adopted to be $R_{\rm V}=3.1$ following \cite{massey09}, while 
a distance modulus to the SMC of 18.91$\pm$0.02 \citep{harries03} was used. 
Using  $R_{\rm V}=2.7$ as in \cite{evans06} would induce changes of less than 5\% in the flux, mostly in the UV, which is on the same order of magnitude or larger than the uncertainties caused by the photometric quantities 
or the distance modulus, and in any case would translate into uncertainties on the luminosities of 0.04 dex or less. 
We found no indications for deviation from this standard $R_{\rm V}$ in our reddening correction, which is
sometimes suggested particularly in regions of strong star formation where $R_{\rm V}$ could be higher \citep[e.g. in 30 Dor, see][]{Bestenlehner11}. 
We note, however, that \cite{maiz12} reported that four stars in a quiescent molecular cloud in the SMC presented significant variations in the extinction law, although all are located within a few pc of each other. 
The strongest variations were found in the 2175 \AA\ bump, and despite these UV variations, the SMC visible-NIR extinction law appears to be more uniform, with no support found for high $R_{\rm V}$.
A somewhat more complex approach had to be adopted to derive the luminosity in three cases, which is discussed in Sect. \ref{bin_sect}.

\item {\bf Effective temperature and surface gravities}:
When optical data were available, we derived an initial estimate of \teff\ with the \hei\ \lb4471 to \heii\ \lb4542 equivalent width ratio, as well as by fitting the line profiles. The \teff\ determination was then
refined with other available \hei\ and \heii\  lines and listed in Table \ref{tab2}. 
Following \cite{bouret03}, \cite{hillier03}, and \cite{heap06}, we also used the ionization ratio of the UV metal lines to estimate the effective temperature. 
This includes for instance the ratios from iron forest
ions\footnote{\feiv\ lines between 1550 and 1700\,\AA, \fev\ lines between 1300 and 1550\,\AA, and \fevi\ lines below 1350 \AA.},  \civ\ \lb1169 
to \ciii\ \lb1176, or \niii\ \lb\lb1183-1185 and \niii\ \lb\lb1748-1752 to \niv\ \lb1718 ratios.
This is particularly important for the hottest stars of the sample, e.g., \object{MPG 355}, where the \hei\ lines become very weak \citep[see][for the use of nitrogen optical lines to constrain 
\teff\ for the earliest O-stars]{rivero12}.
The \civ\ \lb1169  to \ciii\ \lb1176 ratio is especially useful because of its very good sensitivity to \teff\ and weak dependance on gravity, hence we used it as the primary \teff\ diagnostic in the UV.
However, in several cases, we found that effective temperature determinations based on the ionization ratio of helium lines are inconsistent with the ionization balance 
of carbon as predicted by \cmfgen. 
In such cases, we found an offset toward higher \teff\ by about 2000~K compared with those derived from the helium lines, and
we were unable to achieve a good fit to these lines without degrading the agreement with the optical (helium) spectrum. The values listed in Table \ref{tab2} refer to UV-based
\teff\ only for those stars for which no optical spectra were available. For the other stars, the quoted \teff\ correspond to that providing the best fit over the whole FUV-optical spectral range. 

The Stark-broadened wings of \hgamma\ were the primary constraint on \logg, while the other Balmer lines were used as secondary indicators. 
The S/N of our spectra is sufficient to achieve a typical accuracy of about 0.1~dex in \logg. 
For stars with UV spectra only, we adopted a typical value for O-type dwarfs,  \logg = 4.0, because the UV range shows little sensitivity to changes
in surface gravities \citep[but see][for a discussion on using lines from different iron ions in the UV range]{heap06}.

\item {\bf Microturbulence}:
A constant microturbulent velocity of 15 \kms\ was adopted for calculating the atmospheric structure. 
To calculate the emerging spectrum, we used
a radially dependent turbulent velocity \citep[see also][]{hillier98}. 
The microturbulence in the photosphere (\xit\, see Tab. \ref{tab2}) is set such that it reproduces the strength of the \feiv, \fev, and \sv\ lines in the \cosp\ spectra best. 
In the outer region of the wind, the turbulence was adopted to fit the shape and slope of the blue side of the absorption component of the \civ\ 1548-1551 P~Cygni profile. 

\item {\bf Surface abundance} :
We used SMC abundances from \cite{hunter07} for Mg and Si and from \cite{venn99} for Fe. For all other elements but CNO, we adopted the standard solar abundances from \cite{grevesse07}, scaled
down by a factor of 5 (or equivalently by $-$0.7~dex).  The strength of helium lines was used to constrain the helium abundance.  We concentrated on deriving
the carbon, nitrogen, and oxygen abundances.\\

-- {\it Carbon abundance}:
In the UV range, we used \civ\ \lb1169 and \ciii\ \lb1176 as prime indicators for carbon abundances, although these lines proved to be clearly sensitive 
to \teff\ and \vturb\ and for the earliest stars of the sample showed some sensitivity to the wind density.  
For stars with later spectral types,  \ciii\ \lb1247 shows a clear response to abundance variations. 
In the optical, we used \ciii\ \lb4070 and \civ\ \lb\lb5801-5806 when possible. 
On the other hand, we did not include \ciii\ \lb5696 in our analysis. This line is sensitive to the adopted model atom and to UV \ciii\ lines at 386, 574, and 884 \AA\ \citep{martins12b}. 
A fairly large number of energy levels is required for this line to form as 
an emission as observed in late-O stars, and \cite{martins12b} called for extreme caution when interpreting carbon surface abundances derived from this line
\footnote{The influence of Fe metal lines found by \cite{martins12b} should be less than for galactic O stars because the Fe metal abundance is down a factor of five}.  

-- {\it Nitrogen abundance}: 
In the UV, \niii\ \lb\lb1183-1185 and \niii\ \lb\lb1748-1752  are used as primary diagnostics. Together with \niv\ \lb1718, which is only marginally affected by the stellar winds even in the earliest stars, 
these lines are clearly photospheric and provide a very reliable estimate of N/H. 
\niii\ \lb\lb4634-42 and \niv\ \lb4058 are the strongest nitrogen lines in the optical but are often very weak in SMC stars, and were used as secondary diagnostics for nitrogen abundance.
Their formation processes, including their sensitivity
to the background metallicity, nitrogen abundance, and wind-strength, have been discussed extensively by \cite{rivero11, rivero12}.

-- {\it Oxygen abundance}:
We used the \oiv\lb\lb1338-1343 lines as a diagnostic for O/H in the UV, although they sometimes show a blue asymmetry that indicates a contribution from the wind for the earliest stars with 
stronger winds. We also used the \oiii\ \lb\lb1150-1154\,\AA. These lines are on the shortest side of the UV spectra and the flux level there is more uncertain, however. \ov\lb1371 was not used because it is weaker and is most sensitive to the mass-loss rate and clumping parameters. In the optical, we used \oiii\ \lb5592 when possible. 

\item {\bf Wind parameters}:
The wind terminal velocities, \vinf, were estimated from the blueward extension of the absorption part of UV P~Cygni profiles.
The typical uncertainty in our determination of \vinf\ is 100 \kms\ (depending on the maximum microturbulent velocity we adopted).
For stars withour or with undeveloped P~Cygni profiles, we scaled down the wind terminal velocities found in Galactic stars with the same spectral types.
We started from values extracted from Table 1 in \cite{kudritzki00}, and applied a scaling \vinf\ $\propto {\rm Z}^{n}$, with Z/\zsol\  = $1/5$ and an exponent $n =0.13$ \citep{leitherer92}.

Mass-loss rates were derived from the analysis of UV P~Cygni profiles. In contrast to the Galactic case \citep{martins12a}, a unique value of \mdot\ allows a good fit to both the UV lines
and \halpha, although there is a problem with \nv\ \lb1238-1242 in some stars.
The $\beta$ exponent of the wind velocity law was derived from fitting the shape of the P~Cygni profile. 
The clumping parameters, $f$ and $v_{\rm cl}$ were derived in the UV domain following \cite{bouret03, bouret05}. 
Some photospheric lines in the optical presented some sensitivity to the adopted filling factor (and scaled \mdot). For photospheric \hi\ and \he\ lines,
for instance, this is essentially caused by a wind contribution that is weaker in clumped models, which produces deeper absorption than smooth-wind models.
   
A different procedure had to be adopted for stars without wind signatures. Starting from mass-loss rates of Galactic stars of same spectral type, scaled to  Z$_{\rm SMC}$ = $1/5$ \zsol\
with an exponent $m = 0.83$ \citep{mokiem07} such that \mdot\ $\propto {\rm Z}^{m}$, these values were decreased until no P~Cygni profiles were predicted for the aforementioned lines. 
As a consequence, and because we derived the mass-loss rates using homogeneous models, the quoted \mdot\ must be considered as upper limits.

\end{itemize} 

\subsection{Rotation and macro-turbulence}
\label{rot_sect}
To compare the synthetic spectra to observations, we first convolved the theoretical spectra with the appropriate instrumental profiles.
Then, we convolved our synthetic spectra with rotational profiles to account for the projected rotational velocity (\vsini) of the star. 
In the UV range, the \vsini\ is constrained by matching relatively weak iron lines as well as \ciii\ \lb1176, \oiv\lb\lb1338-1343, and \sv\ \lb1502. Helium lines were used in the optical. 
The \ciii\ \lb1176 multiplet is especially useful because of its six close line components. Many spectra show partial or fully resolved
multiplet components, indicating moderate (\vsini\ < 100 \kms) to slow (\vsini\ $\approx$ 20 \kms) rotational velocities, with a typical precision of 10 \kms .
A standard limb-darkening law was adopted for the convolutions with rotational profiles; a discussion about the impact of limb-darkening on some spectral lines of fast rotators is provided in \cite{hillier12}. 
In most cases, rotational rates for the target stars had been measured in \cite{penny09} from a cross-correlation function (CCF) 
analysis performed on \fuse\ spectra. The \vsini\ we derived from a direct comparison to observed photospheric profiles are quoted in Table~\ref{tab2}. 
They agree very well with those from
 \cite{penny09}, aand those by \cite{mokiem06} for the ten stars in common.

Other broadening mechanisms are nowadays routinely included to account for (isotropic) macro-turbulence, i.e. symmetrically distributed, 
deep-seated and highly supersonic velocity fields. \cite{simon10b} showed that a radial-tangential formulation of macro-turbulence fits the data better than 
a pure isotropic Gaussian.
We used the Fourier transform method described in \citet[][and references therein]{simon07}, applied to wind-free lines in optical spectra to assess
the relative contribution of macro-turbulence and rotation to the observed broadening. Stars without optical spectra or with only low-resolution (i.e., 2dF) optical spectra
were not included in this step. For all other stars, we found that \vsini\ from line fitting and/or CCF analysis \citep{penny09} on one hand, and the values obtained from the Fourier transform analysis
on the other hand, agree within the error bars. This agreement suggests  a minor contribution from macro-turbulence. Since the respective contribution of both broadening 
terms is quadratic, macro-turbulent velocity fields from a few \kms\ to several tens of \kms\  may still be present, however. 
\cite{aerts09} emphasized that using the Gaussian formulation of macro-turbulence to fit line profiles may lead to significant under estimation 
of the actual rotational velocity and warned that appropriate expression for the pulsational velocities need to be used instead. 

Therefore, the \vsini\  values quoted in Table~\ref{tab2} should be regarded as lower limits, but we do not expect that this uncertainty significantly alters
our conclusions for the surface abundances. First, the rotational term is always dominant with respect to macro-turbulence.
Second, we note that the stars studied here are dwarfs, and macroturbulent velocities seem to be significantly lower in dwarfs than in O supergiants
\citep[][and references therein]{markova11}.
Third, rotation and macro-turbulence in the adopted description are not expected to affect line equivalent widths, only the line profiles. 

\begin{figure*}[tbp]
   \centering
\includegraphics[scale=0.8, angle=0]{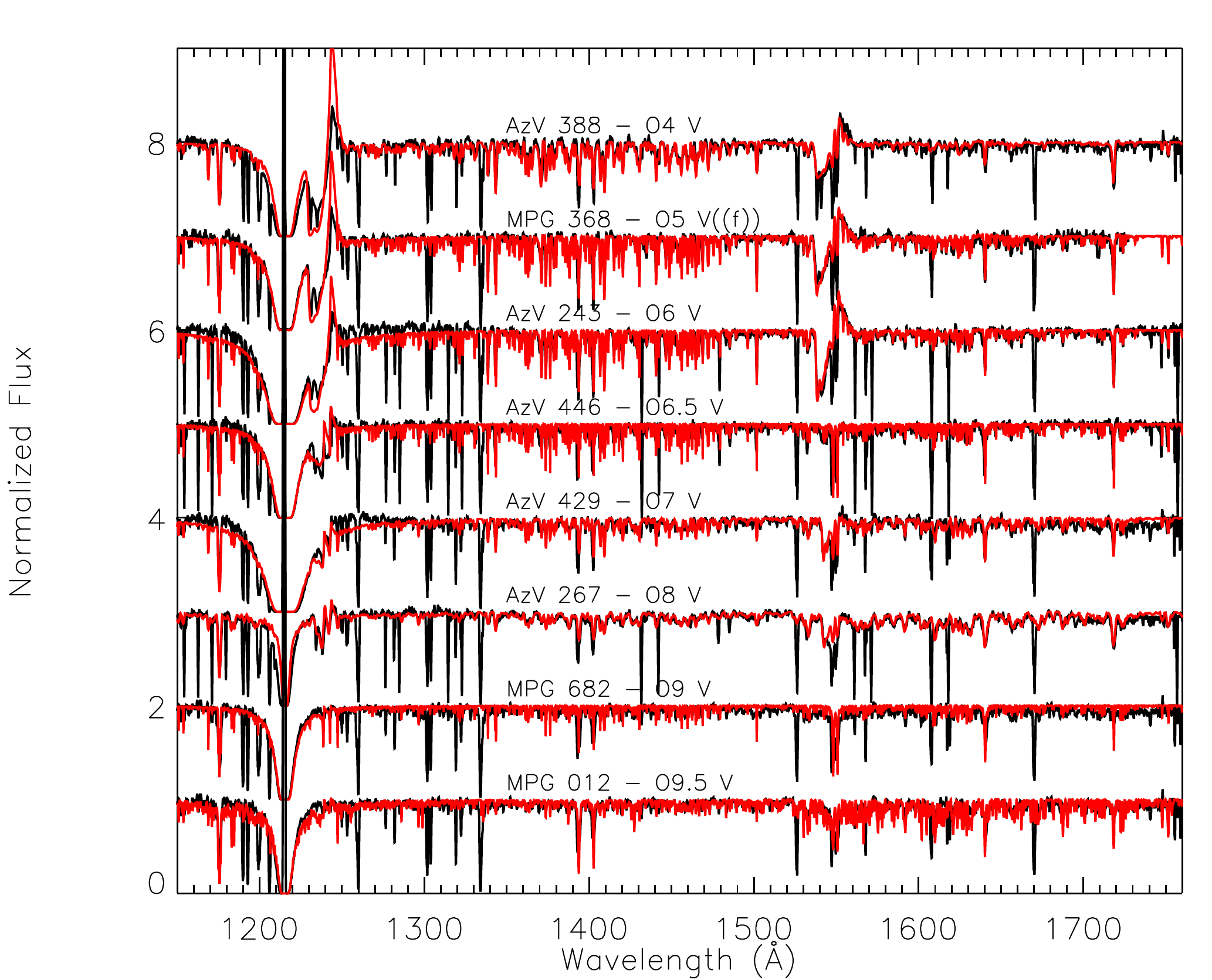}
    \vspace{-0.cm}
      \caption[12cm]{UV spectra for a selection of stars for each spectral type presented in this study, from earlier to later spectral types. Observations (black lines) are shown with
      best-fit model spectra (red lines). 
      The observational sequence is very nicely reproduced by the models (Table \ref{tab2}). Note the strong transition in wind profiles (e.g. \civ\ \lb\lb1548, 1550)
      around spectral type O6 V.    
      }
         \label{o4to09}
   \end{figure*}

\section{Results}
\label{sect_result}
The stellar and wind parameters we derived for the whole sample are gathered in Table~\ref{tab2}. 
The effective gravity derived from the analysis were corrected for the effect of centrifugal forces caused by rotation (see also Sect. \ref{mass_sect}), following the
approach outlined in \cite{repolust04}. These \loggc\  values were used to derive the spectroscopic masses.

Chemical abundances by number relative to hydrogen are listed in Table~\ref{tab3}  
 together with N/C and N/O abundance ratios and the adopted baseline for the abundance analysis. 

Given the complexity of the fitting process over the UV-to-optical spectral range, we elected not to derive formal statistical uncertainties,
nor did we try to estimate the correlation of errors between parameters. Instead, we varied the different parameters until we obtained the solution that provided the 
best fit ``by eye'' to the observations, with more weight being given to primary diagnostics \cite[see Sect. \ref{param_sect}, and][]{bouret12}.
Then, for a more quantitative estimate we varied each parameter independently around this solution and computed the residual (i.e., the difference) between the observed 
and the synthetic spectrum. 
We imposed the condition that this residual must remain within 10\%\ and adopted the range of values that fulfilled this condition as 
error bars on each parameter. The stellar and wind parameters in Table \ref{tab2} are obtained with this procedure. 

As for the uncertainty in the \mbox{CNO} abundance measurements, error bars are based on the fit
quality of the lines listed above (Sect \ref{param_sect}), as estimated from a $\chi^{2}$ procedure as presented for nitrogen in magnetic O-type stars by
\cite{martins12c}. Although this process is not ideal because it neglects correlations between parameters, it provides reasonable representative error bars.

Appendix~B provides a summary of our conclusions for each individual object, with best-model fits.
Fig. \ref{o4to09} presents a selection of UV spectra for stars representative of each spectral type in the sample, together with the corresponding best-fit models.  
Figure \ref{o4to09} shows that the complete observational sequence can be quantitatively reproduced by the models. Important photospheric features such as \feiv\ to \fevi\ ion 
lines or
\mbox{CNO} lines are reproduced and strongly vary as functions of \teff, \vturb, and abundances. Wind profiles are clearly defined for stars earlier than O6 V. Note the steep transition
from strong P~Cygni profile at \civ\ \lb\lb1548-1550 to the characteristic weak-wind profile, occuring at spectral type O6.5 V and later. 
As seen from Fig. \ref{o4to09}, only a small fraction of stars in the present sample exhibit conspicuous wind signatures, which allows a good determination of quantities
such as the mass-loss rates, terminal velocities, and clumping factor (see also Table \ref{tab2}). We investigated these wind quantities (e.g. the wind-momentum-luminosity-relation) but
too few stars provide reliable values for a fruitful discussion. For this reason, we chose not to discuss this in the present paper. 
We are currently investigating the giants and supergiants observed with COS/STIS during our program, and we postpone a complete discussion to a forthcoming, dedicated paper on wind properties at low metallicity.

\begin{figure*}[tbp]
   \centering
\includegraphics[scale=0.51, angle=0]{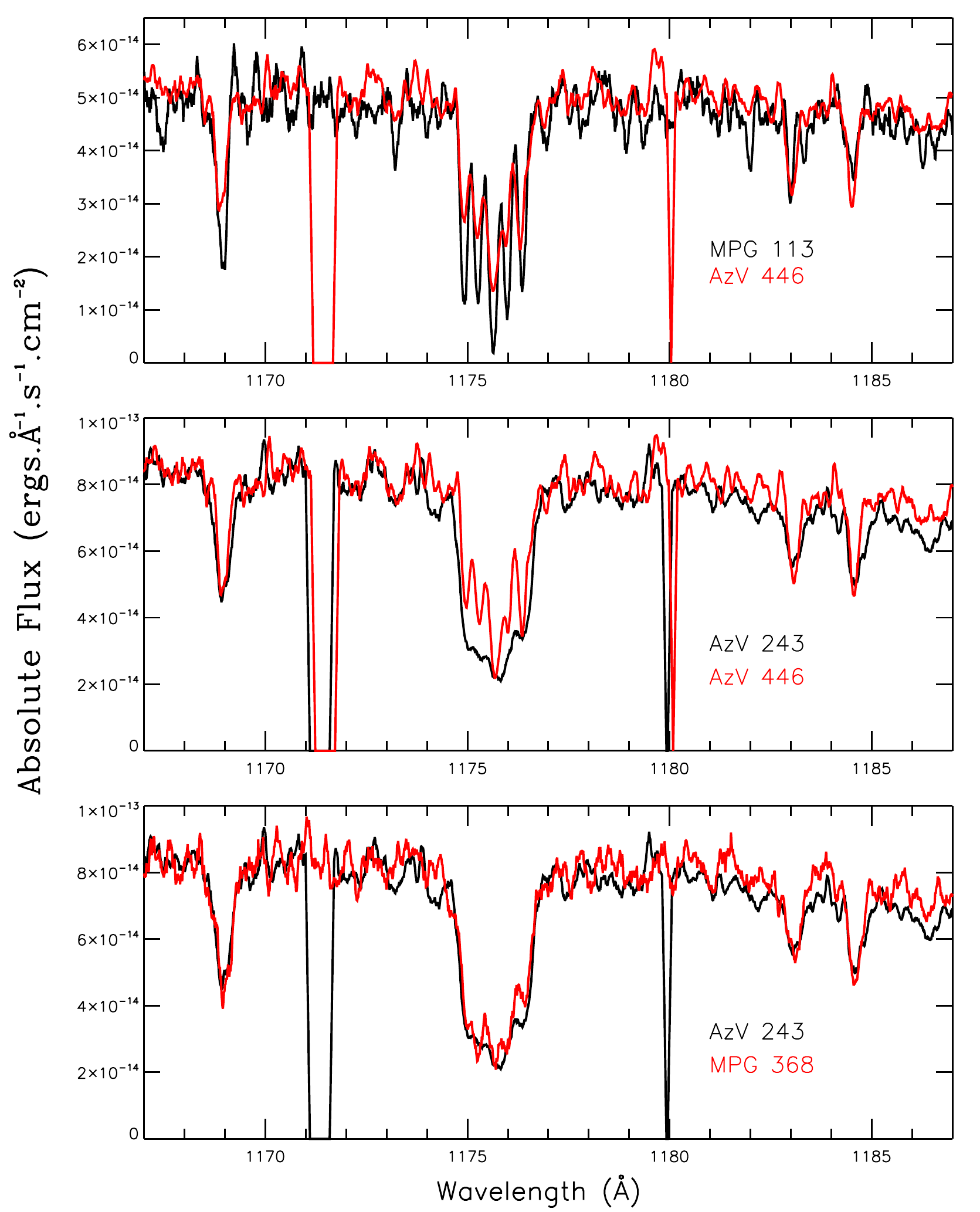}
\includegraphics[scale=0.51, angle=0]{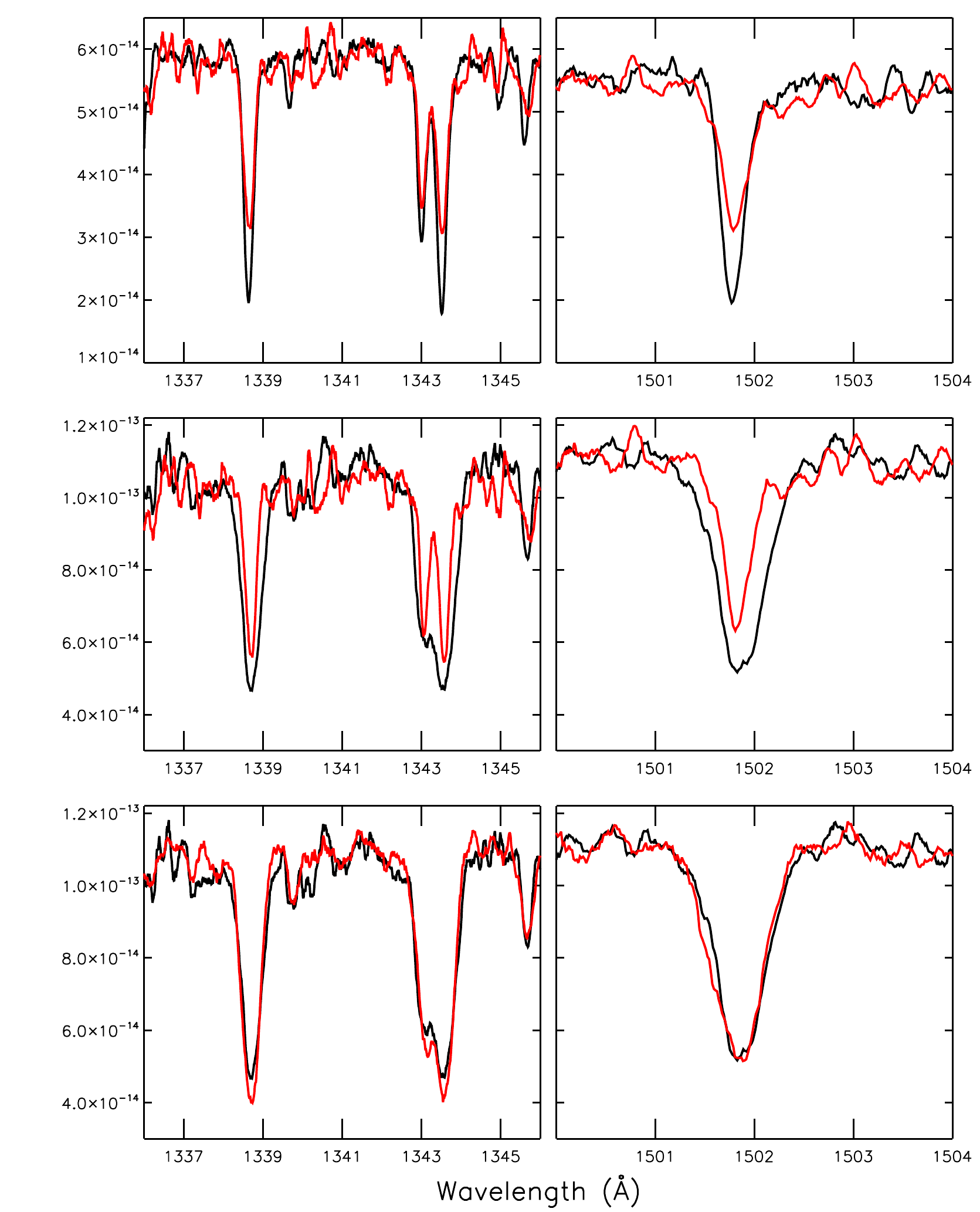}
    \vspace{-0.cm}
      \caption[12cm]{UV spectra for stars for similar fundamental parameters  (Table \ref{tab2}). The profiles of \civ\ \lb1169, \ciii\ \lb1176, and  \niii\ \lb\lb1183-11185
      are shown in the left panels. Profiles of \oiv\  \lb\lb1338-1340 and \sv\ \lb1502 are presented in the right panels.
      }
         \label{abund}
   \end{figure*}

The qualitative behavior of the lines used for abundance determination in the UV is illustrated in Fig.~\ref{abund}. 
We selected stars with very similar fundamental parameters to show the abundance differences. For this comparison, 
the spectra were corrected for reddening (Table \ref{tabone}) and the presence of interstellar \lyalpha\ absorption was taken into account for the plots 
(mostly important for the spectral region where \civ\ \lb1169, \ciii\ \lb1176 and  \niii\ \lb\lb1183-11185 are found). 
The bottom panel of Fig.~\ref{abund} shows \object{AzV 243} and \object{MPG 368}, two stars with very similar parameters, including \vsini\ and micro-turbulence (see Table~\ref{tab2}), as
shown by the similar \sv\ \lb1502 lines in the two stellar spectra. Carbon and oxygen lines (\civ\ \lb1169, \ciii\ \lb1176 and \oiv\  \lb\lb1338-1340) are very 
similar and indicate that carbon and oxygen abundances are the same in both stars.  
The nitrogen abundance on the other hand is slightly higher in \object{MPG 368}, as indicated by the stronger lines (marginally in the case of \niii\ \lb\lb1183-1185
but more clearly for \niv\ \lb1718, not shown here). 
The middle panel shows \object{AzV 243} and \object{AzV 446}. The two stars have also very similar parameters as well, but for \vsini\ and \vturb, which are significantly higher in \object{AzV 243}. 
The \mbox{CNO} lines, although they look different in these stars, are best reproduced with similar abundances. 
Finally, the top panel of Fig.~\ref{abund} compares \object{MPG 113} and \object{AzV 446}, where the \vsini\ values are similar. The effect of the higher \vturb\ in \object{MPG 113} is clearly visible in \sv\ \lb1502. The
\mbox{CNO} line profiles suggest that the surface abundance are different in these two stars: stronger carbon and lower nitrogen in \object{MPG 113}. Although 
 \oiv\lb\lb1338-1343 are sensitive to \vturb, the amplitude of the difference between the lines in the two stars suggests that oxygen abundance is higher
 in \object{MPG 113} as well. We stress that the spectroscopic analysis did not indicate the need for oxygen abundances to differ from the baseline adopted for reference \citep{kurt98} for about half of the sample stars (12 out of 23).
 
\begin{table*}[htbp]
\caption{Fundamental stellar parameters.}
\begin{tabular}{llcccccccccccc}
\hline
\hline
Star				& Sp. Type	& \teff	& \loggc& $\log \frac{L}{L_{\odot}}$ & log(\mdot)& $\beta$	&$f$	& \vinf	&	\xit	& \vsini 	& \msp		& \mev		& Age  \\
                            	& 		& [kK]	& [cgs]	&			     &		 &		&	& \multicolumn{3}{c}{[\kms]}&	 [\msol]	& [\msol]	& [Myr]\\ \hline
\object{MPG 355}	       & ON2III(f$^{*}$)& 51.7	& 4.00	& 6.04	& -6.74  & 0.8    & 0.02       & 2800	& 25		& 120	& 64$\pm13.2$		& $>$70?	& $\geq$ 3?\\		 
\object{AzV177	}		& O4V((f))	& 44.5	& 4.03	& 5.43	& -6.85	 & 1.0    & 0.05       & 2400	& 25		& 220	& 31.9$\pm7.2$  & 38.8$^{+3.5}_{-3.2}$  & 1.6$^{+0.5}_{-0.3}$ \\ 
\object{AzV 388}		& O4V		& 43.1	& 4.01	& 5.54	& -7.00	 & 1.3    & 0.11       & 2100	& 20		& 150	& 41.4$\pm9.3$  & 42.0$^{+4.1}_{-3.5}$  & 2.2$^{+0.5}_{-0.3}$ \\
\object{MPG 324$^{\mathrm{b}}$}	& O4V       	& 42.1	& 4.00	& 5.51	& -6.85	 & 1.0    & 0.07       & 2300	& 15		& 70	& 39.5$\pm9.0$  & 40.0$^{+3.2}_{-3.0}$  & 2.3$^{+0.3}_{-0.3}$\ \\
\object{MPG 368$^{\mathrm{b}}$}	& O6V       	& 39.3	& 3.75	& 5.38	& -7.43  & 1.0    & 0.10       & 2100	& 15		& 60	& 26.0$\pm5.9$  & 33.2$^{+2.7}_{-2.0}$  & 3.0$^{+0.5}_{-0.4}$ \\    
\object{MPG 113}		& OC6Vz		& 39.6	& 4.00	& 5.15	& -8.52  & 0.8    & ...        & 1250	& 10		& 35	& 24.6$\pm5.6$  & 27.8$^{+2.5}_{-2.0}$  & 2.8$^{+0.4}_{-0.4}$ \\
\object{AzV 243}		& O6V		& 39.6	& 3.90	& 5.59	& -7.10	 & 1.0    & 0.05       & 2000	& 15		& 60	& 40.8$\pm9.2$  & 41.0$^{+2.8}_{-2.2}$  & 3.0$^{+0.5}_{-0.5}$ \\
\object{AzV446$^{\mathrm{c}}$}	& O6.5V		& 39.7	& 4.00	& 5.25	& -8.40  & 0.8    & 0.10       & 1400	& 5		& 30	& 28.6$\pm6.5$  & 30.0$^{+2.3}_{-2.1}$  & 3.1$^{+0.5}_{-0.5}$ \\
\object{MPG 356}		& O6.5V		& 38.2	& 4.10	& 4.88	& -8.46  & 0.8    & ...        & 1400	& 5		& 20	& 18.0$\pm4.1$  & 23.1$^{+2.7}_{-2.5}$  & 2.2$^{+0.3}_{-0.2}$ \\
\object{AzV 429$^{\mathrm{c}}$}	& O7V		& 38.3	& 4.01	& 5.13	& -8.40  & 0.8    & 0.10       & 1300	& 10		& 120	& 26.3$\pm6.0$  & 26.5$^{+2.7}_{-2.7}$  & 3.3$^{+0.4}_{-0.3}$ \\
\object{MPG 523}		& O7Vz		& 38.7	& 4.25	& 4.80	& -9.22  & 0.8    & ...        & 1950	& 10		& 50	& 20.1$\pm4.5$  & 23.1$^{+2.6}_{-2.3}$  & 0.5$^{+0.5}_{-0.5}$ \\
\object{NGC346-046}	& O7Vn		& 39.0	& 4.22	& 4.81	& -9.22  & 0.8    & ...        & 1950	& 10		& 300	& 19.0$\pm4.3$  & 23.7$^{+2.6}_{-2.3}$  & 0.3$^{+0.7}_{-0.3}$ \\
\object{NGC346-031}	& O8Vz		& 37.2	& 4.00	& 4.95	& -9.22  & 0.8    & ...        & 1540	& 5		& 25	& 18.6$\pm4.2$  & 23.2$^{+2.5}_{-2.3}$  & 3.3$^{+0.6}_{-0.4}$ \\
\object{AzV 461}		& O8V		& 37.1	& 4.05	& 5.00	& -9.00  & 0.8    & ...        & 1540	& 10		& 200	& 23.8$\pm5.4$  & 23.8$^{+2.5}_{-2.3}$  & 3.6$^{+0.6}_{-0.4}$ \\
\object{MPG 299}		& O8Vn		& 36.3	& 4.25	& 4.64	& -8.52  & 0.8    & ...        & 1540	& 5		& 360	& 18.3$\pm4.1$  &19.5$^{+2.9}_{-2.5}$	& 1.5$^{+0.9}_{-0.7}$ \\
\object{MPG 487$^{\mathrm{a}}$}	& O8V		& 35.8	& 4.00	& 5.12	& -8.52  & 0.8    & ....       & 1540	& 2		& 20	& 31.7$\pm7.1$  & 25.7$^{+2.5}_{-2.0}$  & 4.3$^{+0.7}_{-0.5}$ \\
\object{AzV 267$^{\mathrm{c}}$}	& O8V		& 35.7	& 4.04	& 4.90	& -8.60  & 0.8    & 0.10       & 1250	& 10		& 220	& 21.9$\pm5.0$  & 21.5$^{+2.7}_{-1.9}$   & 4.2$^{+0.6}_{-0.5}$ \\
\object{AzV 468$^{\mathrm{c}}$}	& O8.5V		& 34.7	& 4.00	& 4.76	& -9.15  & 0.8    & ...        & 1540	& 5		& 50	& 15.7$\pm3.5$  & 19.3$^{+2.9}_{-2.2}$  & 4.3$^{+0.9}_{-0.6}$ \\
\object{AzV 148$^{\mathrm{a,c}}$}	& O8.5V		& 32.3	& 4.00	& 4.84	& -8.70  & 0.8    & ...        & 1540	& 12		& 60	& 25.3$\pm5.7$  & 19.0$^{+2.8}_{-2.4}$  & 6.2$^{+0.8}_{-0.6}$ \\
\object{MPG 682}		& O9V		& 34.8	& 4.10	& 4.89	& -9.05  & 0.8    & ...        & 1250	& 5		& 40	& 26.6$\pm6.0$  & 20.8$^{+1.5}_{-1.7}$  & 4.7$^{+0.6}_{-0.6}$ \\
\object{AzV 326$^{\mathrm{a}}$}	& O9V		& 32.4	& 4.03	& 4.81	& -9.15  & 0.8    & ...        & 1250	& 5		& 200	& 25.3$\pm5.7$  & 18.8$^{+2.8}_{-2.5}$  & 6.1$^{+0.8}_{-0.6}$ \\
\object{AzV 189$^{\mathrm{a}}$}	& O9V		& 32.3	& 4.02	& 4.81	& -9.22  & 0.8    & ...        & 1250	& 5		& 150	& 24.8$\pm5.6$  & 18.8$^{+2.8}_{-2.5}$  & 6.1$^{+0.8}_{-0.6}$ \\
\object{MPG 012}		& B0IV	         & 31.0	& 3.65	& 4.93	& -9.30  & 0.8    & ...        & 1250	& 5		& 60	& 17.1$\pm3.9$  & 21.0$^{+2.6}_{-2.1}$  & 6.5$^{+0.5}_{-0.3}$ \\
  \hline
\end{tabular}
 \label{tab2}
 \tablefoot{
\tablefoottext{a}{Stars whose photometry was interpreted as indicating binarity (cf. Sect \ref{bin_sect}) }   
\tablefoottext{b}{Stars showing radial velocity variations (cf. Sect \ref{bin_sect})} 
\tablefoottext{c}{Stars whose \logg\ = 4.0 was adopted, because we lacked of optical spectra.}  \\
\mdot\ are given in \msolyr. Evolutionnary masses (\mev) were derived from the tracks by \cite{brott11a} (see text for more comments).
Uncertainties on \teff\ are $\pm$ 1500~K,  $\pm $ 0.1 dex for \logg\ (except \object{AzV 446}, \object{AzV 429}, \object{AzV 267}, \object{AzV 468} and \object{AzV 148}, $\pm$ 0.2 dex), $\pm $ 0.1 dex on \logL. As for the wind quantities, 
an uncertainty of $\pm$ 0.04 dex was estimated for stars with P~Cygni profiles, while \vinf\ was measured within $\pm$ 100 \kms. 
For stars without \pcyg\ profiles, see Sect. \ref{param_sect} for the procedure we adopted.}
 \end{table*}
   
\section{Discussion}
\label{disc_sect}

\subsection{Binaries and binary fraction}
\label{bin_sect}
In Sect. \ref{param_sect}, we presented the method we used to constrain the stellar luminosity in the models. This method allowed us to reach very satisfactory fits of the global SEDs. Some remaining
discrepancy between models and the flux-calibrated COS spectra is nevertheless observed at wavelengths shorter than the \lyalpha\ line. This discrepancy can be alleviated, to some extent, by adding an extra absorption 
caused by interstellar \lyalpha. Fitting the wings of the observed \lyalpha\ component on the COS spectra allowed us to derive an estimate of the neutral hydrogen column density on each line of sight. 
An additional problem is the uncertain extinction law in the SMC at these short wavelengths.
 
For \object{AzV 326}, \object{AzV 189} and \object{AzV 148}, the luminosities that were derived from the photometry,  
\logL $\approx 5.15$, are 0.3 dex higher than expected for their spectral types O9V and O8.5V compared with stars of the same spectral type, either in the Galaxy \citep[e.g.][]{marcolino09} or
in the SMC \citep[e.g.][]{mokiem06}. 
When used in conjunction with the effective temperatures and the surface gravities obtained from the spectroscopic analysis, we found spectroscopic masses much 
larger than masses derived for stars of similar spectral type and luminosity class, for instance in our Galaxy \citep{martins05b, marcolino09}. 
We stress that we were unable to find any mention of photometric and spectroscopic variability that would cause the observed shift in luminosities. 
A rather straightforward explanation of these high luminosities is that these stars are actually binaries and that both the UV flux in the COS spectra and the optical/NIR photometry show contributions of the two
components of the binary system. 
It is indeed not surprising to find some binaries in our sample, multiplicity being a common, almost ubiquitous feature among massive stars \citep{garcia01, preibisch99, sana11}.

\begin{figure*}[tbp]
   \centering
\includegraphics[scale=0.6, angle=0]{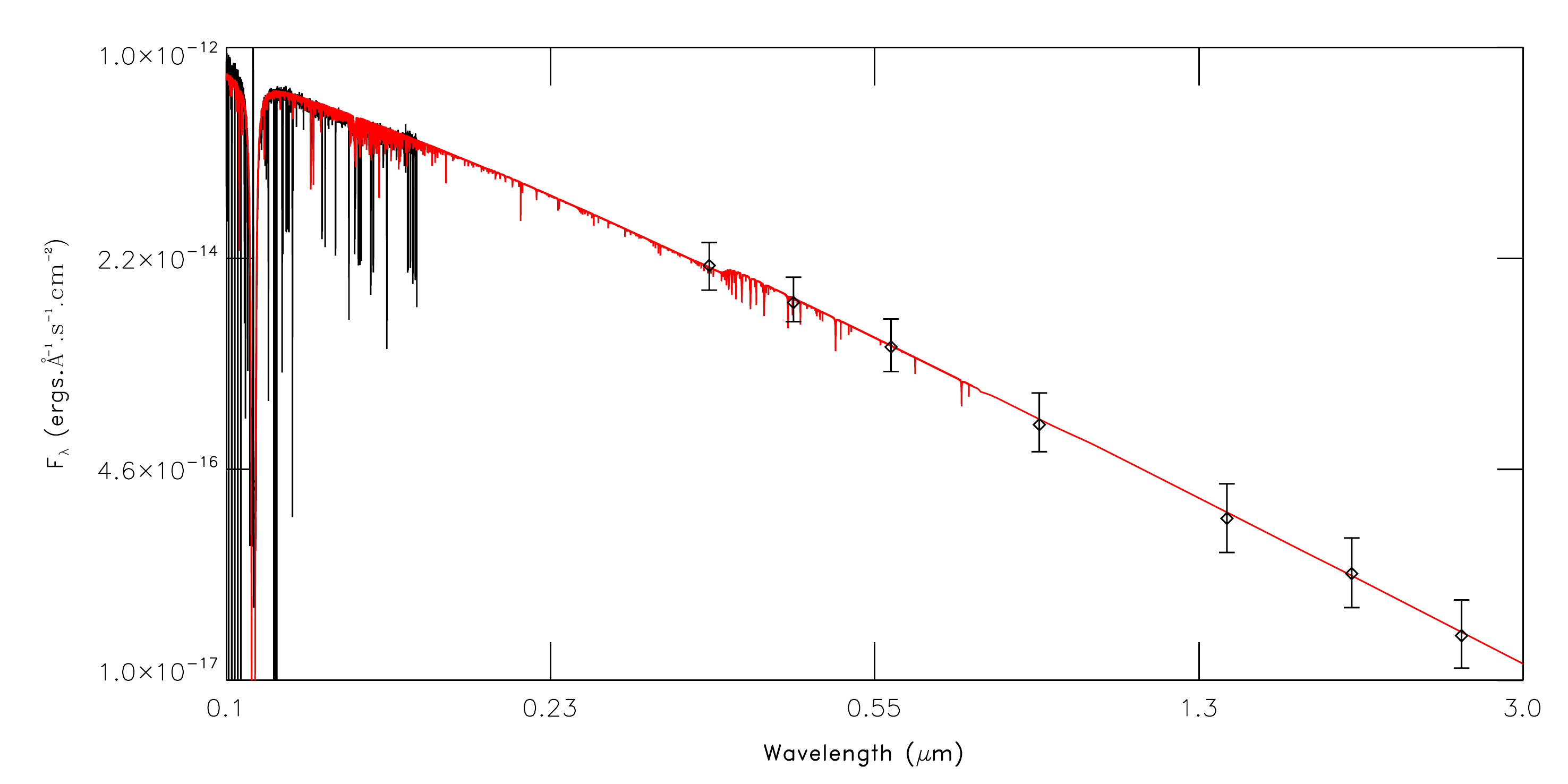}
    \vspace{-0.cm}
      \caption[12cm]{Spectral energy distribution of AzV 189, from FUV (\cosp\ spectrum) to NIR \cite[data from][]{bonanos10}. 
      The fit to the SED with a model with a luminosity derived with the procedure outlined in the text, to take into account the potential binarity of this system, is shown in red.
      We used a distance modulus DM = 18.9 and  E(B-V)$_{\rm Gal} = 0.037 $ and E(B-V)$_{\rm SMC} = 0.1$
      }
         \label{av189_bin}
   \end{figure*}

Furthermore, a detailed check of the Galactic O-type stars catalog by \cite{maiz04} revealed that a significant proportion of late-type binaries are found to be parts 
of couples of stars with the same spectral
type (e.g., O8V + O8V, and O9V + O9V), hence each component contributes about half the observed flux. If we assume a similar behavior in the SMC,
we could then propose the following correction. We divided the observed UV flux and fluxes from UBVIJHK photometry by two on one hand, and adopted 
a similar reduction of the luminosity in the model atmospheres. Since this simple procedure maintains an excellent fit of the overall SEDs (from the FUV to the NIR, see e.g. Fig. \ref{av189_bin}),
it reliably supports the assumption that both stars have similar effective temperatures. This approach resulted in a reduction
of the stellar radii, yielding spectroscopic masses that became then consistent with masses expected for late-type O stars.
Moreover, the luminosities are also compatible with those expected for O8V - O9V Galactic stars  \citep{martins05}.
The luminosities and masses listed in Table~\ref{tab2} are the values obtained with this approach. We emphasize that we do not argue that late-O stars in binary systems are systematically 
in equal mass systems, but simply that the three late O-type stars of our sample that show abnormal luminosities are more easily interpreted in terms of such a scenario. 
Instead, the recent work by \cite{sana12a} for instance shows that low-mass stars might actually dominate the companion mass distribution. Reaching a more general conclusion
on this specific question is beyond the scope of this paper and the sample it is based on. 

Three other stars, \object{MPG 324}, \object{MPG 368}, and \object{MPG 487} have been claimed to be (potential) binaries. \cite{evans06} reported small radial velocity shifts in some lines in the spectra of \object{MPG 324} (NGC 346-007) of about 20 - 30 \kms, 
suggesting it may be a single-lined binary. \cite{massey09} reported a velocity difference of 
50 \kms\ between the \hei\ and \heii\ lines of \object{MPG 368}, which they concluded is a strong indication of binarity. We note that a velocity difference of 70 \kms\ between the UV and optical spectra was also found by \cite{heap06} for this star.  At the COS spectral resolution ($\approx$ 0.1 \AA), any component with radial velocity difference above 50 \kms\ should be detectable, but we found no conspicuous evidence for SB2 binaries in our FUV spectra.

\cite{bouret03} argued that the weak UV lines of \object{MPG 487} might be diluted by one (or several) additional stars in the slit.
From the present modeling, this conclusion was somewhat re-inforced by the failure of fitting the observational SED, from the FUV to the NIR, without down-scaling the optical-NIR photometric fluxes, by 10\% to 20\% over the parameter space we explored.The best fit we obtained (see Table \ref{tab2} for the parameters) implied a reduction 
of 10\% of these fluxes, but we did not need to change the UV flux accordingly. This may be the result of a companion with a significantly later spectral type and a contribution to the total luminosity of only about 
10\%\ of that measured for \object{MPG 487}. 
Because it is cooler, it would mostly contribute to the optical and NIR bands and the lower luminosity  would make the change in the photometry within the observed range. We argue therefore that this suggests that
\object{MPG 487} may be in a binary system. Of course, effects on the luminosity or the SED may result from the presence of several objects in the instrumental aperture, and cannot be unambiguously tied to physical binaries as
for systems identified through radial velocity variations. 
Furthermore, the fact that we are able to match luminosities and SEDs of the other stars with single-star models, does not exclude the possibility that 
these stars have cooler and fainter companions that remain undetected in the SED.
Our study is obviously not based on a complete and homogeneous population of O-type stars in the SMC, because it mixes field stars and stars in the open
cluster \object{NGC~346}. It is biased towards the detection of relatively massive companions, that would contribute most to the total observed flux 
(\object{AzV 326}, \object{AzV 189}, \object{AzV 148}, \object{MPG 487}) or induce radial velocity variations of several tens of \kms\ on early-type objects (\object{MPG 324} and \object{MPG 368}). 
To summarize, although we did not find additional hints for binaries from the modeling analysis, it is not unlikely that there are more multiple objects in our sample than the six stars discussed above.
Our estimate of the binary fraction should therefore be viewed as a lower limit of the actual fraction of multiple systems in the SMC that includes O-type stars. 

With 6 binaries in a sample of 23 stars, the binary fraction would be 26\%. This is consistent with
recent results on the fraction of binaries among massive stars in the Magellanic Clouds. 
The first VLT-FLAMES survey of massive stars reported in \cite{evans06} found a lower limit to the binary fraction for O stars in \object{NGC 346} of 21\% (the total fraction for O-type and 
early-B-type stars being 26\%), while
the investigation of multiplicity in LMC massive stars by \cite{sana12a} found that the observed binary fraction is 35\% but the intrinsic fraction is as high as 51\%, after correcting for observational effects.
 
\subsection{H-R diagram, evolutionary status, and ages}
\label{hr_sect}
\begin{figure*}[tbp]
   \centering
   \includegraphics[width=16.2cm]{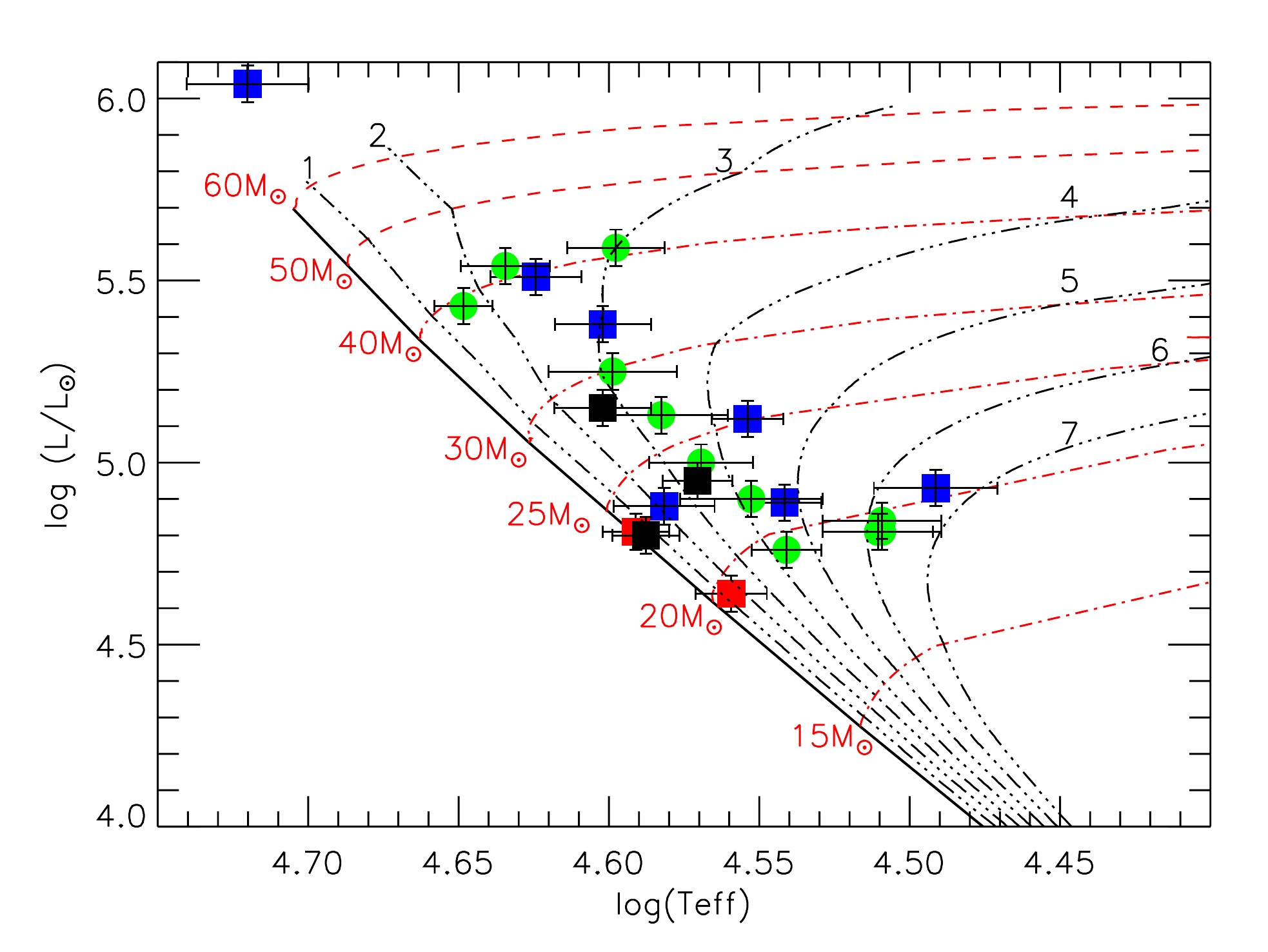}
      \caption{H-R diagram of the sample stars. Evolutionary tracks and isochrones are taken from \cite{brott11a} for an initial rotation rate of 180 \kms. 
      Evolutionary tracks are annotated in solar masses, and isochrones in Myr from 1 to7 Myr. Green circles indicate 
      field stars, while squares indicate stars in \object{NGC 346}. Blue squares show normal luminosity class V stars, while black squares show Vz-class objects (cf. Sect. \ref{sect_vz}).
      Red squares show the position of the two stars that rotate faster than 300 \kms.
              }
         \label{HR_fig}
   \end{figure*}

Using the effective temperature and luminosity from the spectroscopic analysis, we built the H-R diagram shown in Fig. \ref{HR_fig}. 
For this diagram, we also used evolutionary tracks and isochrones from \cite{brott11a} for initial rotational velocities $\approx$ 180 \kms.
The following considerations lead us to select this later value: First, the average \vsini\ of our sample is $\approx$ 115 \kms\ (for a standard
deviation of 1 \kms). We recall that as discussed in Sect. \ref{obs_sect}, part of our sample (stars with \stis\ spectra) was selected 
from slow rotators and therefore the average \vsini\ should be regarded as biased towards lower values. 
Second, from their subsample of (21) unevolved OB stars in the SMC, \cite{mokiem06} concluded that the average \vrot\ of their sample
was 150--180 \kms. These two points made us choose use the \vrot\ of the \cite{brott11a} models which is most consistent with the mean velocity from \cite{mokiem06},
that is, \vrot\  = 180 \kms, although it must be kept in mind that this \vrot\ for evolutionary models is the initial rotational velocity,
without projection effects.
 The error bars on the position of stars in the H-R diagram reflect the uncertainties on the effective temperature and
luminosities from the analysis above.
Fig.~\ref{HR_fig} shows that the stars have ages ranging from younger than 1~Myr to 7~Myr. 
Their evolutionary masses range from 19~\msol\ to 42~\msol\, with the exception of \object{MPG~355} which is apparently more massive than 60~\msol, consistent with its early spectral
type and corresponding high effective temperature and luminosity. 
Our sample consists of field stars as well as stars belonging to the giant H~II region and to the open cluster NGC~346. Overall, the age span of the two groups overlap, but
we note that no stars younger than 1.5~Myr are found in the field sample, suggesting a lack of recent massive star formation in the SMC field.

Of the twelve stars in \object{NGC~346}, two are on or very close to the ZAMS.
A third, rapidly rotating (see below) object is found between the isochrones one and  two~Myr (Table~\ref{tab2}). A second group of five~stars are between two and four~Myr old,  while two objects are younger than 5~Myr. 
The age spread is very consistent with the previous study of \cite{mokiem06}.  Finally, the coolest star of our sample, namely \object{MPG 012}, is older than 6~Myr. 
Our analysis, which includes the \flames\ and \uves\ spectra of \object{MPG~012}, indicates that this star is the oldest of the sample,  significantly older than all analyzed stars in \object{NGC 346}
\citep[see also][]{bouret03}.
The evolved status of \object{MPG~012}  is also supported by its lower surface gravity (as also suggested by the luminosity class IV ascribed to this star) and a high enhancement in surface nitrogen. 
These findings support  the argument of \cite{walborn00} that \object{MPG~012} is most likely not a coeval member of the cluster, which the based on its spatial location outside the nebula, 
a somewhat discrepant radial velocity, and the derived stellar parameters. 

MPG 355 is the hottest and most luminous star of our sample. The optical spectra are devoid of \hei\ lines and we had to rely on alternative diagnostics to constrain the effective temperature
\citep[but see][for a mention to a different spectrum, that potentially shows a very weak \hei\ \lb4471 line]{massey09}. Our parameters for this star agree very well
with those from the recent analysis of O-stars of the earliest spectral types by \cite{rivero12}.
\cite{bouret03} argued that the position of \object{MPG~355} close to the ZAMS could indicate chemically homogeneous evolution due to rotation close to break-up velocity \citep{maeder87}.
Testing different models with high initial masses (50 and 60 \msol) and high initial rotation rates (above 400 \kms), we found that 
the isochrone 3 Myr for models with initial \vrot\ = 400 \kms\ comes closest to the location of MPG 355 in the H-R diagram (see Fig. \ref{HR_400_fig}). 
Still, \object{MPG 355} is above this isochrone and above the 60 \msol, \vrot\ = 400 \kms\ evolutionary tracks, and the main conclusion we can 
reach at this point is that \object{MPG 355} is a very massive object, probably more massive than 70 \msol, but its actual evolutionary status is more
uncertain. Dedicated evolutionary models are needed to reach a better quantitative conclusion regarding its status. We note that a chemically homogeneous evolution was 
concluded to be unlikely in \cite{rivero12} because
 of the low \vsini\ measured for this star. The analysis of the surface abundances support a non-standard evolutionary status for this object, however (see Sect. \ref{ab_sect}).

Overall, there is a trend for the most massive stars in the sample to be younger than less massive stars. 
However, it is not clear how much of this finding is true evidence for a real age sequence with mass or instead is due to a selection effect of the sample, because we selected stars to be primarily dwarfs. 
Indeed, more massive stars have shorter lifetimes and are more quickly identified as evolved, and hence not as dwarfs. In other words, above a certain mass, any massive star identified as 
a dwarf must be somewhat younger than less massive counterparts. For instance, an O star more massive than 40 \msol\ has already moved a long way from the ZAMS 
at an age of 3 Myr (see Fig. \ref{HR_fig}) and any star more massive would appear as a giant or even supergiant, hence it would not have been selected in the sample. 
It could also be a bias from the spectroscopic analysis, as observational results found the same trend for Galactic O-type dwarfs in clusters \citep{martins12a} or in Galactic supergiants (Bouret et al. 2012). 

If true, however, in the specific case of \object{NGC~346} this trend would suggest a scenario in which the most massive stars of the cluster formed last, and after the had formed, they would quench subsequent star formation (as suggested originally in Heap et al. 2006).
This trend can also be seen in the spectroscopic study of \object{NGC~346} by \cite{mokiem06} (see their Fig.~12), although not as clearly. 
When comparing the stars in common with our analysis and the work by \cite{mokiem06},
we note that they obtained effective temperatures systematically higher than ours (1700~K on average). This systematic difference has two important
consequences: {\it i}) their sample is globally shifted toward the ZAMS compared to ours, {\it ii}) most stars lie on a relatively narrow band,
especially in the lower part of the H-R diagram, which is much less sensitive to age because of the smaller separation between isochrones.

An explanation for the observed trend might be that the tracks adopted 
to derive an age for higher and lower masses stars should not have the same rotation rate. If for some reason the initial rotation rate \vrot\ of more massive objects 
turned out to be higher than for lower-mass stars, tracks corresponding to this higher \vrot\ should be used preferentially.     
It is now a strong conclusion from stellar evolution models with rotation that the main-sequence lifetime increases as rotation is faster \citep[e.g.][]{maeder00, brott11a}. 
Age determinations based on non-rotating models can be biased toward younger age, which in contrast means that 
for models with higher rotation rates, older ages would be derived. 
On the other hand, actual differences between age estimates from non-rotating and rotating models with moderate rotation should remain smaller than the error bars resulting from 
the uncertainties on temperature and luminosities for the relatively unevolved objects discussed here.  
In the specific case of \object{NGC~346}, \cite{mokiem06} investigated the impact of using non-rotating versus rotating Geneva models (with initial \vrot\ = 300 \kms\ from \cite{meynet05})
to derive ages of the cluster stars. For stars on the main-sequence, the authors estimated that the age difference may increase up
to 0.5~Myr (see their Fig.~13 and 14). We concur with this conclusion using models without rotation and with initial rotational 
velocities of $\approx$ 200 \kms\ from \cite{brott11a}.  

\subsubsection{Fast rotation}
\label{sect_fast}
The fastest-rotating two stars of the sample, \object{NGC 346-046} and \object{MPG 299}, lie close to or on the ZAMS in the lower part of the H-R diagram in Fig.~\ref{HR_fig}.  
The derived evolutionary masses are 23.7$^{+1.3}_{-1.0}$ \msol\ and 19.5 $^{+1.0}_{-1.0}$ \msol\ and ages of 0.3 Myr and 1.5 Myr, respectively, for models with \vrot\ $\approx$ 180 \kms. 
It is well-known since the work of \cite{vonzeipel24} \citep[see also][]{maeder99} that a fast-rotating star has stronger radiative fluxes at the pole than at the equator. 
Therefore, the observed luminosity, effective temperature, and also the effective gravity of a star depend on its initial angular velocity as well as on the inclination angle. 
In turn, its location on the H-R diagram depends on the same quantities, hence
its age and mass as determined by fitting evolutionary tracks or isochrones. This holds especially for stars that are located near
the zero-age main sequence. A young, unevolved, slowly rotating star may have the same effective temperature and luminosity
as a less massive, fast-rotating, evolved star. 

\begin{figure}[tbp]
   \centering
   \includegraphics[width=9.2cm]{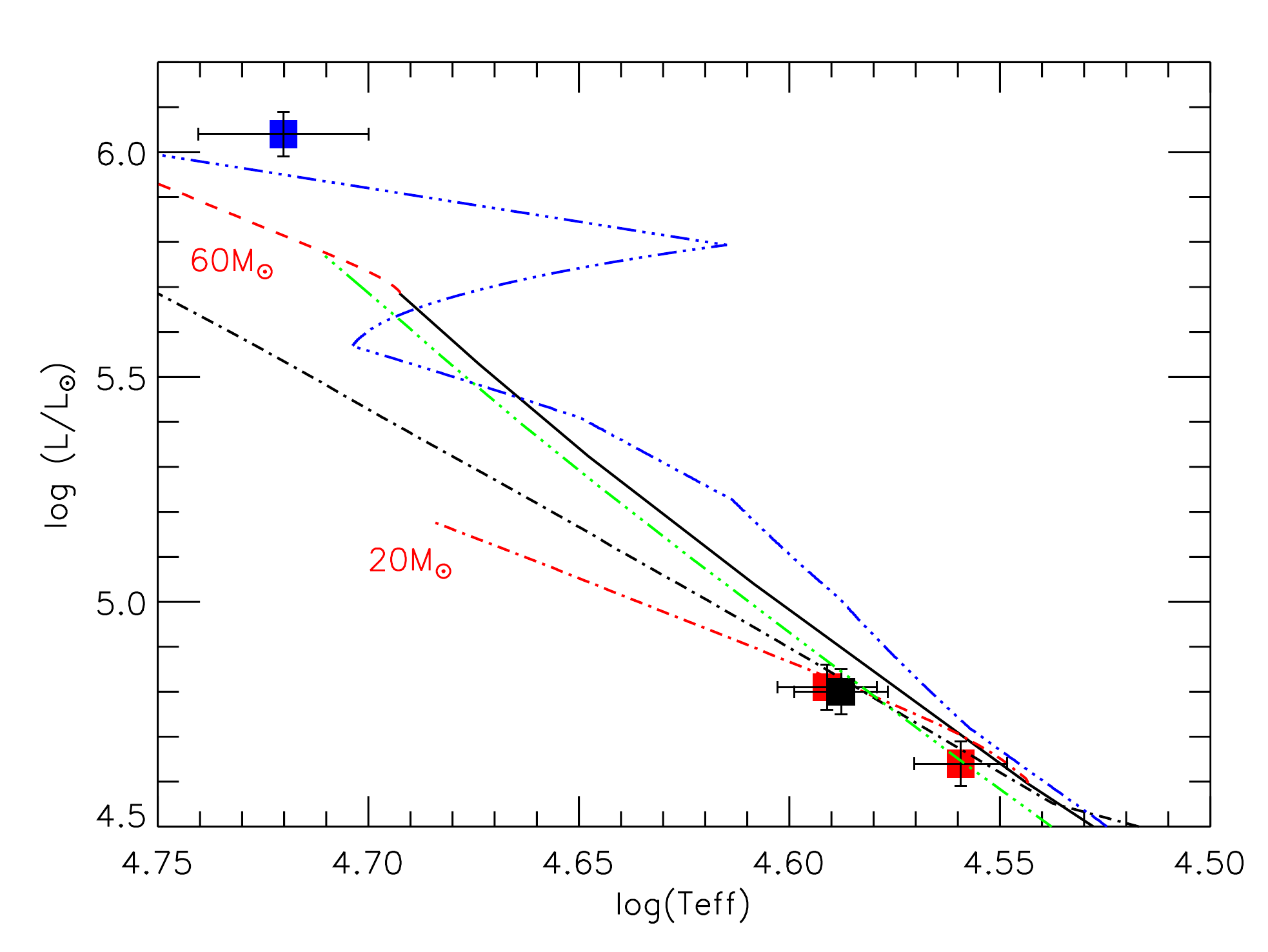}
      \caption{H-R diagram showing evolutionary tracks (red dashed line) with 60 \msol\ and 20 \msol\ for an initial rotation rate of $\approx$ 400 \kms\ \citep{brott11a}. 
            The ZAMS is indicated by the black full line and the isochrone 5 Myr is indicated by the black, dot-dashed line. The green and blue line show
      the isochrones 1 Myr and 3 Myr, respectively. 
      Blue filled square indicates the position of \object{MPG 355}, while the red squares show the positions of the two stars that rotate faster than 300 \kms. The black square  
      indicates the position of \object{MPG 523}.
              }
         \label{HR_400_fig}
   \end{figure}

\cite{brott11a} showed that for fast-rotating stars (with initial \vrot\ = 400 \kms\ or higher),
the location on the H-R diagram stays closer to the ZAMS than for slower rotators of same age. For this extreme rotation, 
mixing is so efficient that the whole internal structure and composition of a star is modified and may lead to homogeneous evolution in which 
the star becomes brighter and hotter and evolves up- and blueward directly in the H-R diagram.
The onset of a blueward bifurcation of the evolutionary tracks caused by chemically homogeneous evolution 
arises around 18 \msol\ at SMC metallicity for stars with \vrot\  $\approx$ 550 \kms\ or higher. 
This strongly suggests that evolutionary tracks with such high rotations rates are better suited to estimating the evolutionary status of \object{NGC 346-046} and \object{MPG 299}.  
In Fig. \ref{HR_400_fig}, we show that the location of \object{NGC 346-046} and \object{MPG 299} on the H-R diagram can be accounted for by various isochrones, with combinations of ages and rotation rates
such as (1 Myr,  \vrot\  $\approx$ 400 \kms) or (5 Myr, \vrot\  $\approx$ 550 \kms). In other words, using isochrones with lower initial \vrot\ leads to a significant underestimation of the age of \object{NGC 346-046} 
and \object{MPG 299}.

It is also argued that rotation induces an enhancement of mass loss, which should affect the ultimate evolution of a star \citep{owocki96, maeder99, maeder00}.
However, we calculated that a contrast of 10\% at most should be observed for the mass-loss rates at the equator and at the pole \cite[see][for a discussion]{maeder00}, 
for the radii, masses, and \vsini\ we derive for \object{NGC 346-046} and \object{MPG 299}, indicating that the actual increase of mass loss by rotation is very marginal for these two stars.  
The net effect of this an increase on the observed spectra is also expected to be negligible, which is also supported by the very good quality of the fits we obtained for the UV and optical spectra
of both stars, indicating no significant deviation from spherical symmetry of their winds. 
The mass-loss rates we derive for \object{NGC 346-046} and \object{MPG 299} are similar to those of stars with same spectral type and much lower \vsini,  which belong
to the range of spectral types where the weak-wind phenomenon is at play. 
In any case, a change by a factor of a few is probably not enough to be detected even in the UV spectra of these stars.
This result implies that these stars lost only a marginal fraction of their initial angular momentum during the course of their evolution. Therefore, 
their rotation rates should not have decreased much and the present measured \vsini\ should reflect the initial values well. 

\cite{mokiem06} found that four SMC O dwarf stars close to or on the ZAMS had an helium-rich surface,
two of them being the fast rotators discussed in this Section. Since helium enrichment is expected for fast rotators, even early in the stellar evolution \citep{meynet00}, \cite{mokiem06} concluded
that the two stars with high helium surface abundance ($Y_{He} = 0.16$) and low \vsini\ were actually fast rotators seen pole-on. The four stars are included in our own work
(\object{MPG 113}, \object{NGC 346-031}, \object{NGC 346-046}, and \object{MPG 299}). Using the surface helium abundance as a measure of stellar age for a given mass for of 
stars with very high initial \vrot, Mokiem et~al.  furthermore derived lower limits on the ages of these stars that were significantly higher than for non-rotating (or slowly-rotating) stars.  
The authors concluded that these stars probably follow a chemically homogeneous evolution.

We used the same \flames\ spectra and supplemented them with \uves\ spectra for the first three stars. However, we found
no evidence of a helium abundance above the initial value, except for \object{MPG~299}. In this case only, we agree with \cite{mokiem06} that an enrichment up to
$Y_{He} = 0.13$ provides a better match to the optical spectrum. Given the high \vsini\ of this star, this is indeed suggestive of chemically homogeneous evolution.
The conclusion of Mokiem and collaborators that \object{MPG~113} and \object{NGC~346-031} are fast rotators cannot be supported from our analysis. 

\subsubsection{The luminosity class Vz}
\label{sect_vz}
Three stars in our sample, \object{MPG 113}, \object{MPG 523}, and \object{NGC 346-031}, are classified as Vz stars \citep{evans06, walborn00}. This luminosity class was 
introduced by \cite{parker92} and \cite{walborn97} for stars whose properties are expected to be characteristic of extreme youth. 
The defining spectroscopic criterion for this class is that at every spectral type, \heii\ \lb4686 absorption is stronger than any other \he\ line. More specifically, this would be the case of  \heii\ \lb4541 at early-O types, but at type O7, the latter  is equal to \hei\ \lb4471, so that \heii\ \lb4686 is stronger than both in a Vz spectrum. At later-O types, however, \heii\ \lb4541 weakens more rapidly with advancing type than \heii\ \lb4686 in normal class-V spectra, such that, a late-O Vz spectrum must have \heii\ \lb4686 stronger than \hei\ \lb4471.
As stated in an illuminating discussion by \cite{walborn09}, this a behavior very likely results from the fact that for objects in the Vz class, a deficit of emission filling in \heii \lb4686 is observed, in contrast to normal class-V stars. This deficit results  from an ``inverse'' Of effect \citep[see e.g.][for a thorough definition]{walborn01}, thus implying that O Vz stars should be subluminous and less evolved. They are therefore expected to be close to or on the ZAMS. 

Of these three stars, only \object{MPG 523} is found to be located on the ZAMS within the error bars; the other two stars are close to the 3~Myr isochrone (see also Table~\ref{tab2}).
In the previous study of \cite{mokiem06}, \object{MPG 523} was clearly to the left of the ZAMS because of a derived \teff\ higher by 3000\,K and only slightly higher \logL\ (by 0.07 dex). 
The strong nebular emission present in the \flames\ spectrum of this star can be interpreted as another hint at the young age of this star. On the other hand, 
Fig. \ref{HR_400_fig} shows that the location of \object{MPG 523} in the H-R diagram is actually consistent with that of a 20 \msol\ star, rotating near 400 \kms\ that is either 1 Myr or 5 Myr old. 
This interpretation is also confirmed by the spectroscopic analysis of \object{MPG 523}, which indicates surface abundances in stark contradiction with the expectation for a young, slowly rotating star. 
From its abundance patterns, \object{MPG 523} would be more compatible with being a more evolved, fast-rotating object. 
\begin{figure}[tbp]
   \centering
   \includegraphics[width=9.2cm]{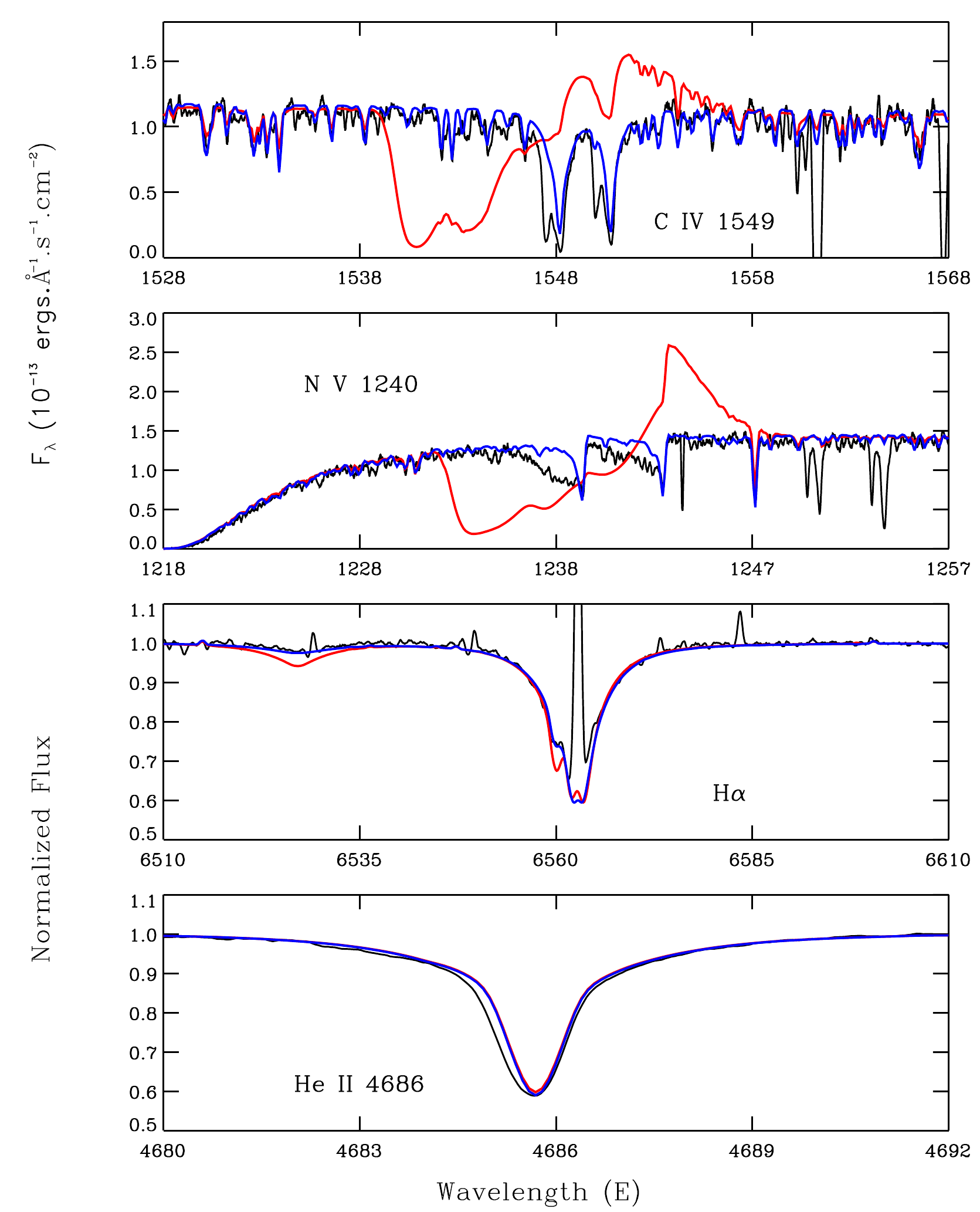}
      \caption{Wind lines of \object{NGC 346-031}. A comparison of our best-fit model (blue line) with a model computed using the parameters provided in 
      \cite{mokiem06} (red line). 
    The mass-loss rates differ by almost two orders of magnitude. The blue side of \ion{N}{V}\,$\lambda$1240 is affected
  by interstellar Ly\,$\alpha$ absorption. We also show the H$\alpha$ line, which proves to be insensitive to mass-loss over a wide range of values.
              }
         \label{Fig_ELS031}
   \end{figure}

The properties of \object{MPG~113} are also very non-standard. 
First, we did not find this star to be close to or on the ZAMS, in contradiction with \cite{mokiem06}. The observed shift in the H-R diagram
is again caused by an effective temperature about 3000\,K higher than our value, while luminosities agree to within 0.05 dex.
We recall that the helium surface abundance of \object{MPG 113} corresponds to that of a normal main-sequence O6V star, with no indication for enrichment ($Y_{He} = 0.16$ is firmly ruled out). 
Second, the carbon and nitrogen contents of \object{MPG 113} are higher than the values measured in H~II regions of the SMC \citep{dufour98},  
resulting in an N/C ratio that is significantly higher than the standard ratio measured in H~II regions \citep[in agreement with][]{bouret03}.  
This does not support the claim by \cite{walborn00} that the N/C abundance ratio in \object{MPG~113} is atypically low to explain its spectral classification
as an OC6 star.
We also note that another ``normal'' O6.5 V star has even higher carbon abundance but is also 
richer in nitrogen, yielding the same N/C ratio, although it is classified as an OC6 star. It seems more likely that this classification
is more complex than a simple abundance criterion and involves several physical parameters such as the effective temperature and the surface gravity. 
 
For the third star, \object{NGC 346-031}, although we found it again to be cooler than did \cite{mokiem06} for very close luminosities (within 0.04 dex), we found that this star does not lie especially close to the ZAMS
and that it is older than 2.5 Myr. \cite{mokiem06} argued that the Vz classification, from a strong \heii\ \lb4686, might come from the high \he\ abundance they derived ($Y_{He} = 0.16$). 
We
needed no such overabundance in our own models to obtain very good fit for this star (see Fig. \ref{Fig_ELS031}). Indeed, using $Y_{He}$ as high as in \cite{mokiem06} degrades the quality of the fit
significantly. On the other hand, we derived a mass-loss rate that is two orders of magnitude lower than that of Mokiem and collaborators, from the modeling of the UV spectrum (Fig. \ref{Fig_ELS031}). The contribution of such a small
\mdot\ to a potential wind-emission that would come and fill in the \heii\ \lb4686 line is drastically reduced.  
We recall that for stars without conspicuous PCygni profiles, we used homogeneous wind models, and that the \mdot\ we quoted are upper limits above which models start producing wind signatures 
not seen in the data. 
We note that the three Vz stars have much smaller \mdot\ in our UV based study than do those from \cite{mokiem06}, which results in all cases from the extreme weakness (or absence) of the observed 
UV stellar-wind profiles. In the temperature range of interest for these stars, this might be the reason why the  \heii\ \lb4686 is observed to be stronger than in normal dwarf stars that experience stronger
wind emission. For later-type stars, a lower temperature would act such that \hei\ lines remain stronger than \heii\ \lb4686 even in cases where weak-wind emission is involved. 

The defining spectroscopic criterion for the Vz class seems to fail in establishing a one-to-one correspondence between an object and a young evolutionary status \citep[but see][for recent findings from VFTS
survey]{walborn13} as inferred from the location in the H-R diagram. The bias toward lower \teff\ (than optical-based \teff) to reproduce the observed ratio of UV carbon lines  
is an obvious cause for the shift of this location away from the  ZAMS. The same trend was already noticed, with almost the same amplitude, for UV-based \teff\ derived with \tlusty\ in \cite{heap06}.  It is therefore
unlikely that this is related to the physical assumptions we made (e.g., including a stellar wind, the adopted micro-turbulence, the atomic data or other issues related to the sphericity of the atmosphere as considered
in \cmfgen). The same trend was also found by \cite{garcia04} from a UV-based analysis, although in a different context. In any case, our modeling suggests that 
more criteria than the strength of \he\ lines should be taken into account to infer the true evolutionary status of a star, including the surface abundance patterns \citep[see also Sect. 4.][for a discussion]{brott11a}.
   
\subsection{Chemical evolution}
\label{ab_sect}
While rotation was claimed to be able to account for the early chemical enrichment of the stellar surface \citep[e.g.][]{maeder00}, 
several observational studies have undermined this conclusion. Investigating the link between surface abundance patterns and rotation,
either in the Galactic context or in the Magellanic Clouds, \cite{hunter07, hunter08, hunter09} concluded that up to 40\% of their sample
presented nitrogen surface abundances that could not be accounted for by rotational mixing in stellar evolution models.
The stars challenging the link between surface abundances and rotation fall into two categories: first, stars with strong
enrichment but slow rotation, and second, fast rotators with a low (if any) surface enrichment. These studies focused on nitrogen
surface abundances of B-type stars up to the supergiants stage. In the core-hydrogen burning phase, slowly rotating nitrogen-rich stars are found for the 
lower metallicity regime of the LMC and SMC \citep{hunter09}, while a relative deficit of such stars is observed in the Galactic case. 

\begin{table*}[htbp]
\centering
\caption{He, C, N, and O abundances by number.}
\begin{tabular}{lcccccc}
\hline
\hline
Star 		& $Y_{He} $	& $\epsilon$(C)	& $\epsilon$(N)	       & $\epsilon$(O)	& N/C	& N/O	\\
\hline
\object{MPG 355}		& 0.09  	& 7.39$\pm0.15$  			& 8.11$\pm0.16$  	         & 8.00$\pm0.18$     &	5.25	& 1.29	\\	     
\object{AzV177	}		& 0.09  	& 7.17$\pm0.16$  			& 8.01$\pm0.22$  		& 7.90$\pm0.30$     &	6.92	& 1.29	       \\ 
\object{AzV 388}		& 0.09  	& 7.39$\pm0.15$  			& 7.96$\pm0.20$  		& 8.05$\pm0.21$     &	3.71	& 0.81		\\
\object{MPG 324}		& 0.09  	& 7.00$\pm0.15$  			& 7.38$\pm0.20$  		& 8.00$\pm0.24$     &	2.40	& 0.24	       \\
\object{MPG 368}		& 0.09  	& 7.20$\pm0.14$  			& 7.70$\pm0.20$ 		& 8.05$\pm0.18$     &	3.16	& 0.45		\\    
\object{MPG 113}		& 0.09  	& 7.52$\pm0.16$  			& 7.40$\pm0.34$  		& 8.30$\pm0.12$     &	0.76	& 0.13		\\
\object{AzV 243}		& 0.09  	& 7.20$\pm0.20$  			& 7.50$\pm0.34$  		& 7.90$\pm0.12$     &	2.00	& 0.40		\\
\object{AzV446	} 		& 0.09  	& 7.20$\pm0.22$  			& 7.48$\pm0.32$  		& 7.98$\pm0.14$     &	2.04	& 0.32		\\
\object{MPG 356}		& 0.09  	& 7.69$\pm0.17$  			& 7.38$\pm0.30$  		& 7.98$\pm0.08$     &	0.49	& 0.25		\\
\object{AzV 429}		& 0.09  	& 7.69$\pm0.17$  			& 7.08$\pm0.29$  		& 7.98$\pm0.22$     &	0.25	& 0.13	       \\
\object{MPG 523}		& 0.09  	& 6.38$\pm0.15$  			& 7.08$\pm0.35$  		& 7.40$\pm0.10$     &	5.00	& 0.48		\\
\object{NGC346-046}		& 0.09  	& 6.68$\pm0.14$  			& 7.56$\pm0.45$  		& 7.98$\pm0.37$     &	7.59	& 0.38	       \\
\object{NGC346-031}		& 0.09  	& 7.69$\pm0.10$  			& 7.08$\pm0.35$  		& 7.98$\pm0.08$     &	0.25	& 0.13		\\
\object{AzV 461}		& 0.09  	& 7.38$\pm0.15$			& 7.38$\pm0.26$  		& 7.98$\pm0.35$     &	1.00	& 0.25	       \\
\object{MPG 299}		& 0.13  	& 7.69$\pm0.15$  			& 7.56$\pm0.44$  		& 8.10$\pm0.33$     &	0.74	& 0.29	       \\
\object{MPG 487}		& 0.15  	& 7.39$\pm0.20$  			& 6.60$\pm0.26$  		& 7.98$\pm0.08$     &	0.16	& 0.04		\\
\object{AzV 267}		& 0.09  	& 7.30$\pm0.14$  			& 7.48$\pm0.30$  		& 7.98$\pm0.30$     &	1.51	& 0.31		\\
\object{AzV 468}		& 0.09  	& 7.30$\pm0.17$			& 6.30$\pm0.25$  		& 7.60$\pm0.14$     &	0.10	& 0.05		\\
\object{AzV 148}		& 0.09  	& 7.69$\pm0.15$ 			& 7.26$\pm0.30$  		& 7.98$\pm0.10$     &	0.37	& 0.19		\\
\object{MPG 682}		& 0.09  	& 7.39$\pm0.14$  			& 6.78$\pm0.25$  		& 7.98$\pm0.14$     &	0.25	& 0.06		\\
\object{AzV 326}		& 0.09  	& 7.08$\pm0.14$  			& 7.38$\pm0.30$  		& 7.98$\pm0.24$     &	2.00	& 0.25	       \\
\object{AzV 189}		& 0.09  	& 7.69$\pm0.15$  			& 7.38$\pm0.30$  		& 7.98$\pm0.21$     &	0.49	& 0.25		\\
\object{MPG012}		& 0.09  	& 7.34$\pm0.16$  			& 7.70$\pm0.21$  		& 8.05$\pm0.16$     &	2.29	& 0.45		\\
\hline
SMC(HII) 	& 0.09		& 7.37							& 6.50				& 7.98		        & 	0.13	 & 0.03	 \\
\hline
\end{tabular}
 \tablefoot{
 For comparison, we list the chemical composition determined by \cite{kurt98} for HII regions in the SMC as used in the evolutionary models
 by \cite{brott11a}. 
By convention log $\epsilon$(X) = 12 + log [X/H].}
\label{tab3}
\end{table*}

This strongly suggests that the global chemical evolution of massive stars, as probed from their surface abundances, is more complex than assumed and
is an intricate combination of several physical parameters, including mass, metallicity, age (evolutionary status) and rotation \citep{maeder09}. 
\cite{martins12a} showed that Galactic O stars in \object{NGC 2244} and \object{Mon~OB1} populated the entire nitrogen abundance -- \vsini\ diagram,
but nicely followed the predictions of evolutionary models in the nitrogen abundance -- luminosity diagram, confirming the arguments of \cite{maeder09}.
Another obvious
complication is binarity. Using population synthesis models, \cite{brott11b} concluded that the observed distribution of \cite{hunter08} could not be 
reproduced by single-star population synthesis. However, they also concluded that the nitrogen-rich slowly rotating stars are unlikely to result from close
binary system evolution, with mass transfer and tidal spin-down. A fraction of this population would end up overpopulating the group of nitrogen-rich
fast rotators, which is not observed. 

\subsubsection{Helium abundances}
\label{He_sect}
For massive stars in the hydrogen-burning phase of their evolution, helium and nitrogen  are produced at the expense of carbon first, as the
star progresses to CN-equilibrium. The relative increase of abundances is expected to be weaker for helium than for nitrogen, however, because the former is already 
abundant in the atmosphere while the latter is not. 
This is confirmed by stellar evolution models with rotation \citep{brott11a}, which predict that helium should
remain fairly constant during the main-sequence phase and that helium transport to the surface is only substantial for the more massive
and fast-rotating stars where mixing processes are more efficient. Quantitatively, the alteration of the helium surface 
abundance remains below 0.2 dex during the H-burning phase for stars with M$_{\star}$ $\leq$ 25 \msol. 
Only fast-rotating models (\vrot\ > 330 \kms) for stars 30 \msol\ and above present modification of the helium surface content on the main-sequence larger than 0.25 dex. 
Our spectroscopic analysis only found the need for helium enrichment in two stars (see Table \ref{tab3}), and for these two stars the enhancement was higher than expected.
Of these two stars, one is indeed a confirmed fast rotator (\object{MPG 299}),  the other a suspected binary (\object{MPG 487}). Overall, the qualitative
and quantitative agreement of our measurements with theoretical predictions is therefore excellent. 

Since the enhancement of the surface helium content is low, a comparison of the predicted versus observed values is far from conclusive for such an early evolutionary phase. 
In principle, a better diagnostic is the N/He ratio, which quickly evolves from the initial value to a value corresponding to the \mbox{CNO} equilibrium.
Stars for which the equilibrium has not been reached will present intermediate values of N/He. Here again, however, the surface enrichment in helium is such that the N/He ration mererly witnesses
the global surface enrichment in nitrogen, which we discuss in more detail in the next sections. 

\subsubsection{Nitrogen abundances and comparisons with B stars}
\begin{figure}[tbp]
   \centering
   \includegraphics[width=9.2cm]{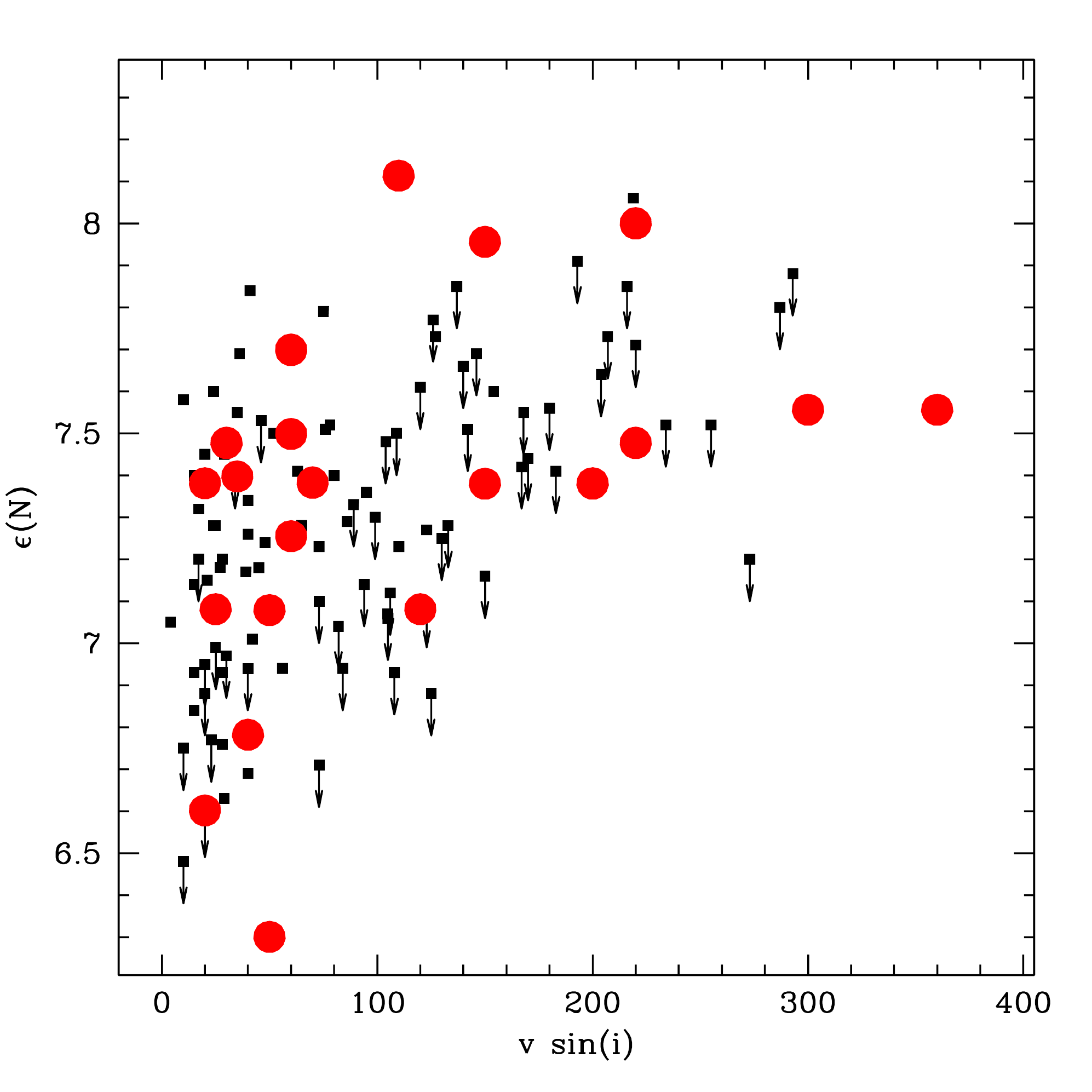}
      \caption{Nitrogen abundance (in units of 12+log(N/H)) versus projected rotational velocity. The black squares indicate the B stars of \cite{hunter09}. 
      The red circles show the O stars analyzed in this study.
              }
         \label{hunterOB}
   \end{figure}

Our results on the surface abundances of O stars in the SMC can be viewed as an extension of the work of \cite{hunter09} toward higher mass stars.
In Fig.~\ref{hunterOB}, we show the $\epsilon (N)$ -- \vsini\ diagram for our sample stars and the B stars (core hydrogen burning sub-sample) of \cite{hunter09}. At first sight,
the distribution of O and B stars in this diagram is the same. 
This is especially true for the low-velocity part of the diagram (\vsini\ $<$ 100 \kms) where actual abundances have been measured in most B stars, not just upper limits (which is the case for fast-rotating B stars).
For all masses, the highest value of $\epsilon(N)$ seems to be about 7.9.
This suggests that the mechanisms responsible for the chemical enrichment of slowly rotating massive stars depends weakly on the mass. In other words,
for any chemically enriched, slowly rotating massive star, the degree of enrichment is controlled by a parameter that is neither the mass nor the rotation rate.
However we cannot exclude that the main-sequence B stars from \cite{hunter09} are older than the O stars of our sample.  Since rotational mixing is expected
to have a stronger effect toward core-hydrogen exhaustion, the observed match of B-type and O-type distributions could be spurious because it may be biased by the fraction of the 
main-sequence lifetime actually covered by each star \citep[see e.g. Fig. 12 in][]{brott11a}.

In the right part of Fig.~\ref{hunterOB} (above 100 \kms), O and B stars also populate the same region. However, it is important to note that the values of $\epsilon(N)$
for the B stars are mostly upper limits. The only safe conclusion one can draw is that O stars are \textit{at least} as enriched as B stars. But one cannot
exclude that O stars are substantially more nitrogen-rich. This result would be consistent with predictions of evolutionary tracks
\citep{maeder00,brott11a,ekstroem11} and the results of \cite{martins12a}, who showed that Galactic O-stars seem to have higher surface nitrogen abundance than B stars.

\begin{figure}[tbp]
   \centering
   \includegraphics[width=9.2cm]{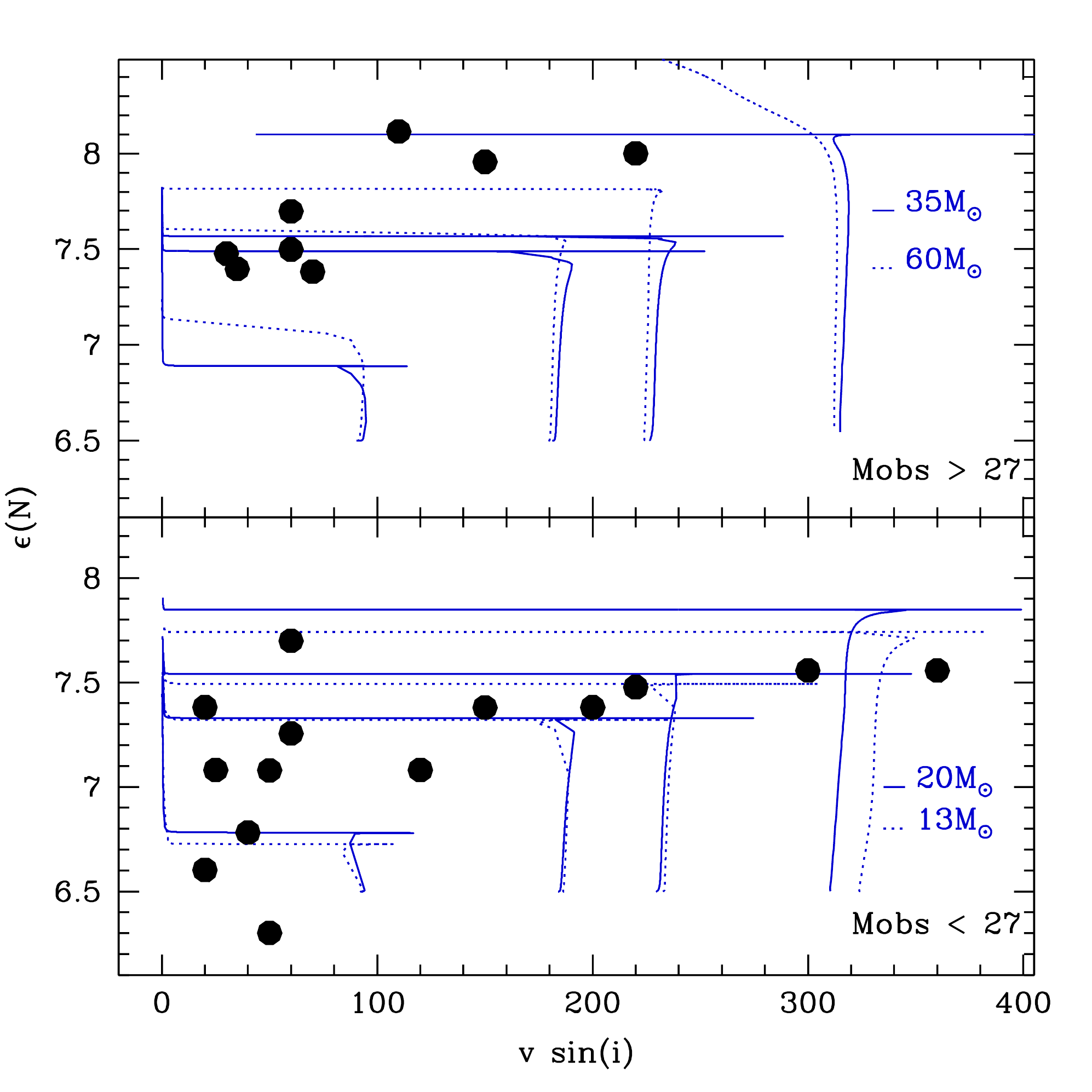}
      \caption{Nitrogen abundance (in units of 12+log(N/H)) versus projected rotational velocity for the O stars analyzed here. The upper (lower) panel shows star swith a mass higher (lower) than 27\msol. Evolutionary tracks for different masses are shown in both panels. They were taken from \cite{brott11a} and have initial rotational velocities $\sim$100, $\sim$200, $\sim$300 and $\sim$400 \kms. A horizontal shift corresponding to a factor $\pi/4$ to account for an average projection
effect has been applied to the theoretical curves.              }
         \label{hunter_smcO}
   \end{figure}

In Fig.\ \ref{hunter_smcO} we compare of the observed surface nitrogen abundance of O stars with predictions of evolutionary tracks. We chose the tracks of \cite{brott11a} since they are currently the only ones computed for different rotational velocities at the SMC metallicity. We divided our sample into two groups: stars below and above 27 \msol, the average mass of our sample. In the lower panel of Fig.\ \ref{hunter_smcO}, we show the low--mass part of our sample. The group of slowly rotating N--rich stars is not reproduced by the tracks, since those tracks reach the low-velocity - high N regime only well beyond the end of the main sequence, represented by the sudden left bifurcation in the tracks. This confirms the well-known results that rotational mixing does not explain the high N content of some of the slowly rotating massive stars. O stars rotating faster than 100 \kms\ can be accounted for by evolutionary tracks provided the stars are all very close to the end of the main sequence. This situation is rather unlikely, and we must therefore conclude that a slightly stronger mixing in the models would be necessary to explain the range of observed N abundances. 

This conclusion applies even more strongly to the high--mass subsample displayed in the upper panel of Fig.\ \ref{hunter_smcO}. In the high--velocity part, there are only three stars, but they all fall well above the main-sequence part of the tracks. Either the evolutionary models do not predict enough mixing, or the stars are already in a post-main-sequence phase. The latter possibility is rather unlikely given their observed properties. It is now well established that very massive stars can still be core-H burning objects and appear as Wolf-Rayet objects \citep{martins08, crowther10}. Here, one would need the exact opposite to explain the strong N enrichment of massive SMC O stars: stars already in a post-core-H-burning phase, but with the appearance of main sequence stars. This plot thus favors the possibility that rotational mixing is not sufficient at high masses in the models of \cite{brott11a}. In this high--mass sample, slowly rotating stars cluster around $\epsilon(N)$ = 7.5, which is not explained by evolutionary models of the main-sequence. Hence, here again, an unidentified mechanism is responsible for the chemical enrichment of those objects. 
We warn about overinterpretating the present results. A rigorous comparison of measured nitrogen abundances with theoretical prediction in a Hunter-like diagram requires population synthesis as performed 
in \cite{brott11b}, which we do not perform in this paper. 

Whatever the exact predictions of the models, another conclusion of our study is that on average, the higher the mass, the stronger the N-enrichment. 
This is clear from Fig.\ \ref{hunter_smcO}, in which we see that above 100\kms, stars with M$<$27 \msol\ have $\epsilon(N)$ below 7.6, while stars with  M$>$27 \msol\ have $\epsilon(N)$  above 7.9. This is a qualitative confirmation of the role of rotational mixing.

\subsubsection{Carbon-to-nitrogen ratio}
\label{CN_sect}
In massive stars, the total mass fraction of \mbox{CNO} remains roughly constant during the \mbox{CNO} cycle. In the very early phases of evolution, 
the CN-equilibrium is achieved first, and nitrogen is enhanced at the expense of carbon.The oxygen content remains roughly constant, and 
somewhat later enters the burning cycle. 
The N/C ratio is an excellent tracer of the chemical evolution of massive stars, because it increases substantially when CNO cycle-processed material appears at the
stellar surface. 

\begin{figure}[tbp]
   \centering
   \includegraphics[width=9.2cm]{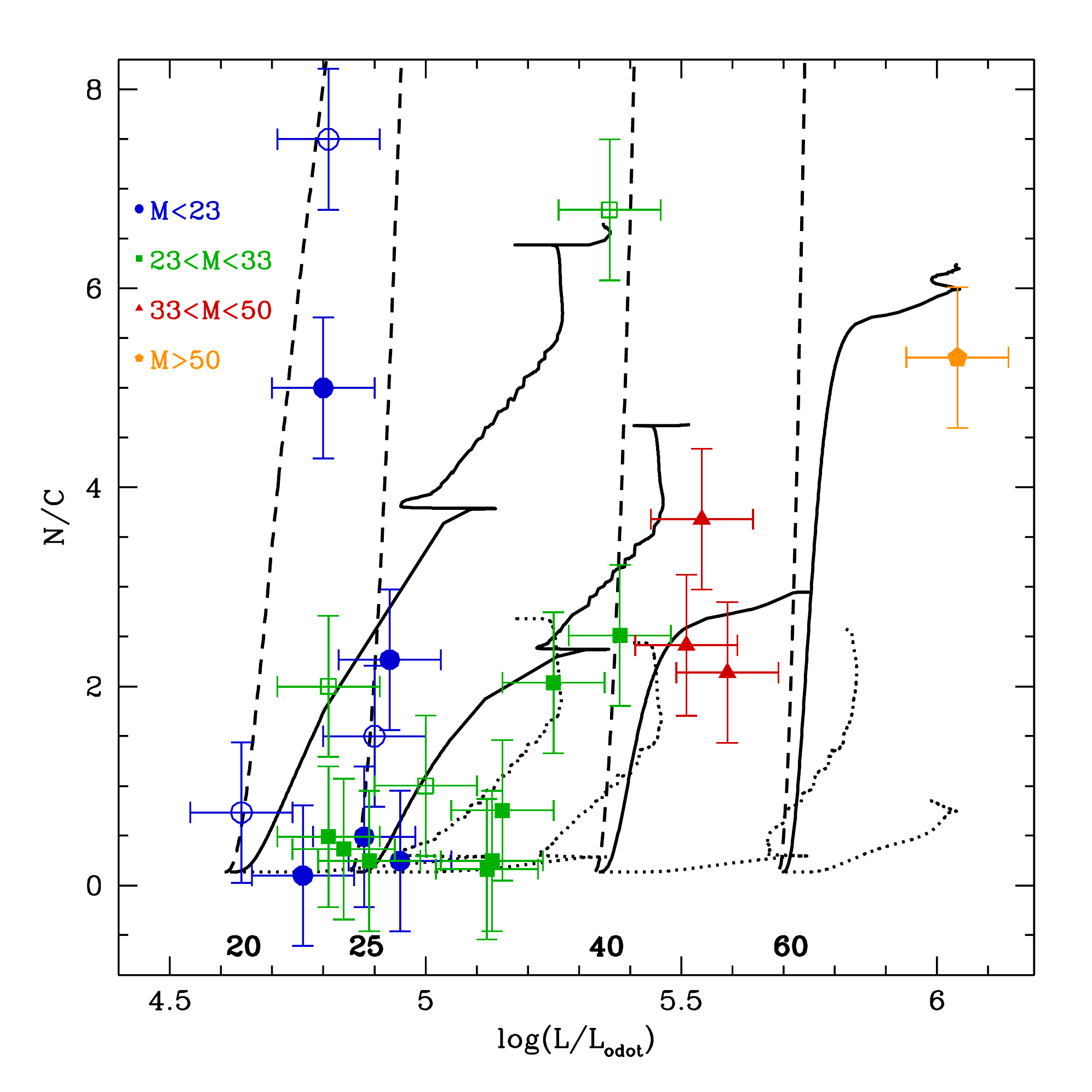}
      \caption{Nitrogen-to-carbon ratio as a function of luminosity. Circles correspond to stars with M $\leq$ 23 \msol, squares indicate stars with $23 <$ \mstar\ $< 33$ \msol, triangles show 
      stars with $33 <$ \mstar\ $< 50$ \msol\ and pentagons show stars with \mstar\ $\geq$ 50 \msol. Open (filled) symbols correspond to stars with \vsini\ $\geq$ 200 \kms\ ($\leq$, resp.). 
      Stellar evolution models taken from
      \cite{brott11a} for \mstar\ = 20, 25, 40 and 60 \msol\ are overplotted: dashed (solid, dotted) lines correspond to models with \vsini\ $\gtrsim$ 400 (300, 100) \kms.
              }
         \label{lum_nc}
   \end{figure}

In Fig. \ref{lum_nc}, we present a comparison between the measured N/C ratios (by number) as a function of stellar luminosity.
Stellar evolution models for \mstar\ = 20, 25, 40, and 60 \msol\ by \cite{brott11a} are also shown. For each mass, we selected three initial rotational velocities (100, 300, and 400 \kms) to cover the range of observed present \vsini\ of our sample stars. 
The amplitude of the N/C ratio is sensitive to differences in rotational velocities, masses, and ages.
We can therefore use this plot to check whether a star is evolving according to the prediction of standard evolutionary tracks.
We note that initial abundances play a critical role, because the relative increase of the nitrogen abundance depends
on the initial amount of carbon available for conversion 
into nitrogen, as measured by N/C ratios. We note in addition that although the absolute value of the initial C, N, and O abundances remains a vigorously debated topic,
especially in the solar case, the nitrogen-to-carbon abundance ratio (N/C = 0.13 in the SMC, see Table \ref{tab3} and Fig.~\ref{lum_nc}) is fairly well established
and may be considered a better reference point for comparison.

The direct comparison between the derived N/C and luminosities and the predictions of stellar evolution leads to the following conclusions:

\begin{enumerate}
\item \mstar\ $<$ 23 \msol: four out of the five filled circles, corresponding to \vsini\ $<$ 200 \kms, can be accounted for by the 25 \msol\ evolutionary tracks with rotational velocities between 100 and 300 \kms. For these objects, normal stellar evolution is therefore able to explain the observed properties. One slowly rotating object (\object{MPG~523}) has N/C=5, much higher than predicted. A high initial rotation (of about 400 \kms) would be needed to reproduce its position in the N/C vs L diagram. Of the three fast rotators (open symbols), two are well reproduced by  the 400 \kms\ track. The latest object (\object{AzV 267}) displays a lower enrichment than expected for its luminosity and \vsini. In conclusion, six out eight stars in this mass range can be accounted for by stellar evolution.

\item 23 $<$ \mstar\ $<$ 33 \msol: of the eight objects with \vsini\ $<$ 200 \kms, six have N/C $<$ 1.0, consistent with the 25 \msol\ evolutionary tracks with initial rotational velocities below 300 \kms. 
The two remaining objects (\object{MPG~368} and \object{AzV~446}) have N/C$\sim$2 and \logL$\sim\,5.3$. The 25 \msol\ evolutionary track with \vsini\ = 300 \kms\ can marginally account for these stars, provided they are very close to the end of the main sequence. Three stars have high rotation rates. Two are reproduced by evolutionary models with \vrot\ = 400 \kms, one \object{AzV~117} has a luminosity and N/C too 
high for the 25 \msol\ track. But its mass (31.9 \msol) is intermediate between the 25 and 40 \msol\ track, the latter being able to reproduce its the position. Consequently, \object{AzV~117} is marginally reproduced by stellar evolution. In total, in the 23--33 \msol\ mass range, eight stars follow normal evolution, and three are marginally explained.

\item 33 $<$ \mstar\ $<$ 50 \msol: the three objects in this mass range are no fast rotators. Two objects (\object{MPG 324} and \object{AzV 243}) fall on the 40 \msol\ evolutionary track with \vrot\ = 300 \kms, near the end of the main sequence. 
\object{MPG 324} is a suspected binary, which obviously might affect its evolution and surface abundances.
One object (\object{AzV~388}) has a higher N/C that cannot be explained by moderate rotation. Its \vsini\ is 150 \kms, higher than that of \object{MPG~324} which has a similar luminosity, gravity, and mass, but a lower \vsini\ (70 \kms). Hence, \object{AzV~388} might have been initially a fast rotator and has now braked to a moderate \vsini. In conclusion, the three objects in this mass range are marginally explained by normal evolution.

\item \mstar\ $>$ 50 \msol: the only star in this range (\object{MPG~355}) is marginally explained by the 60 \msol\ track at 300 \kms\ if the star is close to the end of the main sequence. Given its spectral type (ON2III(f*)), this is a likely possibility.  
The nitrogen abundance derived for this star agrees very well with the value measured by \cite{rivero12} from their analysis of the optical nitrogen lines. These authors found values
in the range [N]=7.98-8.10 (depending on the temperature solution), which is consistent with their conclusion that their method and tool (the \fastwind\ code) for analyzing the earliest O-stars would yield slightly lower abundances than 
those based on \cmfgen. Nevertheless, the nitrogen abundances from both studies agree within the error bars. The N/C ratio is thus much higher than predicted by stellar evolution models with moderate initial rotation. 
Only models with \vrot\ as high as 300 \kms\ can account for the observed ratio. 
At SMC metallicity, a 60 \msol\ (or above) star with such a high rotational velocity may have chemically homogeneous evolution \citep[cf.][]{brott11a}.
\end{enumerate}

In total, 15 out of 23 stars (65\%) are correctly explained by the \cite{brott11a} evolutionary tracks; 6 out of 23 (26\%) are marginally explained, and only 2 objects (9\%) are clear outliers. The bulk of the SMC O dwarf stars follows normal evolution. A fraction of 10 to about 30\% of the SMC O dwarfs deviates mildly or strongly from normal evolution. For comparison, \cite{brott11b} reported that about 40\% of the SMC B stars analyzed by \cite{hunter08} cannot be reproduced by their evolutionary models (the same we used in the present study). Using the same dataset, \cite{maeder09} concluded that the fraction of stars escaping normal evolution was about 20\%. Given the uncertainty concerning both the value of B stars following unusual evolution and the number of true outliers in our sample, we can conclude that the majority of SMC O dwarfs are well explained by single-star evolution with rotation and that a fraction similar to that observed for B stars might need additional physical explanation (such as binarity or magnetism).

 \subsubsection{CNO}
\begin{figure*}[t]
   \centering
   \includegraphics[width=19.cm]{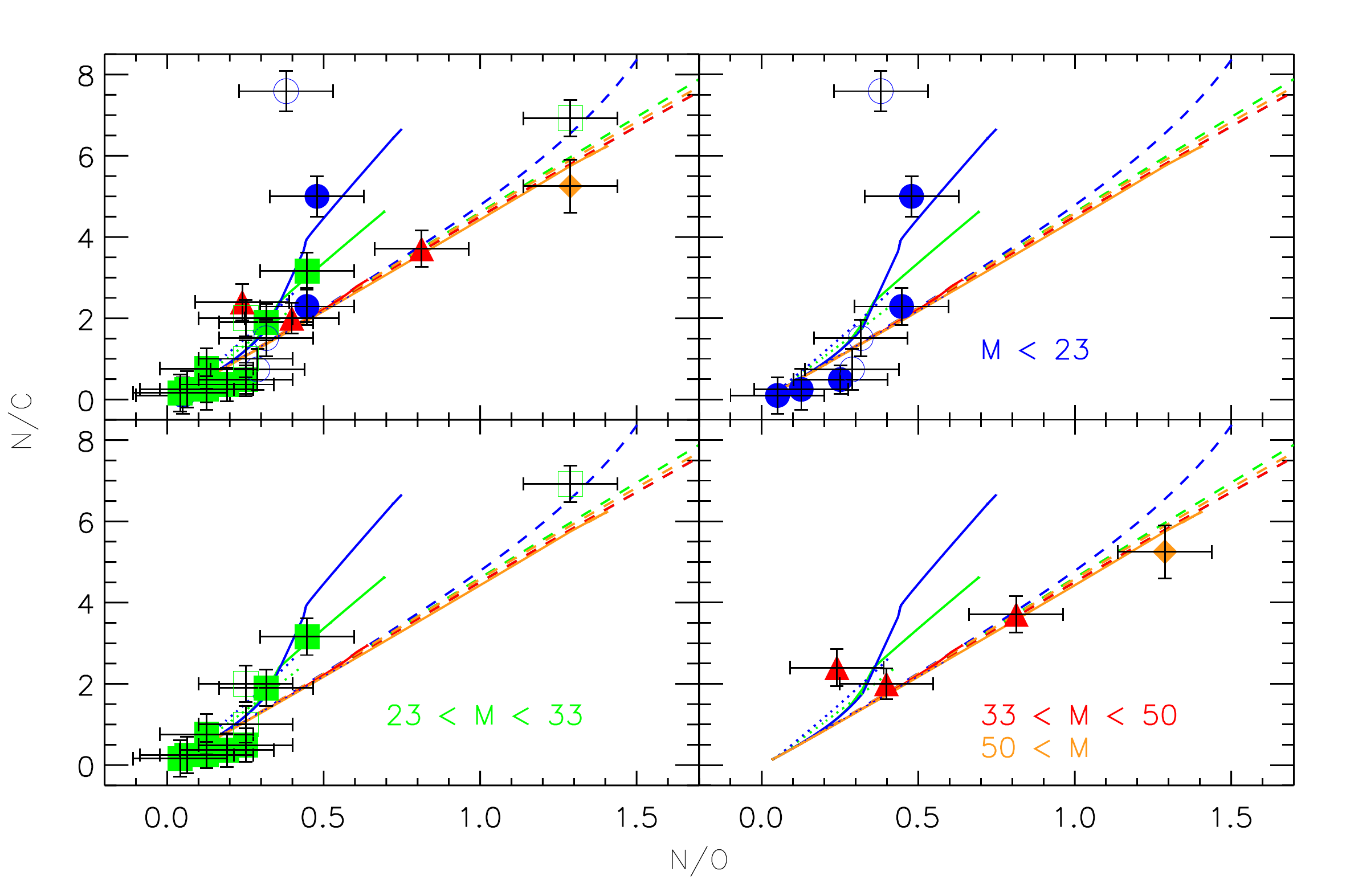}
      \caption{N/C ratio as a function of N/O. The same symbols and color coding as in Fig. \ref{lum_nc} have been adopted.  
      Stellar evolution models according to  \cite{brott11a} for \mstar\ = 20, 25, 40 and 60 \msol\ are overplotted. Here again, the same line styles as in  Fig. \ref{lum_nc} were adopted;
      we also color-scoded each corresponding mass range for clarity. The upper left panel presents the complete sample, the three remaining panels display
      the three mass ranges discussed in the text.
              }
         \label{no_nc}
   \end{figure*}

Achieving the complete CNO-equilibrium requires that hydrogen has been burnt in significant amount in the stellar core.  
Changes in the oxygen surface abundance should therefore be observed preferentially in the later evolutionary stages or in the more massive 
and faster rotating stars.
In Fig. \ref{no_nc}, we compare measured N/C versus N/O ratios with those predicted by stellar evolution models for different masses 
and rotation rates \citep{brott11a}. This plot indicates whether the variations of N/C and N/O follow the trend expected for material
processed through the CNO-cycle. The amplitudes of the departure from the initial abundance ratios are notoriously sensitive to
differences in rotational velocities, masses, and ages. They might also reflect departures of the initial mixture from the standard 
 \mbox{CNO} abundances measured in the SMC interstellar medium \citep{kurt98}, and used in the stellar evolution models. Our results can be summarized as follows:
\begin{enumerate}
\item \mstar\ $<$ 23 \msol: four out of the five filled circles, corresponding to \vsini\ $<$ 200 \kms, can be accounted for by evolutionary tracks with rotational velocities between 100 and 300 \kms\ and masses 20 \msol\  and/or 25 \msol. Their surface properties are therefore compatible with the mixing processes as implement in the evolution models. 
Again, \object{MPG~523} does not follow this trend. The N/C and N/O ratios we measured are compatible with a 20 \msol\ model with 300 \kms\ (but N/O would need to be twice as high to be consistent with the 20 \msol\ model and \vrot\ $\approx$ 400 \kms\ that was found to best accounts for \object{MPG 523} behavior, see Sect. \ref{CN_sect}). The location of \object{MPG 523} 
in Figs. \ref{lum_nc} and \ref{no_nc} strengthens the case that this star is indeed fast-rotating. 
This also implies an evolutionary stage rather later than expected from the H-R diagram in Fig. \ref{HR_fig}, but consistent with a location on isochrone 5 Myr (see Fig. \ref{HR_400_fig}).

Of the three fast rotators (open symbols), two are well reproduced by the 400 \kms\ track, these are \object{AzV 267} and \object{MPG 299}. 
The other object (\object{NGC 346-046}) displays an oxygen depletion much too weak for the measured carbon and nitrogen abundances. To match the prediction of a 20 \msol\  and \vrot\ $\approx$ 400 \kms\ model, O/H would need to be roughly four times lower than measured. Although the number of oxygen lines for abundance determination is limited and the sensitivity/accuracy reduced because of the high \vsini\ of the star, we can rule out such a low abundance. 
At this point, six out eight stars in this mass range can be accounted for by stellar evolution.

\item 23 $<$ \mstar\ $<$ 33 \msol: Seven out of eight objects with \vsini\ $<$ 200 \kms\ display N/O and N/C ratios consistent with the 25 \msol\ evolutionary tracks with initial rotational velocities of about 100 \kms. 
For the one remaining object (\object{MPG~368}), this evolutionary track is able to account marginally for its surface abundance but implies that the star is at the terminal stage of its evolution (the end of helium-core-burning in this grid), 
which can be ruled out from its other observed properties. 25 \msol\ models with faster rotation (\vsini\ $\geq$ 300 \kms) can better account for the surface abundances of MPG 368. This behavior is easier to reconcile with the likely binary status of this object. 
 
Of the three stars with high rotation rates, two are reproduced by evolutionary models with 25 \msol\  and \vsini\ = 300-400 \kms.  The last one (\object{AzV~177}) has an N/C 
too high for any 25 \msol\ track, but N/C and N/O are compatible with a model with 30 \msol\ and \vrot\ $\approx$ 300 \kms, consistent with the conclusion reached in Sect. \ref{CN_sect}. 
Including the present result, we argue that \object{AzV~117} is indeed reproduced by stellar evolution. In total, in the 23--33 \msol\ mass range, all (but one) stars follow normal evolution.

\item 33 $<$ \mstar\ $<$ 50 \msol : Two out of three stars in this mass range, namely  \object{AzV 388} and \object{AzV 243}, fall on the 40 \msol\ evolutionary track with 
\vrot\ $\approx$ 400 \kms\ (\object{AzV 243} is also on the 40 \msol, 300 \kms\ track), although none of them is a fast rotator. 
\object{MPG 324}, on the other hand, is a clear outlier in this diagram for this mass range, possibly related to its binary status. \object{MPG 324} and \object{AzV 388} have very different
N/O and N/C ratios, although they have similar luminosity, gravity, and mass, but their \vsini\ differs by a factor of two. Together with the behavior observed in Fig. 
\ref{lum_nc}, this reinforces the conclusion that \object{AzV~388} initially was a fast rotator and has now braked down to a moderate \vsini. 

\item \mstar\ $>$ 50 \msol: As the only star clearly more massive than 50 \msol\ (but whose real mass is unknown), \object{MPG~355} has N/O and N/C ratios consistent with the 60 \msol\ tracks having \vrot\ $\approx$ 300
\kms. This again strengthens our interpretation throughout this paper that this star is a very massive fast-rotating object in a chemically homogeneous evolution.
\end{enumerate}

In summary, 19 out of 23 stars have surface abundance patterns that are compatible with normal stellar evolution as implemented in the \cite{brott11a} evolutionary tracks. 
Only four objects are clear outliers, one of them is marginally consistent with the expected evolution for the correct mass range although of unlikely evolutionary
status.

For the stars that deviate from stellar evolution predictions, the amplitude of the deviations are generally large enough that 
differences in the initial abundances could not help to reconcile the outputs from stellar evolution models and our measurements.
It is indeed well-known that the slope of the theoretical tracks is very sensitive to the adopted initial mixture (see for instance
Fig. 3 of \cite{przybilla10} and a span of N/O and N/C ratios is expected for different initial chemical composition. 
In our case, all outliers but one (\object{MPG 523}) have $\mbox{C+N+O}$ in good consistency with the values
measured in the H~II regions \citep{kurt98}. We furthermore stress that several stars that follow the normal evolution do present 
some departures from these ISM values, on the other hand . 
However, the scatter around the SMC abundances is small for most objects, which indicates that the initial composition of the stars
is compatible with the ISM values. This rules out a difference from standard ISM values in the initial abundances as the origin for the observed behavior of the outliers stars. 

Alternatively, including magnetic fields in stellar evolution models could change the
speed and amplitude with which abundance patterns are built up at
the surface. A recent study by \cite{meynet11} confirmed that including  magnetic braking yields very different results for the surface abundances, 
which depend on the assumed rotation law inside the star. 
For stars with differential rotation, mixing is faster and stronger, but for stars with solid-body rotation,
mixing is inhibited because the stars spin down rapidly. In this case, surface abundance ratios are lower than in models
without magnetic braking. 
Alternatively, (close) binarity would affect the evolution and likely the surface abundances
of the stars (from mixing and possibly mass transfer), although it would also imply different luminosities, hence evolutionary masses. 
However, a comparison with models tailored for magnetic braking (and/or lower masses) would be required
for more quantitative conclusions. 

The overall properties of the sample stars, including their location in the H-R diagram and surface abundances, indicate that they are observed at different stages
of their chemical evolution, even though they are of very similar age. The underlying drivers of the observed differences are unclear
and could be manyfold -- from enhanced mixing related to binary evolution, or to evolution with rotation and a magnetic field \citep{meynet11}. 

Recent results reported in \cite{sana12a, sana12b} suggest that the binary fraction among massive stars is at least as high as 50\%.
Most of the (potential) binaries we identified follow a normal, single-star-like, evolution. These are of the late spectral type and are expected to
evolve more slowly, which might explain that no signs of binary evolution is visible yet.  
On the other hand,  we note that the most two massive binaries, namely \object{MPG 324} and \object{MPG 368}, do present surface 
abundance properties at odds with normal stellar evolution. Two circumstances may boost this behavior. 
First, because they are more massive, these stars are expected to evolve faster and display signs of chemical evolution more quickly. Second, 
they are potentially close-binary systems (we recall that radial-velocity variations have been found for these stars); this is very likely to impact
the evolution of these objects even more. 

\subsection{Masses}
\label{mass_sect}
Spectroscopic masses have been obtained from line profile fitting, while 
evolutionary masses have been derived by bilinear interpolation on the tracks and isochrones in the H-R diagram.
Since model atmospheres only account for radiation pressure plus gas pressure to balance gravity, spectroscopic masses are correct only if the
centrifugal acceleration is negligible, that is, only for slow rotators.
Otherwise, spectroscopic analyses are expected to underestimate the stellar mass. For this reason, we corrected the spectroscopic
surface gravities for 
centrifugal force induced by rotation, following the approach outlined in \cite{repolust04}.   
Assuming a random distribution of orientations for the rotational axis, the corrected surface gravity is the sum of the effective gravity and a 
correction term where the centrifugal acceleration averaged over the stellar disk is given to a very good approximation 
by the projected centrifugal velocity: $g_{c} = g_{eff} + (v\sin\,i)^{2}/R_{*}$, \\
where $g_{c}$ and $g_{eff}$ are the centrifugally corrected and effective surface gravity, respectively.
In most cases, the correction is small, with $\Delta \log g \leq 0.05$ dex, implying corrections of 10\% or less of the spectroscopic masses. Only stars
with \vsini\ $\geq$ 300 \kms\ -- \object{NGC 346-046} and \object{MPG 299} -- are affected by a larger correction of 0.06 dex and 0.1 dex respectively, implying
a correction of the stellar mass as high as 25\% in the later case. 

\begin{figure}[tbp]
   \centering
   \includegraphics[width=9.2cm]{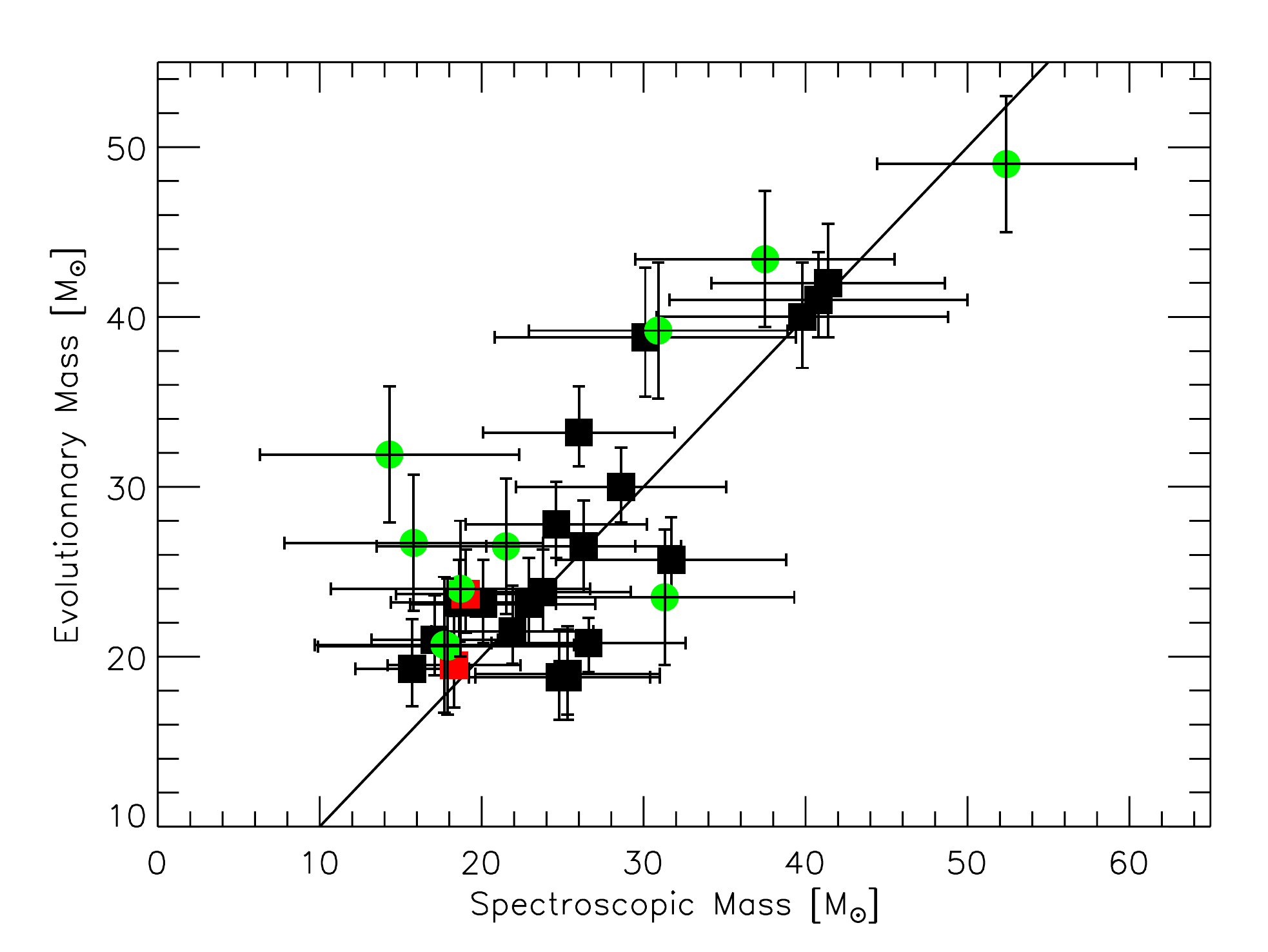}
      \caption{Spectroscopic versus evolutionary masses. The red filled squares show the rotators with \vsini\ $\geq$ 300 \kms\
      (see text). Green symbols indicate masses as determined in \cite{mokiem06} for the stars in 
       common with the present study. \object{MPG 355} is not plotted because its evolutionary mass could not be determined with the 
       model grid we used \citep{brott11a}.
              }
         \label{mass_fig}
   \end{figure}

A comparison of spectroscopic and evolutionary masses as listed in Table~\ref{tab2} is displayed in Fig.~\ref{mass_fig}.
The errors for the evolutionary mass reflect the errors spanned by the stellar luminosity and effective temperature,
respectively, for the adopted set of evolutionary tracks, i.e., for a given initial rotation rate.
With the introduction of rotation in stellar evolution models, it is no longer
possible to assign a unique value for the stellar mass from a  given set ($L_{\star}$, \teff),
and therefore this question translates into larger uncertainties on the evolutionary masses. 
Likewise, the large error bars on spectroscopic masses result from uncertainties on \logg.
Since the stellar distance is known within a fairly good accuracy for SMC stars, the error in \logg\ is indeed 
the critical source of error in the spectroscopic mass, hence in the ratio of the evolutionary mass to the spectroscopic mass.
The uncertainties in the determination of surface gravities are partially rooted in the use of echelle spectra,
whose rectification is difficult, especially in the vicinity of broad \hyd\ lines.  
The accuracy of the surface gravity measurements is considerably improved, however, when optical spectra can be used to constrain \logg\ in the models 
(see also Sect. \ref{param_sect}). This is the case for the 12 stars in \object{NGC~346} and for  \object{AzV 177}, \object{AzV~388} and  \object{AzV 243}.
Four more stars have 2dF spectra, that yield a relatively accurate determination of \logg\  despite a lower spectral resolution.
For the five stars with UV spectroscopy only, we adopted a higher error bars on \logg\ that is indicative of \logg\  scatter for dwarf O stars
around the assumed value of \logg = 4.0. 

Figure~\ref{mass_fig} shows that, within the error bars, most stars follow the 1:1 relation, although three objects still present (i.e.
even within the error bars) the mass-discrepancy first pointed out by \cite{herrero92}. A fourth object is also only marginally 
on the 1:1 relation.  
The study by \cite{mokiem06} also concluded that a mass-discrepancy was present in some SMC stars. They argued that
the discrepancy observed in helium rich stars of their sample was possibly a consequence of fast rotation in these objects. 
Our work shows that one star showing the discrepancy is indeed helium rich, namely MPG 487 but MPG 299, who is the other
helium rich star and is a fast rotator, shows a very good agreement between the spectroscopic and evolutionary masses.  
Furthermore, the other stars showing the mass discrepancy are easily identified as suspected binaries (\object{AzV 326}, \object{AzV 189}, \object{AzV 148}) 
and are not helium enriched objects  (hence not likely to rotate rapidly).
We note that for MPG 355 whose evolutionary masse is not constrained from the grid by \cite{brott11a}, the mass discrepancy could be large,
as pointed out in \cite{rivero12}.

\begin{figure}[tbp]
   \centering
   \includegraphics[width=9.2cm]{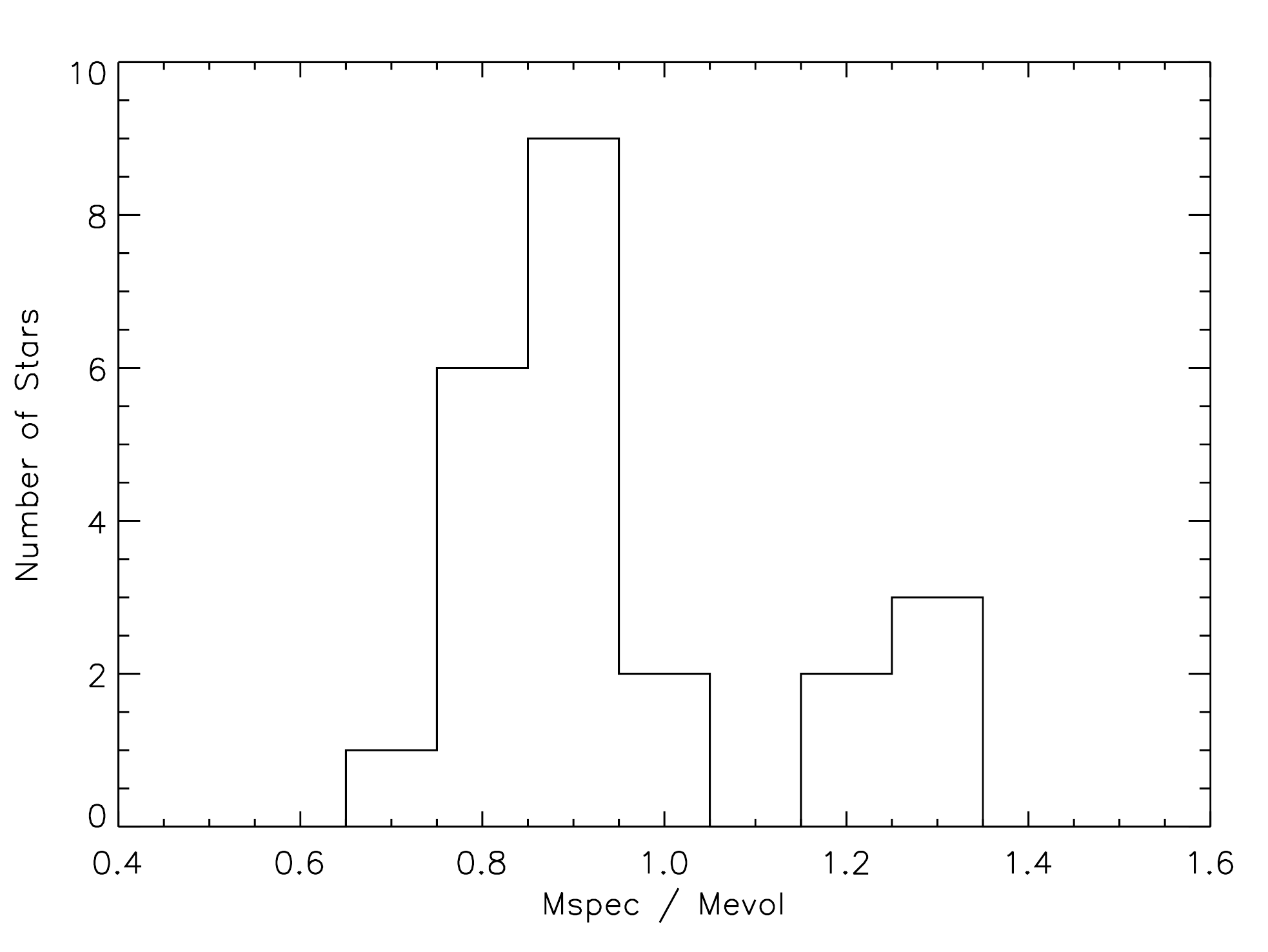}
      \caption{Histogram of the mass-ratio distribution. A bin size of 0.1 is adopted for the plot. Here again, \object{MPG 355} is not included (see text for comment). 
              }
         \label{histo_fig}
   \end{figure}

A different view of our results is provided by the histogram of the mass ratios \msp\ / \mev, which we present in Fig. \ref{histo_fig}. 
The distribution of the ratio divides into two peaks.
In total, five stars present mass ratios higher than 1.2. Four of these stars are clear binaries 
(\object{AzV 326}, \object{AzV 189}, \object{AzV 148}, and \object{MPG 487}) and their masses, hence \msp\ / \mev\ ratios, are affected by larger uncertainties. 
We also note that for MPG 368, also identified as a binary \citep[see also][]{massey09}, the mass ratio deviates from the 1:1 relation by the same amplitude 
as the four later-type binaries. The bias toward highly discrepant ratios for these five stars is therefore probably related to the binary status of these objects. 
We note, on the other hand, that the earliest-type binary, \object{MPG 324} instead presents an \msp\ / \mev\ ratio
that excellently agrees between evolutionary and spectroscopic masses. 
Furthermore, we were not able to identify any issue in the spectroscopic determination of the surface gravity of the fifth star, namely \object{MPG 682}. A lower \logg\ than quoted 
in Table \ref{tab2} is firmly excluded; this would decrease the spectroscopic mass and help reconcile it with the evolutionary mass. 

For the large majority of the sample stars, the distribution is skewed toward values 
lower than unity (with a mean of 0.92 and a standard deviation of 0.08). This indicates that
even including the corrections for rotation applied to spectroscopic masses, the latter are overall smaller than 
evolutionary masses. Obviously, this comment holds for the adopted set of evolutionary tracks only. 
Evolutionary masses (and ages) would be different if we adopted tracks and isochrones with different initial rotation rates. 
The lack of knowledge of the actual rotation rates of the stars 
hampers the use of evolutionary models computed with rotation rates appropriate for each individual object.This translates into larger uncertainties on the actual 
evolutionary masses and the observed skewness of the histogram in Fig. \ref{histo_fig} probably reflects this effect.  
Quantitatively, however, we found that \mev\ for our target stars would be changed by half a solar mass on average if we used
evolutionary models with \vrot\ $\approx$ 100 \kms\ for instance. 
With these models, though, the youngest two stars (\object{NGC 346-046} and \object{MPG 523}) of our sample would be located to the left of the ZAMS,
even within the error bars. For such cases, estimates of evolutionary masses are highly uncertain. We presented  
arguments in Sect. \ref{sect_fast} that showed that \object{NGC 346-046} is potentially best described by a model for a 20 \msol\ star rotating between 400 \kms\ and 500 \kms\ (also consistent
with the chemical N/C pattern, see Sect. \ref{ab_sect}). The derived evolutionary mass would be roughly 3 \msol\ lower than quoted in Table \ref{tab2}, a
perfect match with the spectroscopic mass. We stress furthermore that the same models also account very well for the position of \object{MPG 523} in the H-R diagram, 
while providing a better match to the measured N/C ratio. 


\section{Summary}
\label{conclu_sect}
We have modeled the UV and optical spectra of 23 O-type dwarf stars in the SMC to derive 
their surface and evolutionary properties. Our results can be summarized as follows:

\begin{enumerate}

\item We can derive accurate luminosities from modeling the SEDs of the sample stars. 
We found three objects with luminosities too high for their spectral types. 
We showed that these stars are very likely binaries consisting of two objects of the same or very similar spectral types.
Three other stars were already identified as potential binaries; we confirmed one of them from SED fitting.
The measured fraction of binaries of our sample is 26\%, which agrees well with more systematic dedicated studies of binarity in massive stars.   
 
\item The sample consists of stars in the giant H~II region \object{NGC~346} and field stars. 
From the H-R diagram, we found that the younger stars belong to NGC~346, while fields stars are
2\,Myr old or older. 
The stars in \object{NGC~346} have ages ranging from younger than 1\,Myr to about 6\,Myr (for \object{MPG 012}, the only subgiant of this sample). 

\item We found two fast rotators with \vsini~$\geq$ 300 \kms, \object{MPG~299} and \object{NGC~346-046}. Both stars are 1~Myr old at most,
as indicated by their location on or very close to the ZAMS. 
The offset in the position of these two fast rotators compared to the other stars confirms the predictions of evolutionary models
that fast-rotating stars tend to evolve more vertically in the H-R diagram.

\item Three stars of our sample, \object{MPG~113}, \object{MPG~523} and \object{NGC~346-031}, were originally classified as luminosity-class Vz.
This luminosity class points to massive stars whose spectral properties are 
expected to indicate extreme youth. In contrast to \cite{mokiem06}, our analysis indicates
that only \object{MPG~523} is on the ZAMS, the other two stars are more evolved, even within the error bars. 
For \object{MPG~113}, a difference in the effective temperature of 3,000~K between both studies is the cause of the shift in age.
The case of \object{NGC~346-031} is very interesting because it points to a different interpretation of the 
strength of the \heii\ \lb4686 line, the defining criterion of the Vz class, based on different mass-loss rates and helium abundances. 

\item The discrepancy between spectroscopic and evolutionary masses is reduced but still present. 
For a subsample of 17 of the 23 stars, the mean mass ratio is 0.92, with a standard deviation of $\pm$0.08. 
Binary stars present stronger deviations from the 1:1 relation of  \msp\ / \mev\ (i.e., a deviation of 20\%\ or more), while one single star is the only clear outlier to
the \msp\ = \mev\ relation. 
Obviously, the detailed numerical results of this analysis are valid only for the adopted set of evolutionary tracks and isochrones. Investigation of the impact
of different rotation rates on the evolutionary calculations showed rather limited influence on the mass determination, however, not enough to explain
the observed mass discrepancy. 

\item Photospheric UV lines of \mbox{C} and \mbox{N} were used to derive surface abundances. Our results may be regarded as an extension
toward higher masses of the
work by \cite{hunter09} on B-type stars. Using their published abundances for B stars, we found that the distribution of O and B stars
in the $\epsilon (N)$ -- \vsini\ diagram is the same, especially for the low-velocity part
of this diagram. This suggests that the mechanisms responsible for the chemical enrichment
of slowly rotating massive stars depends weakly on the mass and that O stars are \textit{at least} as enriched as B stars.

\item We compared the measured N/C ratios as a function of stellar luminosity,
and the predictions of the stellar evolution models by \cite{brott11a}. Sixty-five percent of our sample stars are well reproduced by evolutionary models including rotation. 
Twenty-six percent are marginally compatible with these models, and 9\% are clear outliers. The fraction of O dwarfs in the SMC that cannot be explaned by standard evolution is similar to that of B stars, considering the uncertainties in the estimate of the exact fraction of these outliers (between 20 and 40\%). 

\item Finally, a majority of stars, irrespective of their mass, have N/C ratios as a function of N/O that
match the predicted behavior well, which strengthens the physical assumptions for mixing that are 
currently implemented in the evolution models presented in \cite{brott11a}. Two out of the four stars that 
differ from this normal evolution are massive, possibly close, binaries,  
and their surface abundances, should be investigated with ad hoc evolutionary models. The fraction of stars close to the ZAMS that show unexplained
differences to standard evolution is thus 9\%. This fraction is significantly smaller than the $\approx$ 40 \%\ outliers of B-type core-hydrogen-burning stars in the SMC
found by \cite{hunter09} and \cite{brott11b}. Whether or not this reflects the actual trend for O stars needs to be investigated
by extending the present analysis to more evolved objects in the giant and supergiant luminosity classes, such that the compared samples of O- and B-type stars are as
similar as possible.

\end{enumerate}

The different conclusions that we reached regarding the nitrogen surface abundance on one hand and the N/C or/and N/O ratios
on the other hand deserves more investigation. \cite{maeder09} and \cite{brott11b} previously pointed out that the so-called Hunter diagram might not
be the best way to constrain rotational mixing. These authors discussed how the relative influences of binarity, mass loss, metallicity, age, and rotation 
on the chemical evolution of massive stars could blurr the interpretation of this diagram. 
Our results also support this conclusion. We found indeed that a (very) limited fraction of stars do not comply with the predictions of evolution models
with rotation when abundance ratios are used. Taken at face value, they suggest
that the elements abundances taken separately might not be as relevant as abundance ratios when observations are to be compared
with outcomes from stellar evolution models. In other words, the more constraints are obtained on surface abundances (beyond N abundance only), 
the better the agreement between models and observations. 
A larger sample including more evolved objects would allow one to investigate if these results vary along the evolution.
We will address this question in a forthcoming paper where a similar spectral analysis of
SMC giant and supergiant stars will be presented. 

\begin{acknowledgements}
We thank the referee for helpful comments and suggestions.
We are grateful to P. Crowther, C. Evans and P. Massey who provided several optical spectra most useful to this work.
We thank the French Agence Nationale de la Recherche (ANR) for financial support under Programme Blanc MaSiLU. 
This research has made use of the SIMBAD database, operated at CDS, Strasbourg, France. 
STScI is operated by the Association of Universities for Research in Astronomy, Inc., under NASA contract NAS5-26555. 
J.-C. Bouret is indebted to George Sonneborn for his invitation to work at NASA/GSFC when this work was initiated. 
This work was supported by NASA grant NNX08AC146 to the University of Colorado at Boulder.
DJH acknowledges support from  STScI theory grants HST-AR-11756.01.A and  HST-AR-12640.01.A

\end{acknowledgements}

\bibliographystyle{aa}
\bibliography{article2}

\begin{thebibliography}{138}
\expandafter\ifx\csname natexlab\endcsname\relax\def\natexlab#1{#1}\fi

\bibitem[{{Aerts} {et~al.}(2009){Aerts}, {Puls}, {Godart}, \&
  {Dupret}}]{aerts09}
{Aerts}, C., {Puls}, J., {Godart}, M., \& {Dupret}, M.-A. 2009, \aap, 508, 409

\bibitem[{{Allard} {et~al.}(1991){Allard}, {Le Dourneuf}, {Artru}, \&
  {Lanz}}]{AAL91_NIV}
{Allard}, N., {Le Dourneuf}, M., {Artru}, M.-C., \& {Lanz}, T. 1991, \aaps, 91,
  399

\bibitem[{{Azzopardi} \& {Vigneau}(1982)}]{azzopardi82}
{Azzopardi}, M. \& {Vigneau}, J. 1982, \aaps, 50, 291

\bibitem[{{Azzopardi} {et~al.}(1975){Azzopardi}, {Vigneau}, \&
  {Macquet}}]{azzopardi75}
{Azzopardi}, M., {Vigneau}, J., \& {Macquet}, M. 1975, \aaps, 22, 285

\bibitem[{{Ballester} {et~al.}(2000){Ballester}, {Modigliani}, {Boitquin},
  {Cristiani}, {Hanuschik}, {Kaufer}, \& {Wolf}}]{Ballester2000}
{Ballester}, P., {Modigliani}, A., {Boitquin}, O., {et~al.} 2000, The
  Messenger, 101, 31

\bibitem[{{Barnard} {et~al.}(1974){Barnard}, {Cooper}, \&
  {Smith}}]{BCS74_HeI_stark}
{Barnard}, A.~J., {Cooper}, J., \& {Smith}, E.~W. 1974, \jqsrt, 14, 1025

\bibitem[{{Berrington} {et~al.}(1985){Berrington}, {Burke}, {Dufton}, \&
  {Kingston}}]{BBD85_col}
{Berrington}, K.~A., {Burke}, P.~G., {Dufton}, P.~L., \& {Kingston}, A.~E.
  1985, Atomic Data and Nuclear Data Tables, 33, 195

\bibitem[{{Berrington} \& {Kingston}(1987)}]{BK87_HeI_col}
{Berrington}, K.~A. \& {Kingston}, A.~E. 1987, Journal of Physics B Atomic
  Molecular Physics, 20, 6631

\bibitem[{{Bestenlehner} {et~al.}(2011){Bestenlehner}, {Vink}, {Gr{\"a}fener},
  {Najarro}, {Evans}, {Bastian}, {Bonanos}, {Bressert}, {Crowther}, {Doran},
  {Friedrich}, {H{\'e}nault-Brunet}, {Herrero}, {de Koter}, {Langer}, {Lennon},
  {Ma{\'{\i}}z Apell{\'a}niz}, {Sana}, {Soszynski}, \&
  {Taylor}}]{Bestenlehner11}
{Bestenlehner}, J.~M., {Vink}, J.~S., {Gr{\"a}fener}, G., {et~al.} 2011, \aap,
  530, L14

\bibitem[{{Bonanos} {et~al.}(2010){Bonanos}, {Lennon}, {K{\"o}hlinger}, {van
  Loon}, {Massa}, {Sewilo}, {Evans}, {Panagia}, {Babler}, {Block}, {Bracker},
  {Engelbracht}, {Gordon}, {Hora}, {Indebetouw}, {Meade}, {Meixner}, {Misselt},
  {Robitaille}, {Shiao}, \& {Whitney}}]{bonanos10}
{Bonanos}, A.~Z., {Lennon}, D.~J., {K{\"o}hlinger}, F., {et~al.} 2010, \aj,
  140, 416

\bibitem[{{Bouret} {et~al.}(2012){Bouret}, {Hillier}, {Lanz}, \&
  {Fullerton}}]{bouret12}
{Bouret}, J.-C., {Hillier}, D.~J., {Lanz}, T., \& {Fullerton}, A.~W. 2012,
  \aap, 544, A67

\bibitem[{{Bouret} {et~al.}(2005){Bouret}, {Lanz}, \& {Hillier}}]{bouret05}
{Bouret}, J.-C., {Lanz}, T., \& {Hillier}, D.~J. 2005, \aap, 438, 301

\bibitem[{{Bouret} {et~al.}(2003){Bouret}, {Lanz}, {Hillier}, {Heap}, {Hubeny},
  {Lennon}, {Smith}, \& {Evans}}]{bouret03}
{Bouret}, J.-C., {Lanz}, T., {Hillier}, D.~J., {et~al.} 2003, \apj, 595, 1182

\bibitem[{{Brott} {et~al.}(2011{\natexlab{a}}){Brott}, {de Mink}, {Cantiello},
  {Langer}, {de Koter}, {Evans}, {Hunter}, {Trundle}, \& {Vink}}]{brott11a}
{Brott}, I., {de Mink}, S.~E., {Cantiello}, M., {et~al.} 2011{\natexlab{a}},
  \aap, 530, A115+

\bibitem[{{Brott} {et~al.}(2011{\natexlab{b}}){Brott}, {Evans}, {Hunter}, {de
  Koter}, {Langer}, {Dufton}, {Cantiello}, {Trundle}, {Lennon}, {de Mink},
  {Yoon}, \& {Anders}}]{brott11b}
{Brott}, I., {Evans}, C.~J., {Hunter}, I., {et~al.} 2011{\natexlab{b}}, \aap,
  530, A116

\bibitem[{{Busche} \& {Hillier}(2005)}]{busche05}
{Busche}, J.~R. \& {Hillier}, D.~J. 2005, \aj, 129, 454

\bibitem[{{Butler} {et~al.}(1993){Butler}, {Mendoza}, \&
  {Zeippen}}]{BKM93_Mg_seq}
{Butler}, K., {Mendoza}, C., \& {Zeippen}, C.~J. 1993, Journal of Physics B
  Atomic Molecular Physics, 26, 4409

\bibitem[{{Cardelli} {et~al.}(1989){Cardelli}, {Clayton}, \&
  {Mathis}}]{cardelli89}
{Cardelli}, J.~A., {Clayton}, G.~C., \& {Mathis}, J.~S. 1989, \apj, 345, 245

\bibitem[{{Cochrane} \& {McWhirter}(1983)}]{CM83_lith}
{Cochrane}, D.~M. \& {McWhirter}, R.~W.~P. 1983, \physscr, 28, 25

\bibitem[{{Crowther} {et~al.}(2002){Crowther}, {Hillier}, {Abbott}, \&
  {Fullerton}}]{crowther02}
{Crowther}, P.~A., {Hillier}, D.~J., {Abbott}, J.~B., \& {Fullerton}, A.~W.
  2002, \aap, 392, 653

\bibitem[{{Crowther} {et~al.}(2010){Crowther}, {Schnurr}, {Hirschi}, {Yusof},
  {Parker}, {Goodwin}, \& {Kassim}}]{crowther10}
{Crowther}, P.~A., {Schnurr}, O., {Hirschi}, R., {et~al.} 2010, \mnras, 408,
  731

\bibitem[{{Cunto} {et~al.}(1993){Cunto}, {Mendoza}, {Ochsenbein}, \&
  {Zeippen}}]{Topbase93}
{Cunto}, W., {Mendoza}, C., {Ochsenbein}, F., \& {Zeippen}, C.~J. 1993, \aap,
  275, L5

\bibitem[{{Dekker} {et~al.}(2000){Dekker}, {D'Odorico}, {Kaufer}, {Delabre}, \&
  {Kotzlowski}}]{Dekker2000}
{Dekker}, H., {D'Odorico}, S., {Kaufer}, A., {Delabre}, B., \& {Kotzlowski}, H.
  2000, in Society of Photo-Optical Instrumentation Engineers (SPIE) Conference
  Series, Vol. 4008, Society of Photo-Optical Instrumentation Engineers (SPIE)
  Conference Series, ed. {M.~Iye \& A.~F.~Moorwood}, 534--545

\bibitem[{{Dimitrijevic} \& {Sahal-Brechot}(1984)}]{DS84_Hei}
{Dimitrijevic}, M.~S. \& {Sahal-Brechot}, S. 1984, \aap, 136, 289

\bibitem[{{Dufour} \& {Kurt}(1998)}]{dufour98}
{Dufour}, R.~J. \& {Kurt}, C.~M. 1998, in ESA Special Publication, Vol. 413,
  Ultraviolet Astrophysics Beyond the IUE Final Archive, ed. {W.~Wamsteker,
  R.~Gonzalez Riestra, \& B.~Harris}, 469

\bibitem[{{Ekstr{\"o}m} {et~al.}(2012){Ekstr{\"o}m}, {Georgy}, {Eggenberger},
  {Meynet}, {Mowlavi}, {Wyttenbach}, {Granada}, {Decressin}, {Hirschi},
  {Frischknecht}, {Charbonnel}, \& {Maeder}}]{ekstroem11}
{Ekstr{\"o}m}, S., {Georgy}, C., {Eggenberger}, P., {et~al.} 2012, \aap, 537,
  A146

\bibitem[{{Evans} {et~al.}(2004){Evans}, {Howarth}, {Irwin}, {Burnley}, \&
  {Harries}}]{evans04}
{Evans}, C.~J., {Howarth}, I.~D., {Irwin}, M.~J., {Burnley}, A.~W., \&
  {Harries}, T.~J. 2004, \mnras, 353, 601

\bibitem[{{Evans} {et~al.}(2006){Evans}, {Lennon}, {Smartt}, \&
  {Trundle}}]{evans06}
{Evans}, C.~J., {Lennon}, D.~J., {Smartt}, S.~J., \& {Trundle}, C. 2006, \aap,
  456, 623

\bibitem[{{Evans} {et~al.}(2005){Evans}, {Smartt}, {Lee}, {Lennon}, {Kaufer},
  {Dufton}, {Trundle}, {Herrero}, {Sim{\'o}n-D{\'{\i}}az}, {de Koter},
  {Hamann}, {Hendry}, {Hunter}, {Irwin}, {Korn}, {Kudritzki}, {Langer},
  {Mokiem}, {Najarro}, {Pauldrach}, {Przybilla}, {Puls}, {Ryans}, {Urbaneja},
  {Venn}, \& {Villamariz}}]{evans05}
{Evans}, C.~J., {Smartt}, S.~J., {Lee}, J., {et~al.} 2005, \aap, 437, 467

\bibitem[{{Fernley} {et~al.}(1987){Fernley}, {Seaton}, \& {Taylor}}]{FST87_Hei}
{Fernley}, J.~A., {Seaton}, M.~J., \& {Taylor}, K.~T. 1987, Journal of Physics
  B Atomic Molecular Physics, 20, 6457

\bibitem[{{Garc{\'{\i}}a} \& {Mermilliod}(2001)}]{garcia01}
{Garc{\'{\i}}a}, B. \& {Mermilliod}, J.~C. 2001, \aap, 368, 122

\bibitem[{{Garcia} \& {Bianchi}(2004)}]{garcia04}
{Garcia}, M. \& {Bianchi}, L. 2004, \apj, 606, 497

\bibitem[{{Grevesse} {et~al.}(2007){Grevesse}, {Asplund}, \&
  {Sauval}}]{grevesse07}
{Grevesse}, N., {Asplund}, M., \& {Sauval}, A.~J. 2007, \ssr, 130, 105

\bibitem[{{Harries} {et~al.}(2003){Harries}, {Hilditch}, \&
  {Howarth}}]{harries03}
{Harries}, T.~J., {Hilditch}, R.~W., \& {Howarth}, I.~D. 2003, \mnras, 339, 157

\bibitem[{{Haser} {et~al.}(1998){Haser}, {Pauldrach}, {Lennon}, {Kudritzki},
  {Lennon}, {Puls}, \& {Voels}}]{haser98}
{Haser}, S.~M., {Pauldrach}, A.~W.~A., {Lennon}, D.~J., {et~al.} 1998, \aap,
  330, 285

\bibitem[{{Heap} {et~al.}(2006){Heap}, {Lanz}, \& {Hubeny}}]{heap06}
{Heap}, S.~R., {Lanz}, T., \& {Hubeny}, I. 2006, \apj, 638, 409

\bibitem[{{Heger} \& {Langer}(2000)}]{heger00}
{Heger}, A. \& {Langer}, N. 2000, \apj, 544, 1016

\bibitem[{{Herrero} {et~al.}(1992){Herrero}, {Kudritzki}, {Vilchez}, {Kunze},
  {Butler}, \& {Haser}}]{herrero92}
{Herrero}, A., {Kudritzki}, R.~P., {Vilchez}, J.~M., {et~al.} 1992, \aap, 261,
  209

\bibitem[{{Hillier}(1987)}]{Hil87a}
{Hillier}, D.~J. 1987, \apjs, 63, 947

\bibitem[{{Hillier} {et~al.}(2012){Hillier}, {Bouret}, {Lanz}, \&
  {Busche}}]{hillier12}
{Hillier}, D.~J., {Bouret}, J.-C., {Lanz}, T., \& {Busche}, J.~R. 2012, \mnras,
  426, 1043

\bibitem[{{Hillier} {et~al.}(2003){Hillier}, {Lanz}, {Heap}, {Hubeny}, {Smith},
  {Evans}, {Lennon}, \& {Bouret}}]{hillier03}
{Hillier}, D.~J., {Lanz}, T., {Heap}, S.~R., {et~al.} 2003, \apj, 588, 1039

\bibitem[{{Hillier} \& {Miller}(1998)}]{hillier98}
{Hillier}, D.~J. \& {Miller}, D.~L. 1998, \apj, 496, 407

\bibitem[{{Hubeny} \& {Lanz}(1995{\natexlab{a}})}]{hubeny95}
{Hubeny}, I. \& {Lanz}, T. 1995{\natexlab{a}}, \apj, 439, 875

\bibitem[{{Hubeny} \& {Lanz}(1995{\natexlab{b}})}]{HL95_TLUSTY}
{Hubeny}, I. \& {Lanz}, T. 1995{\natexlab{b}}, \apj, 439, 875

\bibitem[{{Hunter} {et~al.}(2009){Hunter}, {Brott}, {Langer}, {Lennon},
  {Dufton}, {Howarth}, {Ryans}, {Trundle}, {Evans}, {de Koter}, \&
  {Smartt}}]{hunter09}
{Hunter}, I., {Brott}, I., {Langer}, N., {et~al.} 2009, \aap, 496, 841

\bibitem[{{Hunter} {et~al.}(2008){Hunter}, {Brott}, {Lennon}, {Langer},
  {Dufton}, {Trundle}, {Smartt}, {de Koter}, {Evans}, \& {Ryans}}]{hunter08}
{Hunter}, I., {Brott}, I., {Lennon}, D.~J., {et~al.} 2008, \apjl, 676, L29

\bibitem[{{Hunter} {et~al.}(2007){Hunter}, {Dufton}, {Smartt}, {Ryans},
  {Evans}, {Lennon}, {Trundle}, {Hubeny}, \& {Lanz}}]{hunter07}
{Hunter}, I., {Dufton}, P.~L., {Smartt}, S.~J., {et~al.} 2007, \aap, 466, 277

\bibitem[{{K{\"o}hler} {et~al.}(2012){K{\"o}hler}, {Borzyszkowski}, {Brott},
  {Langer}, \& {de Koter}}]{kohler12}
{K{\"o}hler}, K., {Borzyszkowski}, M., {Brott}, I., {Langer}, N., \& {de
  Koter}, A. 2012, \aap, 544, A76

\bibitem[{{Krti{\v c}ka}(2006)}]{krticka06}
{Krti{\v c}ka}, J. 2006, \mnras, 367, 1282

\bibitem[{{Kudritzki} \& {Puls}(2000)}]{kudritzki00}
{Kudritzki}, R.~P. \& {Puls}, J. 2000, \araa, 38, 613

\bibitem[{{Kurt} \& {Dufour}(1998)}]{kurt98}
{Kurt}, C.~M. \& {Dufour}, R.~J. 1998, in Revista Mexicana de Astronomia y
  Astrofisica, vol. 27, Vol.~7, Revista Mexicana de Astronomia y Astrofisica
  Conference Series, ed. {R.~J.~Dufour \& S.~Torres-Peimbert}, 202--+

\bibitem[{{Kurucz}(2009)}]{Kur09_ATD}
{Kurucz}, R.~L. 2009, in American Institute of Physics Conference Series, Vol.
  1171, American Institute of Physics Conference Series, ed. {I.~Hubeny,
  J.~M.~Stone, K.~MacGregor, \& K.~Werner}, 43--51

\bibitem[{{Kurucz} \& {Peytremann}(1975)}]{KP75_linelist}
{Kurucz}, R.~L. \& {Peytremann}, E. 1975, SAO Special Report, 362

\bibitem[{{Lanz} \& {Hubeny}(2003)}]{lanz03}
{Lanz}, T. \& {Hubeny}, I. 2003, \apjs, 146, 417

\bibitem[{{Leibowitz}(1972)}]{Lei72_CIV}
{Leibowitz}, E.~M. 1972, Journal of Quantitative Spectroscopy and Radiative
  Transfer, 12, 299

\bibitem[{{Leitherer} {et~al.}(1992){Leitherer}, {Robert}, \&
  {Drissen}}]{leitherer92}
{Leitherer}, C., {Robert}, C., \& {Drissen}, L. 1992, \apj, 401, 596

\bibitem[{{Lemke}(1997)}]{Lem97_stark}
{Lemke}, M. 1997, \aaps, 122, 285

\bibitem[{{Lennon} \& {Burke}(1994)}]{LB94_col}
{Lennon}, D.~J. \& {Burke}, V.~M. 1994, \aaps, 103, 273

\bibitem[{{Liang} \& {Badnell}(2011)}]{LB11_Li_like}
{Liang}, G.~Y. \& {Badnell}, N.~R. 2011, \aap, 528, A69

\bibitem[{{Luo} \& {Pradhan}(1989)}]{LP89_C_seq}
{Luo}, D. \& {Pradhan}, A.~K. 1989, Journal of Physics B Atomic Molecular
  Physics, 22, 3377

\bibitem[{{Luo} {et~al.}(1989){Luo}, {Pradhan}, {Saraph}, {Storey}, \&
  {Yu}}]{LPS89_OIII_phot}
{Luo}, D., {Pradhan}, A.~K., {Saraph}, H.~E., {Storey}, P.~J., \& {Yu}, Y.
  1989, Journal of Physics B Atomic Molecular Physics, 22, 389

\bibitem[{{Maeder}(1987)}]{maeder87}
{Maeder}, A. 1987, \aap, 178, 159

\bibitem[{{Maeder}(1999)}]{maeder99}
{Maeder}, A. 1999, \aap, 347, 185

\bibitem[{{Maeder} \& {Meynet}(2000)}]{maeder00}
{Maeder}, A. \& {Meynet}, G. 2000, \araa, 38, 143

\bibitem[{{Maeder} \& {Meynet}(2001)}]{maeder01}
{Maeder}, A. \& {Meynet}, G. 2001, \aap, 373, 555

\bibitem[{{Maeder} {et~al.}(2009){Maeder}, {Meynet}, {Ekstr{\"o}m}, \&
  {Georgy}}]{maeder09}
{Maeder}, A., {Meynet}, G., {Ekstr{\"o}m}, S., \& {Georgy}, C. 2009,
  Communications in Asteroseismology, 158, 72

\bibitem[{{Ma{\'{\i}}z Apell{\'a}niz} \& {Rubio}(2012)}]{maiz12}
{Ma{\'{\i}}z Apell{\'a}niz}, J. \& {Rubio}, M. 2012, \aap, 541, A54

\bibitem[{{Ma{\'{\i}}z-Apell{\'a}niz}
  {et~al.}(2004){Ma{\'{\i}}z-Apell{\'a}niz}, {Walborn}, {Galu{\'e}}, \&
  {Wei}}]{maiz04}
{Ma{\'{\i}}z-Apell{\'a}niz}, J., {Walborn}, N.~R., {Galu{\'e}}, H.~{\'A}., \&
  {Wei}, L.~H. 2004, \apjs, 151, 103

\bibitem[{{Marcolino} {et~al.}(2009){Marcolino}, {Bouret}, {Martins},
  {Hillier}, {Lanz}, \& {Escolano}}]{marcolino09}
{Marcolino}, W.~L.~F., {Bouret}, J.-C., {Martins}, F., {et~al.} 2009, \aap,
  498, 837

\bibitem[{{Markova} {et~al.}(2011){Markova}, {Markov}, {Puls},
  {Sim{\'o}n-D{\'{\i}}az}, \& {Herrero}}]{markova11}
{Markova}, N., {Markov}, H., {Puls}, J., {Sim{\'o}n-D{\'{\i}}az}, S., \&
  {Herrero}, A. 2011, Bulgarian Astronomical Journal, 17, 54

\bibitem[{{Martins} {et~al.}(2012{\natexlab{a}}){Martins}, {Escolano}, {Wade},
  {Donati}, {Bouret}, \& {Mimes Collaboration}}]{martins12c}
{Martins}, F., {Escolano}, C., {Wade}, G.~A., {et~al.} 2012{\natexlab{a}},
  \aap, 538, A29

\bibitem[{{Martins} \& {Hillier}(2012)}]{martins12b}
{Martins}, F. \& {Hillier}, D.~J. 2012, \aap, 545, A95

\bibitem[{{Martins} {et~al.}(2008){Martins}, {Hillier}, {Paumard},
  {Eisenhauer}, {Ott}, \& {Genzel}}]{martins08}
{Martins}, F., {Hillier}, D.~J., {Paumard}, T., {et~al.} 2008, \aap, 478, 219

\bibitem[{{Martins} {et~al.}(2012{\natexlab{b}}){Martins}, {Mahy}, {Hillier},
  \& {Rauw}}]{martins12a}
{Martins}, F., {Mahy}, L., {Hillier}, D.~J., \& {Rauw}, G. 2012{\natexlab{b}},
  \aap, 538, A39

\bibitem[{{Martins} {et~al.}(2005{\natexlab{a}}){Martins}, {Schaerer}, \&
  {Hillier}}]{martins05}
{Martins}, F., {Schaerer}, D., \& {Hillier}, D.~J. 2005{\natexlab{a}}, \aap,
  436, 1049

\bibitem[{{Martins} {et~al.}(2005{\natexlab{b}}){Martins}, {Schaerer},
  {Hillier}, {Meynadier}, {Heydari-Malayeri}, \& {Walborn}}]{martins05b}
{Martins}, F., {Schaerer}, D., {Hillier}, D.~J., {et~al.} 2005{\natexlab{b}},
  \aap, 441, 735

\bibitem[{{Massey}(2002)}]{massey02}
{Massey}, P. 2002, \apjs, 141, 81

\bibitem[{{Massey} {et~al.}(1989){Massey}, {Parker}, \& {Garmany}}]{massey89}
{Massey}, P., {Parker}, J.~W., \& {Garmany}, C.~D. 1989, \aj, 98, 1305

\bibitem[{{Massey} {et~al.}(2005){Massey}, {Puls}, {Pauldrach}, {Bresolin},
  {Kudritzki}, \& {Simon}}]{massey05}
{Massey}, P., {Puls}, J., {Pauldrach}, A.~W.~A., {et~al.} 2005, \apj, 627, 477

\bibitem[{{Massey} {et~al.}(2009){Massey}, {Zangari}, {Morrell}, {Puls},
  {DeGioia-Eastwood}, {Bresolin}, \& {Kudritzki}}]{massey09}
{Massey}, P., {Zangari}, A.~M., {Morrell}, N.~I., {et~al.} 2009, \apj, 692, 618

\bibitem[{{McLaughlin} \& {Bell}(2000)}]{MB00_NeIII_col}
{McLaughlin}, B.~M. \& {Bell}, K.~L. 2000, Journal of Physics B Atomic
  Molecular Physics, 33, 597

\bibitem[{{Mendoza}(1983)}]{Men83_col}
{Mendoza}, C. 1983, in IAU Symposium, Vol. 103, Planetary Nebulae, ed.
  {D.~R.~Flower}, 143--172

\bibitem[{{Meynet} {et~al.}(2011){Meynet}, {Eggenberger}, \&
  {Maeder}}]{meynet11}
{Meynet}, G., {Eggenberger}, P., \& {Maeder}, A. 2011, \aap, 525, L11

\bibitem[{{Meynet} \& {Maeder}(2000)}]{meynet00}
{Meynet}, G. \& {Maeder}, A. 2000, \aap, 361, 101

\bibitem[{{Meynet} \& {Maeder}(2005)}]{meynet05}
{Meynet}, G. \& {Maeder}, A. 2005, \aap, 429, 581

\bibitem[{{Mihalas} {et~al.}(1975){Mihalas}, {Heasley}, \& {Auer}}]{MHK75_prog}
{Mihalas}, D., {Heasley}, J.~N., \& {Auer}, L.~H. 1975, NASA STI/Recon
  Technical Report N, 76, 30128

\bibitem[{{Mokiem} {et~al.}(2006){Mokiem}, {de Koter}, {Evans}, {Puls},
  {Smartt}, {Crowther}, {Herrero}, {Langer}, {Lennon}, {Najarro}, {Villamariz},
  \& {Yoon}}]{mokiem06}
{Mokiem}, M.~R., {de Koter}, A., {Evans}, C.~J., {et~al.} 2006, \aap, 456, 1131

\bibitem[{{Mokiem} {et~al.}(2007){Mokiem}, {de Koter}, {Vink}, {Puls}, {Evans},
  {Smartt}, {Crowther}, {Herrero}, {Langer}, {Lennon}, {Najarro}, \&
  {Villamariz}}]{mokiem07}
{Mokiem}, M.~R., {de Koter}, A., {Vink}, J.~S., {et~al.} 2007, \aap, 473, 603

\bibitem[{{Muijres} {et~al.}(2012){Muijres}, {Vink}, {de Koter}, {M{\"u}ller},
  \& {Langer}}]{muijres12}
{Muijres}, L.~E., {Vink}, J.~S., {de Koter}, A., {M{\"u}ller}, P.~E., \&
  {Langer}, N. 2012, \aap, 537, A37

\bibitem[{{Nahar}(1998)}]{Nah98_Ophot}
{Nahar}, S.~N. 1998, \pra, 58, 3766

\bibitem[{{Nahar} \& {Pradhan}(1993)}]{NP93_sil}
{Nahar}, S.~N. \& {Pradhan}, A.~K. 1993, Journal of Physics B Atomic Molecular
  Physics, 26, 1109

\bibitem[{{Najarro} {et~al.}(2011){Najarro}, {Hanson}, \& {Puls}}]{najarro11}
{Najarro}, F., {Hanson}, M.~M., \& {Puls}, J. 2011, \aap, 535, A32

\bibitem[{{Najarro} {et~al.}(2006){Najarro}, {Hillier}, {Puls}, {Lanz}, \&
  {Martins}}]{NHP06_HeI}
{Najarro}, F., {Hillier}, D.~J., {Puls}, J., {Lanz}, T., \& {Martins}, F. 2006,
  \aap, 456, 659

\bibitem[{{Nussbaumer} \& {Storey}(1979{\natexlab{a}})}]{NS79_NIII}
{Nussbaumer}, H. \& {Storey}, P.~J. 1979{\natexlab{a}}, \aap, 71, L5

\bibitem[{{Nussbaumer} \& {Storey}(1979{\natexlab{b}})}]{NS79_Inter}
{Nussbaumer}, H. \& {Storey}, P.~J. 1979{\natexlab{b}}, \aap, 74, 244

\bibitem[{{Nussbaumer} \& {Storey}(1983)}]{NS83_LTDR}
{Nussbaumer}, H. \& {Storey}, P.~J. 1983, \aap, 126, 75

\bibitem[{{Nussbaumer} \& {Storey}(1984)}]{NS84_CNO_LTDR}
{Nussbaumer}, H. \& {Storey}, P.~J. 1984, \aaps, 56, 293

\bibitem[{{Owocki} {et~al.}(1996){Owocki}, {Cranmer}, \& {Gayley}}]{owocki96}
{Owocki}, S.~P., {Cranmer}, S.~R., \& {Gayley}, K.~G. 1996, \apjl, 472, L115+

\bibitem[{{Parker} {et~al.}(1992){Parker}, {Garmany}, {Massey}, \&
  {Walborn}}]{parker92}
{Parker}, J.~W., {Garmany}, C.~D., {Massey}, P., \& {Walborn}, N.~R. 1992, \aj,
  103, 1205

\bibitem[{{Peach} {et~al.}(1988){Peach}, {Saraph}, \& {Seaton}}]{PSS88_LI_seq}
{Peach}, G., {Saraph}, H.~E., \& {Seaton}, M.~J. 1988, Journal of Physics B
  Atomic Molecular Physics, 21, 3669

\bibitem[{{Penny} \& {Gies}(2009)}]{penny09}
{Penny}, L.~R. \& {Gies}, D.~R. 2009, \apj, 700, 844

\bibitem[{{Preibisch} {et~al.}(1999){Preibisch}, {Balega}, {Hofmann},
  {Weigelt}, \& {Zinnecker}}]{preibisch99}
{Preibisch}, T., {Balega}, Y., {Hofmann}, K.-H., {Weigelt}, G., \& {Zinnecker},
  H. 1999, \na, 4, 531

\bibitem[{{Przybilla} {et~al.}(2010){Przybilla}, {Firnstein}, {Nieva},
  {Meynet}, \& {Maeder}}]{przybilla10}
{Przybilla}, N., {Firnstein}, M., {Nieva}, M.~F., {Meynet}, G., \& {Maeder}, A.
  2010, \aap, 517, A38+

\bibitem[{{Puls} {et~al.}(2006){Puls}, {Markova}, {Scuderi}, {Stanghellini},
  {Taranova}, {Burnley}, \& {Howarth}}]{puls06}
{Puls}, J., {Markova}, N., {Scuderi}, S., {et~al.} 2006, \aap, 454, 625

\bibitem[{{Puls} {et~al.}(2008){Puls}, {Vink}, \& {Najarro}}]{puls08}
{Puls}, J., {Vink}, J.~S., \& {Najarro}, F. 2008, \aapr, 16, 209

\bibitem[{{Repolust} {et~al.}(2004){Repolust}, {Puls}, \&
  {Herrero}}]{repolust04}
{Repolust}, T., {Puls}, J., \& {Herrero}, A. 2004, \aap, 415, 349

\bibitem[{{Rivero Gonz{\'a}lez} {et~al.}(2012){Rivero Gonz{\'a}lez}, {Puls},
  {Massey}, \& {Najarro}}]{rivero12}
{Rivero Gonz{\'a}lez}, J.~G., {Puls}, J., {Massey}, P., \& {Najarro}, F. 2012,
  \aap, 543, A95

\bibitem[{{Rivero Gonz{\'a}lez} {et~al.}(2011){Rivero Gonz{\'a}lez}, {Puls}, \&
  {Najarro}}]{rivero11}
{Rivero Gonz{\'a}lez}, J.~G., {Puls}, J., \& {Najarro}, F. 2011, \aap, 536, A58

\bibitem[{{Runacres} \& {Owocki}(2002)}]{runacres02}
{Runacres}, M.~C. \& {Owocki}, S.~P. 2002, \aap, 381, 1015

\bibitem[{{Sana} {et~al.}(2012{\natexlab{a}}){Sana}, {de Koter}, {de Mink},
  {Dunstall}, {Evans}, {Henault-Brunet}, {Maiz Apellaniz}, {Ramirez-Agudelo},
  {Taylor}, {Walborn}, {Clark}, {Crowther}, {Herrero}, {Gieles}, {Langer},
  {Lennon}, \& {Vink}}]{sana12a}
{Sana}, H., {de Koter}, A., {de Mink}, S.~E., {et~al.} 2012{\natexlab{a}},
  ArXiv e-prints

\bibitem[{{Sana} {et~al.}(2012{\natexlab{b}}){Sana}, {de Mink}, {de Koter},
  {Langer}, {Evans}, {Gieles}, {Gosset}, {Izzard}, {Le Bouquin}, \&
  {Schneider}}]{sana12b}
{Sana}, H., {de Mink}, S.~E., {de Koter}, A., {et~al.} 2012{\natexlab{b}},
  Science, 337, 444

\bibitem[{{Sana} {et~al.}(2011){Sana}, {James}, \& {Gosset}}]{sana11}
{Sana}, H., {James}, G., \& {Gosset}, E. 2011, \mnras, 416, 817

\bibitem[{{Sana} {et~al.}(2006){Sana}, {Rauw}, {Naz{\'e}}, {Gosset}, \&
  {Vreux}}]{sana06}
{Sana}, H., {Rauw}, G., {Naz{\'e}}, Y., {Gosset}, E., \& {Vreux}, J. 2006,
  \mnras, 372, 661

\bibitem[{{Sch\"oning} \& {Butler}(1989{\natexlab{a}})}]{SB89errata_stark}
{Sch\"oning}, T. \& {Butler}, K. 1989{\natexlab{a}}, \aaps, 79, 153

\bibitem[{{Sch\"oning} \& {Butler}(1989{\natexlab{b}})}]{SB89sup_stark}
{Sch\"oning}, T. \& {Butler}, K. 1989{\natexlab{b}}, \aaps, 78, 51

\bibitem[{{Sch\"oning} \& {Butler}(1989{\natexlab{c}})}]{SB89_stark}
{Sch\"oning}, T. \& {Butler}, K. 1989{\natexlab{c}}, \aap, 219, 326

\bibitem[{{Seaton}(1987)}]{Sea87_OP}
{Seaton}, M.~J. 1987, Journal of Physics B Atomic Molecular Physics, 20, 6363

\bibitem[{{Shamey}(1969)}]{Shamey69PhDT........47S}
{Shamey}, L.~J. 1969, PhD thesis, UNIVERSITY OF COLORADO AT BOULDER.

\bibitem[{{Sim{\'o}n-D{\'{\i}}az} \& {Herrero}(2007)}]{simon07}
{Sim{\'o}n-D{\'{\i}}az}, S. \& {Herrero}, A. 2007, \aap, 468, 1063

\bibitem[{{Sim{\'o}n-D{\'{\i}}az} {et~al.}(2010){Sim{\'o}n-D{\'{\i}}az},
  {Herrero}, {Uytterhoeven}, {Castro}, {Aerts}, \& {Puls}}]{simon10b}
{Sim{\'o}n-D{\'{\i}}az}, S., {Herrero}, A., {Uytterhoeven}, K., {et~al.} 2010,
  \apjl, 720, L174

\bibitem[{{Tayal}(1997)}]{Tay97_SIII_col}
{Tayal}, S.~S. 1997, \apj, 481, 550

\bibitem[{{Tr{\"a}bert} {et~al.}(1999){Tr{\"a}bert}, {Gwinner}, {Knystautas},
  {Tordoir}, \& {Wolf}}]{TGK99_CII_NIII}
{Tr{\"a}bert}, E., {Gwinner}, G., {Knystautas}, E.~J., {Tordoir}, X., \&
  {Wolf}, A. 1999, Journal of Physics B Atomic Molecular Physics, 32, L491

\bibitem[{{Tully} {et~al.}(1990){Tully}, {Seaton}, \&
  {Berrington}}]{TSB90_Be_seq}
{Tully}, J.~A., {Seaton}, M.~J., \& {Berrington}, K.~A. 1990, Journal of
  Physics B Atomic Molecular Physics, 23, 3811

\bibitem[{{Venn}(1999)}]{venn99}
{Venn}, K.~A. 1999, \apj, 518, 405

\bibitem[{{Vink} {et~al.}(2001){Vink}, {de Koter}, \& {Lamers}}]{vink01}
{Vink}, J.~S., {de Koter}, A., \& {Lamers}, H.~J.~G.~L.~M. 2001, \aap, 369, 574

\bibitem[{{von Zeipel}(1924)}]{vonzeipel24}
{von Zeipel}, H. 1924, \mnras, 84, 665

\bibitem[{{Walborn}(2001)}]{walborn01}
{Walborn}, N. 2001, in ASP Conf.\ Ser., Vol. 242, Eta Carinae and Other
  Mysterious Stars: The Hidden Opportunities of Emission Spectroscopy, ed.
  {T.~R.~Gull, S.~Johannson, \& K.~Davidson} (San Francisco: ASP), 217--+

\bibitem[{{Walborn}(2009)}]{walborn09}
{Walborn}, N.~R. 2009, {Optically observable zero-age main-sequence O stars:},
  ed. {Livio, M.~\& Villaver, E.} (Cambridge: Cambridge University Press),
  167--177

\bibitem[{{Walborn} \& {Blades}(1997)}]{walborn97}
{Walborn}, N.~R. \& {Blades}, J.~C. 1997, \apjs, 112, 457

\bibitem[{{Walborn} \& {Howarth}(2000)}]{walborn00}
{Walborn}, N.~R. \& {Howarth}, I.~D. 2000, \pasp, 112, 1446

\bibitem[{{Walborn} {et~al.}(2000){Walborn}, {Lennon}, {Heap}, {Lindler},
  {Smith}, \& {Evans}}]{walbornSMC}
{Walborn}, N.~R., {Lennon}, D.~J., {Heap}, S.~R., {et~al.} 2000, \pasp, 112,
  1243

\bibitem[{{Walborn} {et~al.}(2004){Walborn}, {Morrell}, {Howarth}, {Crowther},
  {Lennon}, {Massey}, \& {Arias}}]{walborn04}
{Walborn}, N.~R., {Morrell}, N.~I., {Howarth}, I.~D., {et~al.} 2004, \apj, 608,
  1028

\bibitem[{{Walborn} \& {Nichols-Bohlin}(1987)}]{walborn87}
{Walborn}, N.~R. \& {Nichols-Bohlin}, J. 1987, \pasp, 99, 40

\bibitem[{{Walborn} \& {Panek}(1984)}]{walborn84}
{Walborn}, N.~R. \& {Panek}, R.~J. 1984, \apjl, 280, L27

\bibitem[{{Walborn} {et~al.}(2013){Walborn}, {Sana}, {Taylor},
  {Sim{\'o}n-D{\'{\i}}az}, \& {Evans}}]{walborn13}
{Walborn}, N.~R., {Sana}, H., {Taylor}, W.~D., {Sim{\'o}n-D{\'{\i}}az}, S., \&
  {Evans}, C.~J. 2013, in Astronomical Society of the Pacific Conference
  Series, Vol. 465, Astronomical Society of the Pacific Conference Series, ed.
  L.~{Drissen}, C.~{Rubert}, N.~{St-Louis}, \& A.~F.~J. {Moffat}, 490

\bibitem[{{Witt} \& {Gordon}(2000)}]{witt00}
{Witt}, A.~N. \& {Gordon}, K.~D. 2000, \apj, 528, 799

\bibitem[{{Zhang} {et~al.}(1994){Zhang}, {Graziani}, \&
  {Pradhan}}]{ZCO94_B_like}
{Zhang}, H.~L., {Graziani}, M., \& {Pradhan}, A.~K. 1994, \aap, 283, 319

\bibitem[{{Zhang} \& {Pradhan}(1997)}]{ZP97_FeIV_col}
{Zhang}, H.~L. \& {Pradhan}, A.~K. 1997, \aaps, 126, 373

\end{thebibliography}

\begin{appendix}
\label{appendice}

\section{Atomic data}

We provide below a brief description of the atomic data used in the present set of calculations. For many atomic species different atomic data sets are available. In the present calculations we did not always use the latest data set, although test calculations show that the choice of other data sets would not affect our conclusions. 

For H we used collision strengths adapted from \cite{MHK75_prog}, which are similar to those used in {\sc tlusty} \citep{HL95_TLUSTY}. For the Stark broadening of optical hydrogen lines we used the results from \cite{Lem97_stark}; for the optical lines they give similar results as are found in \cite{SB89_stark}. Collision rates for He\,{\sc i} were taken from \cite{BK87_HeI_col}, while photoionization cross-sections are ``below-resonance''  fits to the calculations of \cite{FST87_Hei}. For He\,{\sc ii} we used the Stark-
broadening calculations of \cite{SB89_stark,SB89sup_stark,SB89errata_stark}. For He\,{I} $\lambda$4471 and $\lambda$4922 we used data from \cite{BCS74_HeI_stark}, for $\lambda$4389 and $\lambda$4026 the data was taken from \cite{Shamey69PhDT........47S} (kindly supplied by Keith Butler), while for other He\,{i} lines we used data from \cite{DS84_Hei}. Other data for H and He are discussed in \cite{Hil87a}.

C\,{\sc iv} oscillator strengths were taken from \cite{Lei72_CIV} while the photoionization data were taken from \cite{PSS88_LI_seq} and \cite{Lei72_CIV}. 
C\,{\sc iii} oscillator strengths, photoionization data, and dielectronic data were taken from P.~J. Storey (private communication), while collision rates for the lowest six terms were taken from \cite{BBD85_col}.

 N\,{\sc v} oscillator strengths for $n < 6$ were taken from \cite{LB11_Li_like}, the photoionization data were taken from \cite{PSS88_LI_seq}, and collisional data for the 2s-2p, 2s-3 and 2p-3 transitions were taken from \cite{CM83_lith}. N\,{\sc iv} oscillator strengths and photoionization cross-sections were computed in \cite{TSB90_Be_seq} as part of the opacity project \citep{Sea87_OP,Topbase93}. The identification of the 2s\,4s\,$^3$S and 2p\,3p\,$^3$S states were switched \citep[see][]{AAL91_NIV}. Oscillator strengths for N\,{\sc iv} forbidden (and semi-forbidden) lines were taken from \cite{NS79_Inter},  while for oscillator strengths for low-temperature  dielectronic recombination transitions, we followed \cite{NS83_LTDR,NS84_CNO_LTDR} (private communication). Transition probabilities for the N\,{\sc iii} intercombination lines were taken from \cite{NS79_NIII} which agree reasonably well with the experimental values found in the recent study of \cite{TGK99_CII_NIII}.
 
O\,{\sc v} oscillator strengths were taken from \cite[][private communication]{NS83_LTDR,NS84_CNO_LTDR}. Collision rates for the six lowest terms of O\,{\sc v} were taken from \cite{BBD85_col}. O\,{\sc iv} oscillator strengths were computed as part of the opacity project \citep{Sea87_OP} and were obtained from {\sc topbase} \citep{Topbase93}.   Intercombination data for O\,{\sc iv} is from the compilation of \cite{Men83_col}.  O\,{\sc iv} photoionization data were taken from \cite{Nah98_Ophot} and were obtained through {\sc norad}\footnote{http://www.astronomy.ohio-state.edu/$\sim$nahar/nahar\_radiativeatomicdata/}. O\,{\sc iv} collision rates were taken from \cite{ZCO94_B_like}. O\,{\sc iii} oscillator strengths and photoionization data were taken from \cite{LPS89_OIII_phot}.  Collisional data for the four lowest terms in O\,{\sc iii} were taken from the compilation of \cite{Men83_col}.

Oscillator strengths for Ne\,{\sc iii} and Ne\,{\sc iv} were taken from the compilation of \cite{KP75_linelist} and the opacity project \citep{Sea87_OP}. The Ne\,{\sc iii} and Ne\,{\sc iv} photoionization cross-sections were obtained from the opacity project \citep{Sea87_OP,Topbase93}. Collisional data for the four lowest terms in Ne\,{\sc v} were taken from \cite{LB94_col}, collisional data for the three lowest terms  of Ne\,{\sc iv} were taken from the compilation of \cite{Men83_col}, while collisional data for Ne\,{\sc iii} were taken from \cite{MB00_NeIII_col}. 

Oscillator strengths and photoionization cross-sections for  N\,{\sc iii}, Ne\,{\sc v} \citep{LP89_C_seq}, Si\,{\sc iii} \citep{BKM93_Mg_seq}, Si\,{\sc iv}, S\,{\sc iii} \citep{NP93_sil}, and S\,{\sc iv} were obtained from {\sc topbase} \citep{Topbase93}, and were  computed as part of the opacity project \citep{Sea87_OP}. Intercombination data for S\,{\sc iv} were taken from the compilation of \cite{Men83_col} while collision strengths between the $^2$P$^o$ and $^4$P levels were taken from the compilation of \cite{Men83_col}. Collisional data for the four lowest terms for S\,{\sc iii} were taken from \cite{Tay97_SIII_col}.
 
Oscillator strengths for Fe\,{\sc iv}, Fe\,{\sc v}, and Fe\,{\sc vi} were computed by \cite{Kur09_ATD} and were obtained (before 2001) from Bob Kurucz's website\footnote{http://kurucz.harvard.edu\/}. The oscillator strengths of two Fe\,{\sc iv} lines that overlap with the He\,{\sc i} resonance line near 58.4 nm were reduced by a factor of 10 \citep[see][]{NHP06_HeI}. Photoionization cross-sections for Fe\,{\sc iv}, Fe\,{\sc v}, Fe\,{\sc vi} and Fe\,{\sc vii} were computed as part of the opacity project \citep{Sea87_OP} and were obtained from {\sc topbase} \citep{Topbase93}. Fe\,{\sc iv} collisional data is from \cite{ZP97_FeIV_col}.

\section{Best fits}
In this appendix we give more details of the parameter determination of each individual star and compare our best-fit models to the optical and UV spectra.
Absolute fluxes in UV spectra are provided after correcting for reddening (see Sect. \ref{param_sect} and Table \ref{tabone}) and are 
expressed in ergs\,.\,\AA$^{-1}$.\,s$^{-1}$.\,cm$^{-2}$ 

\subsection{\object{MPG 355}\ - {\rm ON2~III (f$^{*}$)}}
\label{m355_sect}
\object{MPG 355} was already studied extensively in the UV and optical in \cite{bouret03}, to which we refer for a thorough discussion of the \stis\ spectrum. 
More recently, \cite{massey09} obtained a new optical spectrum with a higher S/N of MPG 355; they detected a very weak \he\ \lb4471 (their Fig. 7) 
that is not visible in our own spectrum. This biased \cite{bouret03} toward a higher \teff. 

On the other hand, \hei\ \lb5876 is not detected in our high S/N \uves\ spectrum while this line would be expected
at \teff\ = 50,000~K in a test model we computed. 
Furthermore, several lines of \mbox{CNO} elements argue in favor of a higher effective temperature \teff\ = 52,500~K.  
This is the case of \civ\ \lb1169 and \ciii\ \lb1176, where the intensity of individual lines (the \ciii\ line is not detected in the
\stis\ spectrum) is best fitted for such a \teff. Other lines include \nv\ \lb\lb4604-4620  and \niv\ \lb6381, where both the intensity of individual lines 
and line ratio of these successives ionization stages argue for the adopted \teff. 
Finally, we found that emission from incoherent electron scattering in the wings of \heii\ \lb4686 is predicted somewhat too strong and that a higher 
effective temperature (55,000 K) would solve this problem. Overall, we are quite confident that the effective temperature of MPG 355 
as quoted in \cite{bouret03} and in the present paper is correct. The recent results by \cite{rivero12} also strengthen this conclusion and the
other quantities derived here.

Compared with our previous study, we only changed the nitrogen surface abundance to better fit the  \niv\ \lb6381, still maintaining a very good fit to other
lines from nitrogen ions. 

\begin{figure*}[tbp]
\includegraphics[scale=0.51, angle=0]{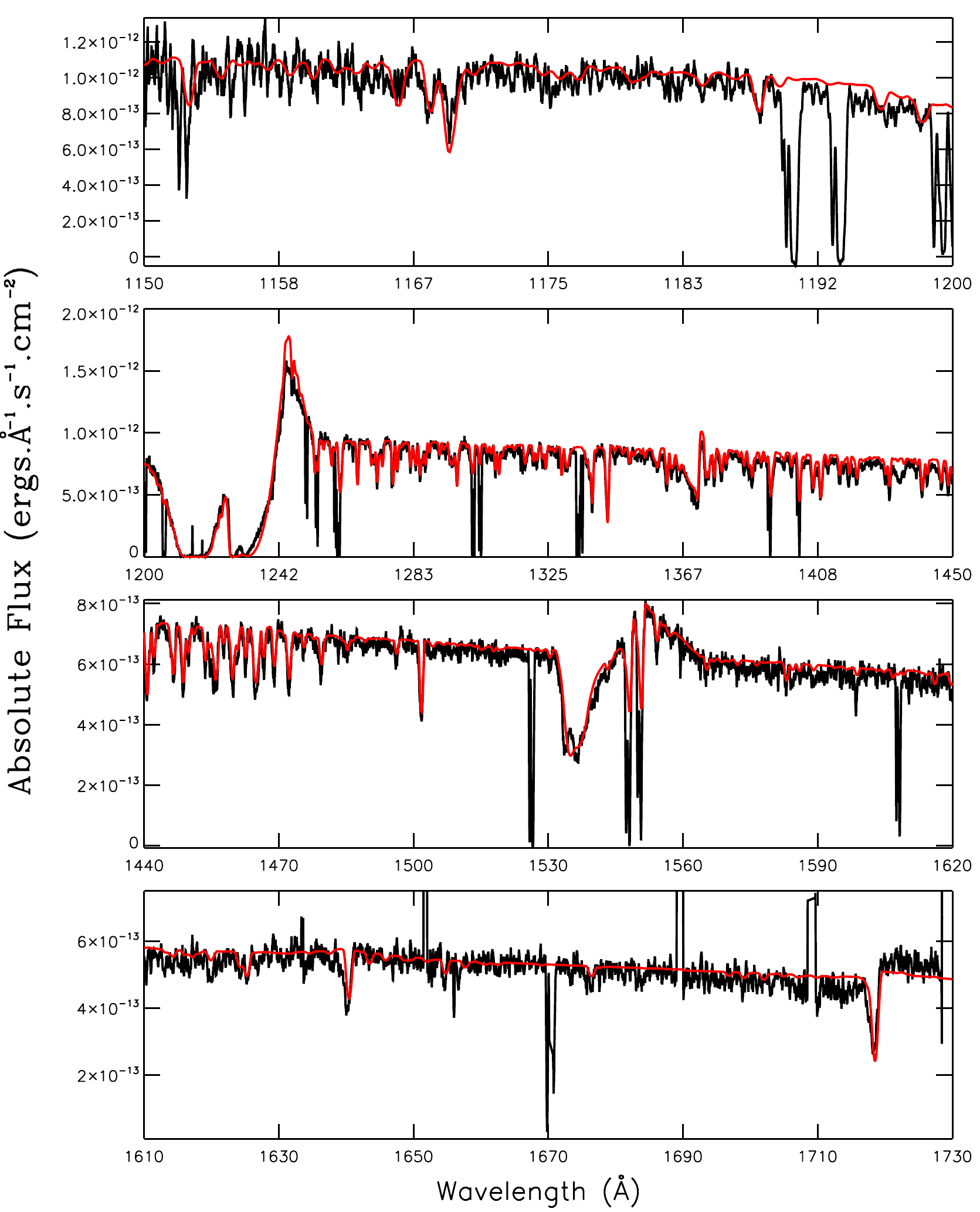}
\includegraphics[scale=0.51, angle=0]{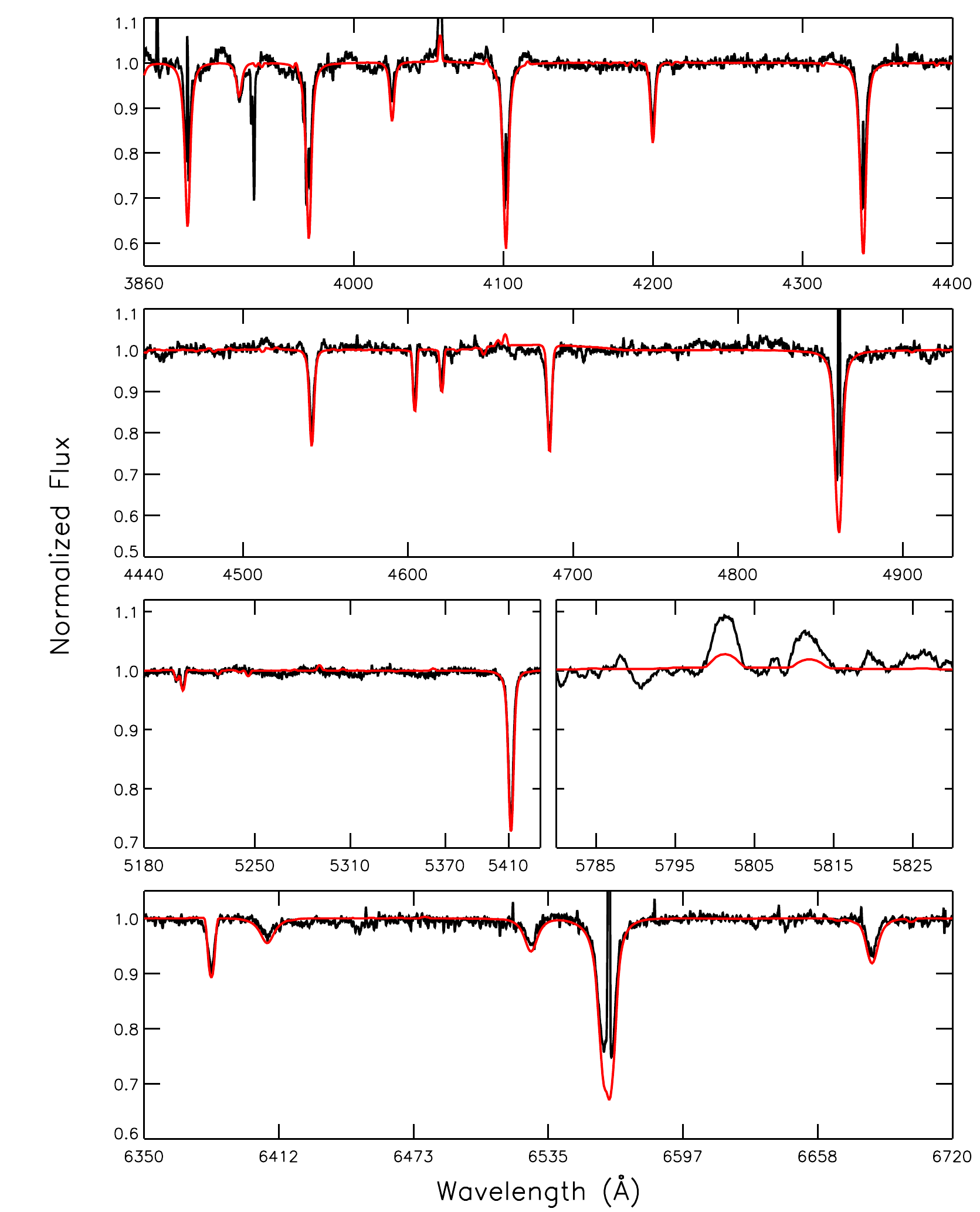}
      \caption[12cm]{Left: Best-fit model for \object{MPG 355} (red line) compared with the \cosp\ spectrum (black line). Right: Best-fit model for \object{MPG 355} (red line) compared with the \aat\ 
      and \uves\ spectra (black line).}
         \label{Fig_mpg355}
   \end{figure*}

 \subsection{\object{AzV 177}\  - {\rm O4~V}}
Detailed modeling of the optical spectrum of \object{AzV 177} was presented by \cite{massey05} and \cite{rivero12}. 
Significant differences are found from these studies that concern photospheric quantities such as the surface gravity or the helium abundance, and
the wind parameters. The 0.2 dex difference in \loggc\ is very surprising since we used the same optical spectrum (kindly provide by P. Massey)
and derived a similar \teff.
Our modeling of this star relies on the FUV \cosp\ spectrum, which offers a better grasp on the wind properties. We also
have several \mbox{CNO} lines at our disposal to constrain the surface abundances. The line ratio \civ\ \lb1169
to \ciii\ \lb1176 argues for a \teff\ slightly lower than in \cite{massey05} and \cite{rivero12}, although both estimates would agree within error bars. 
The mass-loss rates also match within a factor of two, which is a very good agreement given the clumping filling factor we adopted. 
The wind velocity exponent is slightly higher in our case ($\beta = 1.1$ rather than 0.8), a result mostly driven by the fit to the 
\civ\ resonance doublet. We note that this higher $\beta$ additionally precludes using He/H=0.15 from \cite{massey05} and \cite{rivero12}, because \heii\ 4686
would show strong emission wings (already too strong here), which are not observed.

\begin{figure*}[tbp]
\includegraphics[scale=0.51, angle=0]{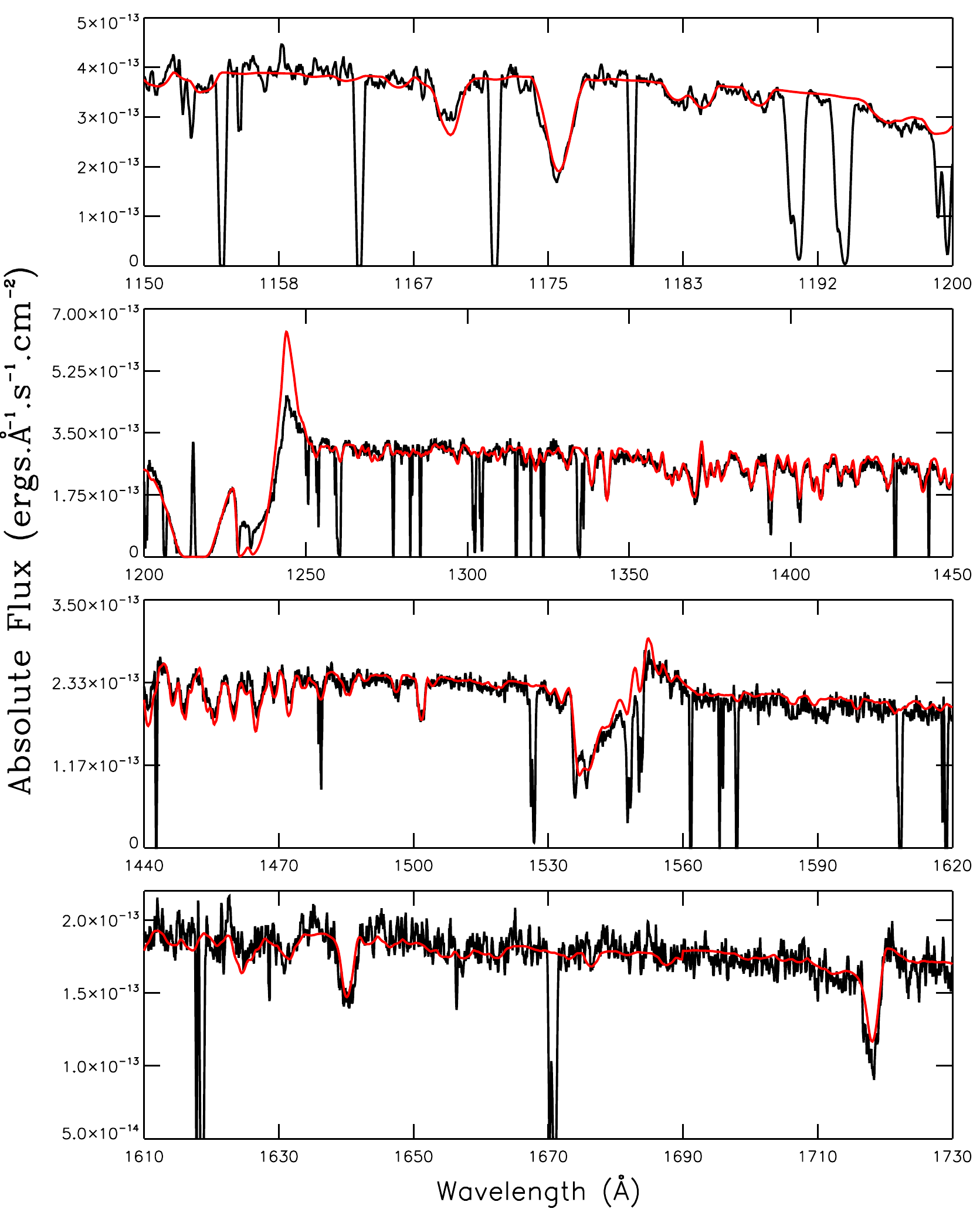}
\includegraphics[scale=0.51, angle=0]{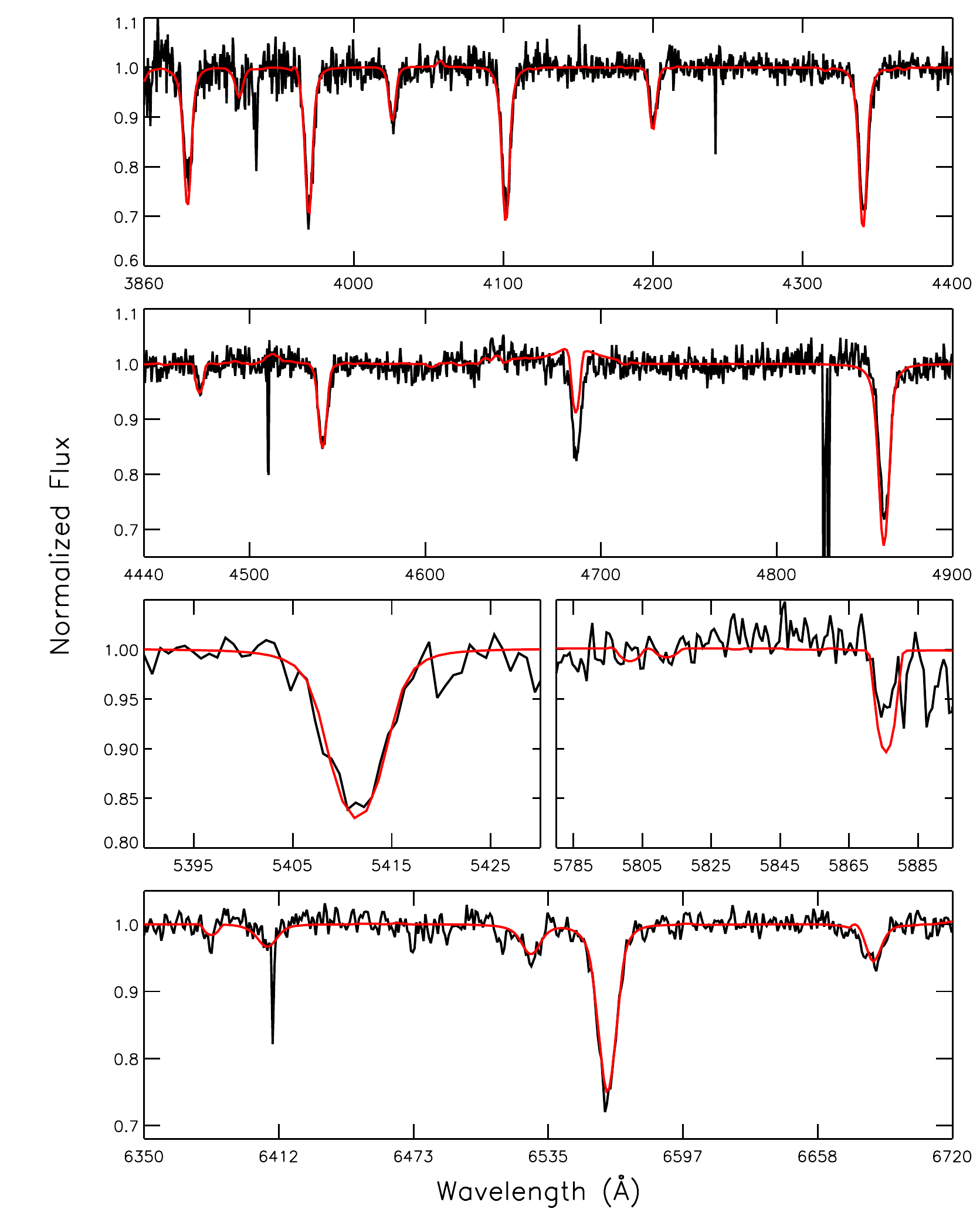}
      \caption[12cm]{Left: Best-fit model for \object{AzV 177} (red line) compared with the \cosp\ spectrum (black line). Right: Best-fit model for \object{AzV 177} (red line) compared with the 
      {\it CTIO} spectrum (black line). }
         \label{Fig_av177}
   \end{figure*}
  
\subsection{\object{AzV 388}\  - {\rm O4~V}}
\object{AzV 388} was included in the sample of SMC stars investigated in \cite{mokiem06}. We used the same optical spectrum,
provided by P. Crowther. The fit to this spectrum is excellent throughout the whole spectral range, including UV. The derived parameters
agree very well with those listed in \cite{mokiem06}, well within the error bars. The only major discrepancy is the 
mass-loss rate, which we found to be a factor 3.3 times lower than their value which is based mostly on \halpha. 
Still, the derived clumping filling factor $f = 0.11$ implies that our mass-loss rate would be scaled up by the exact
same factor, if the wind were assumed to be homogeneous. This last assumption is incompatible with the \ov\ \lb1371
line profile, however.
\begin{figure*}[tbp]
\includegraphics[scale=0.51, angle=0]{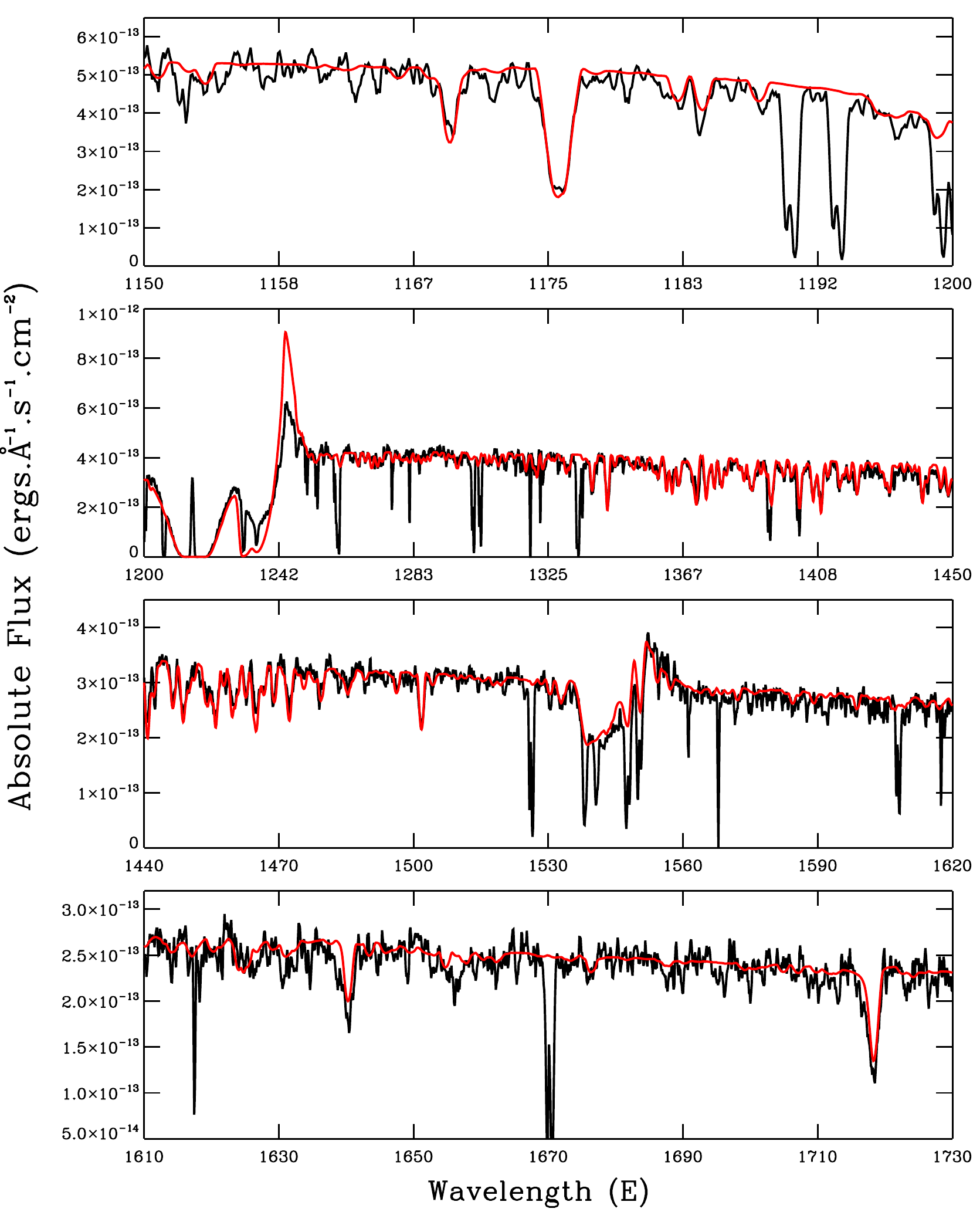}
\includegraphics[scale=0.51, angle=0]{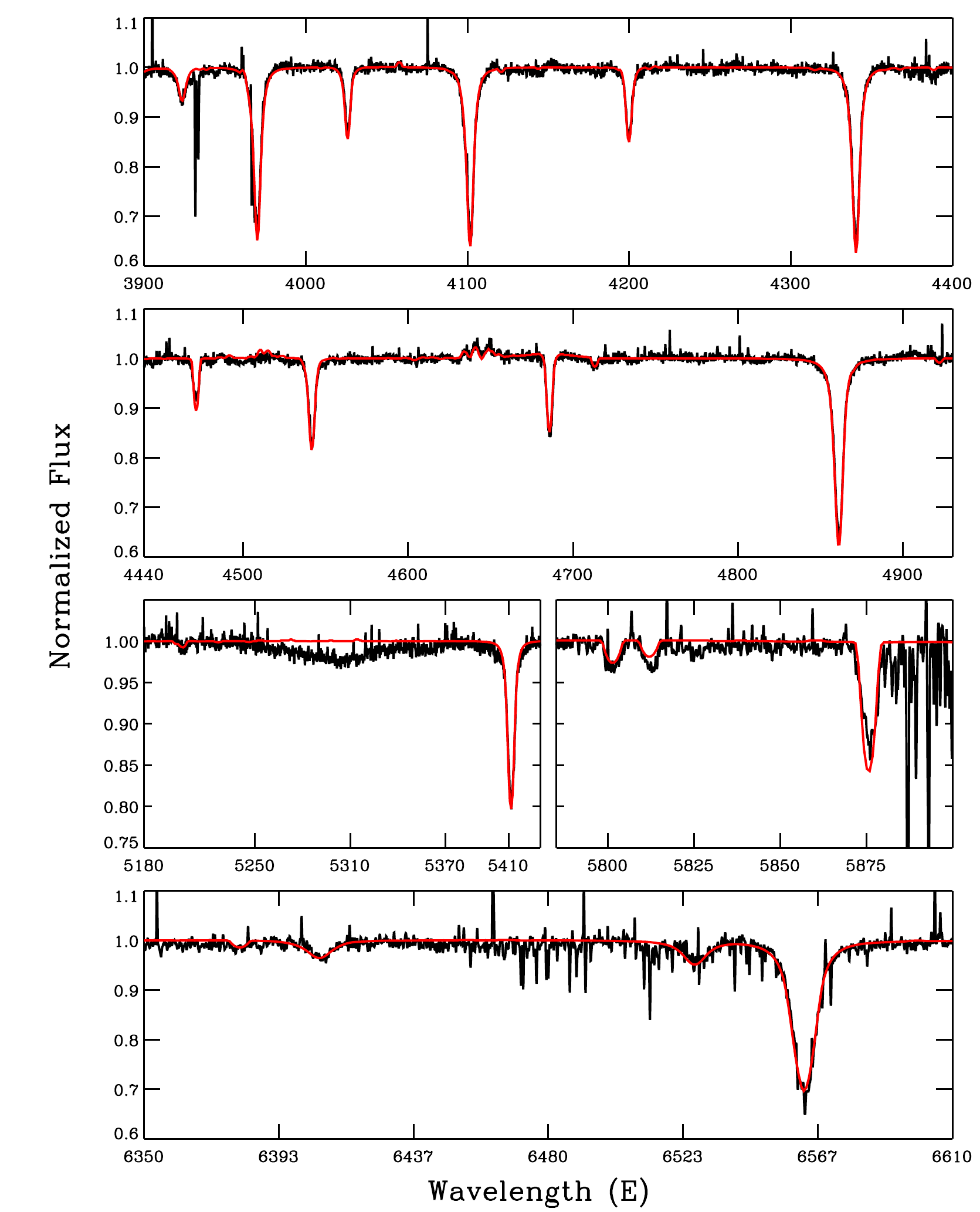}
\includegraphics[scale=1., angle=0]{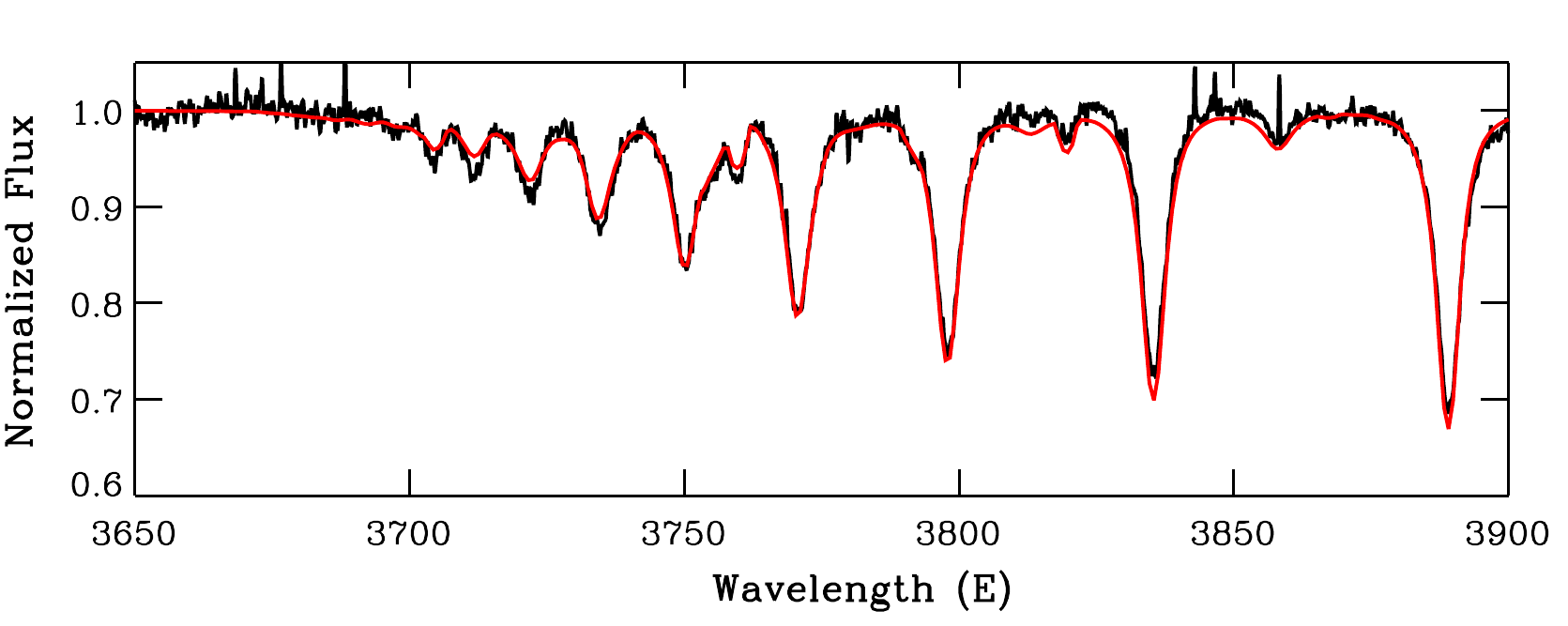}
      \caption[12cm]{Best-fit model for \object{AzV 388} (in red) compared with observed spectra (in black): the \cosp\ spectrum (upper left) and the \uves\ spectrum (upper right).
      A comparison with the \uves\ spectrum near the Balmer series is shown in the bottom panel.}
         \label{Fig_av388_2}
   \end{figure*}

\subsection{\object{MPG 324}\  - {\rm O4~V ((f))}}
\object{MPG 324} was studied in the UV and optical (\aat\ spectrum) in \cite{bouret03} and \cite{heap06} and was more recently included in two other studies by \cite{mokiem06}
and \cite{massey09}, both of them using the code \fastwind\ for the modeling of the optical spectrum. 
Using this time the same optical spectrum (\flames) as in \cite{mokiem06}, supplemented with our own \uves\ spectrum, the fundamental
and wind parameters we derived differ only moderately from our initial study. \teff\ was increased by 600~K (a change driven by the \hei\ 5876 line
observed with very good S/N ratio with \uves\ and no apparent nebular contamination) 
and agrees reasonably with other estimates for this star. The only significant difference with published results is
the projected rotation velocity, which we find to be two times lower (\vsini\ = 70 \kms\  vs 150 \kms) than quoted \cite{massey09}. Such a high value seems
inconsistent with the UV spectrum, however, especially with the \ciii\ \lb1176  multiplet. 
We finally note that from small velocity shifts observed in some lines, \cite{evans06} concluded that MPG 324 was probably a single-lined binary.  
\begin{figure*}[tbp]
\includegraphics[scale=0.51, angle=0]{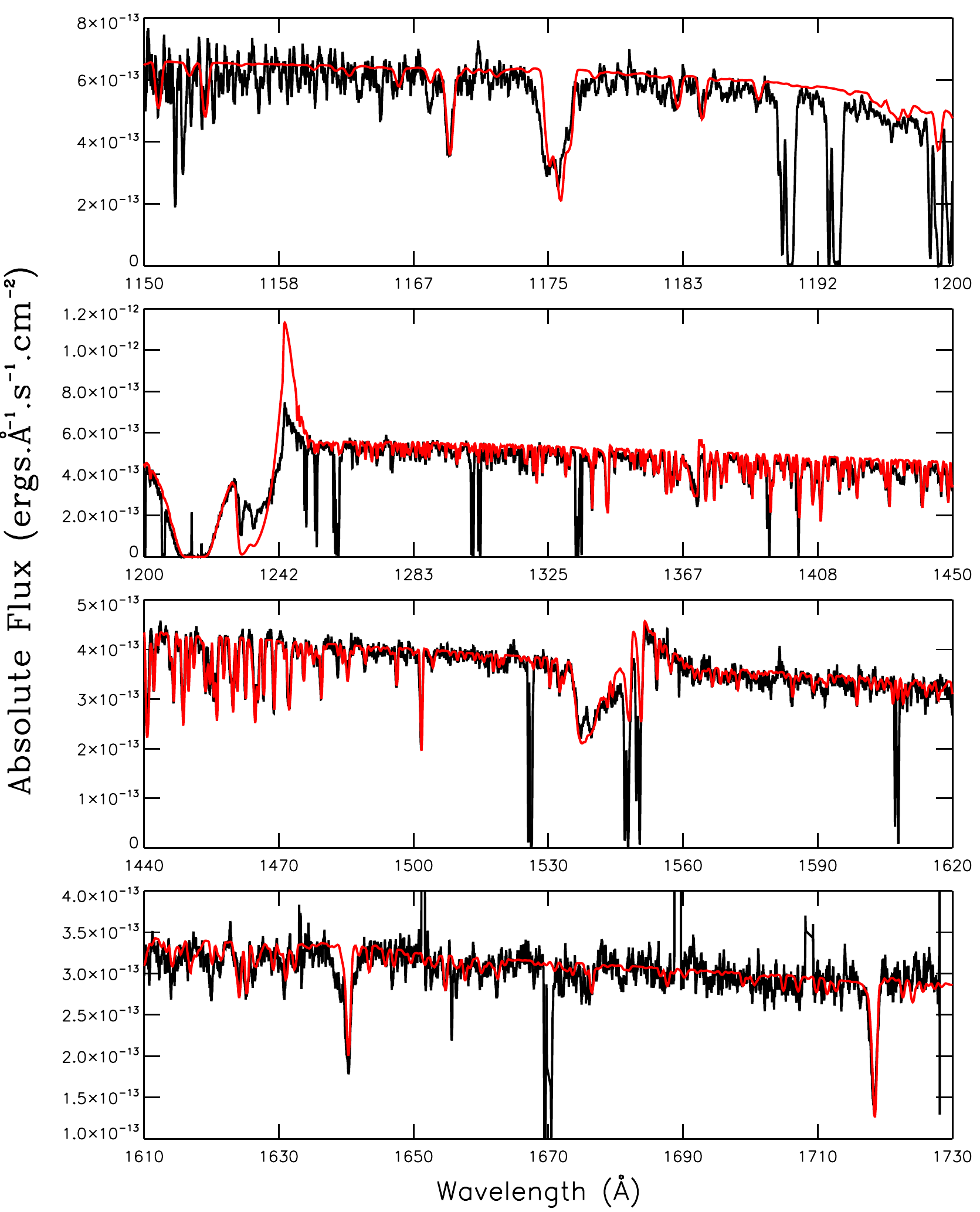}
\includegraphics[scale=0.51, angle=0]{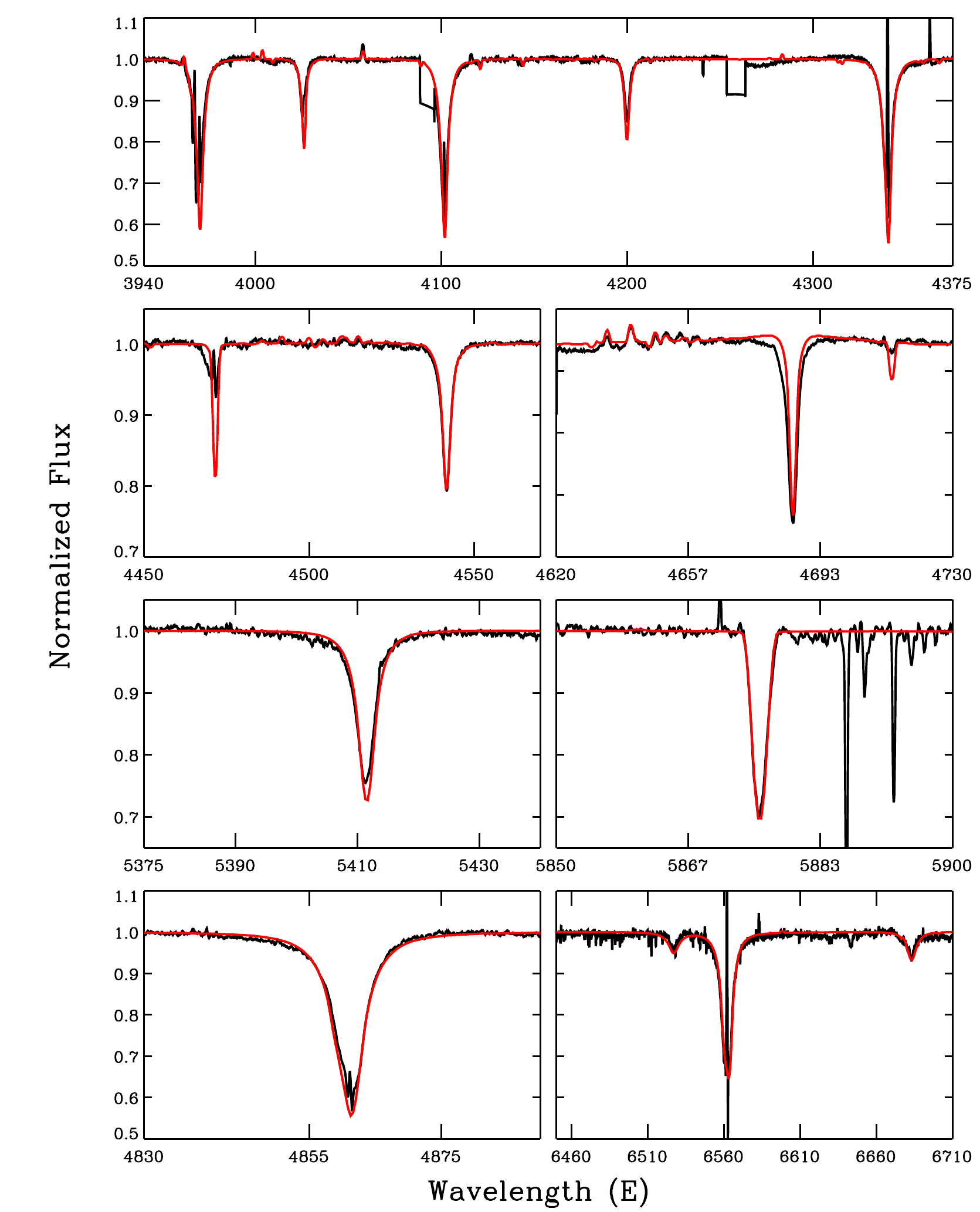}
      \caption[12cm]{Left: Best-fit model for \object{MPG 324} (red line) compared with the \stis\ spectrum (black line). Right: Best-fit model for \object{MPG 324} (red line) compared with the 
      \flames\ and \uves\ spectra (black line).}
         \label{Fig_mpg324}
   \end{figure*}
\subsection{\object{MPG 368}\  - {\rm O5~V ((f+))}}
\begin{figure*}[tbp]
\includegraphics[scale=0.51, angle=0]{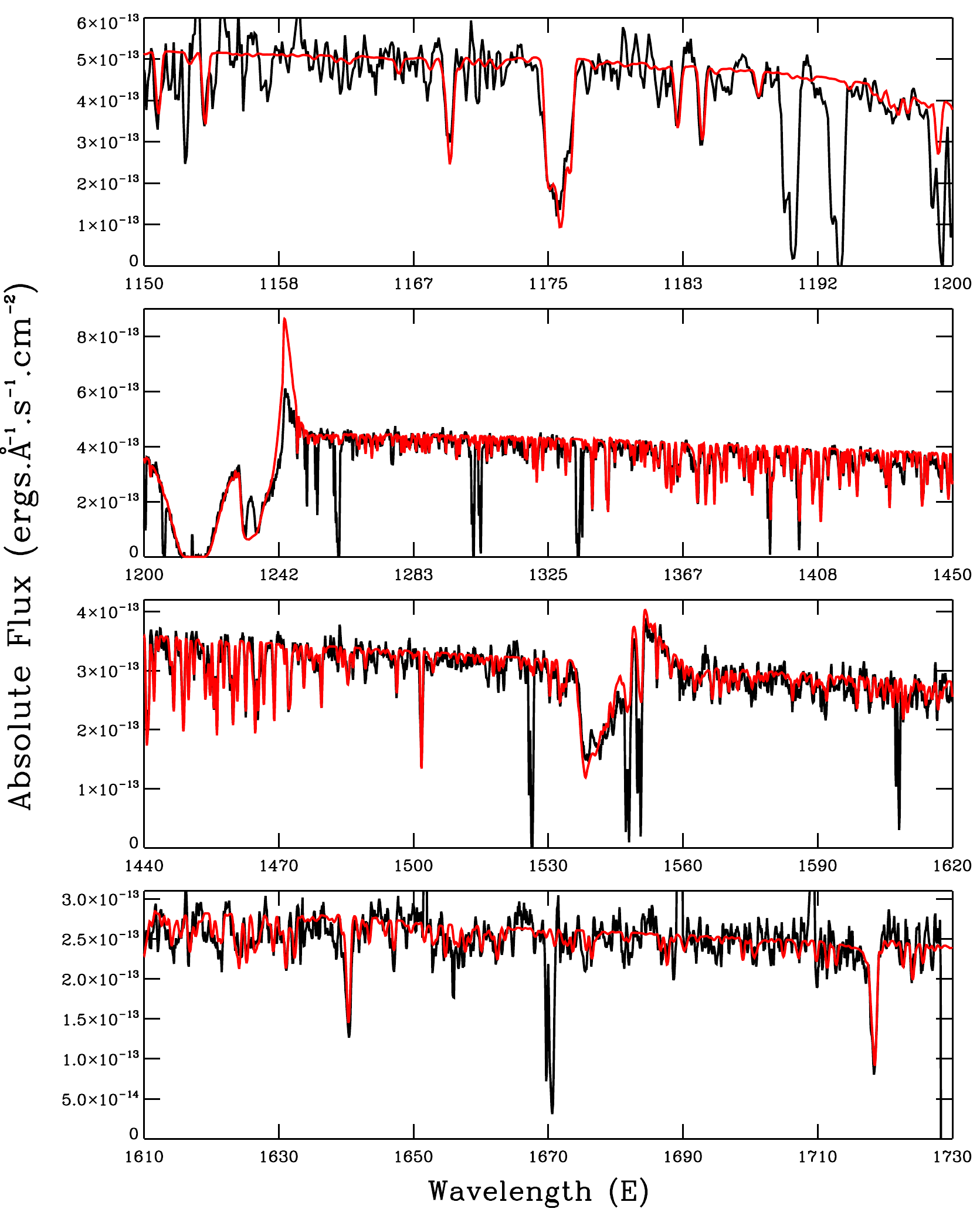}
\includegraphics[scale=0.51, angle=0]{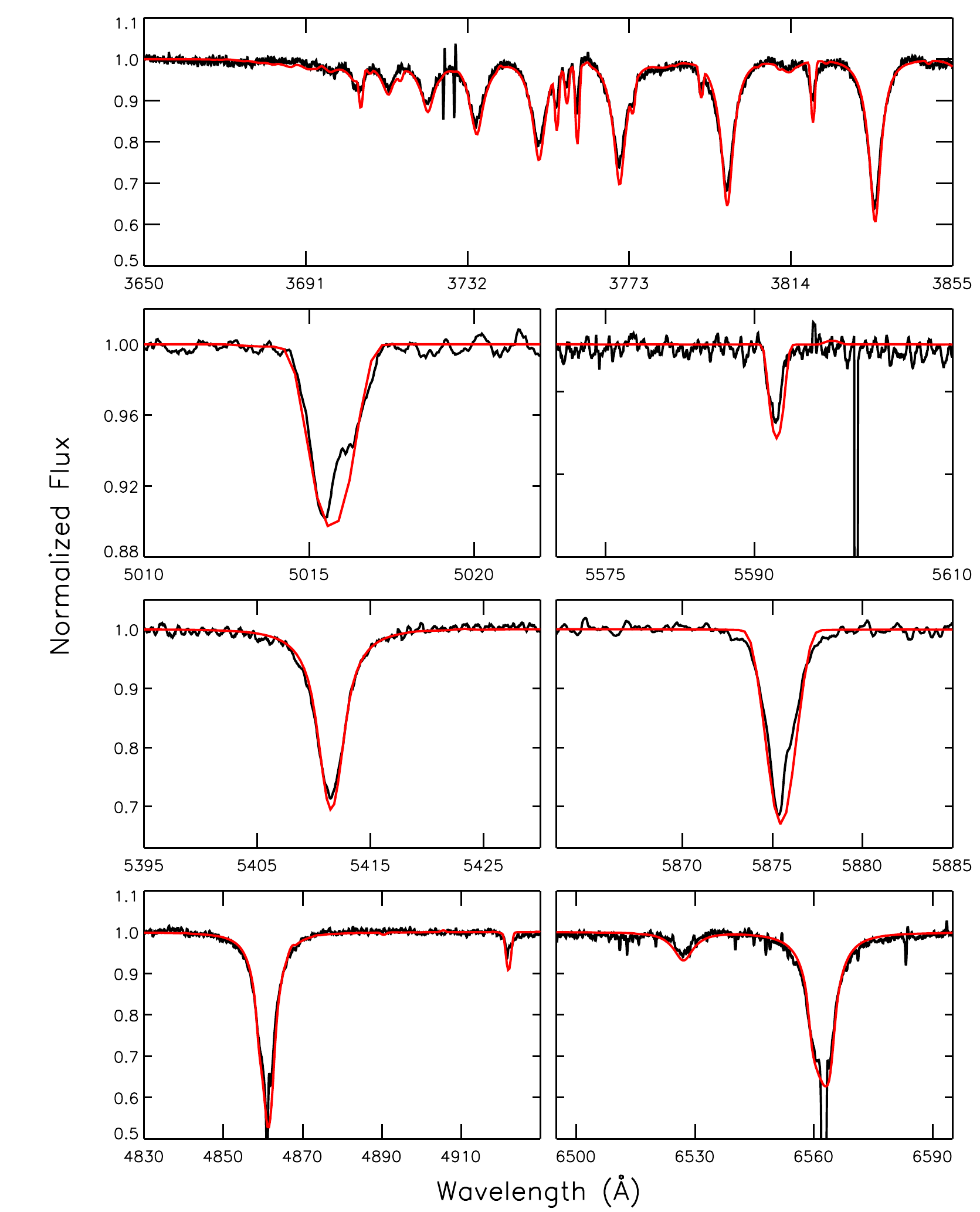}
      \caption[12cm]{Left: Best-fit model for \object{MPG 368} (red line) compared with the \stis\ spectrum (black line). Right: Best-fit model for \object{MPG 368} (red line) compared with the 
      \uves\ spectrum (black line).}
         \label{Fig_mpg368}
   \end{figure*}

\object{MPG 368} was studied in the UV and optical (\aat\ spectrum) in \cite{bouret03} and \cite{heap06}. 
Based on their own optical spectrum, \cite{massey09} confirmed the interpretation of \cite{heap06} that this star is a binary. 
A new optical spectrum obtained with \uves\ shows the \hei\ lines as composites of two components (see Fig. \ref{Fig_mpg368}).
Higher ionization lines seem to be unaffected, either in UV or optical, suggesting that the other component of this
system is a late-O or B-type star. 
The fit to the observations with our model proves to be very reasonable, confirming that the companion of \object{MPG 368} contributes only marginally 
to the wind line spectrum. 
\subsection{\object{MPG 113}\ - {OC6 Vz}}
\object{MPG 113} was extensively studied in \cite{bouret03}, \cite{heap06}, and \cite{mokiem06}. Our best-fit model has the same parameters as quoted
in our first study but for the mass-loss rate, which we slighlty increased to 3\eix\ \msolyr\ as a result of an improved \logLX\ ratio used 
in the new model (too high in the former study). Our \teff\ is still 3000~K lower than found in \cite{mokiem06},
and although we used the same \flames\ spectrum as they did, we did not need to increase the helium abundance to fit the helium lines. 
This difference might come from a bias toward lower temperatures to reproduce the observed ratio of lines \civ\ \lb1169 / \ciii\ \lb1176 in our case. 
For the adopted carbon abundance (based on these UV lines), the optical lines of \ciii\ \lb\lb4647-50-52 were predicted slightly too strong, while
\ciii\ \lb5696 was slightly too weak. Both are nevertheless predicted in emission, as observed. We note that the \niii\ \lb\lb4634-4640 lines are neither
observed nor predicted in our model. The resulting nitrogen-to-carbon ratio is above solar (see Table \ref{tab3}), like other stars of the O6 class, and 
as discussed in Sect. \ref{hr_sect}, the reason for the OC6 classification of \object{MPG 113} is more complex than a simple abundance criterion. 

\begin{figure*}[tbp]
\includegraphics[scale=0.54, angle=0]{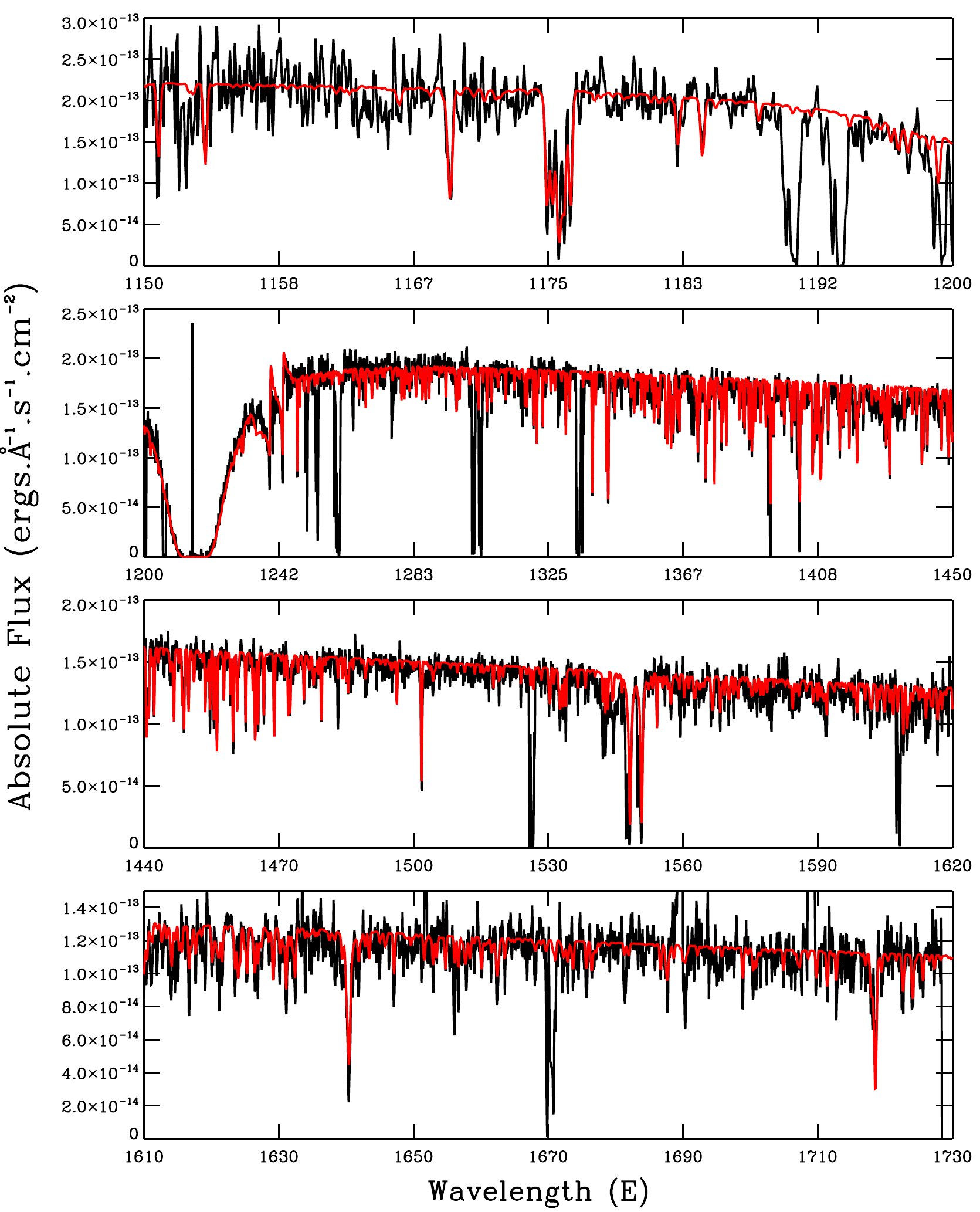}
\includegraphics[scale=0.54, angle=0]{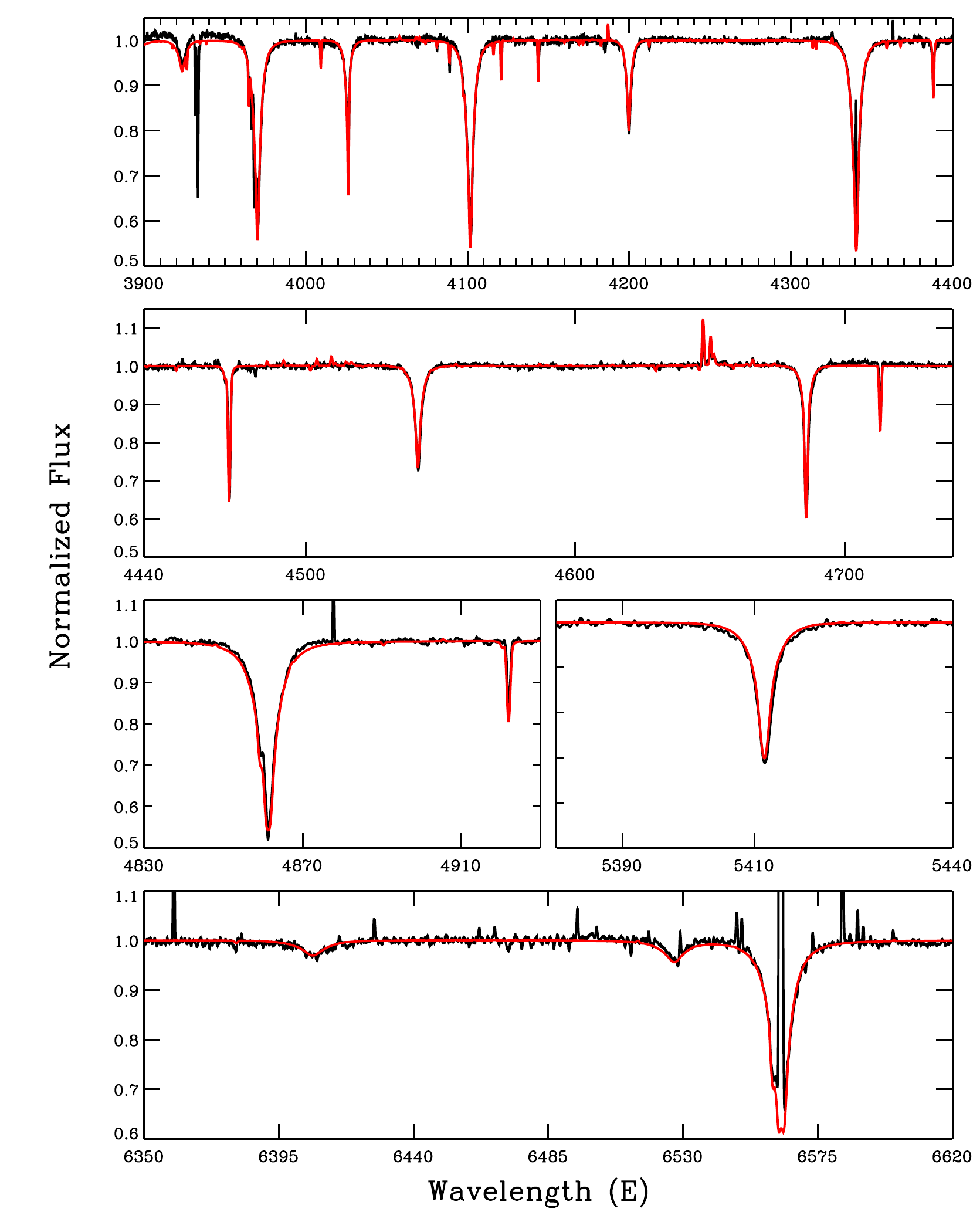}
\includegraphics[scale=1., angle=0]{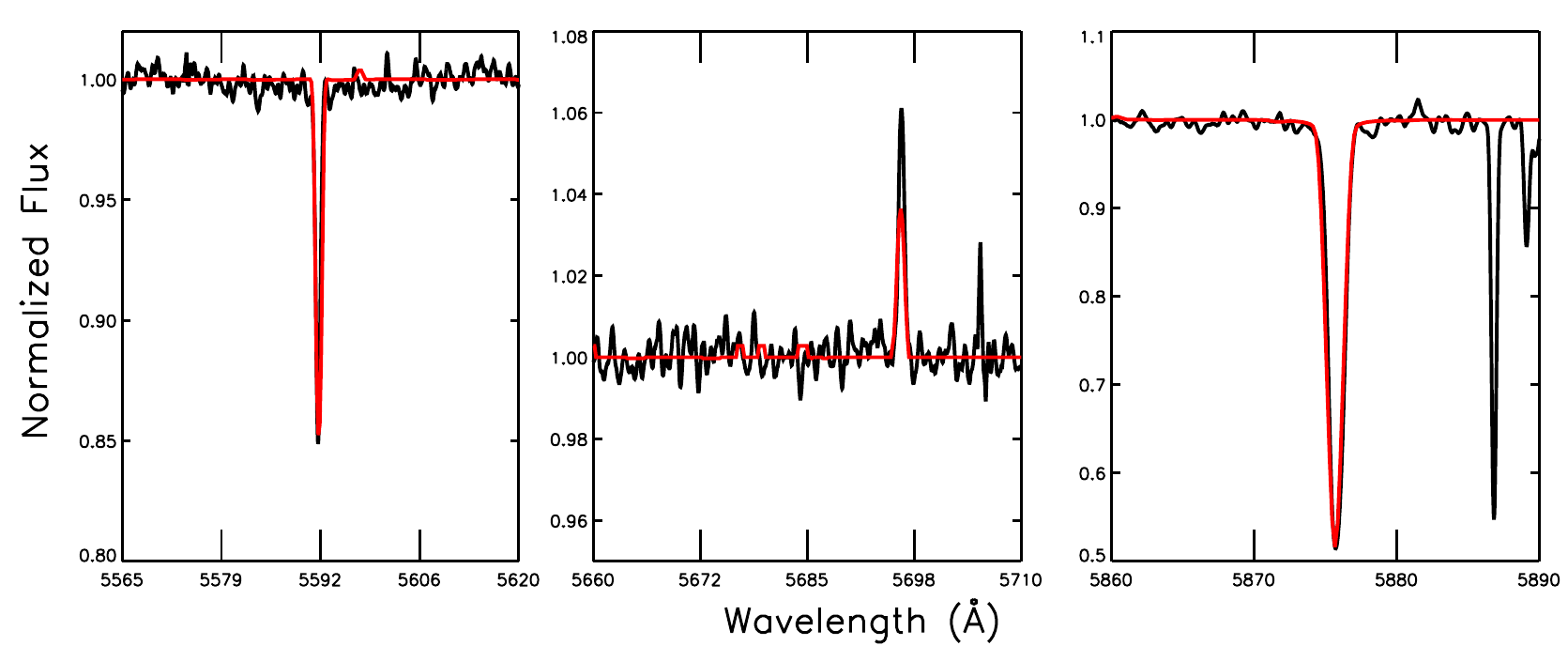}
    \caption[12cm]{Best-fit model for \object{MPG 113} (in red) compared with observed spectra (in black): the \stis\ spectrum (upper left) and the \flames + \uves\ spectra (upper right).
      The bottom plot shows a comparison with the \uves\ spectrum for spectral regions around \oiii\ \lb5592 (left), \ciii\ \lb5696 (middle), and \hei\ \lb5876 (right).}
               \label{Fig_m113_1}
   \end{figure*}
\subsection{\object{AzV 243}\ - {\rm O6 V}}
Of the O6 V stars of our sample, \object{AzV 243} has the largest luminosity, consistent with its photometry. The luminosity we derive 
agrees well with the one quoted in \cite{mokiem06}, but we found this star to be significantly cooler (by 2500~K) than in this study. 
With its higher luminosity, \object{AzV 243} does not fall into the weak-wind regime, which is also confirmed by its \cosp\ spectrum, showing clear
PCygni profiles at \nv\ and \civ\ resonance doublet. The wind acceleration parameter $\beta\ = 1$ is intermediate between the value $\beta\ = 0.7$ 
given in \cite{haser98} and the value $\beta\ = 1.37$ in \cite{mokiem06}. The latter is prone to higher uncertainty, however, because it was derived from the \halpha\ line mostly, 
which poorly responds  to wind parameters for \mdot\ $\le$ 1.\evii\ \msolyr\ as is the case here. 
\begin{figure*}[tbp]
\includegraphics[scale=0.51, angle=0]{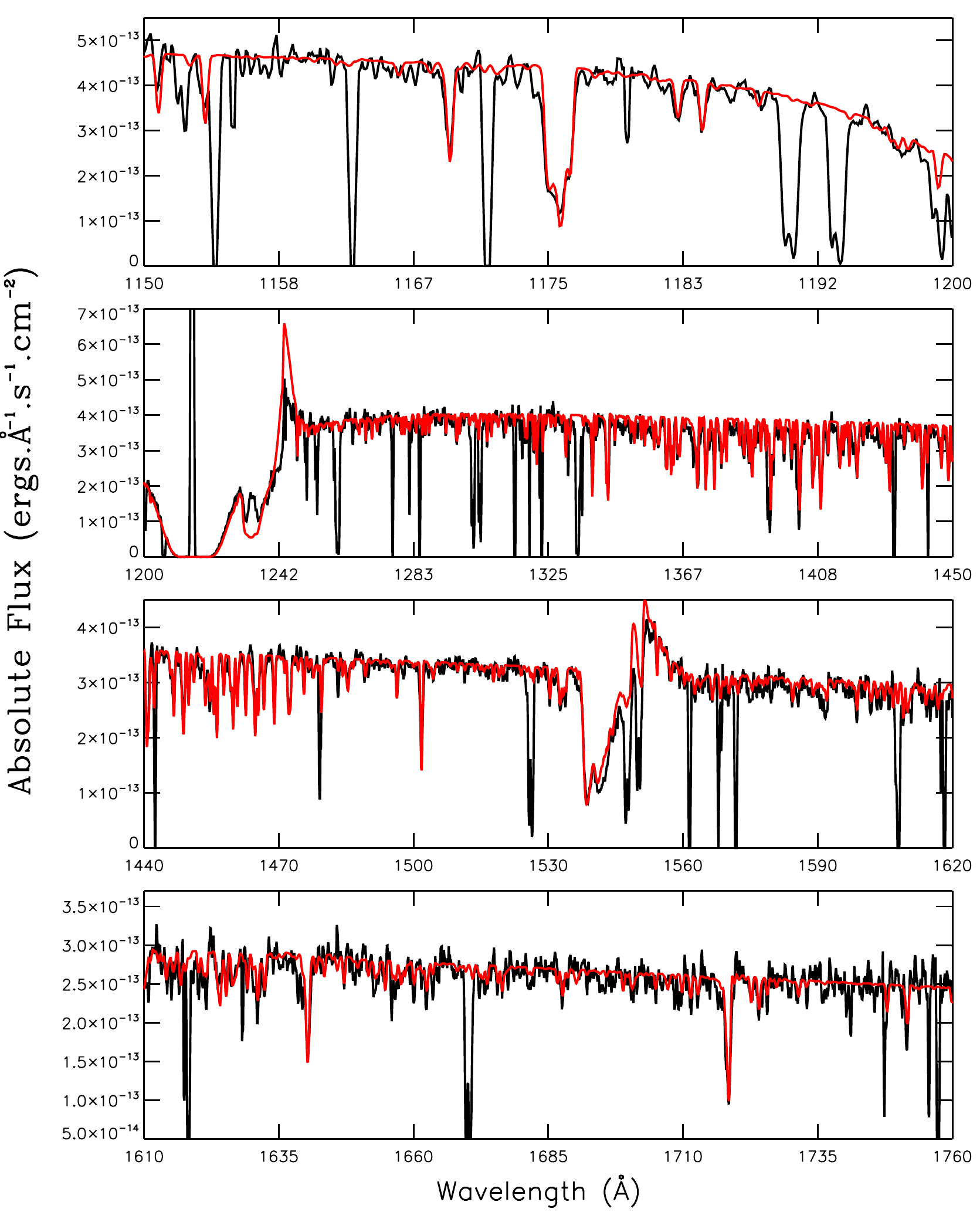}
\includegraphics[scale=0.51, angle=0]{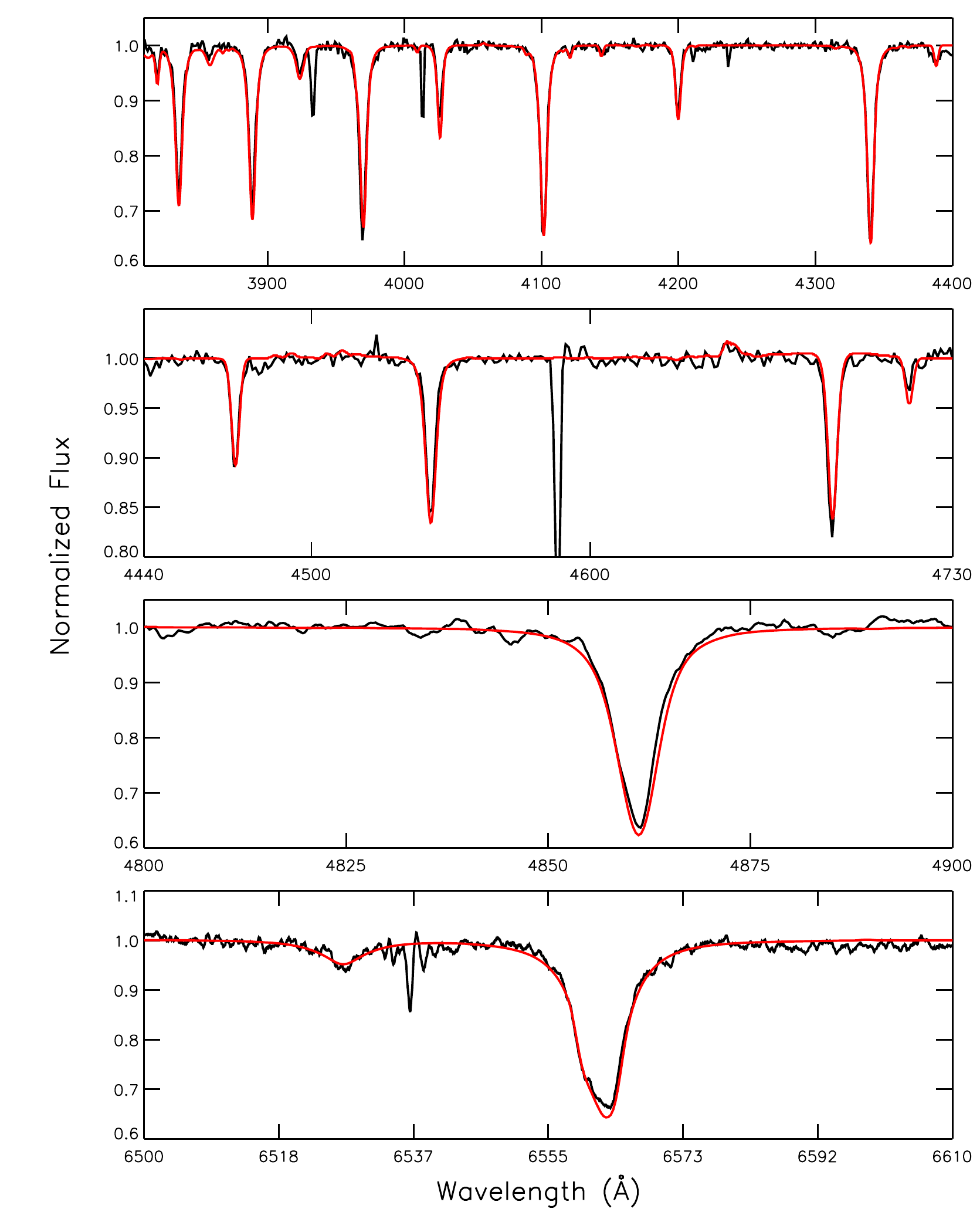}
      \caption[12cm]{Left: Best-fit model for \object{AzV 243} (red line) compared with the \cosp\ spectrum (black line). Right: Best-fit model for \object{AzV 243} (red line) compared with the 
       \flames\ spectrum (black line).}
         \label{Fig_av243}
   \end{figure*}	
\subsection{\object{AzV 446}\ - {\rm O6.5 V}}
\object{AzV 446} was studied in \cite{massey05}. Based on the line widths of \hei\ \lb4471 and \heii\ \lb4542, these authors noted that this star could be
a binary although the absolute magnitude of \object{AzV 446} was consistent with a single star of luminosity class V. The very good agreeement
of their model with the observed optical spectrum also supported the single-star interpretation. Our modeling of the flux-calibrated UV spectrum
of \object{AzV 446} additionally confirms this (the SED of our model also matches the photometry up to NIR). The fundamental parameters of our best model
marginally differ from those in \cite{massey05} but for the wind parameters. From the UV only, though, we were unable to constrain the surface gravity 
and helium abundance to a good accuracy and simply adopted standard values for a young main-sequence star. We note that \cite{massey05}
measured $y$ = 0.15, surprisingly high for such a star. 
From the very weak \civ\ resonance profile, \mdot\ = 1\evii\ \msolyr\ is ruled out. However, this profile, although weak, is clearly seen, as is 
the \nv\ \lb\lb1248-1250 profile, showing a discrete absorption component near the blue edge of the line. We also note that the
\civ\ \lb\lb1548-1550 resonance line presents a blueward extension that is clearly sensitive to $f$. 
We find that these two wind profiles are best 
reproduced when \mdot\ = 4\eix\ \msolyr,  $f = 0.1$, and \vinf\ = 1400 \kms. 
\begin{figure*}[tbp]
\includegraphics[scale=1., angle=0]{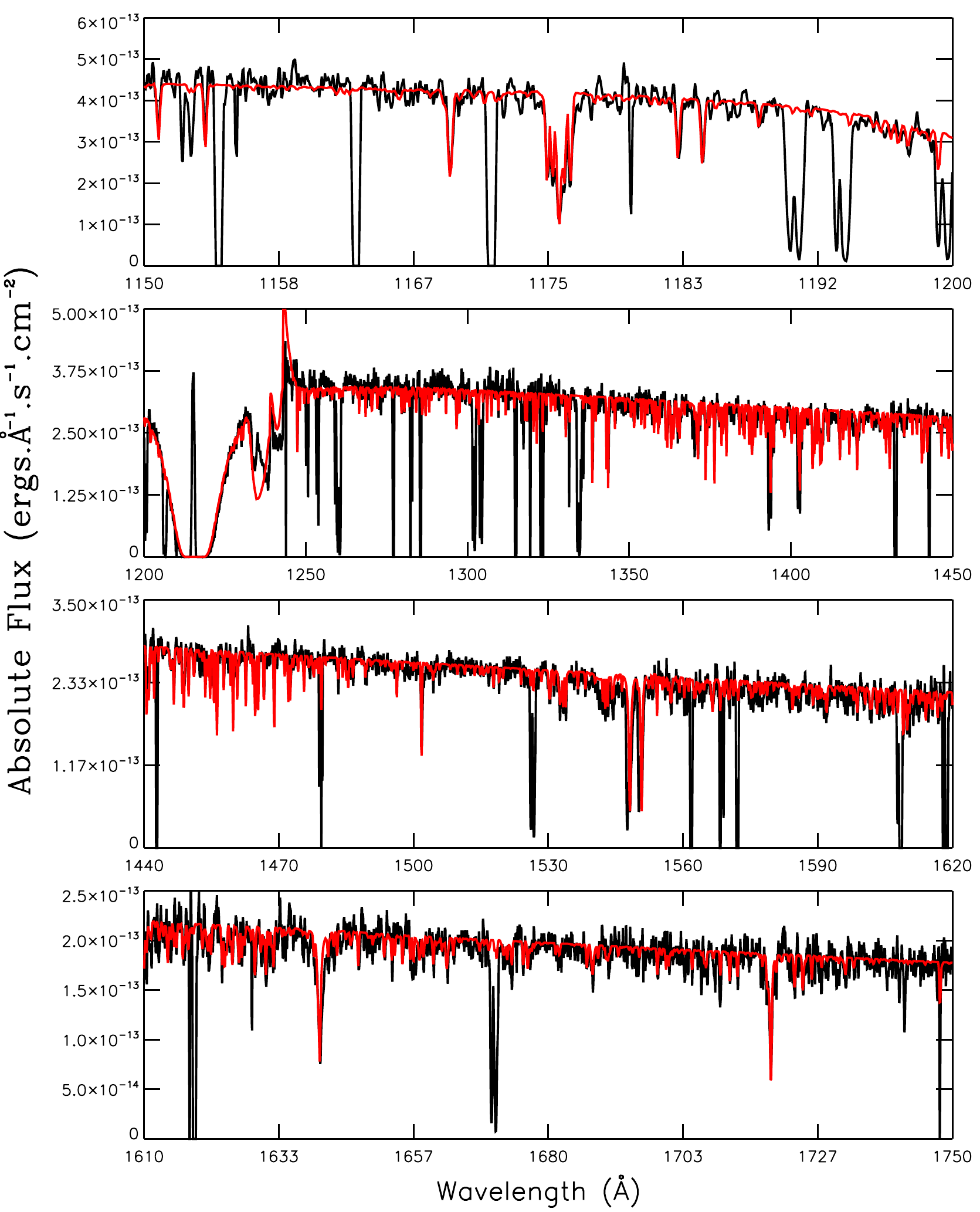}
      \caption[12cm]{Left: Best-fit model for \object{AzV 446} (red line) compared with the \cosp\ spectrum (black line).}
         \label{Fig_av446}
   \end{figure*}	   
\subsection{\object{MPG 356}\  - {\rm O6.5~V }}
This star was modeled in \cite{heap06}, although in the UV only. Using the \uves\ spectrum we obtained for this star, we derived a 
higher temperature and surface gravity than \cite{heap06}, although still within their error bars. Our luminosity is also lower
(0.4 dex) but consistent with the complete photometric dataset, from UV to NIR. 
From the photospheric lines of carbon and nitrogen in the UV spectrum, we derived surface abundances significantly different
from those in \cite{heap06}. The carbon abundance is higher than measured in H~II region of the SMC or even than a solar
carbon abundance scaled down by a factor 0.2 corresponding to the global SMC metallicity. This abundance is consistent with the UV lines
and with the \ciii\ \lb5696 line in the \uves\ spectrum. \cite{heap06} found a moderate (if any) nitrogen enhancement from \niii\ \lb\lb1182-1184, while
we derived an enrichment by up to a factor of seven (compared to the adopted baseline), from the same \stis\ spectrum. The N/C ratios, on the
other hand, agree in both studies. Keeping the carbon and nitrogen abundances fixed, we constrained the mass-loss rate from \nv\ \lb\lb1248-1250 
mostly, because the \civ\ resonance doublet shows no clear wind profile. An upper limit on the mass-loss rate \mdot\ $\le$ 3.5\eix\ \msolyr\ was found. 

\begin{figure*}[tbp]
\includegraphics[scale=0.51, angle=0]{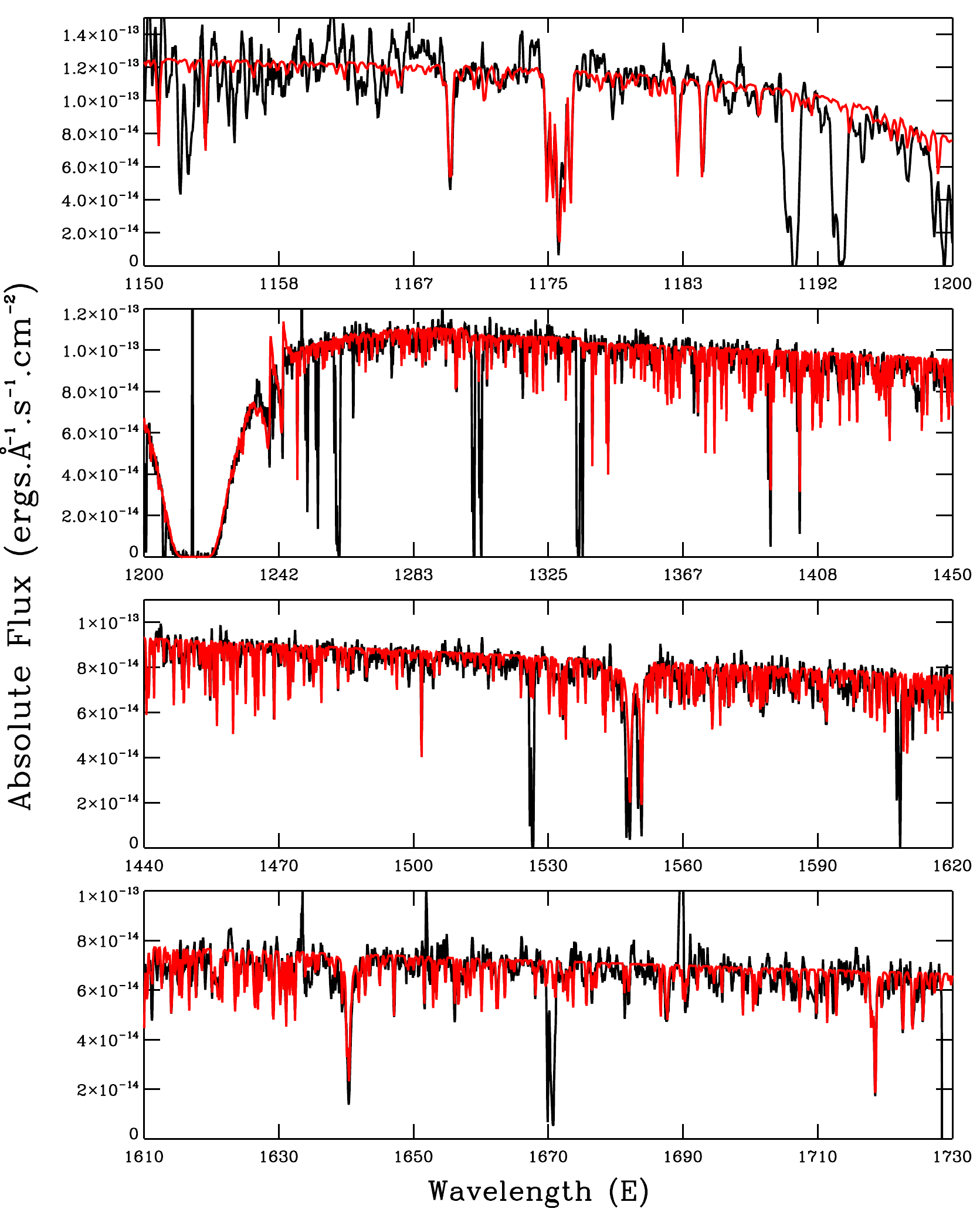}
\includegraphics[scale=0.51, angle=0]{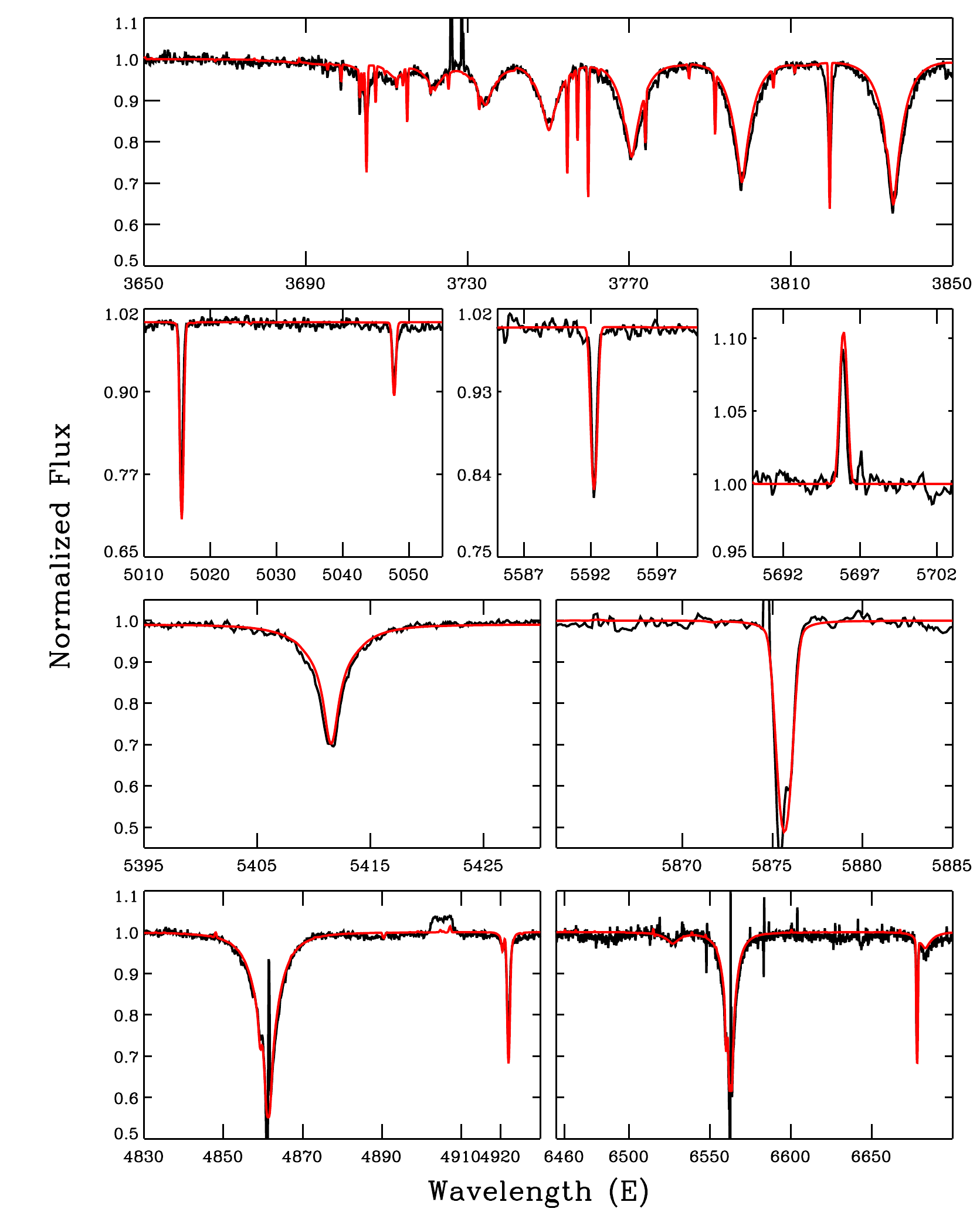}
      \caption[12cm]{Left: Best-fit model for \object{MPG 356} (red line) compared with the \stis\ spectrum (black line). Right: Best-fit model for \object{MPG 356} (red line) compared with the 
       \uves\ spectrum (black line).}
         \label{Fig_mpg356}
   \end{figure*}	
\subsection{\object{AzV 429}\  - {\rm O7~V }}
\object{AzV 429} is substantially more luminous than the other O7~V stars of the sample (see below and Table \ref{tab2}). It is also probably
more evolved, as indicated by its position in the H-R diagram. Having only the UV spectrum for this star, our estimate of the surface
gravity is based on the \fev\ to \feiv\ index from \cite{heap06} and suffers from higher uncertainty. 
The surface abundances of carbon and nitrogen are such that the N/C ratio is one of the lowest of the whole sample and only twice the
interstellar value. Nitrogen is enhanced, but the carbon abundance is higher than listed in \cite{kurt98}. 
The \civ\ \lb\lb1548-1550 doublet presents a weak but well-defined wind profile. The best fit to this profile is obtained using \mdot\ = 4.0\eix\ \msolyr\ and $f$ = 0.05.  

   \begin{figure*}[tbp]
\includegraphics[scale=1., angle=0]{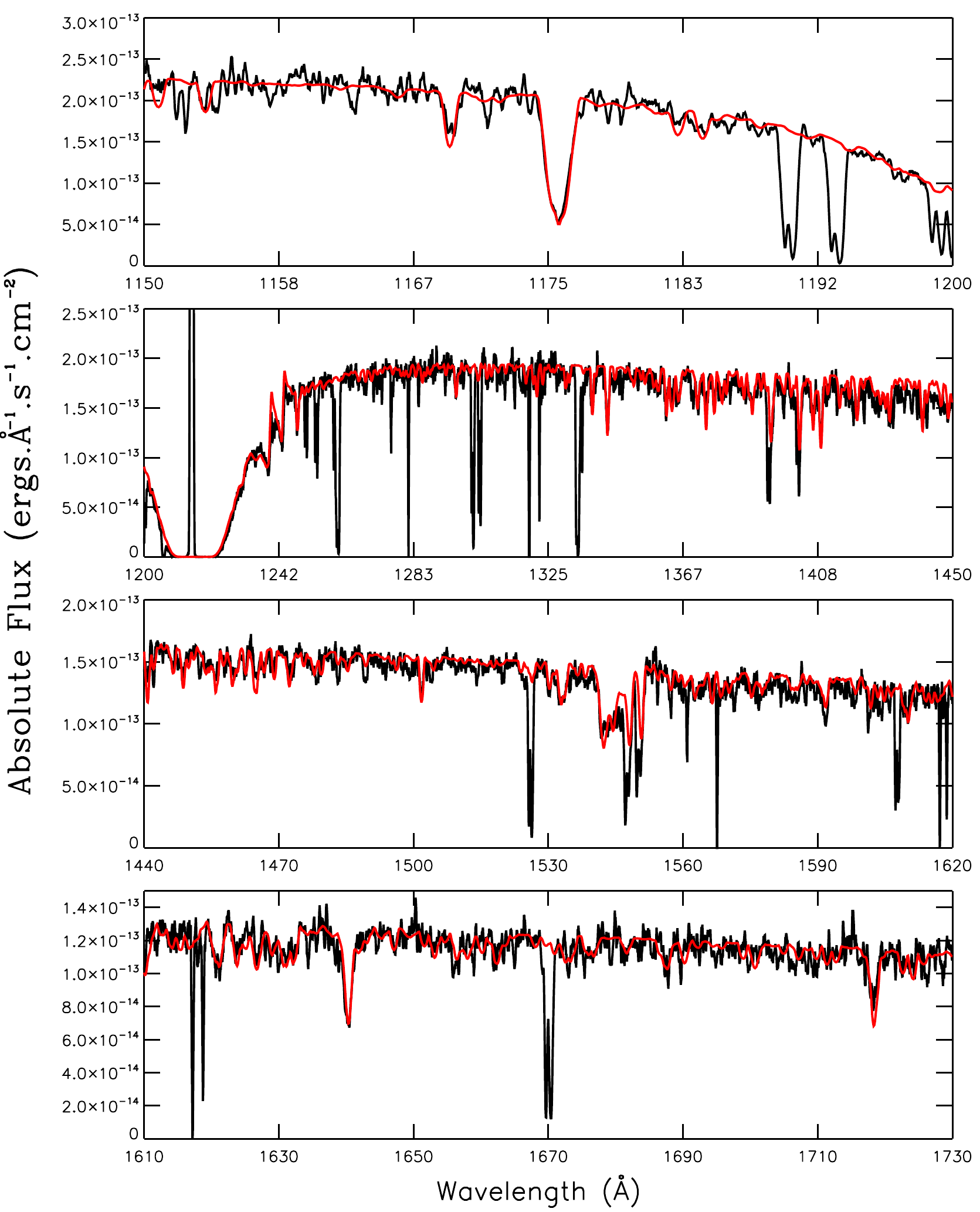}
      \caption[12cm]{Best-fit model for \object{AzV 429} (red line) compared with the \cosp\ spectrum (black line). }
         \label{Fig_av429}
   \end{figure*}
\subsection{\object{MPG 523}\  - {\rm O7~Vz }}
Our \uves\ spectrum of \object{MPG 523} is not as strongly contaminated with nebular emission as was the \flames\ spectrum used in \cite{mokiem06}. 
Both spectra can be best fitted with \teff\ = 38,700~K and a relatively high surface gravity, \logg\ = 4.25. The fit to the helium
lines is good except for \hei\ \lb5876, which is too broad in our model, while the observed profile also presents a warp on its blue side. 
The \vsini\ is moderate (50 \kms) and consistent with the UV spectrum as well. This star is not listed as a potential or known binary, nor do we find it
overluminous for its spectral type. We were unable to find a satisfactory explanation for this profile. 
The \civ\ \lb\lb1548-1550 shows up as photospheric absorption, and the upper limit on the mass-loss rate
is  \mdot\ = 6\eix\ \msolyr, 

Depending on the \vrot\ used for stellar evolution models, the location of \object{MPG 523} in the H-R diagram indicates contradictory ages, either very young (with an age younger
than 1 Myr) or 
much older (with an age of about 5 Myr).
On the other hand, the carbon and nitrogen surface abundances are characteristic of significantly more evolved status (carbon
depleted and nitrogen enriched). 
The N/C ratio is one of the highest of the whole sample. We note that the \civ\ \lb1169 is too weak in our model, however, but could not
be matched by increasing the carbon abundance without degrading the fit to \ciii\ \lb1176. The weak \ciii\ \lb\lb4647-50-52 supports the low
value for the carbon abundance.   

\begin{figure*}[tbp]
\includegraphics[scale=0.51, angle=0]{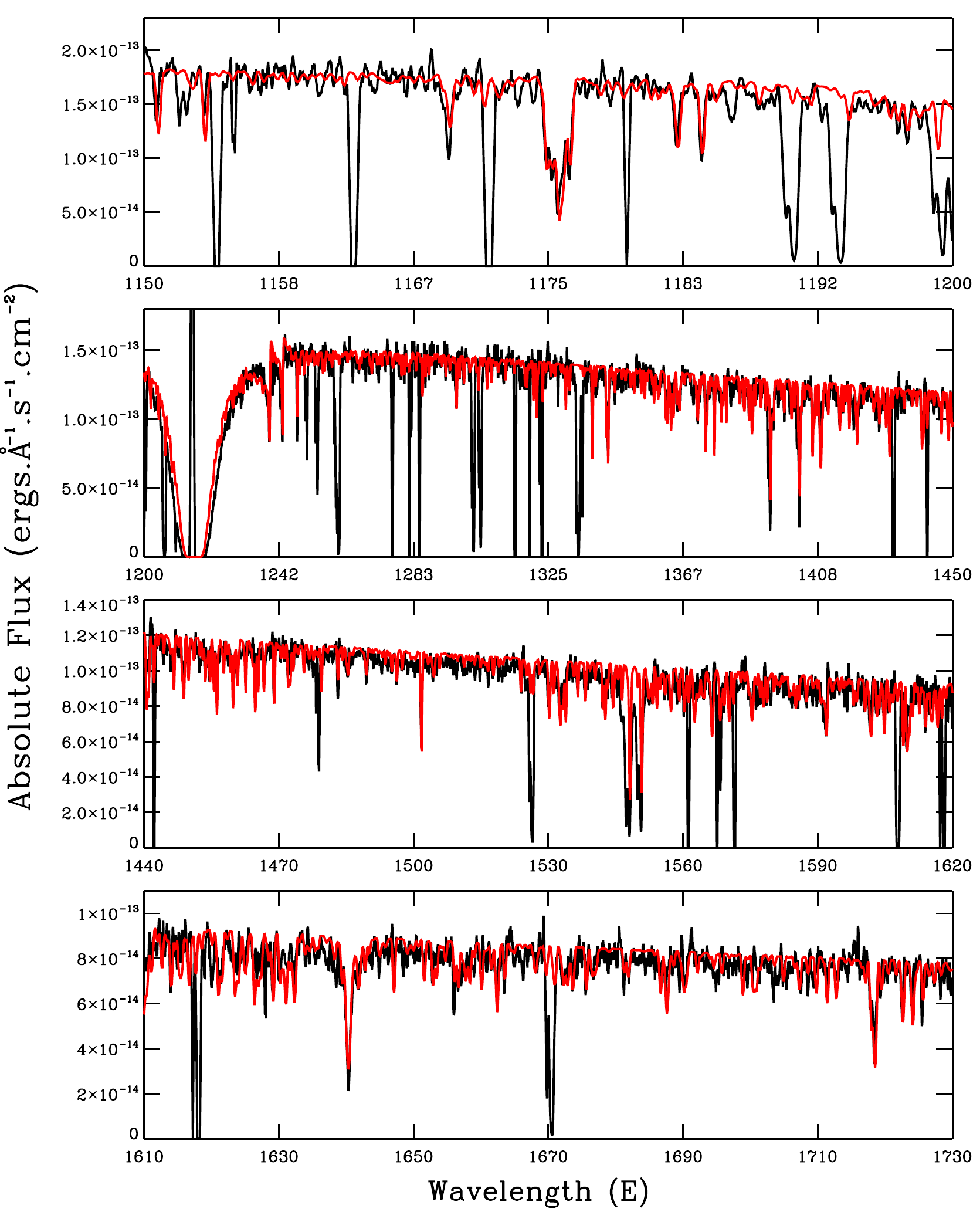}
\includegraphics[scale=0.51, angle=0]{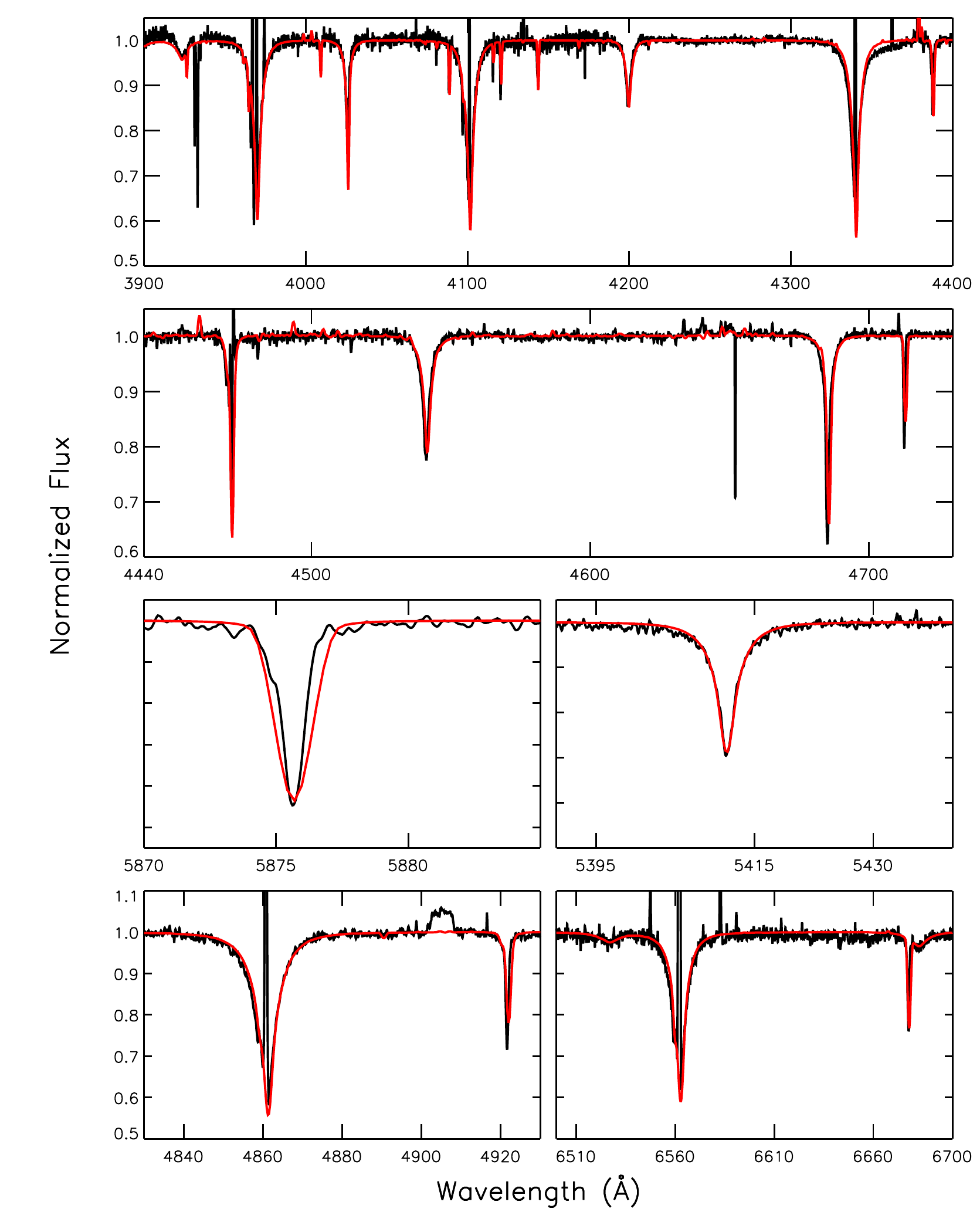}
      \caption[12cm]{Left: Best-fit model for \object{MPG 523} (red line) compared with the \stis\ spectrum (black line). Right: Best-fit model for \object{MPG 523} (red line) compared with the 
       \uves\ spectrum (black line).}
         \label{Fig_mpg523}
   \end{figure*}	
\subsection{\object{NGC 346-046}\  - {\rm O7~Vn }}
The luminosity and effective temperature we derived for \object{NGC 346-046} indicates that this star is very close to the ZAMS, possibly leftward
of the ZAMS within the error bars. 
Consistent with its spectral classification and with the results by \cite{mokiem06}, this star has a very high rotation, the second highest 
of the sample, with \vsini\ = 300 \kms. This strongly suggests that this evolutionary status of this star is to a large extent influenced
by its rotation, possibly to the point of following chemically homogeneous evolution.  
However, our modeling does not indicate any departure from the initial helium abundance, which is inconsistent with such an evolution
\citep[and with][]{mokiem06}. The nitrogen-to-carbon ratio, though, is the highest of the whole sample. 
The \civ\ resonance doublet is present as a pure absorption profile, from which we derive an upper limit
 of \mdot\ = 6\eix\ \msolyr\ for the mass-loss rate. 

\begin{figure*}[tbp]
\includegraphics[scale=0.51, angle=0]{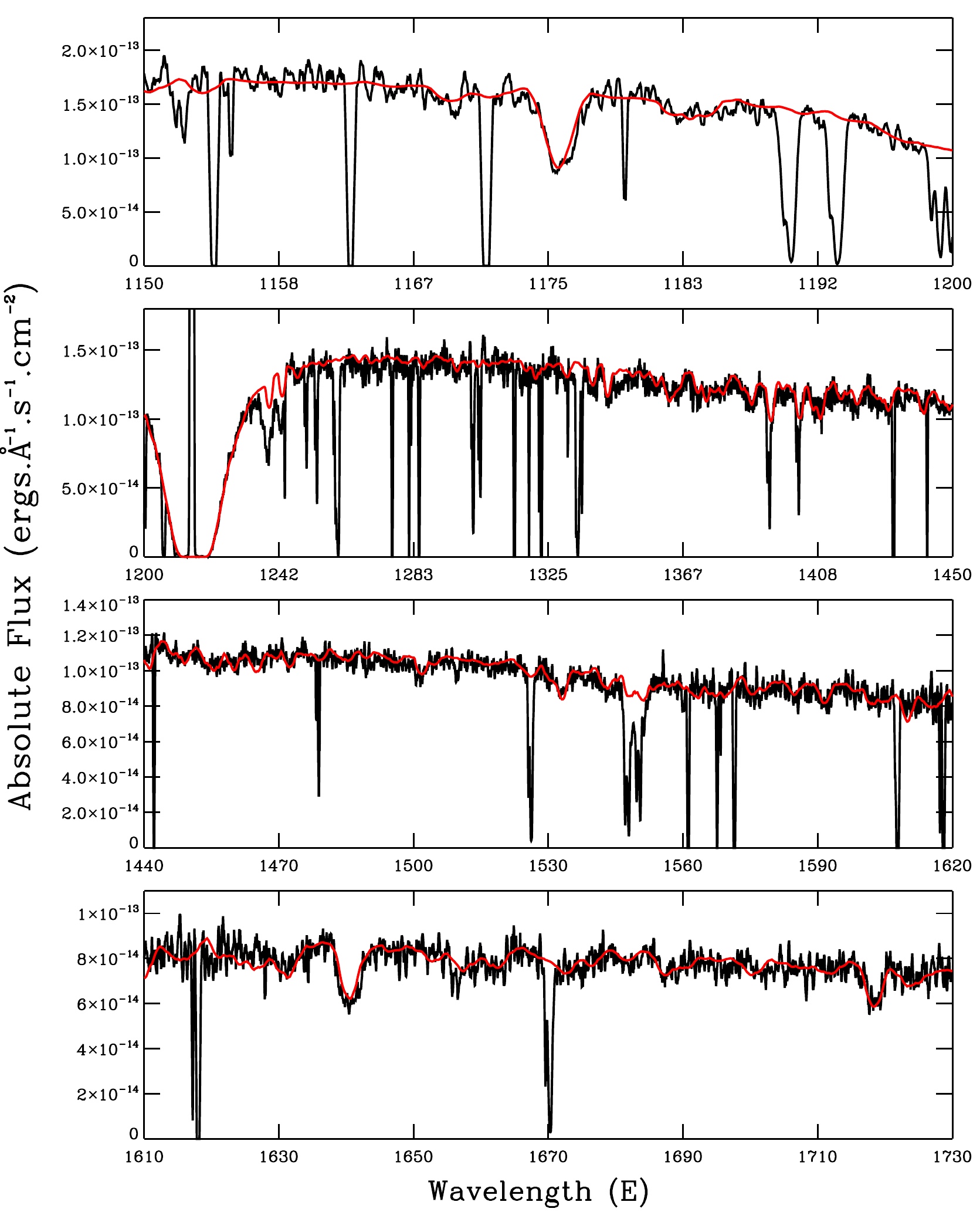}
\includegraphics[scale=0.51, angle=0]{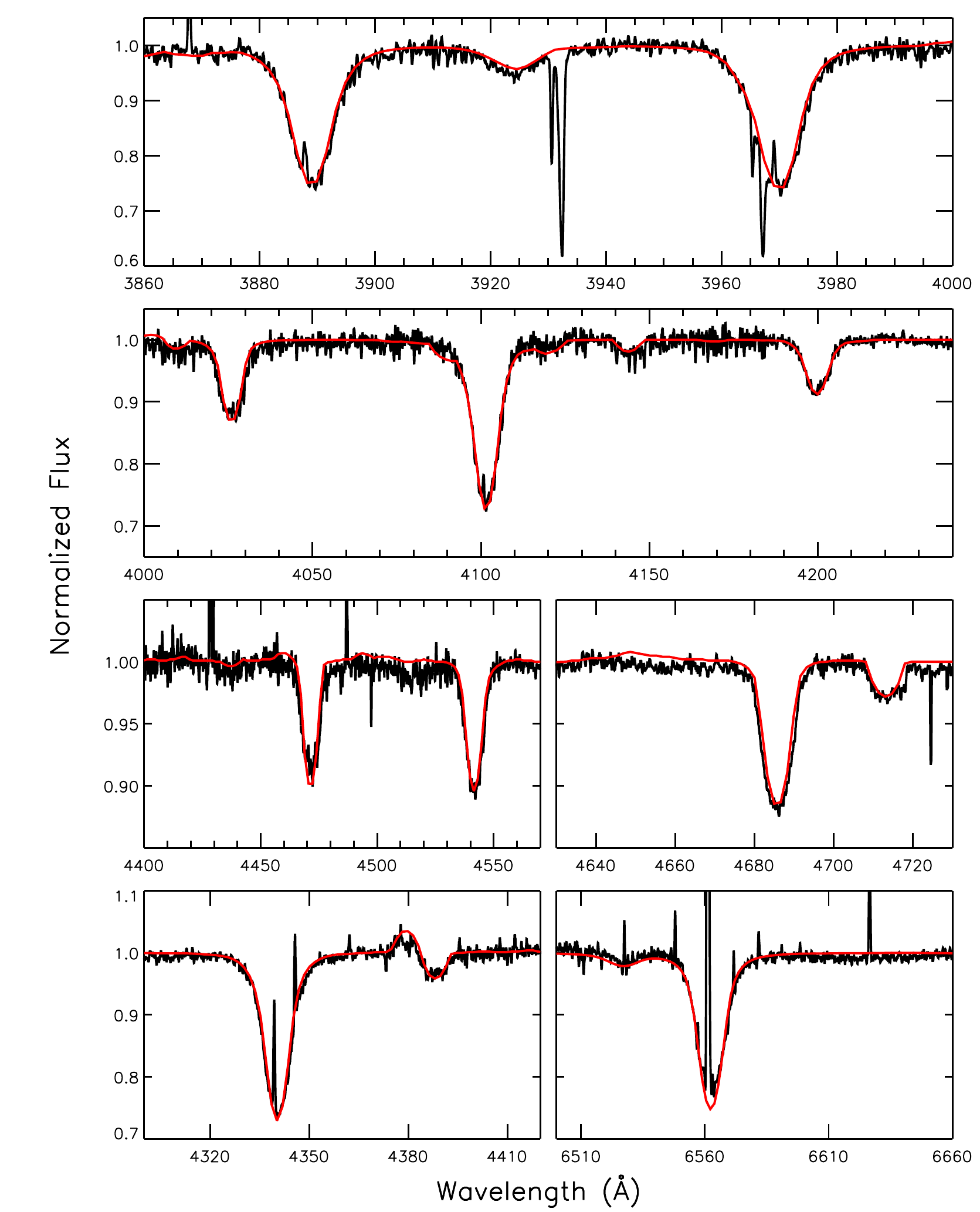}
      \caption[12cm]{Left: Best-fit model for \object{NGC 346-046} (red line) compared with the \cosp\ spectrum (black line). Right: Best-fit model for \object{NGC 346-046} (red line) compared with the \flames\ spectrum (black line).}
         \label{Fig_els046}
   \end{figure*}	
\subsection{\object{NGC 346-031}\  - {\rm O8~Vz }}
\object{NGC 346-031} is the last member of the Vz class of our sample. We achieved a very good fit of both the UV and optical spectra and 
derived fundamental parameters such that its age is above 3 Myr, in contradiction with the expected youth of Vz class-members. 
In contrast to \cite{mokiem06}, we did not find it necessary to increase the helium abundance, but instead found that the very low mass-loss
rate we derived can explain the spectral classification in this parameter range (see Sect. \ref{sect_vz}). 

\begin{figure*}[tbp]
\includegraphics[scale=0.54, angle=0]{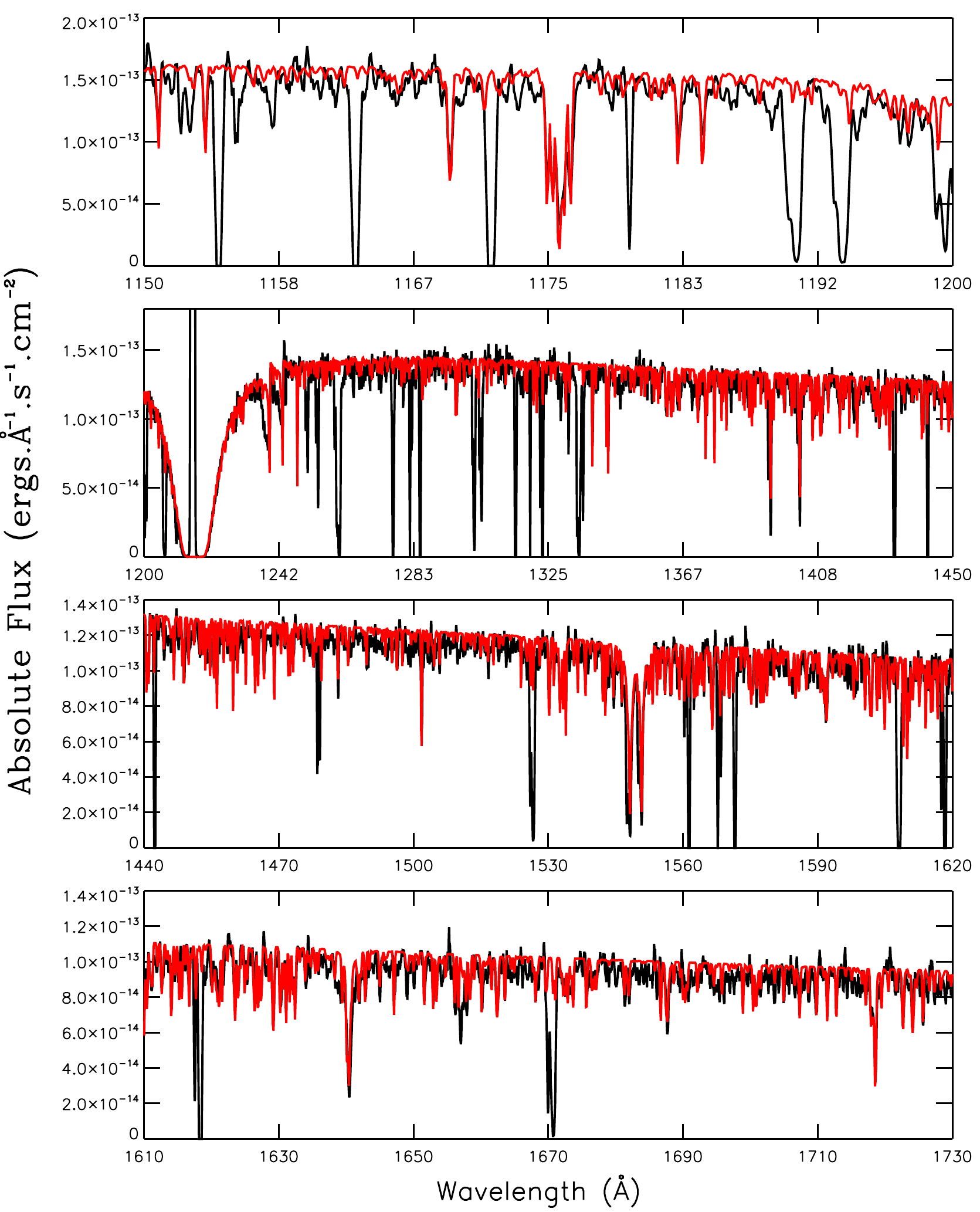}
\includegraphics[scale=0.54, angle=0]{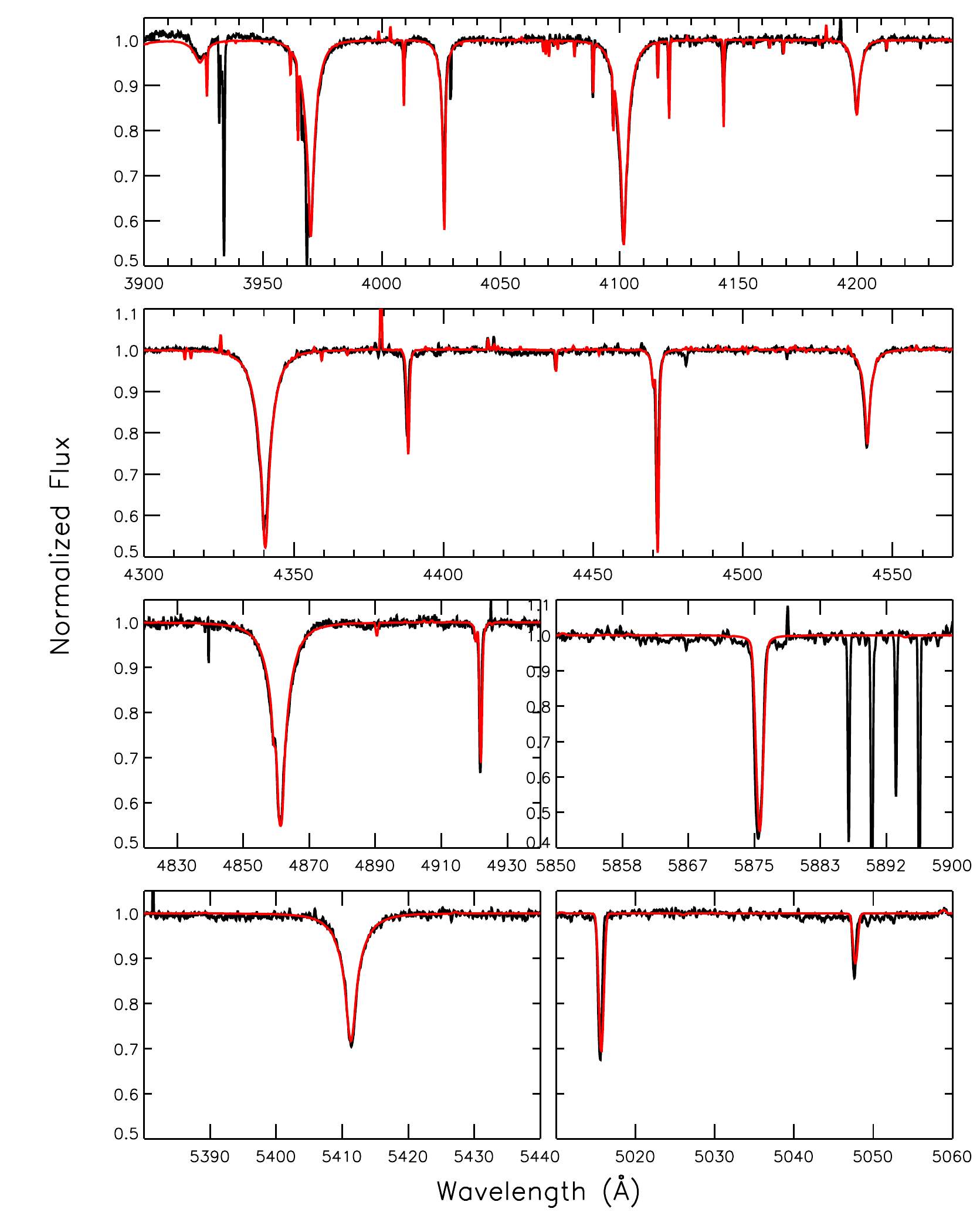}
 \includegraphics[scale=1., angle=0]{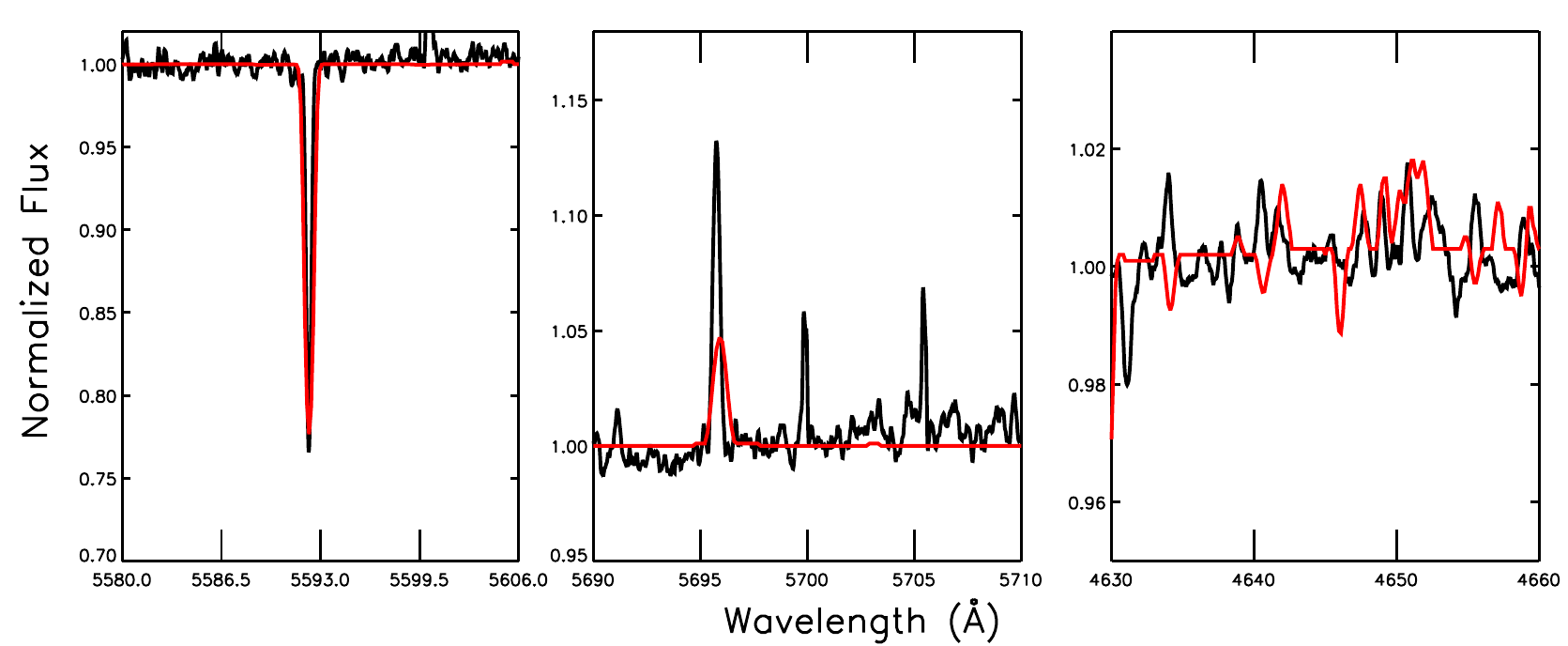}
      \caption[12cm]{Best-fit model for \object{NGC 346-031} (in red) compared with observed spectra (in black): the \cosp\ spectrum (upper left) and the \flames spectrum (upper right).
      The bottom plot shows a comparison with the \uves\ and \flames\ spectra for spectral regions around \oiii\ \lb5592 (left), \ciii\ \lb5696 (middle), and \niii\ - \ciii\  features (right).}
         \label{Fig_els031_2}
   \end{figure*}
\subsection{\object{AzV 461}\  - {\rm O8~V}}
The fundamental and wind parameters we derived for \object{AzV 461} are close to those of \object{NGC 346-031}, but for the micro-turbulence and rotational velocity.  
For the effective temperature we derived from the UV analysis (see Sect. \ref{param_sect}), \hei\ \lb4471 and 
\heii\ \lb4686 are too weak, while \heii\ \lb4200 and \heii\ \lb4542 are well reproduced. Since the optical (2dF) spectrum is at lower resolution and S/N,
we mostly relied on the \cosp\ spectrum to derive the fundamental parameters.
The very close \teff\ and luminosities of \object{AzV 461} and \object{NGC 346-031} imply very close evolutionary status. Consistent with the higher \vsini, we derived a higher nitrogen enrichment and carbon
depletion in \object{AzV 461}. 

\begin{figure*}[tbp]
\includegraphics[scale=0.51, angle=0]{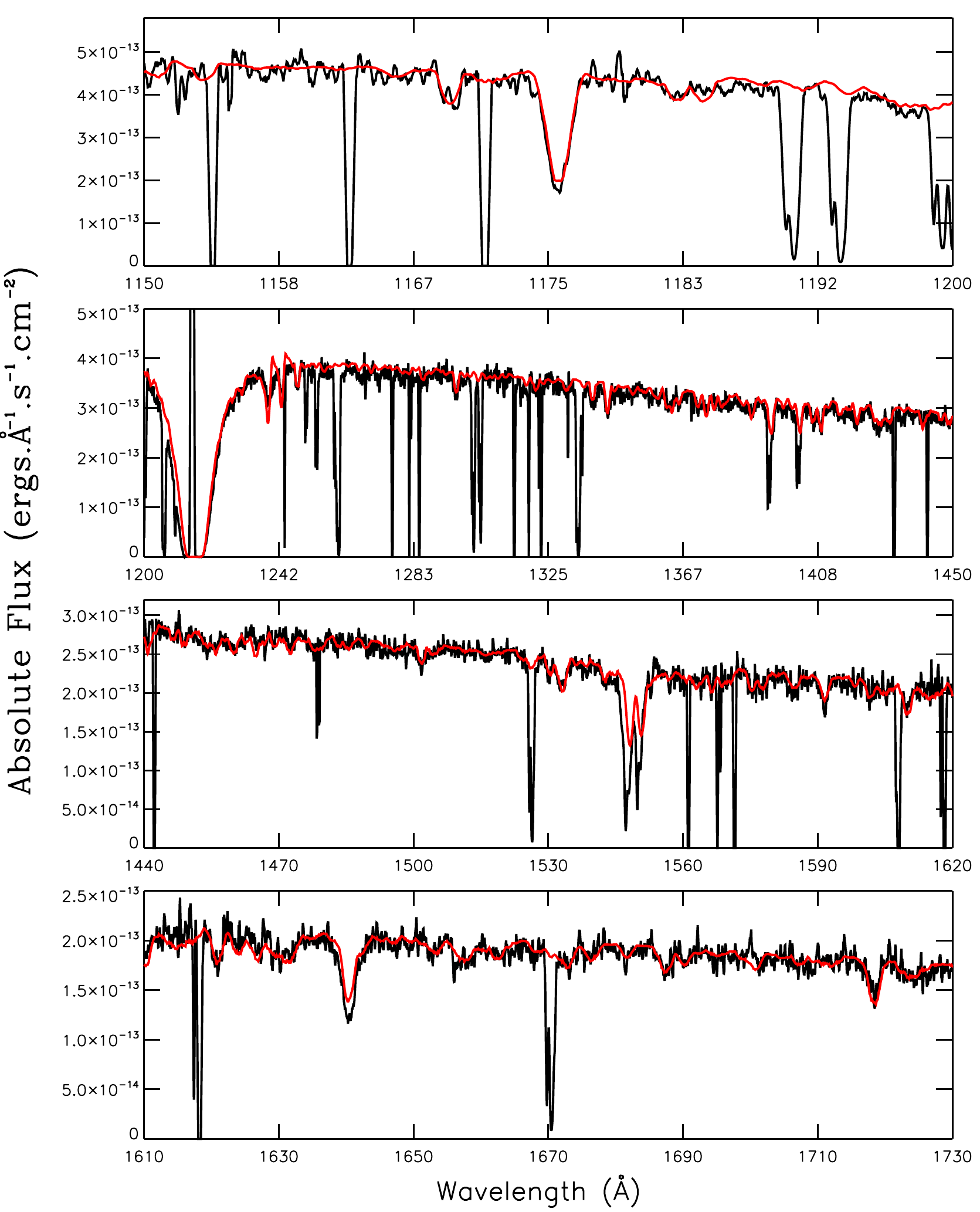}
\includegraphics[scale=0.49, angle=0]{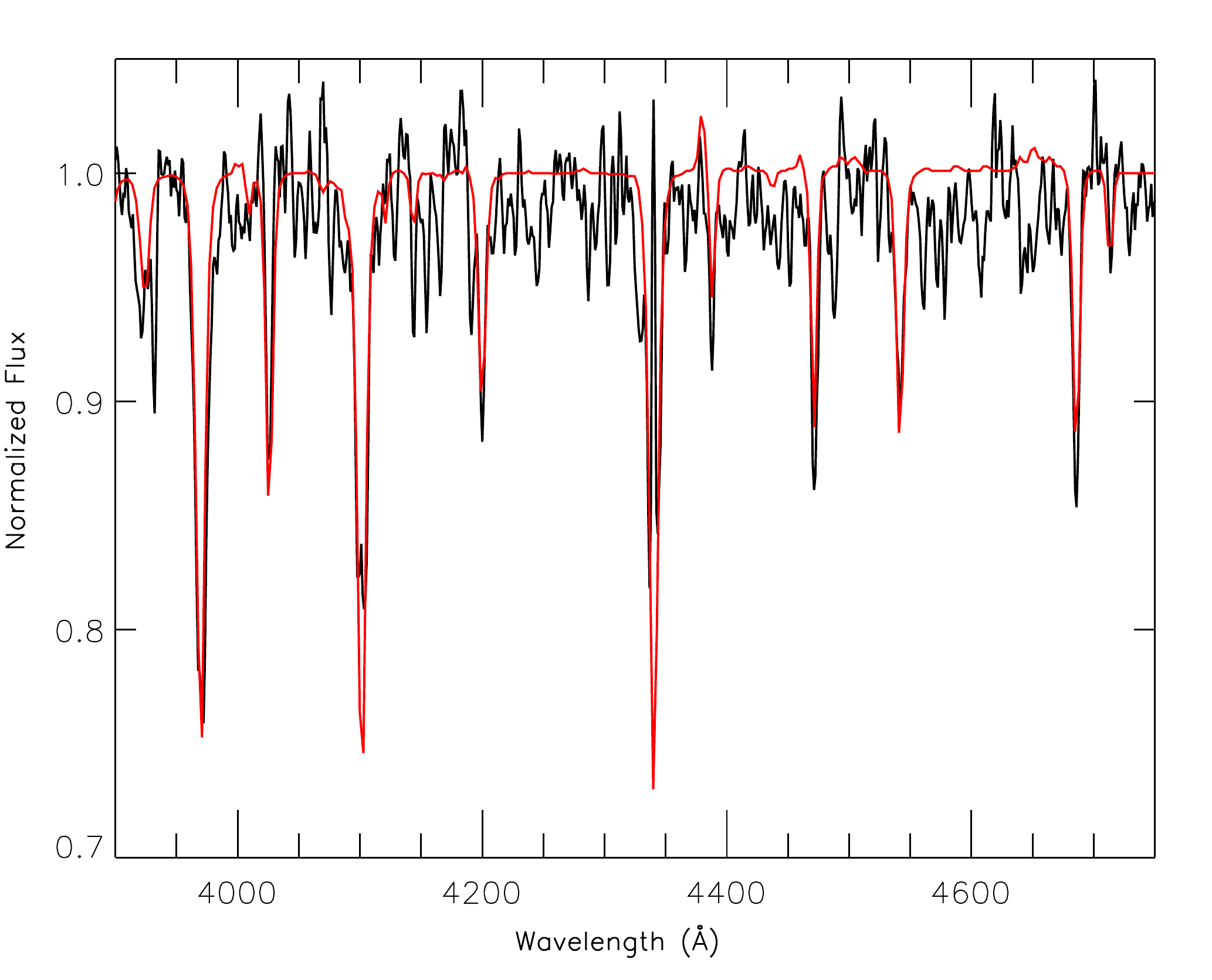}
      \caption[12cm]{Left: Best-fit model for \object{AzV 461} (red line) compared with the \cosp\ spectrum (black line). Right: Best-fit model for \object{AzV 461} (red line) compared with the 
      2dF spectrum (black line).}
         \label{Fig_av461}
   \end{figure*}	
\subsection{\object{MPG 299}\  - {\rm O8~Vn / O9 V}}
With \vsini\ = 360 \kms, \object{MPG 299} is the fastest rotator of our sample. The analysis of the \flames\ spectrum yields parameters in good agreement within the error bars 
with those by \cite{mokiem06}. We confirm their conclusion that helium is enriched in \object{MPG 299}. Despite the significant centrifugal correction to the surface gravity, we found
that the spectroscopic mass of \object{MPG 299} agrees very well with its evolutionary mass. Although the location of \object{MPG 299} in the H-R diagram indicates 
that it is a young object, we note that evolutionary tracks with initial \vrot\ higher than 360 \kms\ should be considered in this specific case (see Sect. \ref{sect_fast}).
For \object{MPG 299}, the nitrogen abundance is such that higher \mdot\ would produce too strong a PCygni profile at the \nv\ resonance line, which is not observed.

\begin{figure*}[tbp]
\includegraphics[scale=0.51, angle=0]{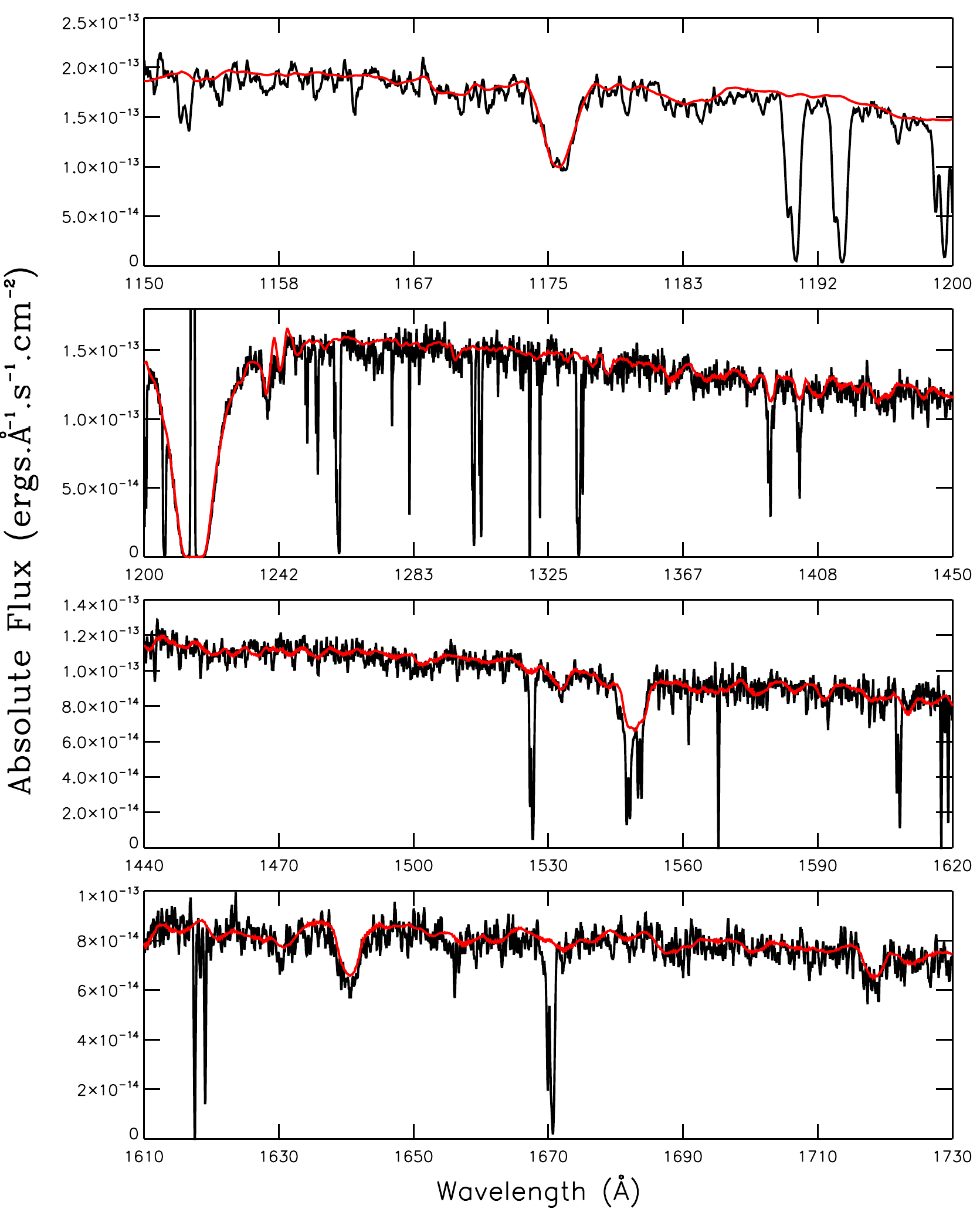}
\includegraphics[scale=0.51, angle=0]{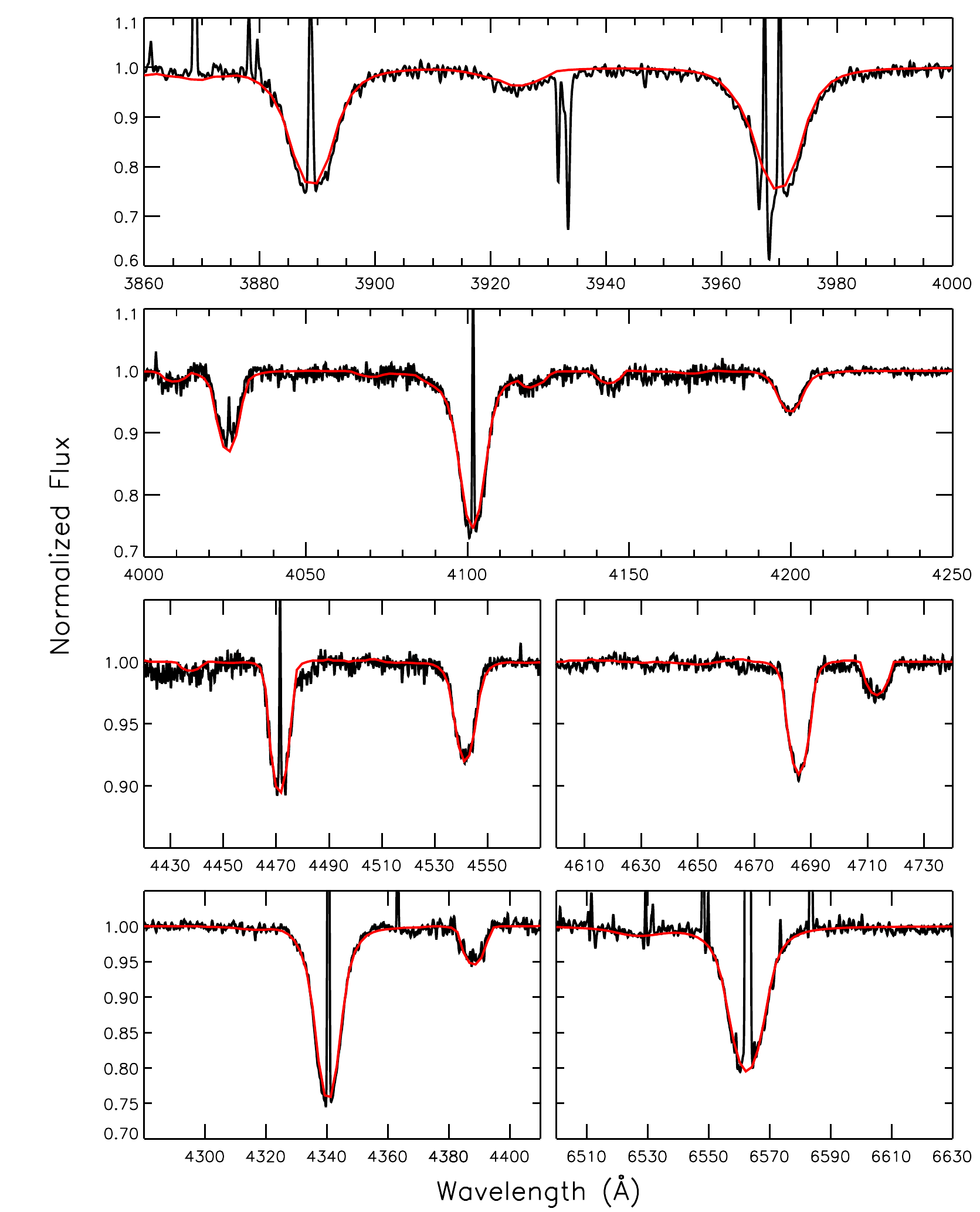}
      \caption[12cm]{Left: Best-fit model for \object{MPG 299} (red line) compared with the \cosp\ spectrum (black line). Right: Best-fit model for \object{MPG 299} (red line) compared with the 
      \flames\ spectrum (black line).}
         \label{Fig_mpg299}
   \end{figure*}	
   
\subsection{\object{MPG 487}\  - {\rm O8~V}}
\object{MPG 487} was originally studied in the FUV and optical in \cite{bouret03}, then \cite{heap06}. Subsequently, \cite{massey09} used their own optical
spectrum to provide a different analysis of this star. They argued that the cooler \teff\ we derived (along with Heap et al. 2006)
was caused by a wrong correction of the moonlight continuum contribution that contaminates the \aat\ spectrum (same comments hold for \object{MPG 368}). 
Such a contribution was indeed present and mis-corrected in our initial study. On the other hand, the claim by
\cite{massey09} that we were unable to obtain a satisfactory fit to the optical without assuming an unlikely low value for the metallicity, is wrong, as we demonstrate here. 
The metallicity was derived on 
the basis of the \stis\ spectrum analysis \cite[see also][]{bouret03}, which is not affected by moonlight contamination. The derived value allows the best fit to
the very weak lines of iron ions present on this \stis\ spectrum. Assuming higher metallicity would lead to an even lower \teff\ than the one we derived, hence increasing
the discrepancy with the \cite{massey09} results. Moreover, the value quoted by these authors for the projected rotation rate of \object{MPG 487} (\vsini\ = 160 \kms) 
simply cannot be true (a typo, we assume). 
The lines of this star, whether in the FUV or in the optical are really narrow, indicating a \vsini\ that must be 20 \kms\ or even lower (see e.g. the \ciii\ \lb1176
lines).

The SED of this star furthermore suggests that \object{MPG 487} is a binary with a rather late-type companion (cf. Sect. \ref{bin_sect}). This conclusion was already mentioned in 
\cite{bouret03},
although it was not as firmly established as it is now. 

Finally, our new \uves\ spectrum of \object{MPG 487} (cf. Sect. \ref{obs_sect}) does not suffer from moonlight contamination and has an excellent S/N ratio. 
We still find the same value of \teff\ = 35 kK that we derived in \cite{bouret03}. The surface gravity is higher than in \cite{bouret03} and more compatible with
the value from \cite{massey09}; we used the full line-set from the Balmer series to constrain this parameter.  
We also derived a non-standard abundance pattern for this star, with nitrogen being depleted while carbon seems enhanced compared
with the standard chemical composition of HII regions of the SMC by \cite{kurt98}.  

\begin{figure*}[tbp]
\includegraphics[scale=0.51, angle=0]{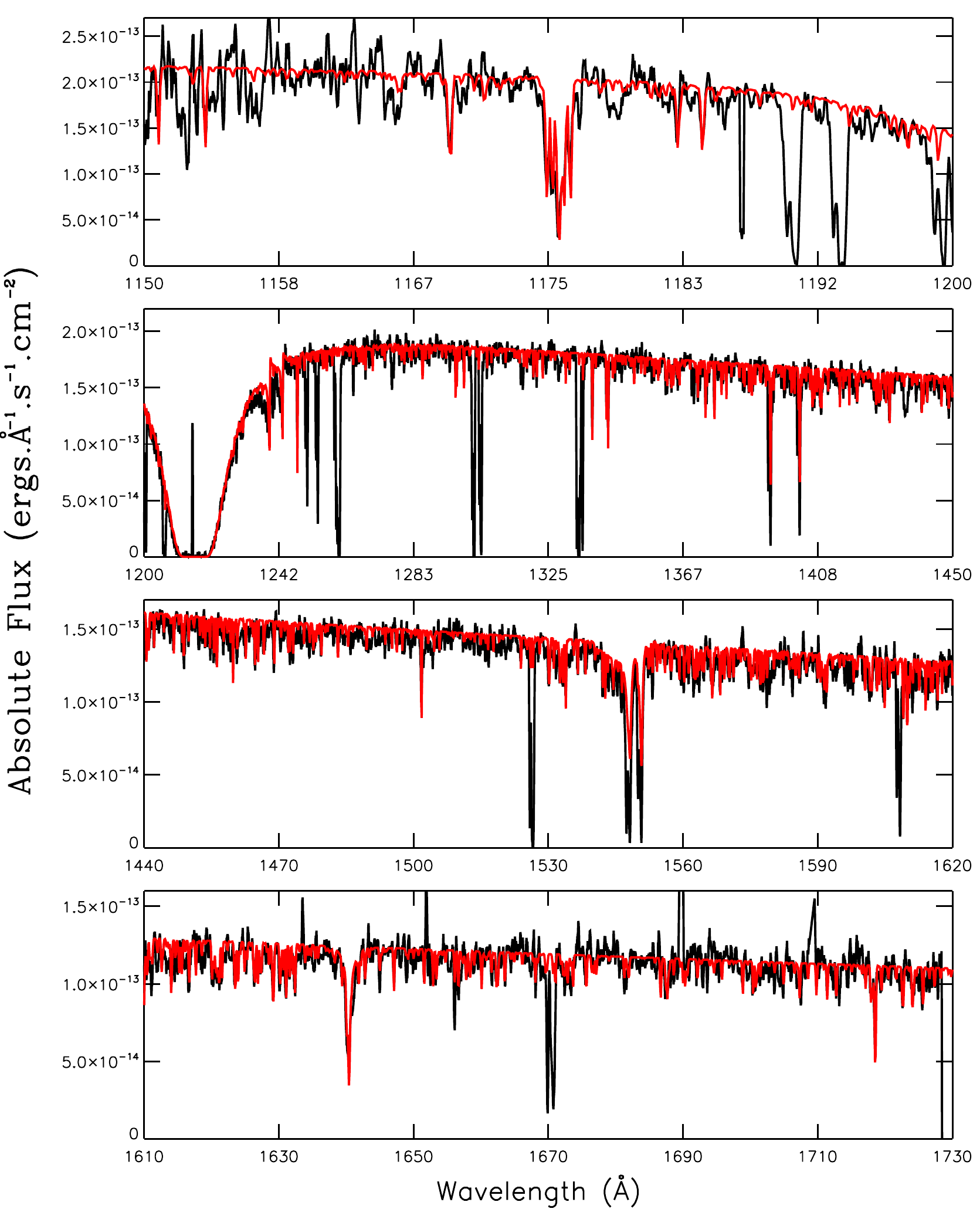}
\includegraphics[scale=0.51, angle=0]{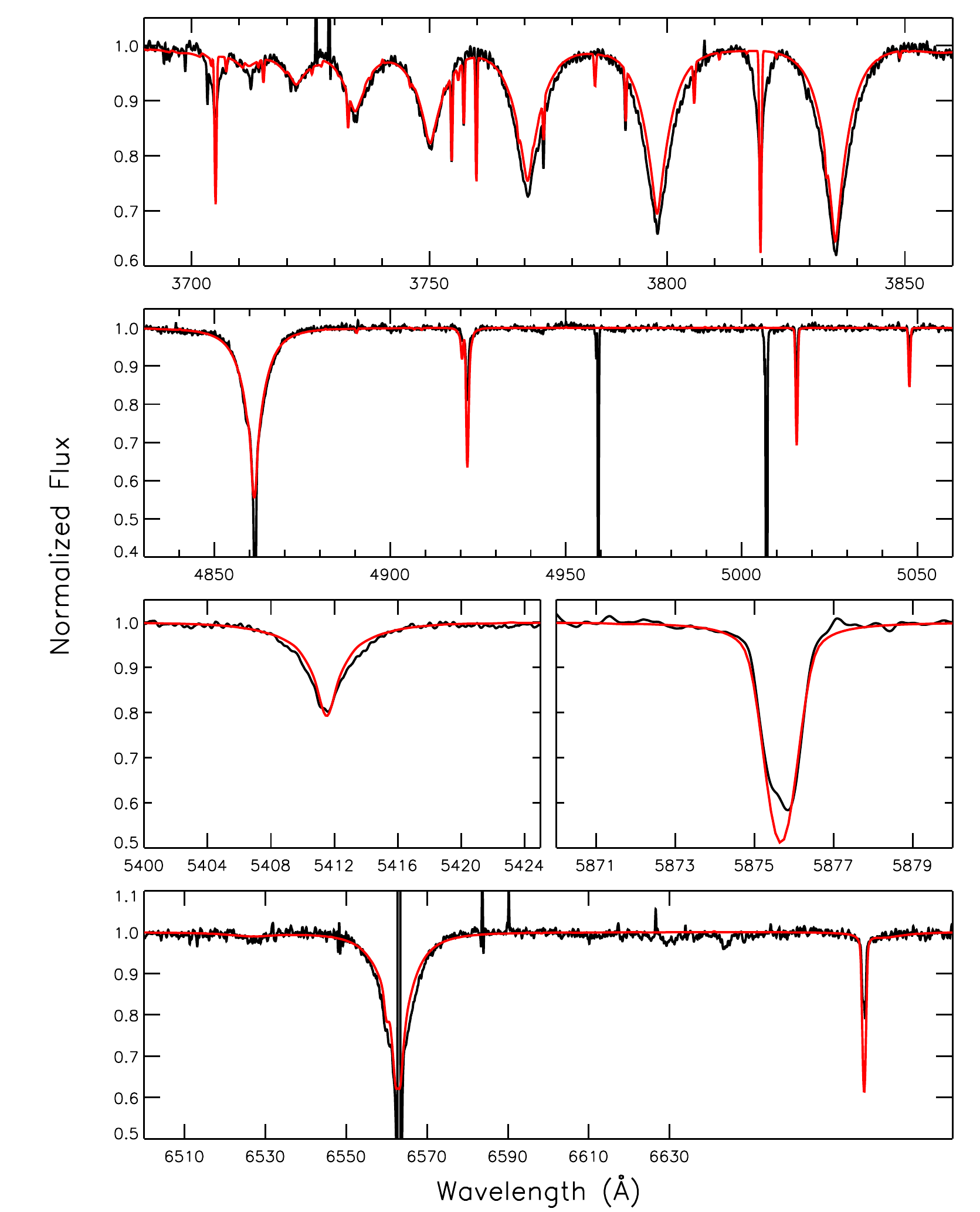}
      \caption[12cm]{Left: Best-fit model for \object{MPG 487} (red line) compared with the \cosp\ spectrum (black line). Right: Best-fit model for \object{MPG 487} (red line) compared with the 
      \uves\ spectrum (black line).}
         \label{Fig_mpg487}
   \end{figure*}	
\subsection{\object{AzV 267}\  - {\rm O8~V}}
With \vsini\ =  220 \kms, \object{AzV 267} is another fast rotator of the sample. Because we only had the UV \cosp\ spectrum for this star, we constrained the effective temperature from ratios between iron ions 
and \civ\ \lb1169 to \ciii\ \lb1176. Although it is a later-type O star, the \civ\ UV resonance doublet presents a clear absorption extending to the blue that we were able to use to constrain 
the mass-loss rate and the clumping filling factor. The fit to the \nv\ resonance doublet is very good for the derived wind parameters. The terminal velocity we adopted to fit the lines is lower
than for the other stars of the same type (i.e., with no wind detection). 

\begin{figure*}[tbp]
\includegraphics[scale=1, angle=0]{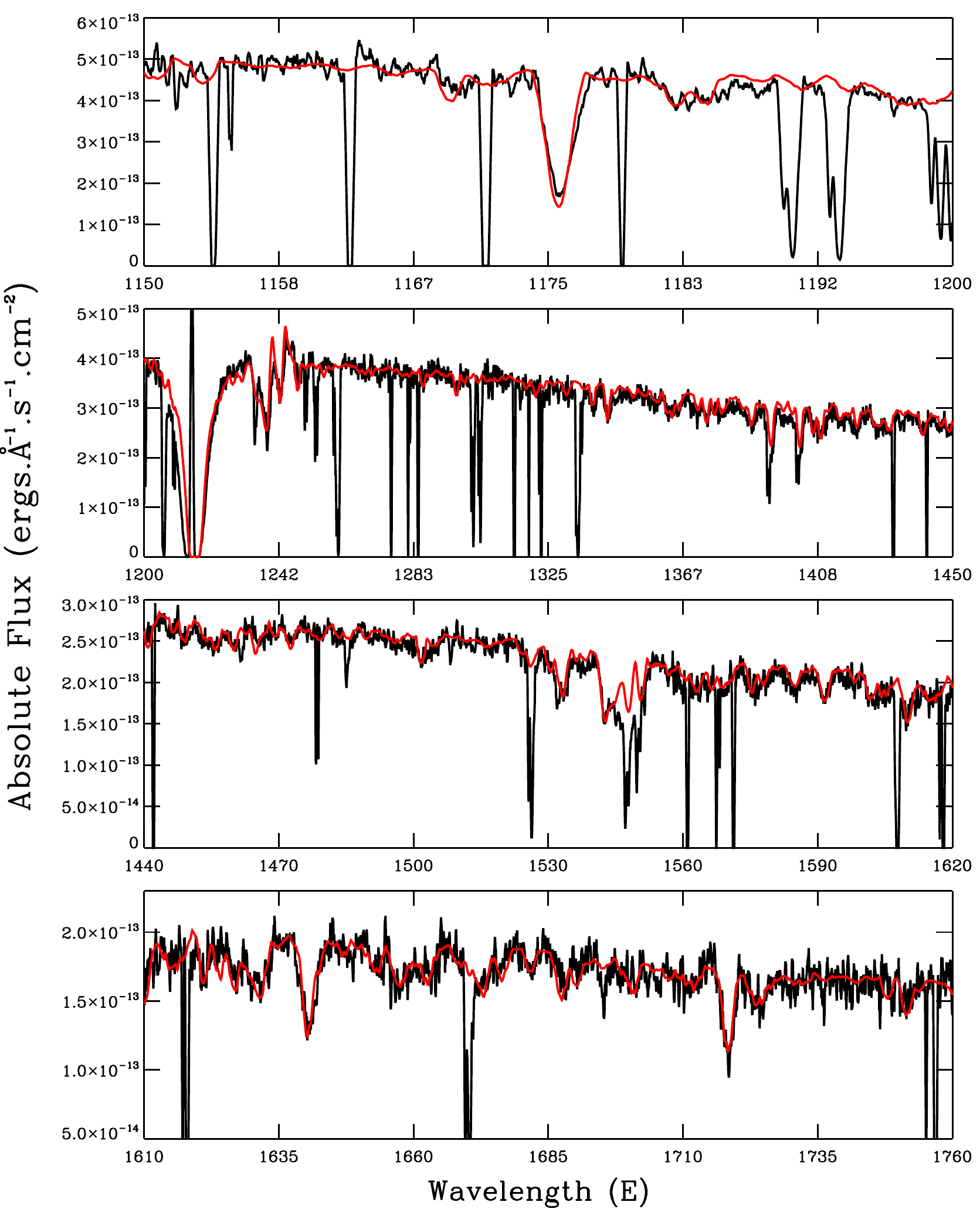}
      \caption[12cm]{Left: Best-fit model for \object{AzV 267} (red line) compared with the \cosp\ spectrum (black line).}
         \label{Fig_av267}
   \end{figure*}	

 \subsection{\object{AzV 468}\  - {\rm O8.5~V}}
Fundamental parameters were derived for the \cosp\ spectrum, except for the surface gravity, which we adopted as \logg\ = 4.0, because 
no reliable diagnostics are present in the UV range. 
Because of the moderate \vsini\ of this star, photospheric iron and carbon lines are clearly defined and were used to constrain \teff\ to a good accuracy. 
The \cosp\ spectrum of AzV 468 shows no trace of wind, either at \civ\ or at \nv\ resonance doublets. The derived mass-loss rate is indeed very low and is merely
an upper limit.

\begin{figure*}[tbp]
\includegraphics[scale=1, angle=0]{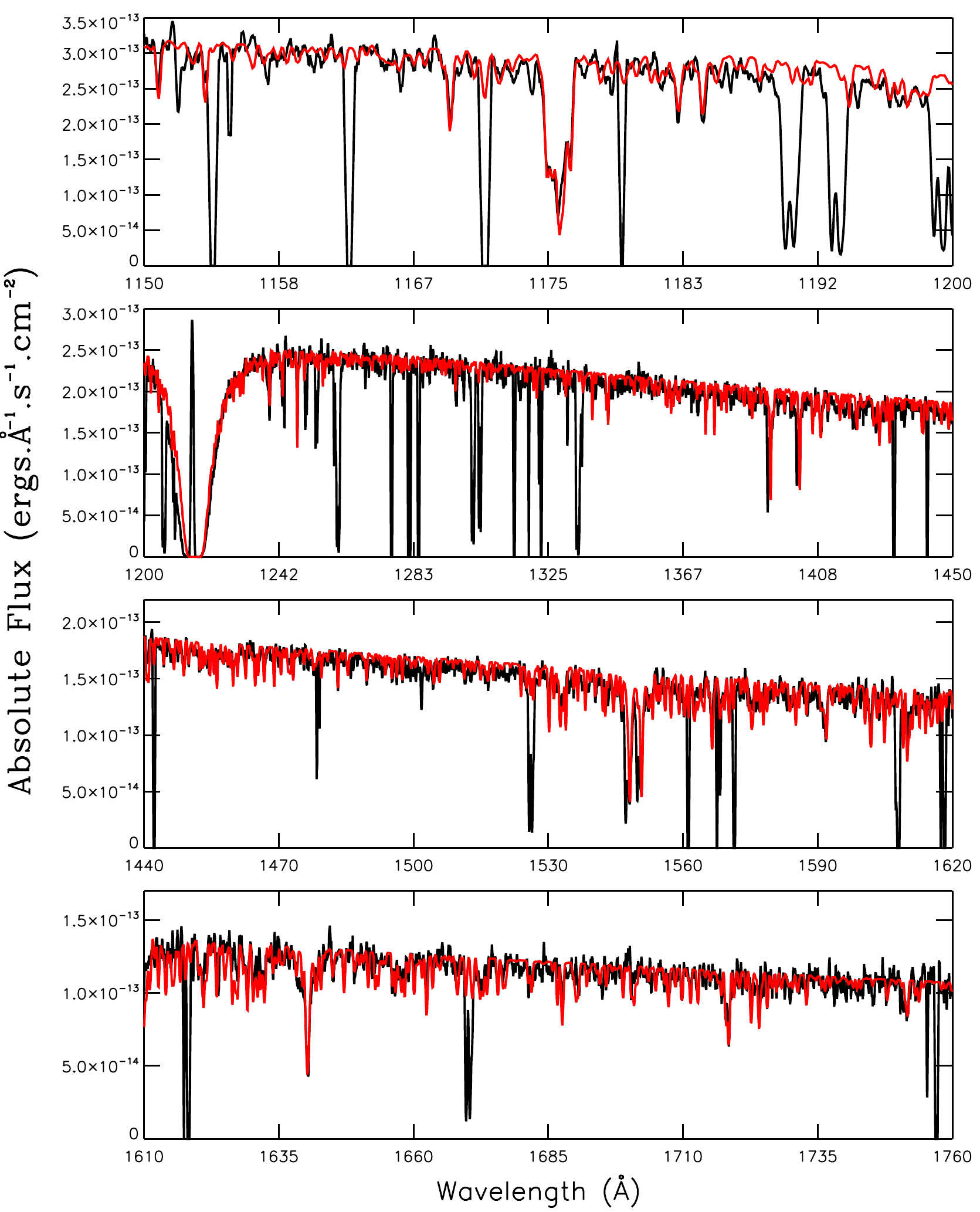}
      \caption[12cm]{Left: Best-fit model for \object{AzV 468} (red line) compared with the \cosp\ spectrum (black line).}
         \label{Fig_av468}
   \end{figure*}	
\subsection{\object{MPG 682}\  - {\rm O9~V}}
The fit to the \cosp\, \uves\ and \flames\ spectra of \object{MPG 682} is excellent, including the weak lines
of \niii\ \lb\lb4634-42. Our \teff\ is cooler by 2000~K than the value reported in \cite{mokiem06}, while \logL\ are very similar. 
The wings of the hydrogen lines (including the Balmer series below 3860 \AA) indicate that \logg\ = 4.1, which yields a spectroscopic mass
close to 27 \msol, significantly higher than the evolutionary mass we derived from the H-R diagram ($\approx 21$ \msol). The surface gravity 
quoted in \cite{mokiem06} is even higher (4.2 dex), but consistent with ours, within the error bars. 
The absence of UV wind profiles confirm that this star is in the weak-wind regime, as expected from its spectral type and luminosity class. Our mass-loss
rate is about two orders of magnitude lower than the mass-loss rate in \cite{mokiem06}. 

\begin{figure*}[tbp]
\includegraphics[scale=0.54, angle=0]{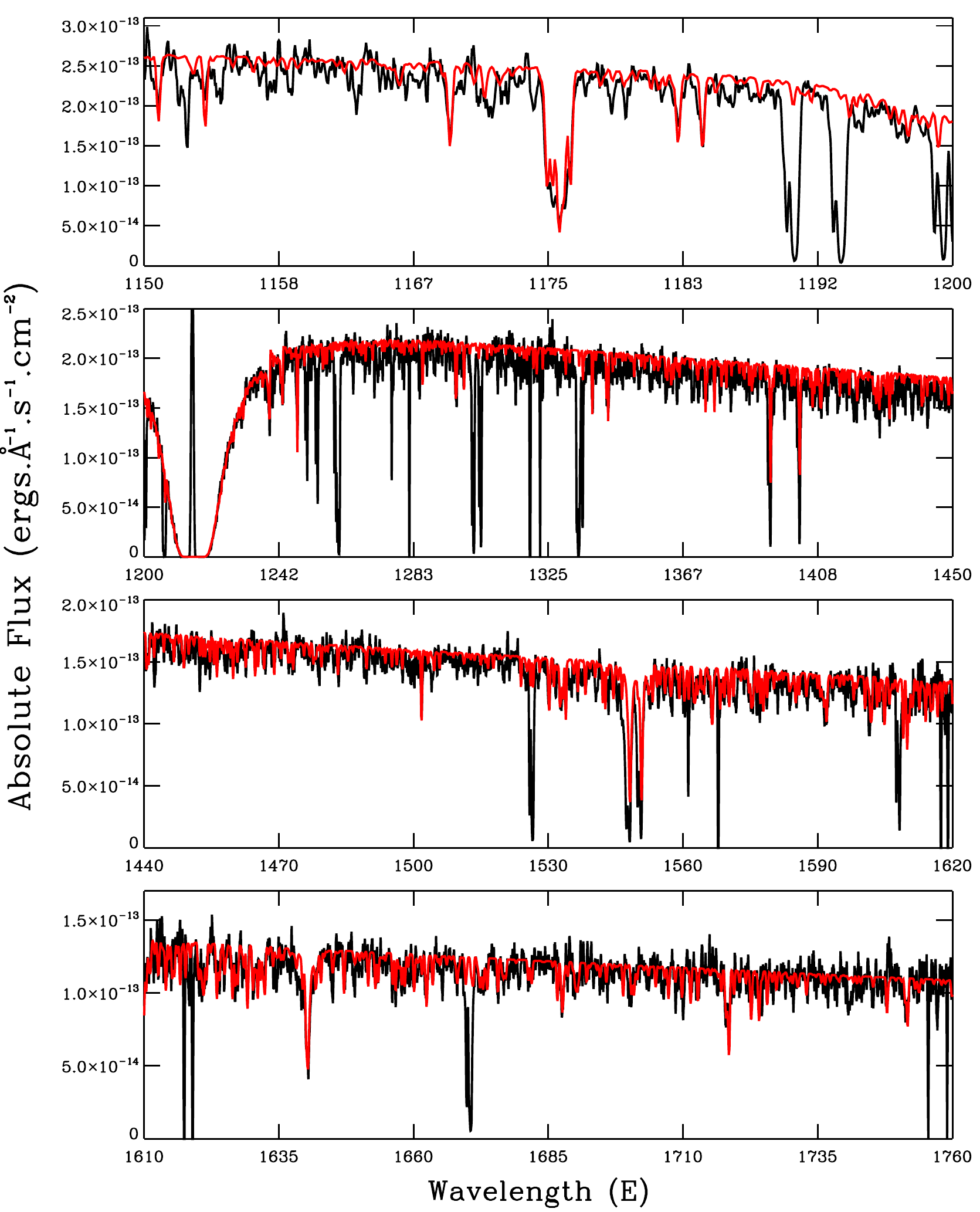}
\includegraphics[scale=0.54, angle=0]{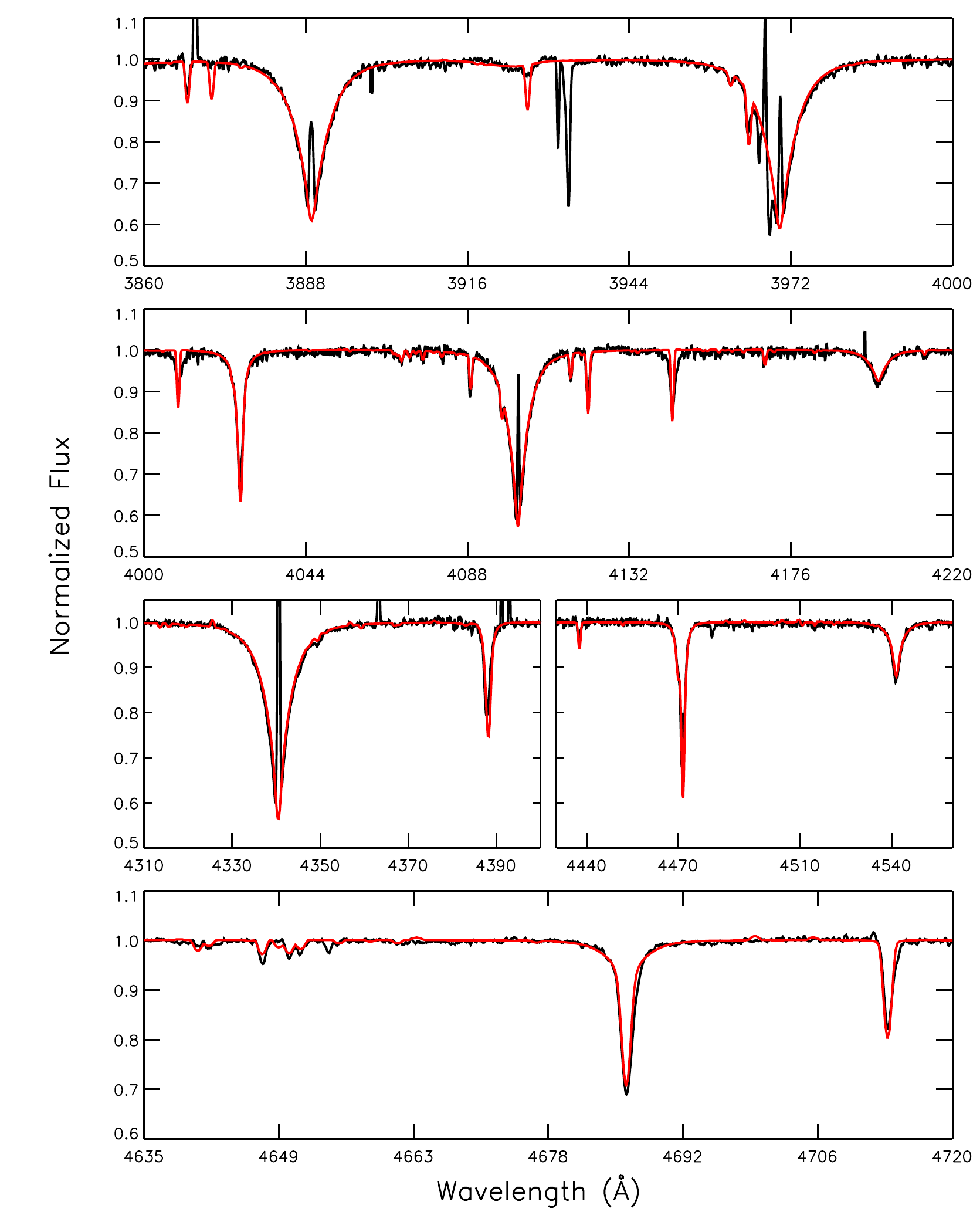}
\includegraphics[height=5in,width=7in, angle=0]{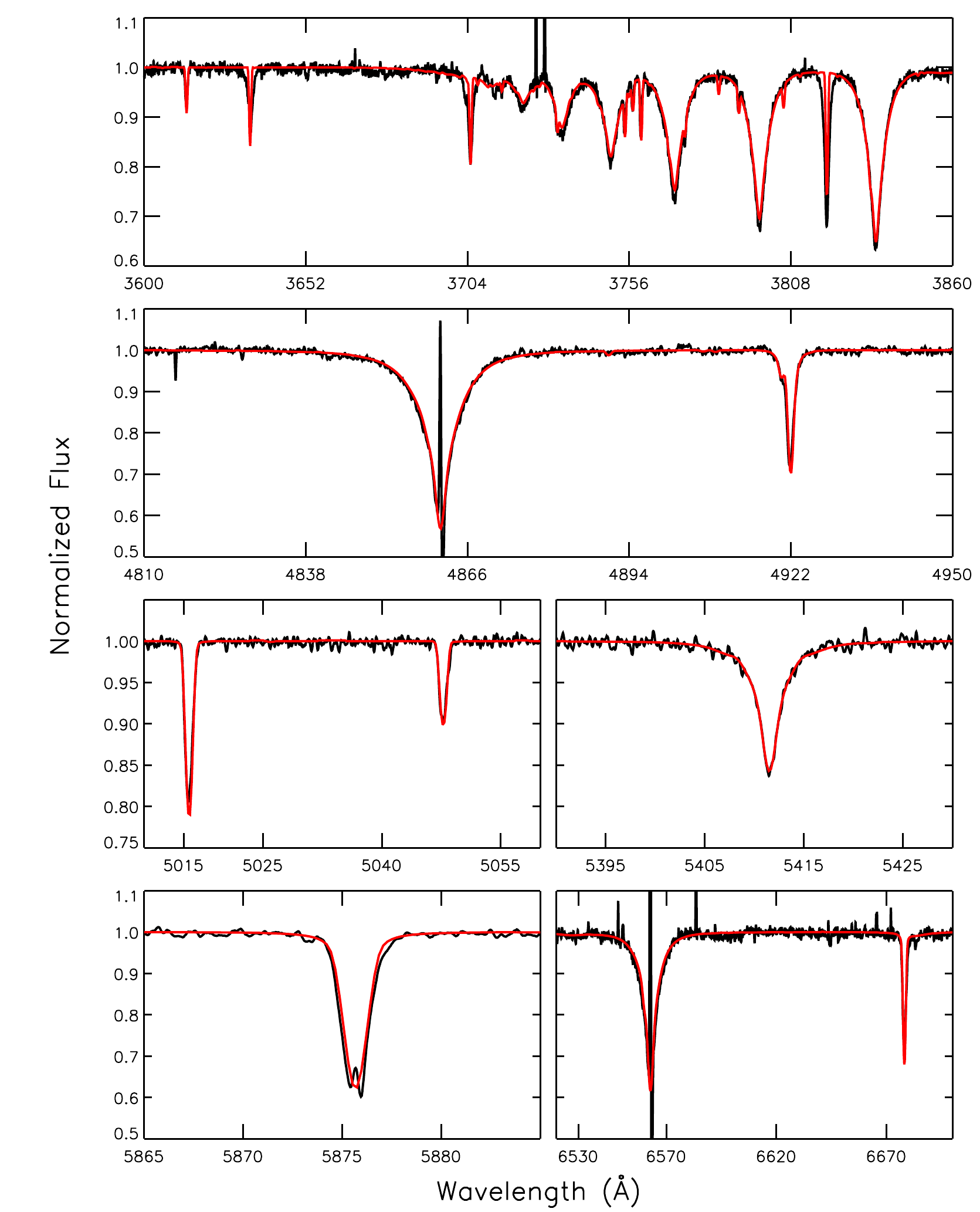}
      \caption[12cm]{Best-fit model for \object{MPG 682} (in red) compared with observed spectra (in black): the \cosp\ spectrum (upper left) and the \flames spectrum (upper right).
      The bottom plot shows a comparison with the \uves\ spectrum.}
         \label{Fig_mpg682_2}
   \end{figure*}	
 \subsection{\object{AzV 326}\  - {\rm O9~V}}
\begin{figure*}[tbp]
\includegraphics[scale=0.51, angle=0]{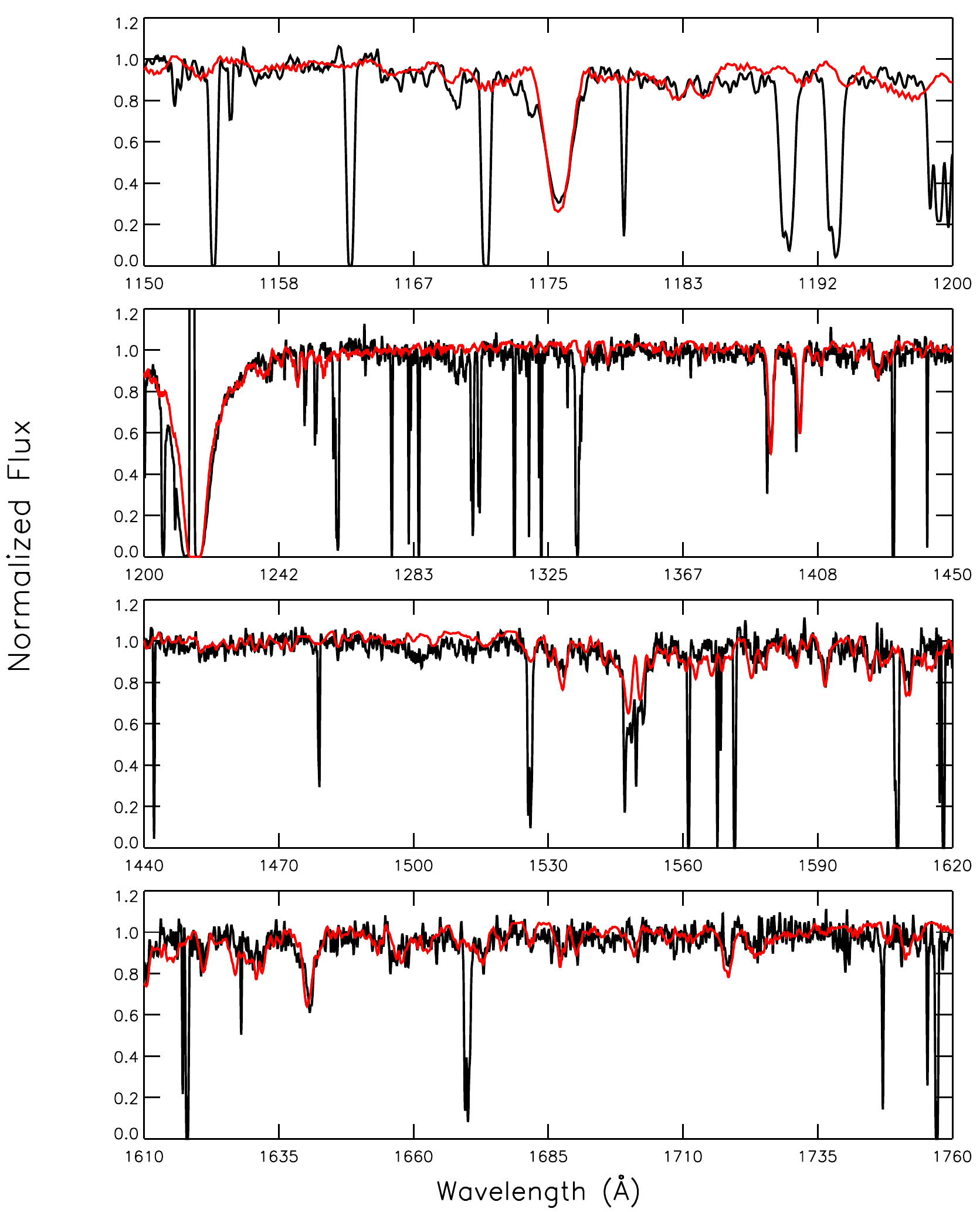}
\includegraphics[scale=0.49, angle=0]{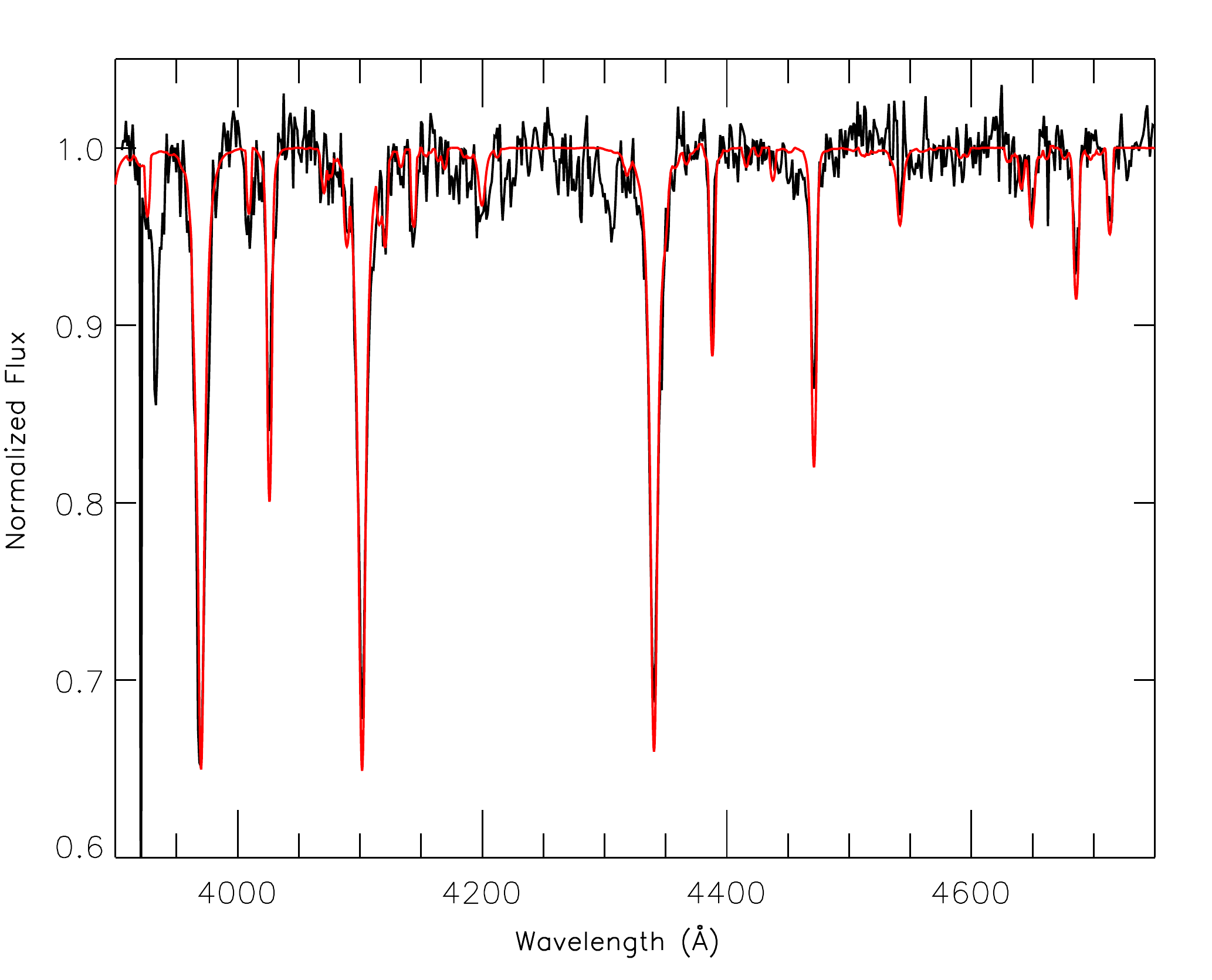}
\caption[12cm]{Left: Best-fit model for \object{AzV 326} (red line) compared with the \cosp\ spectrum (black line). Right: Best-fit model for \object{AzV 326} (red line) compared with the 
      2dF spectrum (black line).}
         \label{Fig_av326}
   \end{figure*}

\object{AzV 326} is very likely a multiple star (see the discussion in Sect. \ref{bin_sect}). However, even when we corrected for a possible binary component, there were
some remaining disagreement in the fit to the \cosp\ SED and we finally chose to normalize the spectrum, which we present in Fig. \ref{Fig_av326}). 
The fit to the individual spectral line from UV to the lower resolution 2dF spectrum in the optical is very good and we are confident that the derived
 parameters provide a meaningful representation of the primary component of this (yet hypothetical) system.  

  \subsection{\object{AzV 189}\  - {\rm O9~V}}
Here again, we are dealing with a potential binary. The correction applied assuming that this system consists of two identical components  yields an excellent fit to the 
whole SED from UV to NIR with a single parameter set. For the same parameters, the fit to the optical 2dF spectrum is also very good. This spectrum was used to constrain 
the surface gravity. For the derived \logg\, the spectroscopic mass is higher than the evolutionary mass. This could indicate that the adopted luminosity is too low, 
although a change by 0.1dex
in \logg\ would also reconcile the two masses and is not excluded given the lower S/N and spectral resolution of the 2dF spectrum. 

\begin{figure*}[tbp]
\includegraphics[scale=0.51, angle=0]{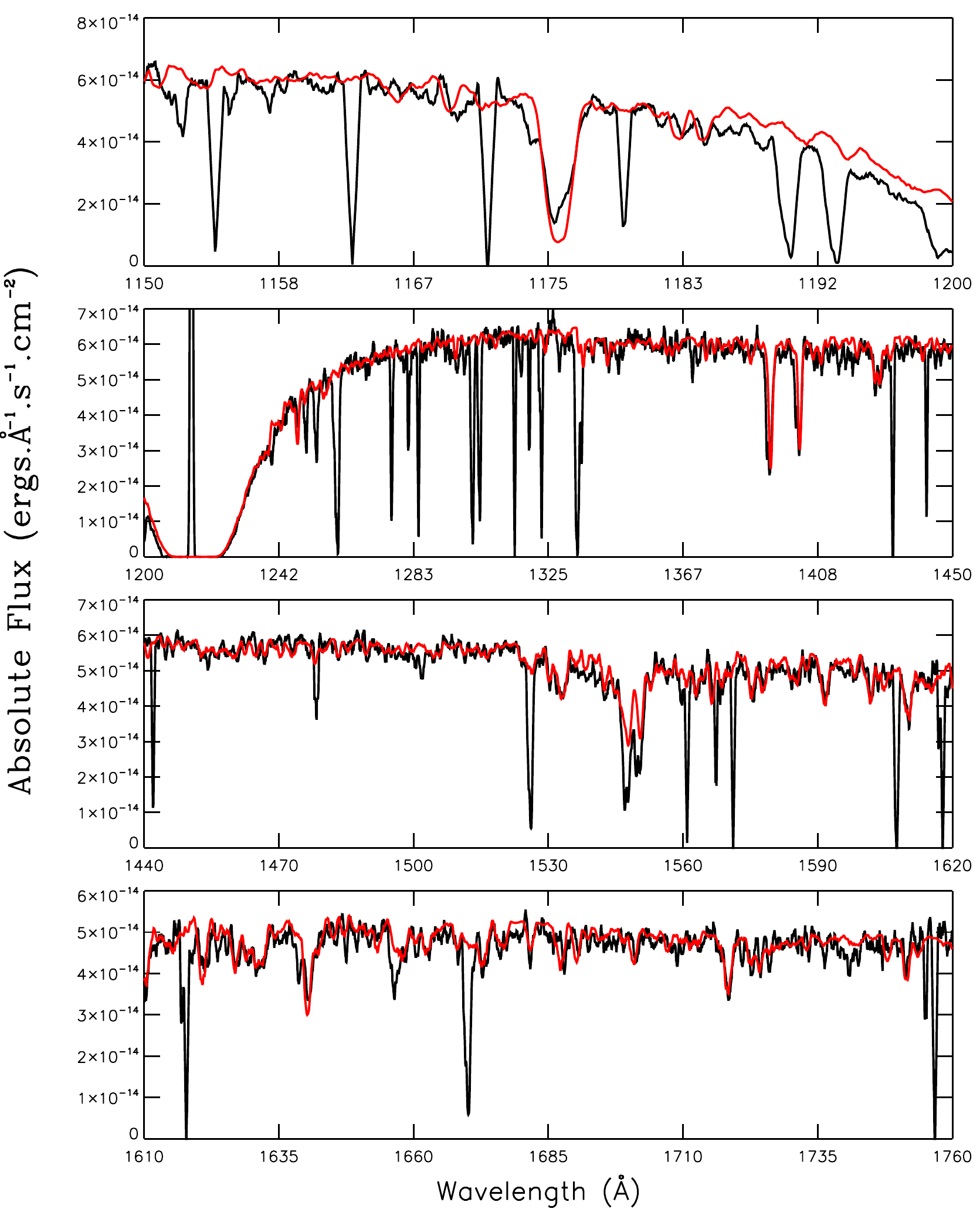}
\includegraphics[scale=0.49, angle=0]{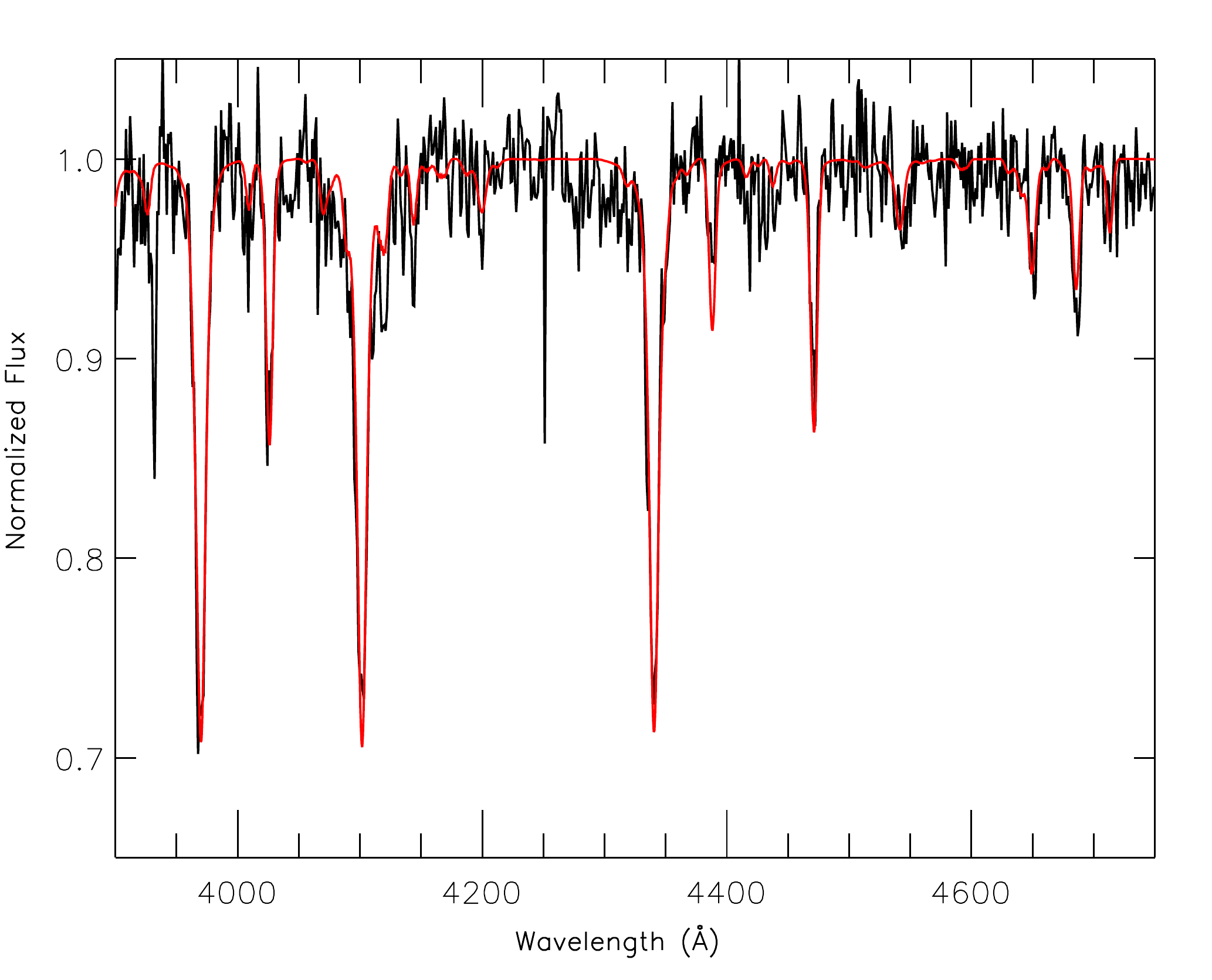}
\caption[12cm]{Left: Best-fit model for \object{AzV 189} (red line) compared with the \cosp\ spectrum (black line). Right: Best-fit model for \object{AzV 189} (red line) compared with the 
      2dF spectrum (black line).}
         \label{Fig_av189}
   \end{figure*}	
 \subsection{\object{AzV 148}\  - {\rm O8.5~V}}
\object{AzV 148} is the third star with  photometry and UV flux unusual for its spectral type, and which would correspond to a luminosity typical of O6~V.  
The lower \vsini\ for this star allowed us to derive accurate \teff\ from ionization ratios, despite the lack of optical spectrum. The value we derived is consistent
with those of the stars with spectral type O9~V. Surface gravity, on the other hand, was adopted as \logg\ = 4.0, which translates into a spectroscopic mass significantly 
higher than the evolutionary mass. Like in \object{AzV 189}, though, a change by 0.1dex in \logg\ would be enough to reconcile the two estimates. 
The \civ\ \lb\lb1548-1550 line shows some blue extension, blueward of the interstellar components. The profile, together with the \nv\  resonance profiles, 
is well reproduced by adopting a mass-loss rate three times higher than for the other O9~V stars of the sample. 

\begin{figure*}[tbp]
\includegraphics[scale=1, angle=0]{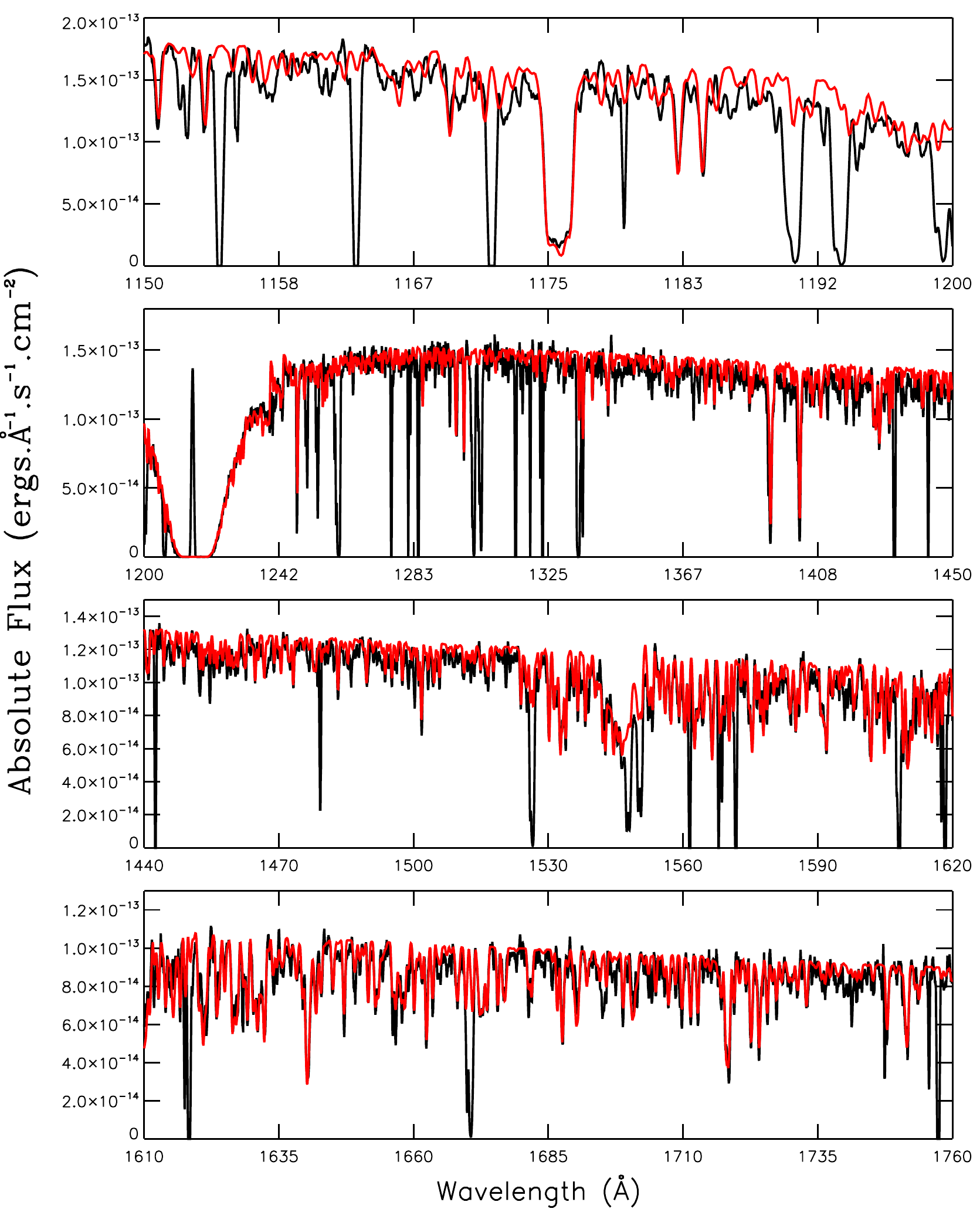}
\caption[12cm]{Left: Best-fit model for \object{AzV 148} (red line) compared with the \cosp\ spectrum (black line).}
         \label{Fig_av148}
   \end{figure*}	
\subsection{\object{MPG 012}\  - {\rm O9.5~V / B0 IV}}
This star was previously studied in the UV and optical in \cite{bouret03} and in the optical in \cite{mokiem06}. 
The higher S/N ratios of the \uves\ and \flames\ spectrum led us to revise \logg\ (0.05 dex) slightly upward compared
to \cite{bouret03}, while \teff\ was unchanged. The fit to the three spectra is excellent except for \hei\ \lb5876; the synthetic profile is not as
strong and broad as observed. This behavior cannot be explained at this point. 
The star is still the coolest and lowest gravity object of the whole sample. 
The fundamental parameters suggest a clearly evolved status for this object, as indicated by its age, the highest by far among the 
stars in \object{NGC 346}. The surface abundances are also typical of more advanced evolution, as noted in \cite{bouret03}. 
Because it is at the outskirts of this H~II region, this object might not be coeval with this region, as previously suggested by
\cite{walborn00}, \cite{bouret03}, and \cite{mokiem06}. 

\begin{figure*}[tbp]
\includegraphics[scale=0.54, angle=0]{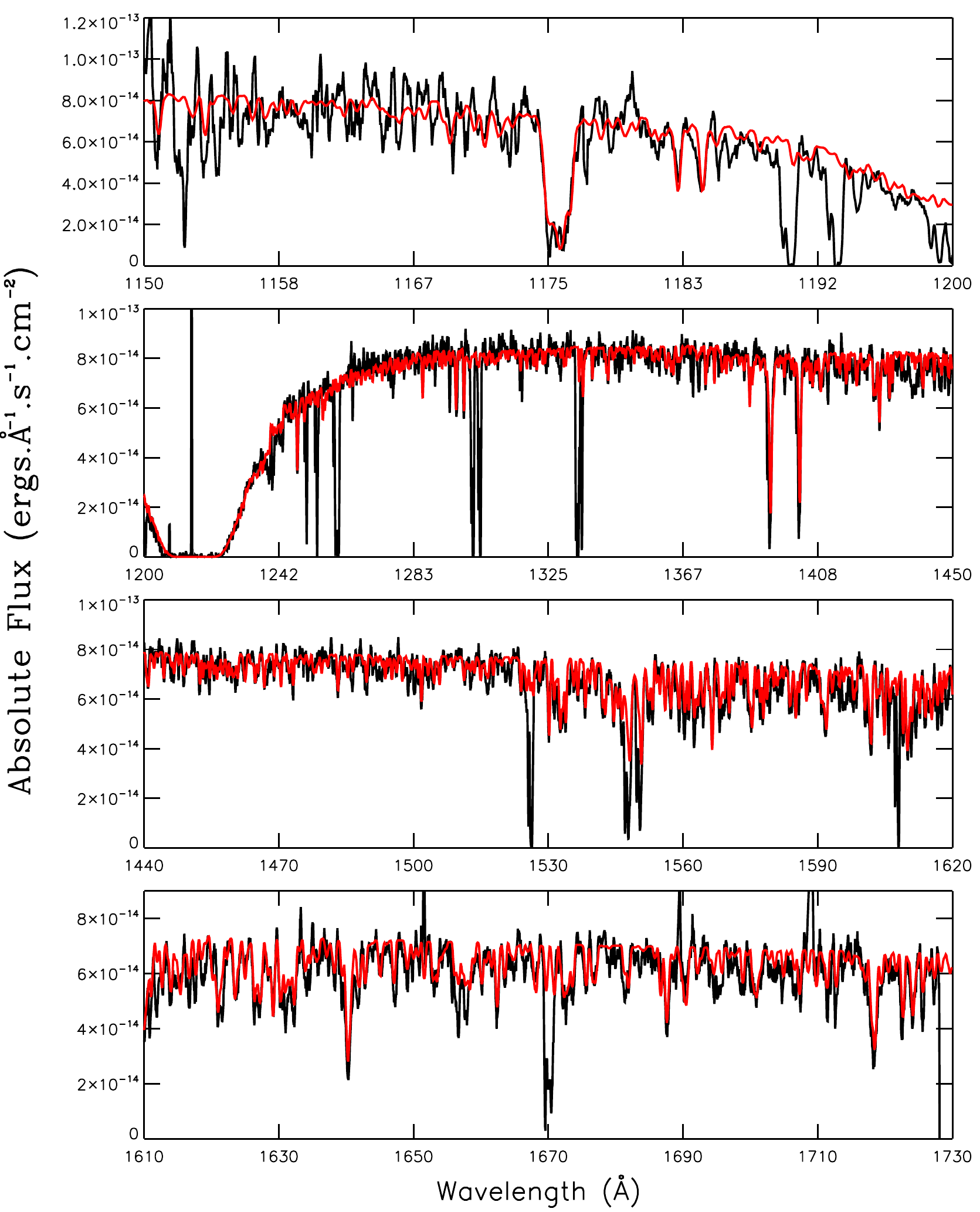}
\includegraphics[scale=0.54, angle=0]{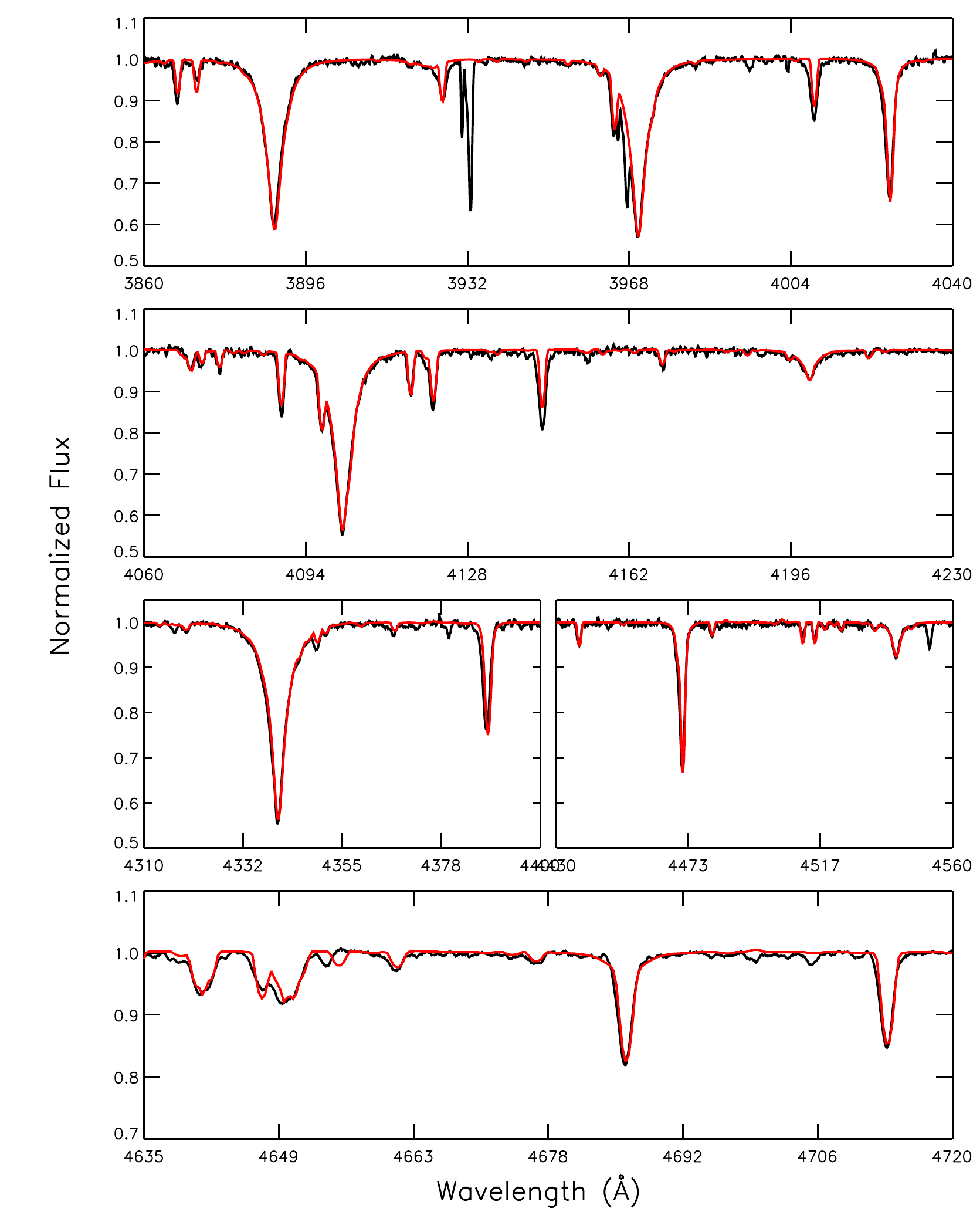}
\includegraphics[height=5in,width=7in, angle=0]{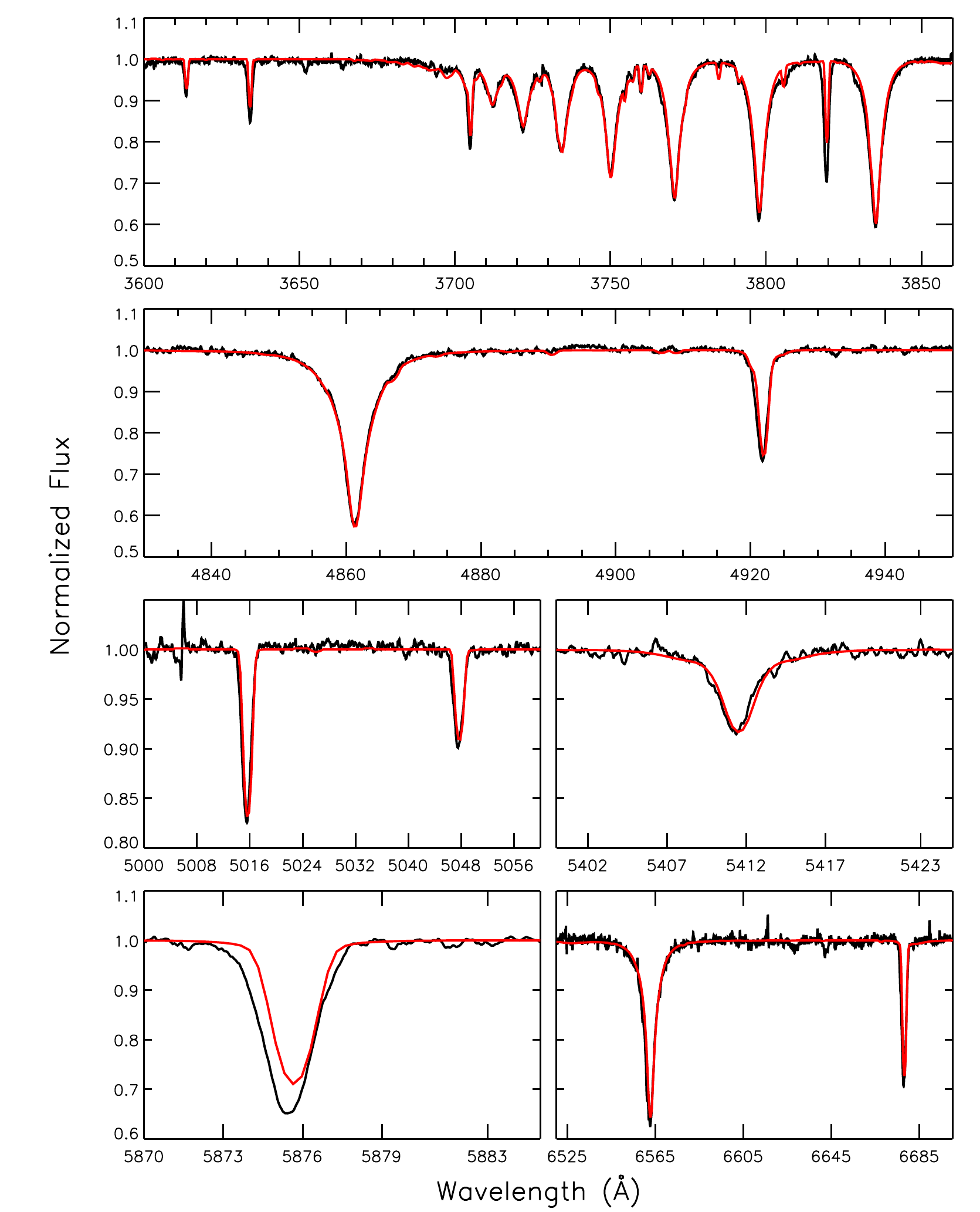}
      \caption[12cm]{Best-fit model for \object{MPG 012} (in red) compared with observed spectra (in black): the \stis\ spectrum (upper left) and the \flames\ spectrum (upper right).
      The bottom plot shows a comparison with the \uves\ spectrum.}
         \label{Fig_mpg012_2}
   \end{figure*}	 

\end{appendix}


\end{document}